# Development of Semantics-Based Distributed Middleware for Heterogeneous Data Integration and its Application for Drought

by

---

## ADEYINKA KABIR AKANBI

Submitted in fulfilment of the requirements for the degree

## DOCTOR OF PHILOSOPHY
## IN INFORMATION TECHNOLOGY

In the

Faculty of Engineering, Built Environment and Information Technology:

Department of Information Technology

**Central University of Technology, Free State**

**Promoter: Prof. Muthoni Masinde**

Co-Promoter: Prof. Yali Woyessa

2020

# Copyright Notice

This doctoral thesis is to be used for only academic or non-commercial research purposes. The information contained in this thesis is to be published with acknowledgement of the source.

This doctoral thesis is published by the Central University of Technology, Free State, in terms of a non-exclusive licence granted to CUT by the author.

# Disclaimer

The presentation of this thesis contains colour images meant to present the results visually in a more understandable form. While printing or viewing the thesis in greyscale (black and white) is possible, it is recommended that for clarity, the thesis is viewed (or printed) in full-colour format.

# Dedication

This thesis is dedicated to all my family members for their kind support and to my unborn children.

# Declaration

This research as presented in this thesis is my original work and has not been presented for any other university award. Knowledge derived from other sources has been clearly indicated, with acknowledgement and reference to the literature.

This study was conducted and completed under the guidance of Professor Muthoni Masinde, Department of Information Technology at Central University of Technology, Free State, South Africa, and co-supervised by Professor Yali Woyessa, Department of Civil Engineering at Central University of Technology, Free State, South Africa.

**Adeyinka Kabir Akanbi**

Signature: ______________________

Date: _________________________

In our capacity as the supervisors of this thesis, we certify that the above statements are true to the best of our knowledge.

**Professor Muthoni Masinde**

Signature: ______________________

Date: _________________________

**Professor Yali Woyessa**

Signature: ______________________

Date: _________________________

# Preface and Acknowledgements

This thesis is the result of a PhD research study carried out from May 2015 to May 2019 at the Department of Information Technology, Faculty of Engineering, Built Environment and Information Technology, Central University of Technology, Free State, South Africa. In the words of Miguel de Cervantes Saavedra (1614): "*For a man to attain to an eminent degree in learning, costs him time, watching, hunger, nakedness, dizziness in the head, weakness in the stomach, and other inconveniences.*" However, the task of completion of this thesis would not have been possible without the support of several individuals to whom I would like to express my deepest gratitude.

First and foremost, I would like to express my whole-hearted gratitude to my main supervisor, Prof Muthoni Masinde for always believing in me since inception, directing and supervising my research work. Your constant encouragement at every step of the journey – from idea conception till the end is highly appreciated. Without your timely support, problem-solving skills, enthusiasm, responsibility towards students and expert guidance during my research work, this thesis would not have been possible. I believe that the invaluable knowledge and experience have gained from you will always guide me forward in the future. I would also like to thank my co-supervisor, Prof Yali Woyessa for being a pillar of support with words of encouragement and upliftment.

I would like to thank my friends and colleagues – those that made the University home away from home for me, starting with my brothers from another mother – Mr Jacob Adedeji and Mr Olugbenga Abejide for the support and help that I have received from them whenever needed during this research period. My gratitude is extended to Dr Bankole Awuzie for his insightful talks on research ideas whenever the need arises. My sincere gratitude goes to Professor Fidelis Emuze, HOD, Department of Built Environment, and Mrs Mpho Mbeo, Faculty Officer, Faculty of Engineering Built Environment and Information Technology, CUT, for her timely advice and support during my PhD work.

My profound and unreserved gratitude goes to Almighty God for his mercies and loving kindness over my life. I am indebted to my parents: Hon Adebowale Akanbi and Mrs Ibijola Akanbi for their unconditional moral and financial support, understanding, belief and love; without them this research would have been impossible. I am also grateful to my loving wife, Mrs Adedoyin Kareemat Akanbi and my daughter Adeshewa Rahmatalah Damilola Akanbi for their love, affection and understanding during this PhD study.

Last but not the least, I would like to acknowledge my brothers: Ademola Akanbi, Adewale Akanbi, Adekunle Akanbi, Yusuf Akanbi and my sisters: Jadesola Akanbi, Ayobami Akanbi and Tomiwa Akanbi (*of loving memory*); my uncles: Mr. MG Akanbi, Mr Binyamin Yusuf, Mr Leke Akanbi, Mr Abiodun Akanbi; my aunts: Mrs Kuburat Fawole, Ms Iyabo Ibrahim,  other family and friends for their understanding and love.

<div align="right">

Adeyinka Kabir Akanbi

October 2019

Bloemfontein, Free State

South Africa

</div>

# Abstract

Drought is a complex environmental phenomenon that affects millions of people and communities all over the globe and is too elusive to be accurately predicted. This is mostly due to the scalability and variability of the web of environmental parameters that directly/indirectly causes the onset of different categories of drought. Since the dawn of man, efforts have been made to uniquely understand the natural indicators that provide signs of likely environmental events. These indicators/signs in the form of indigenous knowledge system have been used for generations. Also, since the dawn of modern science, different drought prediction and forecasting models/indices have been developed which usually incorporate data from sparsely located weather stations in their computation, producing less accurate results – due to lack of the desired scalability in the input datasets.

The intricate complexity of drought has, however, always been a major stumbling block for accurate drought prediction and forecasting systems. Recently, scientists in the field of ethnoecology, agriculture and environmental monitoring have been discussing the integration of indigenous knowledge and scientific knowledge for a more accurate environmental forecasting system in order to incorporate diverse environmental information for a reliable drought forecast. Hence, in this research, the core objective is the development of a semantics-based data integration middleware that encompasses and integrates heterogeneous data models of local indigenous knowledge and sensor data towards an accurate drought forecasting system for the study areas of the KwaZulu-Natal province of South Africa and Mbeere District of Kenya.

For the study areas, the local indigenous knowledge on drought gathered from the domain experts and local elderly farmers, is transformed into rules to be used for performing deductive inference in conjunction with sensors data for determining the onset of drought through an automated inference generation module of the middleware. The semantic middleware incorporates, *inter alia*, a distributed architecture that consists of a streaming data processing engine based on *Apache Kafka* for real-time stream processing; a rule-based reasoning module; an ontology module for semantic representation of the knowledge bases. The plethora of sub-systems in the semantic middleware produce a service(s) as a combined output – in the form of drought forecast advisory information (DFAI). The DFAI as an output of the semantic middleware is disseminated across multiple channels for utilisation by policy-makers to develop mitigation strategies to combat the effect of drought and their drought-related decision-making processes.

# List of Acronyms

AHP – Agro-hydropotential

AI – Artificial Intelligence

API – Application Programming Interface

BFO - Basic Formal Ontology

CEP – Complex Event Engine

CF – Certainty Factor

CIQ – Customer Information Quality

CLI – Command Line Interface

CSV – Comma-separated values

CUT - Central University of Technology

DAML – Digital Asset Modelling Language

dB – Document Database

DCSP – Distributed Stream Computing Platforms

DDL – Data Definition Language

DE – Domain Expert

DEWS – Drought Early Warning System

DFAI – Drought Forecast Advisory Information

DL – Descriptive Logic

DML – Data Manipulation Language

DOLCE – Descriptive Ontology for Linguistic and Cognitive Engineering

DRIC – Departmental Research and Innovation Committee

EDI – Effective Drought Index

EDXL-SitRep – Emergency Data Exchange Language Situation Reporting

EP – Event Processing

EPL – Event Processing Language

ESTemd – Event STream Processing Engine for Environmental Monitoring Domain

EWS – Early Warning Systems

FAO – Food and Agriculture Organisation

FEWS-Net – Famine Early Warning System

FG – Functional Group

FGDC – American Federal Geographic Data

FOL – First Order Logic

FR – Functional Requirement

FRIC – Faculty Research and Innovations Committee

GIEWS – Global Information and Early Warning System

GML – Geography Markup Language

GUI – Graphical User Interface

HEWS – Humanitarian Early Warning Service

HTTP – Hyper Text Transfer Protocol

IK – Indigenous Knowledge

IKF – Indigenous Knowledge Forecasts

IKF – Indigenous Knowledge Forecasts

IKON – Indigenous Knowledge on Drought Domain ONtology

IKS – Indigenous Knowledge System

IKSDC – Indigenous Knowledge System Data Collection

IKSDC – Indigenous Knowledge System Data Collection

IoT – Internet of Things

IS – Information Systems

IS – Information Systems

ITIKI – Information Technology with Indigenous Knowledge

JESS – Java Expert Shell Script

JSON – JavaScript Object Notation

JSON-LD – JSON for Linked Data

KAON – Karlsruhe Ontology

KBE – Knowledge Base Editor

KSQL – Kafka Structured Query Language

LODE – Live OWL Documentation Environment

MAPE – Mean Absolute Percentage Error

ME – Mean Error

MQTT – Message Queuing Telemetry Transport

NFR – Non-Functional Requirement

NIEM – National Information Exchange Model

O&M – Observations & Measurements

OBO – Open Biological and Biomedical Ontology

ODK – Open Data Kit

OGC – Open Geospatial Consortium

OWL – Web Ontology Language

OWLViz – Web Ontology Language Visualiser

PDSI – Palmer Drought Severity Index

PiECE – Pilot, Exploratory and Confirmatory Experiments

PL – Propositional Logic

RB-DEWES – Rule-based Drought Early Warning Expert System

RBES – Rule-based Expert System

RC – Research Contribution

RDF – Resource Descriptive Framework

RDFS – Resource Description Framework Schema

REST – Representational State Transfer

RH – Research Hypothesis

RIF – Rule Interchange Format

RO – Research Objective

RQ – Research Question

RSE – Root Square Error

SB-DEWS – Semantics-based Drought Early Warning Systems

SBDIM – Semantic-based Data Integration Middleware

SOA – Service Oriented Architecture

SOAP – Simple Object Access Protocol

SPI – Standard Precipitation Index

SQL – Structured Query Language

SSN – Semantic Sensor Network

SSN – Semantic Sensor Network

SUMO – Suggested Upper Merged Ontology

SWSI – Surface Water Supply Index

TAHMO – Trans-African Hydro-Meteorological Observatory

TAHMO – Trans-African Hydro Meteorological Observatory

UID – Unique Identifier

UML – Universal Modelling Language

UNISDR – United Nations International Strategy for Disaster Reduction

UoD – Universe of Discourse

USSD – Unstructured Service Supplementary Data

UUID – Unique Universal Identifier

UUID –Unique Universal Identifier

VOWL – Visual Ontology Web Language

W3C – World Wide Web Consortium

WFP – World Food Programme

WSDC – Wireless Sensor Data Collection

WSDC – Wireless Sensor Data Collection

WSN – Wireless Sensor Networks

XML – Extensible Markup Language

# Glossary

- *Axioms*: A form of assertions (including rules) that can be used to generate inference and reasoning.

- *Blob storage*: An optimised storage for storing massive amounts of unstructured data, such as text or binary data. Users or client applications can access blobs via URLs, REST API.

- *Case study*: An in-depth study of an event in a selected area using selected individuals.

- *CEP Engine* (Complex Event Processing): An event processing engine that combines data from multiple sources to infer events or patterns that suggest more complicated circumstances.

- *Certainty Factor*: A numerical value that expresses the extent to which, based on a given set of evidence, a given conclusion should be accepted.

- *Data analysis:* The interrogation of acquired data to come up with summaries and trends in the study variable.

- *Event*: An event is an occurrence taking place at a determinable time and place, with or without the participation of human agents.

- *Focus groups:* A selected group of expert/respondents.

- *Heavyweight Ontology*: An ontology is an ontology with a higher level of expressivity, with the capability to perform formal reasoning.

- *JSON*: An open-standard file format that uses human-readable text to transmit data objects consisting of attribute–value pairs and array data types.

- *Judgmental sampling*: The use of prior knowledge to select respondents to research questions.

- *Legacy System*: A computer system, programming or application software that is outdated or that can no longer receive support and maintenance.

- *Lightweight Ontology*: A directed graph whose nodes represent concepts.

- *Measurand*: A quantity intended to be measured.

- *Mind Map*: An illustration showing the interconnection of thoughts towards achieving an objective.

- *OGC*: An international consortium of industry, academic and government organisations who collaboratively develop open standards for geospatial and location services.

- *Open-ended questions*: A set of questions to which respondents are free to give their own responses.

- *Pilot/Pre-Test Study*: A trail study to gauge the adequacy of research tools and redefine questionnaires.

- *Population:* The set of all people in the communities' studies.

- *Qualitative research*: Research focusing on descriptive data and responses.

- *Quantitative research*: Research focusing on a number of responses.

- *Research design:* A plan for conducting research.

- *Rule*: In knowledge representation, an IF-THEN structure that relates given information or facts in the IF part to some action in the THEN part.

- *Sample:* A subset of a population.

- *Sigfox*: A global network that makes it simple to connect devices anywhere in the world.

- *Structured interview:* A set of predefined questions to guide researchers and respondents in answering of questions.

- *Subsumption*: A subsumption relation is "is-a-superclass-of" and the converse of "is-a", "is-a-subtype-of" or "is-a-subclass-of", defining which objects are classified by which class in a hierarchical format.

- *Validity:* The degree of a result to reflect the meaning of a tested variable.

- *W3C*: The main international standards organisation for the World Wide Web.

# Table of Contents

# List of Figures

# List of Tables

# List of Publications

As outcome of this research, the author of this thesis is the first author of the current published papers listed below.

**Paper A:**  Akanbi, A. K. & Masinde, M. (2015, December). Towards semantic integration of heterogeneous sensor data with indigenous knowledge for drought forecasting. In Proceedings of the Doctoral Symposium of the 16th International Middleware Conference (p. 2). ACM.

**Paper B:**  Akanbi, A. K. & Masinde, M. (2015, December). A Framework for Accurate Drought Forecasting System Using Semantics-Based Data Integration Middleware. In International Conference on e-Infrastructure and e-Services for Developing Countries (pp. 106-110). Springer, Cham.

**Paper C:**  Akanbi, A. K. & Masinde, M. (2018, May). Semantic Interoperability Middleware Architecture for Heterogeneous Environmental Data Sources. In 2018 IST-Africa Week Conference (IST-Africa) (pp. Page-1). IEEE.

**Paper D:**  Akanbi, A. & Masinde, M. (2018). IKON-OWL: Using Ontologies for Knowledge Representation of Local Indigenous Knowledge on Drought. In proceedings of the 24th Americas Conference on Information Systems (AMCIS 2018), New Orleans, Louisiana, US.

**Paper E:**  Akanbi, A. K. & Masinde, M. (2018, August). Towards the Development of a Rule-based Drought Early Warning Expert Systems using Indigenous Knowledge. In 2018 International Conference on Advances in Big Data, Computing and Data Communication Systems (icABCD) (pp. 1-8). IEEE.

# CHAPTER ONE

# INTRODUCTION

## 1.1. Background Information and Motivation

The past half-century has witnessed rapid advancement in various areas of Information Communication and Technology (ICT) (Atzori, Iera & Morabito, 2010), with smart environments now representing the next evolutionary development step in the home and environmental monitoring systems. The notion of an intelligent environment evolves from the definition of ubiquitous systems. According to Van der Veer and Wiles (2008), it promotes the idea of "*a physical world that is richly and invisibly interwoven with sensors, actuators, displays, and computing element seamlessly in the everyday objects of our lives, and connected through a continuous network.*" Enabling technologies needed for the realisation of this concept is multifaceted and most especially involves wireless communication, algorithm design, multi-layered software architecture (middleware), event-processing engines, sensors, semantic web, knowledge graphs, and adaptive control, amongst others. Currently, the integration of all these technologies has inherent challenges, mostly due to heterogeneity in ubiquitous components. The expectation that networks of heterogeneous smart devices and services can be integrated to form an interoperable information system is driving the need for broad agreement or solutions on data integration and interoperability across software boundaries (Berners-Lee, Hendler & Lassila, 2001).

In this PhD research, the focus is on environmental monitoring domain, with drought forecasting and prediction as a case study. Droughts are currently ranked number one (Guha-Sapir, Vos, Below & Ponserre, 2012) in terms of negative impacts, compared to other natural disasters such as floods, hurricanes, earthquakes and epidemics. They are now more rampant, severe and have become synonymous with Sub-Saharan Africa, where they are a significant contributor to the acute food insecurity in the region (Guha-Sapir *et al.*, 2012). Though this is not different from other areas of the world, the uniqueness of the problem in the Sub-Saharan Africa countries is the ineffectiveness of the drought monitoring and predicting tools in use in these countries. Droughts are very difficult to predict; they creep slowly and last longest of all-natural phenomena. The complex nature of droughts from onset to termination has made it acquire the title "*the creeping disaster*" (Mishra & Singh, 2010). The greatest challenge is designing a prediction and forecasting systems which can track information about the '*what*', '*where*' and '*when*' of environmental phenomena and the representation of the various dynamic aspects of thereof (Peuquet & Duan, 1995). The representation of such phenomena requires a better understanding of the 'process' that leads to the 'event'. For example, a soil moisture sensor is used to measure the property, soil moisture. The measured property can also be influenced by the temperature heat index measured over the observed period. This makes an accurate prediction based on these sensor values almost impossible without understanding the semantics and relationships that exist between these various properties.

Technological advancement in Wireless Sensor Networks (WSN) and Internet of Things (IoT) has facilitated efficient monitoring of environmental properties irrespective of the geographical location. However, in current IoT/WSN solutions, environmental parameters are measured using heterogeneous sensors that are mostly distributed in different locations. Further, different abstruse terms and vocabulary in most cases are used to denote the same observed property, thereby leading to data heterogeneity (Kuhn, 2005; Akanbi & Masinde, 2015a; Devaraju, 2005) with different data representation formats and communication protocol. However, effective forecasting and prediction of a complex environmental phenomenon such as drought involve combining diverse data sources (for example, sensor data, weather station data, geospatial data, satellite imagery, indigenous knowledge) for accurate forecasting information – which might still not be fool-proof.

Moreover, in order to increase the level of accuracy of drought forecasting and prediction systems, scientists in the field of anthropology, conservation biology and agriculture have been discussing the possibility of integrating indigenous knowledge on drought with scientific drought prediction knowledge (Ludwig, 2016). Furthermore, research (Mugabe *et al.*, 2010; Masinde, 2015) on indigenous knowledge (IK) on droughts has pointed to the fact that local IK in a geographic area can imply the likely occurrence of a drought event over time (Sillitoe, 1998), for example, worms like *Sifennefene* worms and plants like the *Mugaa* tree in Kenya could indicate drier or wetter conditions. However, researchers have often focused on differences between knowledge systems. Recent debates about how knowledge integration will benefit the weather forecasting/prediction domain cannot be overemphasised (Ludwig, 2016; Fogwill, Alberts & Keet, 2012). Indigenous knowledge (IK) has been in existence since the dawn of civilisation but seems to have been forgotten and currently on its way to its extinction, although development of new scientific knowledge is rapid, beneficial and well-documented. IK, on the other hand, is oral, scattered and unstructured knowledge and used by local indigenous people in certain geographic locations from generations to generations (Masinde, 2015). The possible integration of this ancient method with modern methods is significant, but will not be possible until full knowledge representation of the domain is fully achieved.

Many local communities and tribal farmers in Africa (and indeed, elsewhere in the world) have developed their intricate native systems of natural indicators for prediction. The Indigenous Knowledge System (IKS) is also used for local-level environmentally related decision-making in many rural communities as opposed to scientific knowledge. A typical example of the indigenous knowledge is the local IK on drought, which comprises the use of a variety of natural indicators associated with the environment for drought forecasting and prediction (Masinde, 2015). The local farmers in the community have relied on the IKS and their experience on drought for their farming decision-making. The indicators for the indigenous knowledge are from elderly farmers observations and years of use – making these farmers IK experts of that locality. Integrating this knowledge with modern drought forecasting models will increase the accuracy level currently hampered by the variability of scientific weather data (observation/simulation data) and the difficulty in achieving the desired level of scalability (Díaz *et al.*, 2015, Reid *et al.*, 2005).

Although modern scientific knowledge and methods have dominated the drought forecasting and prediction sphere, Fogwill *et al.* (2012) argue that modern science and technology with the help of indigenous knowledge will increase the level of accuracy. Hence, achieving the curation of quality vocabularies that will facilitate the detailed understanding of the natural indicators associated with drought forecasting in the local indigenous domain is essential. Studies such as the natural behaviour and ecological interactions between different species of insects and animals in a particular region can be used to infer drought forecast accurately and importantly in developing an accurate drought early warning system for the region. The most important method of collecting data on behaviour and ecological interaction is through detailed observation (Krebs & Davies, 2009). These observations, known by the IK experts, are shared orally. The data include the names of the animal and plant species, their relationships, and their behavioural tendencies due to weather changes.

Hence, this study envisages a very large unstructured knowledge base that captures how the weather influences the natural indicators, and the ecological interactions between different species of animals/plants with the environment for generating inference. However, due to lack of vocabulary standardisation brought about by heterogeneity and the use of local terminology and languages, analysts face significant challenges when attempting to analyse and integrate the indigenous knowledge data with the scientific knowledge base. This can be solved by attributing semantic annotation and representation of the IK using an ontology. Analysing the ecological interaction using ontology will provide descriptive and explanatory knowledge that will be useful in weather forecasts and climate predictions. The formal representation of indigenous knowledge, therefore, promises "*access to a large amount of information and experience that has been previously ignored, or treated as mysticism*" (Ludwig, 2016). The knowledge, with its empirically derived emphasis on the natural world, can provide scientifically testable insights into drought forecasting (Manyanhaire, 2015).

Considering the aforementioned, it can be concluded that the key to improving the accuracy of forecasting a drought event is the understanding of 'space-time' interactions of variables with processes, ontology representation of the domain and semantic integration of the heterogeneous sensor data with indigenous knowledge for efficient drought forecasting. Eventually, this leads to the processing and integration of a large amount of heterogeneous data from multiple sources. These factors encouraged the researcher to study and implement efficient ways to achieve heterogeneous data integration, interoperability for purpose of generating a more accurate drought prediction, and forecasting inference in environmental monitoring domain through a mediator-based system.

## 1.2. Problem Statement

In order to achieve heterogeneous data integration and interoperability in the environmental monitoring domain, semantic levels interoperability offers the technologies needed for enabling the same meaning to an exchange piece of data to be shared by communicating nodes, are currently lacking. This can be achieved through the representation of the data in a machine-readable format using knowledge representation and automated reasoning for accurate predictions and forecast. Moreover, modern sensory and legacy devices in communication systems were open systems built using the manufacturer's unique data and communication standards and thus require common semantics-level interoperability solutions.

The following problems and hypothesis were identified as a major bottleneck for the utilisation of semantic technologies for drought forecasting:

a) *The current lack of ontology-based middleware for the semantic representation of environmental data and processes*:
   - Hypothesis: Ontological modelling of key concepts of environmental phenomena such as an object, state, process and event can ensure the drawing of accurate inferences from the sequence of processes that lead to an event. At presently, what is missing is an environmental ontology with well-defined vocabularies that allow the explicit representation of the process, events and also attach semantics to the participants in the environmental domain. The developed semantic middleware prototype will enhance efficient integration and interoperability of heterogeneous data, facilitate ease of communication of weather/drought data/information between different platform/domain through standardised semantic annotation, and generate a more accurate drought forecast and prediction inference from the data inputs.

b) *Lack of semantic integration of heterogeneous data sources for accurate environmental forecasting*:

- Hypothesis: An environmental monitoring system made up of interconnected heterogeneous weather information sources such as sensors, mobile phones, conventional weather stations and IK could improve the accuracy of environmental forecasting by providing environmental data streams required to be semantically represented for seamless data integration with existing indigenous knowledge. Local indigenous knowledge of drought is relevant to contextualise the occurrence of a climate event in the area under study based on the ecological integration of the natural indicators. This integration will improve the accuracy of drought prediction.

c) *Lack of effective drought forecasts communication and dissemination channels*:

- Hypothesis: There is a lack of effective dissemination channels for drought forecasting information across various channels for utilisation by policymakers or analysts. For example, drought forecast information should be available in a standardised format that could be accessed through an application programming interface (APIs) for dissemination via notifications hubs, smart apps, documents (dB), or cloud repository for offline analysis.

## 1.3. Research Questions and Objectives

To solve the above research problems, the following research questions were taken into consideration for the heterogeneous data integration and interoperability:

a) **RQ1**: To what extent does the adoption of knowledge representation and semantic technology in the development of a middleware enable seamless sharing and exchange of data among heterogeneous IoT entities?

b) **RQ2**: What are the main components of an implementation framework/architecture that employs a distributed middleware for the implementation of a heterogeneous data drought early warning systems (DEWS)?

c) **RQ3**: What method is currently suitable to predict drought event given a combination of heterogeneous sensor data with indigenous knowledge on drought for an accurate drought forecasting system?

In order to answer the above questions, the main objective of this research was laid out as follows "to develop a semantic middleware for heterogeneous data integration and interoperability – using local indigenous knowledge on drought and wireless sensor data". This overall objective was demarcated using the following sub-objectives:

a) **RO1**: To identify aspects of indigenous knowledge used for drought prediction by three selected communities in Kenya and South Africa.

b) **RO2**: To develop an IoT framework for the use of WSN in environmental monitoring and use it to collect relevant data.

c) **RO3**: To develop a distributed semantics-based data integration middleware framework for heterogeneous data integration and generation of accurate inference.

d) **RO4**: To use relevant ontology to represent and integrate the indigenous knowledge identified in RO1 and sensor data collected in RO2.

e) **RO5**: To develop a drought early warning system application prototype and use it to test and evaluate RO4.

## 1.4. The Solution Approach – A Case Study

In this research, to test the solution's applicability and validity, a case study is considered (Benbasat, Goldstein & Mead, 1987). The case study investigated is drought forecasting and prediction in the environmental monitoring domain. It is used to study the heterogeneous data integration, interoperability of services as well as to develop and evaluate the proposed solution. The approach is based on five distributed *functional groups* (FG) of the middleware: *data acquisition*, *data storage*, *stream analytics*, *inference engine* and d*ata publishing*. The first FG – *data acquisition* was achieved through the adoption of ITIKI framework (Masinde, 2015). The *data storage FG* was based on cloud-based data storage infrastructure, while the *stream analytics FG* was used for real-time stream processing of the sensor data using a complex event processing engine (CEP engine) based on open source *Apache Kafka*. Furthermore, the *inference engine FG* is used for computing various forecasts and achieved through distributed services with an inference engine as the core. On the other hand, the *data storage FG* is used to disseminate the output using a standardised format. The solution was tested and validated using actual indigenous knowledge and weather data acquired in two study areas in Kenya and South Africa.

## 1.5. Limitation of the Research Scope

The focus of this research thesis is restricted to drought forecasting and prediction in the environmental monitoring domain. It does not focus on the verification and validation of the local indigenous knowledge acquired in the areas under study because this was not possible within the time frame of this study. Besides, although indigenous knowledge on drought was used to test the semantic middleware, the comprehensive collection of all the indigenous knowledge on drought is outside the scope of this project. Finally, this research did not consider or aim to develop appropriate security mechanisms to secure communication channels or data transmissions in the system.

## 1.6. Significance and Contributions of the Study

This research has made a significant contribution to the scientific knowledge through the novel approach of performing heterogeneous data integration using semantic technologies for the environmental monitoring domain. The main contributions of this thesis are summarised as follows.

The thesis presents a semantics-based data integration middleware framework that addresses the challenges of heterogeneous data integration and interoperability. This framework facilitated the semantic representation of the data sources eliminating data heterogeneity and created a model with a unified data format.

In this thesis, a domain ontology called **I**ndigenous **K**nowledge on Drought Domain **ON**tology (**IKON**), was developed for the local indigenous knowledge on drought. This ontology provides a machine-readable format of the domain. This domain ontology is based on Descriptive Ontology for Linguistic and Cognitive Engineering (DOLCE), and available in Resource Descriptive Framework (RDF) and Web Ontology Language (OWL) format.

A complete and tested implementation of semantic middleware for the integration and interoperability of heterogeneous data sources for drought forecasting and the prediction was presented. A method for real-time processing of environmental monitoring sensor data channelled through a streaming platform was also developed and presented.

A rule-based drought early warning expert (RB-DEWES) sub-system that could be implemented as a standalone system with customisable Graphical User Interface (GUI) for end users was developed, implemented and presented, and the implementation of a more accurate semantics-based drought early warning systems (DEWS) based on the semantic middleware for the study area was presented.

## 1.7. Evaluation Criteria

To evaluate this research, each objective was tested against the research outcomes. The case studies used were adopted to evaluate the research objectives as a measure of the quality and reliability in the form of verification and validation (V&V). The verification involves evaluating the research project to ensure it satisfies the research objectives; and the validation involves using necessary validation metrics to quantify the research processes.

## 1.8. Structure of the Thesis

The thesis is divided into eight chapters; besides the current chapter, the other chapters are as follows:

a) **Chapter Two** presents a comprehensive literature review of the concepts and technologies relevant to this thesis. It explains the use of local indigenous knowledge on drought, drought forecasting and prediction concepts, including related works that have been conducted by researchers.

b) **Chapter Three** presents the methodology followed in executing this research; it also presents the semantic middleware framework. The main aim of this chapter is to explain the research methodologies and presents the overview of the distributed middleware which comprised five different *functional groups,* all working in an orchestrated manner towards achieving the aim and objectives of this research.

c) **Chapters Four, Five, and Six** are dedicated to the *functional groups.* That is, Chapter Four explains the implementation of *Data acquisition FG* for heterogenous data (structured and unstructured collection. Chapter Five covers knowledge modelling and representation of the data sources using semantic annotation and representation in a machine-readable language. It presents the developed domain ontology for local indigenous knowledge on drought and sensor data. Chapter Six focuses on the automated reasoning systems of the semantic middleware. It presents the *Inference Engine FG* and *Stream Analytics FG* of the semantic middleware. Also presented were the inference engine and automated reasoners, including the developed GUI prototype for utilising the reasoners.

d) **Chapter Seven** presented the implementation and data pipelining of the distributed semantic middleware. Discussion on the results, including an evaluation of the developed middleware prototype against the user requirements, and usability evaluation was presented.

e) **Chapter Eight** concludes the thesis by briefly summarising the contributions of the research work and evaluation of the research against the research objectives. The concluding remarks and future research direction were presented.

# CHAPTER TWO

# BACKGROUND AND RELATED WORK

## 2.1. Introduction

This chapter is divided into two parts: background and related work. The first part starts with presenting the overview of the background challenges that necessitate this research using a case study approach. A detailed overview of the theoretical concepts and technological ideas employed in achieving the main contribution of this thesis is also presented. The second part presents the related works carried out by other researchers towards solving the research challenges identified in Chapter One of this thesis.

## 2.2. Background

The very heterogeneity of data presents challenges by hampering the full realisation of heterogeneous data integration and services interoperability potentials (Kuhn, 2005). These challenges are due to the lack of ability to combine multiple data residing in different autonomous information silos for effective use. This is because of incompatible data exchange or representation format. Data integration has been a decades-old issue, from legacy systems to modern information systems, with the goal being to combine disparate sets of data into meaningful information. Currently, with the rise of the Internet of Things (IoT)-enabled devices, different devices are generating a large scale of heterogeneous data sets at an unprecedented level with the challenge becoming grimmer than ever.

In a typical IoT realm, billions of day-to-day things ranging from physical to virtual objects/devices are joining online networks. Enabling technologies needed for identification, sensing and communication drive the success of IoT, include the internet itself, as well as sensors and communication modules. WSNs is a critical component for achieving IoT; it provides the sensing capabilities to collect information about the physical environment without any pre-set physical infrastructure (Akanbi & Masinde, 2015b). This results in extensive amounts of heterogeneous data that could be presented in a seamless and easily interpretable form.

The IoT provides the ability to remotely sense objects using a wireless network infrastructure (Atzori *et al.*, 2010). This has the potential of creating opportunities for more direct integration of physical objects with computer-based systems, resulting in improved accuracy, better analytics and understanding. However, building an IoT application requires the integration of multiple components in such array of sensors (with communications modules) and networks. However, considering each component's use of different underlying proprietary technology or data representation formats, the challenges continue.

Despite the above challenges, the applications of IoT technologies are numerous and diverse, as IoT solutions are increasingly adopted in virtually all aspects of life. The effective impact of IoT technologies transgresses the unit-value created by individually connected products. Instead, the extensive functionality of integrated IoT products creates an intelligent system. For instance, in the environmental monitoring domain, a connected sensor may become part of a farm equipment system, which could include, for example among other things, a sprinkler system, hydroponics, manure spreaders, or actuators that monitors various key environmental parameters. Moreover, integration or the combination of multiple systems or sub-systems may lead to systems of systems, providing insightful analytics (Bartolomeo, 2014).

The main enabling factor for this promising paradigm of IoT is, however, the interoperability of several technologies, seamless data sharing and ease of communication. Conceptually, the most viable solution that can facilitate this effective data sharing and integration is the representation of data in a machine-readable format, i.e. transformation of the data into semantically annotated data with detailed metadata representation for seamless communication between resources/things irrespective of the domain (Kuhn, 2009; Guarino, Oberle & Staab, 2009). This enabling factor is the thrust of this thesis – it focuses on the integration of heterogeneous data sources for drought forecasting and prediction.

The process of turning the heterogeneous data that in form of local indigenous knowledge on drought, as well as WSN data, into useful information, involves a series of five steps. These are: (1) knowledge representation of the domain, (2) semantic representation of the data, (3) integration of the knowledge sources, (4) automated inference generation systems, (5) data analytics, and information utilisation. This realisation further depends on a multitude of distributed services with semantic referencing of data at the core to enable ease of communication and interoperability among different things irrespective of the domain. This scenario is implemented in the form of a drought early warning systems that integrates heterogeneous data sources using available technology seamless data integration and services interoperability.

## 2.2.1. The Concept of Drought

Drought is a naturally occurring climate phenomenon that impacts human and environmental activity globally and is considered to be one of the costliest and most widespread of natural disasters (Smith & Katz, 2013; Below, Grover-Kopec & Dilley, 2007). In terms of negative impacts, droughts are currently ranked number one (Wilhite & Glantz, 1985, Guha-Sapir *et al.*, 2012). Compared to other natural disasters such as floods, hurricanes, earthquakes and epidemics, droughts are very difficult to predict; they creep slowly and last longest. According to Espinoza *et al.* (2011), drought qualifies as a hazard because it is a natural incidence of erratic occurrence but of recognisable recurrence and as a disaster.

Drought is a result of precipitation deficiency, which causes disruption of the water supply to the natural and agricultural ecosystems (Mohamed, 2011). However, drought is a natural environmental phenomenon, and its recurrence in susceptible areas is almost inevitable (Mishra & Desai, 2006; Gana, 2003). However, lack of definite characteristics of drought is a major dilemma for the scientific and policy-making community and is preventing a detailed understanding of the drought phenomenon. The absence of an accurate and precise definition of drought has been an obstacle to understanding drought, which has led to indecision and inaction on the part of the individuals concerned managers, policy-makers (Wilhite & Glantz, 1985; Wilhite *et al.*, 1986).

Drought can be termed as a normal, recurrent feature of climate, which is sometimes rare and occurs randomly. The occurrence of drought in a particular geographic area varies from region to regions. Drought is a creeping disaster; it occurs when there is less than normal precipitation over an extended period of time, usually a season or more (Dea & Scoones, 2003). The lack of or reduced water input causes water shortages to various activities that require water in the ecosystem such as agricultural irrigation, animal use, and other (Edossa, Woyessa & Welderufael, 2014). Drought can also occur when the temperature is higher than normal for a sustained period; this results in higher evapotranspiration (i.e., evaporation and transpiration) than the precipitation. The increase in the evaporation cycle makes water vapour in the air for precipitation but contributes to drying over some land areas, leaving less moisture in the soil. However, drought is not a disaster for nature itself. It is a sequence of events that causes drought, which makes forecasting and predicting it quite complicated without the area of necessary data with the desired level of variability and scalability (Edossa, Woyessa & Welderufael, 2016).

It is generally said that there is no universally accepted definition of drought. In 1965, Palmer came to this conclusion that "*Drought means various things to various people depending on their specific interest. To the farmer, drought means a shortage of moisture in the root zone of his crops. To the hydrologist, it suggests below average water levels in the streams, lakes, reservoirs, and the like. To the economist, it means a shortage which affects the established economy*" (Palmer, 1965). Irrespective of the accepted definition, drought can be classified separately as meteorological, agricultural, hydrological, and socioeconomic drought (Wayne, 1965). In this thesis, drought is generally referred to as based on the conceptual definition provided by Palmer (1965).

### 2.2.1.1    Causes of Drought

The causes of drought are multi-faceted, as, with many environmental phenomena: there is never only one, but multiple causes. Therefore, in order to understand the phenomenon, it must be treated as a manifestation of several factors (Welderufael, Woyessa & Edossa, 2013). While drought may usually be caused by common environmental parameters, such as weather systems and the like, there must be a detailed understanding of the ecological interaction indicating its likely onset. In modern times, drought is forecast using forecasting models and indices with environmental parameters such as soil moisture level, temperature, rainfall level, evapo-transpiration, all of which help in determining the severity of the drought on a larger scale. For example, Figure 2-1 depicts the drought and flood prediction modelling output representation for the African continent. However, in ancient times, these environmental parameters were observed through natural indicators or signs, which helped to understand the onset of drought at a lower level of scalability.

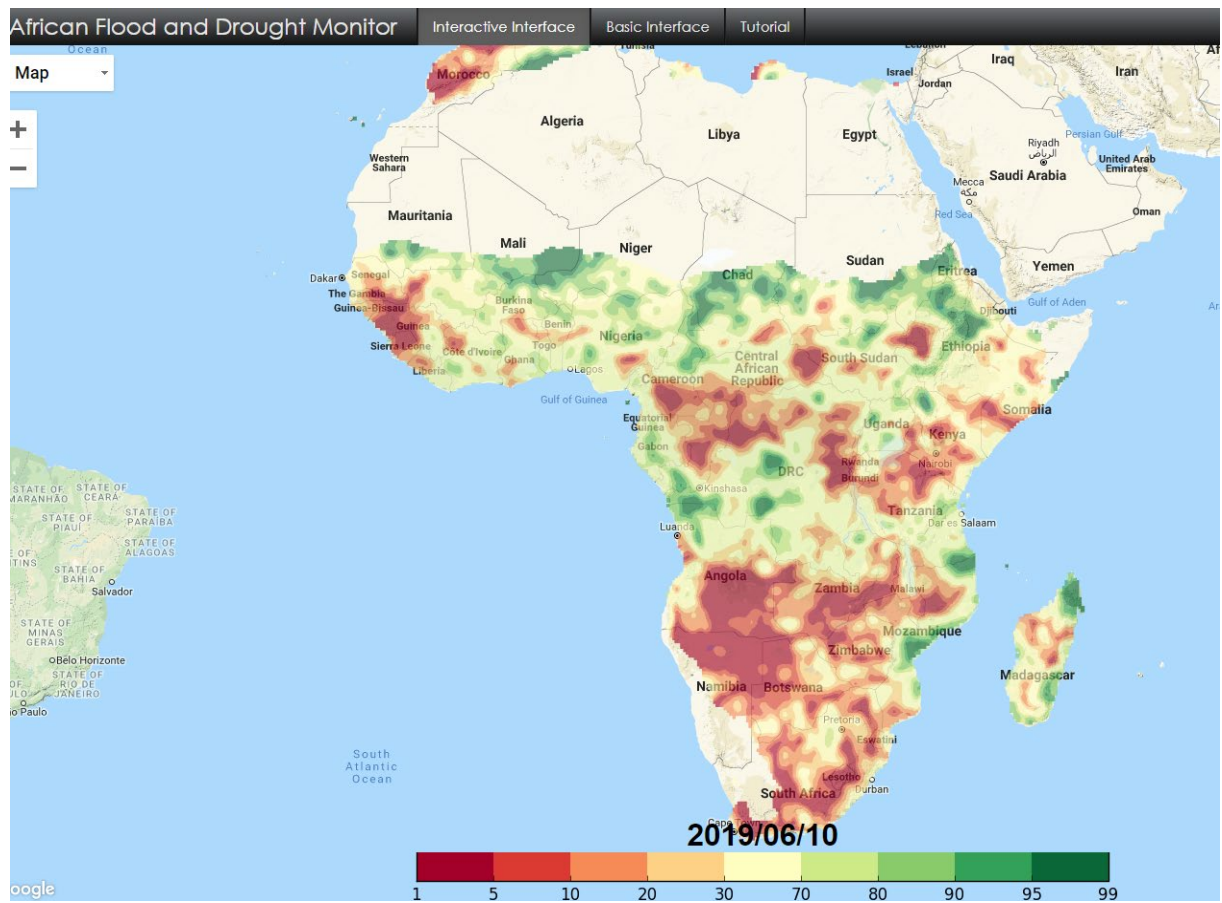

Figure 2- 1: The colour shows the severity of droughts index, red – significant positive trend (towards drier conditions) and green – significant negative trend (towards wetter conditions) (*Source: African Flood and Drought Monitor*)

### 2.2.1.2    Impacts, Cost, and Complexity of Droughts

Drought is a slow-onset natural hazard that has meaningful impact on many sectors of the economy, with the resulting impact exceeding the area experiencing physical drought. This compound effect exists because water is crucial to society's survivability. The complexity of impacts is largely caused by the primordial dependence on water directly or indirectly.

Drought impacts can, however, be classified as either direct or indirect (Mishra & Desai, 2006). This classification is borne out of the impact assessment of the drought and the resulting consequences on humans and the environment. The direct impacts include crop loss, deforestation, the risk of reduced water levels, fire, and damage to animal and fish habitat. The effect of these direct impacts consequently leads to the indirect impacts (Wilhite, Svoboda & Hayes, 2007). For example, lack of crop growth ultimately leads to scarcity and increase in the price of agro-foods and commodities.

Hypothetically, drought prediction tools could be used to determine drought development patterns as early as possible and provide information to farmers and policy makers to develop mitigation strategies to reduce the negative effect.

### 2.2.1.3 Drought Prediction Model and Indices

In modern methods of drought prediction and forecasting, all categories of drought are based on drought severity indices for prediction or modelling. According to Wilhite (2007), the severity of a drought is determined by the drought duration and probability distribution of the drought variables. It is therefore, of great importance to consider temporal parameters with different categories of drought.

Meteorological drought is a result of precipitation deficit and duration of the period, simply expressed in terms of lack of rainfall in relation to some average amount and duration of the drought period (Ceglar, 2008). The severity is defined in the form of indices such as the Palmer Drought Severity Index (PDSI), Standardised Precipitation Index (SPI) and Surface Water Supply Index (SWSI).

Agricultural drought on its own refers to an insufficient soil moisture level to meet the plant needs for water during the vegetation period (Ceglar, 2008). The main assessment of agricultural drought requires the calculation of water balance on a weekly time scale during the growing season. The severity of agricultural drought can be calculated using indices such as the Agro-hydropotential (AHP), Moisture Availability Index, Dry day Sequences, Generalized Hydrologic Model, and Crop Moisture Index.

Hydrological drought occurs after a longer period of precipitation deficit, caused by periods of lack of rain or shortfall on surface and subsurface water supply. It is common understanding that lack of precipitation has a consequent effect on groundwater, soil moisture, snowpack, and streamflow, which led to the development of the Standardised Precipitation Index (SPI) (McKee, Doesken, & Kleist, 1993).

Each category of drought has a specialised type of drought-forecasting indices; however, in this research, the focus will be on the use of the Effective Drought Index (EDI). The EDI encompasses hydrological, agricultural and meteorological drought. Moreover, it is different from the rest of the indices due to the fact that it calculates drought on a daily basis.

## Effective Drought Index (EDI)

The EDI is an agricultural, meteorological and hydrological drought index developed by Byun and Wilhite (1999) and addresses the shortcomings of the SPI. It is used to calculate 30 years mean effective precipitation (EP) and mean effective precipitation (MEP) for each calendar year. The Deviation of EP (DEP) is a measure of the difference between EP and MEP. When DEP is negative, it indicates 'dryer than average' (Byun & Wilhite, 1999).

$$EPi = \sum_{n-1}^{i}[(\sum_{m-1}^{n} Pm)/n] \quad \text{(Equation 2-1)}$$

$$DEP_n = EP_n - MEP_n \quad \text{(Equation 2-2)}$$

$$EDI_n = DEP_n \ / \ SD \ (DEP_n) \quad \text{(Equation 2-3)}$$

Where $EP_i$ represents the valid accumulations of precipitation of each day, accumulated for $n$ days, $P_m$ is the precipitation for $m$ days, $m = n$. In Equation 1, if $m/n = 365$, then, EP is the precipitation for the calender year divided by 365. $DEP_n$ in Equation 2 represents a deviation of $EP_n$ from the mean of $EP_n$ (MEP) – typically a 30-year average of the EP. $EDI_n$ in Equation 3 represents the Effective Drought Index, calculated by dividing the DEP by the standard deviation of DEP – $SD \ (DEP_n)$ for the specified period. After the calculation, the output is associated with different categories of the EDI (Table 2-1). Therefore, the categorisation of the drought phenomenon has to do with the computed values of the EDI. The table below itemises the different categories of drought based on the EDI value.

Table 2- 1: Classification of drought categories using EDI (*Source: Wilhite, 1999*).

| Drought Classes | Criterion |
|---|---|
| Extreme Drought | EDI ≤ 2.0 |
| Severe drought | -2.0 ≤ EDI ≤ -1.5 |
| Moderate drought | -1.5 ≤ EDI ≤ -1.0 |
| Near normal drought | -1.0 ≤ EDI ≤ 1.0 |

### 2.2.1.4     Local Indigenous Knowledge on Drought

Local indigenous knowledge is the knowledge, ways of life, methods and practices of indigenous and local communities around the world. This knowledge mostly called indigenous knowledge (IK) is developed and harnessed over years and centuries. Local knowledge is transmitted orally from generation to generation (Masinde, 2015). However, it is mostly shared in the form of folklore, proverbs, teachings, cultural values, beliefs, rituals, and local language. It is widely adopted in the local community and applied day-to-day activities such as agricultural practices, food preparation, natural resources management, education, and a host of other activities in rural communities (Warren, Brokensha & Slikkerveer, 1991).

Indigenous knowledge (IK) as local knowledge is unique to a given culture, society or tribe. Such knowledge is passed down from generation to generation (Simpson, 2000). Indigenous knowledge has value in the local geographical area and has become valuable for scientists for a better understanding of the rural localities. The application of IK in drought forecasting involves the utilisation of local knowledge on local weather and climate. This local knowledge is assessed, interpreted and predicted by locally observed indicators and experiences using combinations of plant, animals, insects and meteorological and astronomical indications (Boef, Amanor, Wellard & Bebbington, 1993).

Research in the field of IK is aimed at explicitly understanding the connections between local people's understandings and practices and those of scientific knowledge, notably in the environmental monitoring domain and agriculture (Brokensha, Warren & Werner 1980; Warren & Cashman, 1988; Wamalwa, 1989). In recent years, more efforts are necessary on way to accurately forecast drought through the use of every relevant available heterogeneous data source (Akanbi & Masinde, 2015b). This is necessary to mitigate the disastrous effect of drought in a particular geographic area using data sources or knowledge models that offer the required level of scalability and variability for accurate drought predictions.

## Nature of Indigenous Knowledge on Drought Forecasting

Indigenous knowledge is similar to scientific knowledge in that both attempt to make sense of the world and to render it understandable to the human mind. These knowledge bases are based on observations and generalisations derived from those observations. According to Berkes, Folke and Gadgil (1994), IK differs from scientific knowledge in its:

a) reliance on qualitative information;
b) lack of empirical facts;
c) reliance on experimental trial-and-error, rather than on systematic experiments;
d) lack of interest in building theoretical framework.

However, it appears that IK differs from scientific knowledge in being moral, spiritual, holistic and intuitive, with large social context. The major strength of IK lies in long time-series of observations on a geographical area. The veracity of the knowledge is based on long time-series as opposed to short time-series over a large area. The two kinds of data may be incompatible, but could be complementary when fully integrated. There is great potential value in a historical series of observations about particular areas based on knowledge passed from generation to generation provided the geographical area has not been drastically perturbed.

The local community has developed this local indigenous knowledge system (IKS) over the years from their understanding of the environment and used it for forecasting based on the variance of different natural indicators (Masinde & Bagula, 2010). These are used to increase the validity of the rainy season indicators. This category of indicators is used to forecast short-term (in hours or days) trends. IK forecasting is based on observing historical trends; this is one of the IK principles whose reliability is currently under threat due to the increased severity and frequency of droughts over the last decades across the entire world (Mutua *et al.*, 2011). IK on drought forecasting in most indigenous communities falls into six general categories: (1) seasonal patterns; (2) behavioural properties of animals, insects and birds; (3) astronomical; (4) meteorological; (5) human nature and behaviour; and (6) behaviour of plants/trees (Masinde, 2015).

## Indigenous Drought Forecasts in African Communities

According to some studies (Ziervogel & Opere 2010; Murphy *et al.*, 2011; Ajibade & Shokemi, 2003; Luseno *et al.*, 2003; Roncoli 2006; ISDR, 2006; Roncoli, Orlove,

Kabugo & Waiswa, 2011; Mercer, Kelman, Taranis & Suchet-Pearson, 2010), most African communities observe natural indicators such as clouds, wind and lightning; others watch the behaviour of livestock, wildlife, local flora, the ecological indicators interactions as early warning signs to predict the environment based on their local IK. They also observed changing seasons as well as lunar cycles (shape or position of the moon and patterns of stars). Other examples are: (1) mating of animals as a sign of plenty of rains to come (Roos, Chigeza & Van Niekerk, 2010; Masinde, 2012); (2) wind direction before rainfall (Masinde, 2015; Ajibade & Shokemi, 2003).

## Identification of Local Indigenous Knowledge on Drought Indicators

The concept of a local indigenous knowledge system is based on several ecological interactions and observations in the environment called indicators. These local so-called indicators serve as pointers to the likely occurrence of an environmental phenomenon in a pre-/post-observational scenario. The local indicators for the indigenous knowledge on drought are categorised according to the astronomical, meteorological, mythological and behavioural weather indicators (Table 2-2) (Masinde & Bagula, 2011; Masinde, 2015; Mwagha & Masinde, 2016; Mugabe *et al.*, 2010).

Table 2- 2: Categorisation of Local Indigenous Knowledge on Drought (*Source: Masinde, 2015*).

| Indigenous Knowledge Category | Category of Local Indicators by Property |
|---|---|
| Astronomical | Sighting of the moon, sighting of the stars, phases of the moon, clearness of the night sky, cloud levels, sun brightness. |
| Meteorological | Knowledge of the seasons, weather patterns, rain, temperature, humidity, precipitation, dryness, windy, cloudy. |
| Behaviours of birds | Flocking of birds, sighting of the birds |

| Behaviours of insects | Presence and occurrence of insects after environmental events. |
|---|---|
| Behaviours of animals | The weight of animals, the sighting of animals |
| Behaviours of floral and non-floral plants | Withering, flowering, growth, fruiting. |

Through the application of knowledge modelling and representation, each local indicator category has a comprising object(s) and corresponding attributes that exhibit the local indicator. An example instance, the flocking of the *Phezukomkhono* bird – a migratory bird sighted seasonally in the area under study. The notion of classification of the indicators based on the exhibited properties is necessitated for proper classification and the purpose of defining the taxonomy. The properties of the moon, for example, varies between the full/half to visible/dark moon transition, the properties for the classification of the object – the moon would be a full moon, half-moon. (Mwagha, 2017).

The combination(s) of several of these local indigenous indicators observations scenarios have a meaningful interpretation for forecasting drought in the local indigenous knowledge systems of the area under study and help to achieve the desired level of scalability in improving the accuracy of drought prediction and forecasting.

## 2.2.1.5. Indigenous Knowledge versus Modern Science on Droughts

Since the advent of modern science, drought management strategies are largely based on modern knowledge or technology at the expense of indigenous knowledge systems. Environmental phenomena such as droughts are complex and given various challenges in scientific weather and climate forecasting, such as lack of the desired level of scalability, indigenous knowledge (IK) is proposed to complement modern scientific knowledge (Masinde & Bagula, 2010). Collectively, this heterogeneous knowledge base represents a dynamic and localised information dataset that can support most rural communities to adapt to the changing and varying climates (Nyong, Adesina & Elasha, 2007).

The advanced modern technologies of weather forecasting and predictions are still elusive (Luseno *et al.*, 2003; Mugabe *et al.*, 2010; Masinde, 2015). Implementing modern drought prediction technologies such as weather stations, IoT monitoring systems, WSN solutions are still a costly affair for most African countries due to the associated cost challenges for implementation and maintenance.

### 2.2.1.6. Application of IoT/WSN for Drought Forecasting and Prediction

The basic idea behind the IoT paradigm is the interconnectivity of various generic objects to be integrated into a unified framework. According to Atzori *et al.*, (2010), 'Internet of Things' means the integration of various internet-enabled heterogeneous interconnected devices or objects for effective data sharing and machine-to-machine communication. With the advancement of technology, the significant potential of WSN has facilitated its use in environmental monitoring and habitat monitoring systems (Masinde, Bagula & Muthama, 2012).

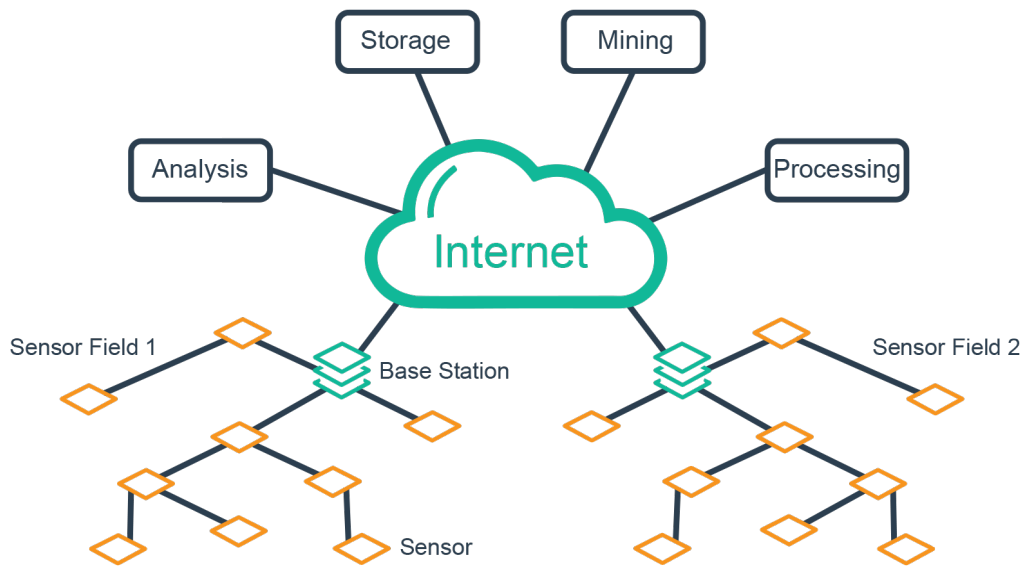

Figure 2- 2: WSN Network (*Source: Author*).

WSNs are networks of interconnected sensors that monitor environmental phenomena in geographic space irrespective of the topographical location (Figure 2-2). They have become an invaluable component of realising an IoT-based environmental monitoring system; they form the 'd*igital skin*' through which to '*sense*' and collect the context of the surroundings and provides information on the process leading to environmental phenomena such as drought and weather changes (Akanbi & Masinde, 2015b). However, these environmental properties are measured by various heterogeneous sensors of different modalities in distributed locations making up the WSN, using different terms in most cases to denote the same observed property.

Moreover, with these potentials, lack of unique addressing and semantic representation of sensor data are one of the most important bottlenecks hampering the realisation of the effective IoT visions and objectives, this is closely followed by security. This is due to different manufacturers, using different data languages; resulting in data formats that are incompatible with each other (Akanbi & Masinde, 2015b), causing a lack of seamless data integration and use. Traditionally, the easiest way to address interoperability is to define standards (Kosanke, 2006). Several standards have been created to cope with the data heterogeneities. Examples are the Sensor Markup Language (SensorML) (http://www.opengeospatial.org/standards) and  Observations and Measurements Encoding Standard, WaterML (Valentine, Taylor & Zaslavsky, 2012, and American Federal Geographic Data (FGDC) Standard (https://www.fgdc.gov/metadata).

These standards provide sensor data to a predefined application in a standardised format and hence do not solve data heterogeneity. The promising technology to tackle these problems of heterogeneity and integration of ubiquitous data sources is semantic technologies. Semantic technologies have a stronger approach to interoperability than contemporary standards-based approaches (Oberle, 2004; Akanbi & Masinde, 2015b). It creates knowledge representation models that are general to allow meaningful information exchange among machines through detailed semantic referencing of metadata. It utilises Resource Description Framework (RDF) and Ontology Web Language (OWL) for seamless data sharing and integration in an event-driven way and adopted for use in this thesis towards achieving heterogeneous data integration for effective drought forecasting and prediction systems.

### 2.2.1.7. Drought Early Warning Systems

Drought Early Warning System (DEWS) is a variant of Early Warning Systems (EWS) for drought disaster management, forecasting with necessary mitigation strategies (Wilk, Andersson, Graham, Wikner, Mokwatlo & Petja, 2017). According to UNISDR (2009) "*Early warning is a major element of disaster risk reduction.*" The adoption of an early warning system can prevent loss of life and reduce the impacts of disastrous events. However, the effectiveness of early warning systems is tantamount to the active participation of people and communities at risk; monitoring of the risk via accurate warning systems; dissemination and communication of warning systems and adequate response capability or mitigation plans (UNISDR, 2009; Rogers & Tsirkunov, 2011). These four key elements of EWS depicted in Figure 2-3 is based on the Hyogo Framework for Action (UNISDR, 2005), which was adopted by the World Conference on Disaster Reduction in Hyogo, Japan in 2005. The development of an intelligent drought forecasting and decision support systems is important to achieving the key element of an EWS highlighted under the Hyogo Framework (Leonard, Johnston, Paton, Christianson, Becker, & Keys, 2008). The thesis proposed the development of a semantics-based drought early warning systems (SB-DEWS) to address the important key elements of an EWS.

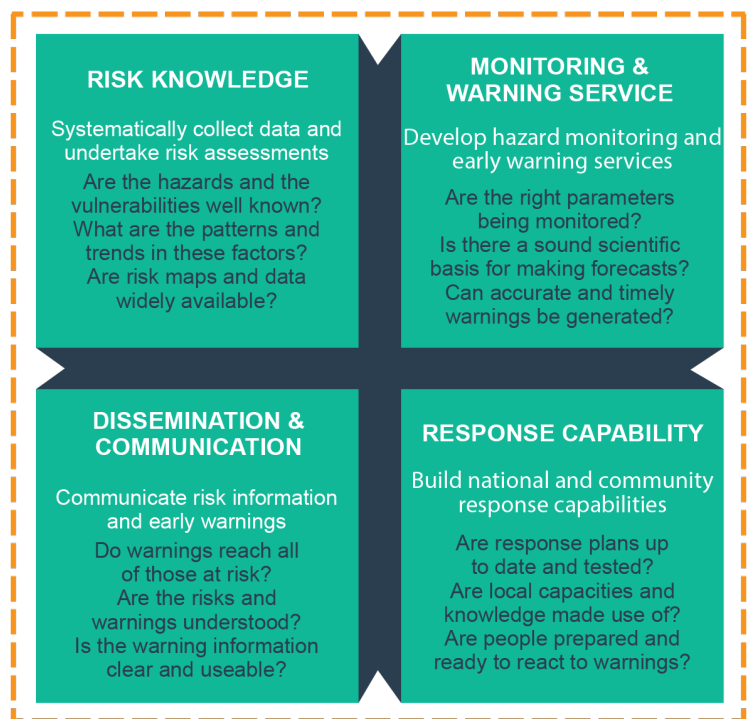

Figure 2- 3: Key element of an early warning system (*Source: Rogers & Tsirkunov, 2011*)

Current systems that address droughts are multi-faceted, and drought forecasting is not the main functionality of the systems. Examples of such systems are the Famine Early Warning System (FEWS-Net) (Verdin, Funk, Senay & Choularton, 2005), which provides monthly famine and droughts reports on in Eastern Africa. There is no one single early warning system (known to the author) dedicated to tackling droughts in Africa. Other such systems described by (Rashid, 2009) are Global Information and Early Warning System on Food and Agriculture (GIEWS) by Food and Agriculture Organisation (FAO), and Humanitarian Early Warning Service (HEWS) by World Food Programme (WFP). At national levels, the U.S. Drought Monitor is the best-known drought early warning system, while the most relevant (to this research) system is the East Asian drought monitoring system that makes use of the Effective Drought Index to describe the spatial and temporal distribution of droughts in East Asia (Oh, Kim, Choi & Byun, 2010).

### 2.2.2. Semantics-based Drought Early Warning Systems (SB-DEWS)

A semantics-based drought early warning system (SB-DEWS) is a form of an (EWS) specifically tailored for the provision of timely, accurate and effective drought forecasting information through semantic integration of heterogeneous data sources, that allows generation of deductive inference from an understanding of *'space-time'* interactions of environmental variables in the form of *rules*.

In this case study of SB-DEWS, the indigenous knowledge on drought is collected through the various data collection tools from the IK experts; the data are analysed to determine the patterns of the hazard, effects, and the vulnerability in the area under study. The knowledge is gathered, and facts in the form of *rules* are identified. The *rules* identified are used to create the risk assessment and indicators or signs of potential occurrence. Natural drought indicators in the form of *rules*, and ecological interaction in the form of events, obtained from the domain experts of the study area are semantically represented and integrated to predict future occurrence using advanced technological solutions using a stream processing engine and an inference engine module.

The inferred warnings outputs called Drought Forecast Advisory Information (DFAI) is disseminated through multiple communication channels via notification hubs, mobile USSD services, web apps, logic apps etc. The disseminated DFAI information is interpreted by the policymakers who are the intended target for the outputs.

### 2.2.2.1. Semantic Technology

Semantic technology consists of a set of methods and tools for discovering in-depth relationships within varied categorised data sets (Sheth & Ramakrishnan, 2003). This technology ensures the discovery of meaning (semantics) within data. The goal of Semantic technology is to make the machine to understand the data by encoding of semantics with the data through the use of machine-readable languages to represent a data or knowledge base (Domingue, Fensel & Hendler, 2011).

The structure of semantic technology is based on the Semantic Web Stack. This stack illustrates the architecture of the semantic technology from the semantic representation of knowledge up to the application in Semantic Web (WEB 3.0). The Semantic Web initiative is mostly a collaborative movement led by international standards body; the World Wide Web Consortium (W3C). It promotes intelligent data formats on the World Wide Web. By encouraging the inclusion of semantic content in web pages, the Semantic Web aims at converting the current web, dominated by unstructured and semi-structured documents into a "*web of data*" to ensure integration and interoperability (Sheth & Ramakrishnan, 2003). Figure 2-4 represents the semantic web stack.

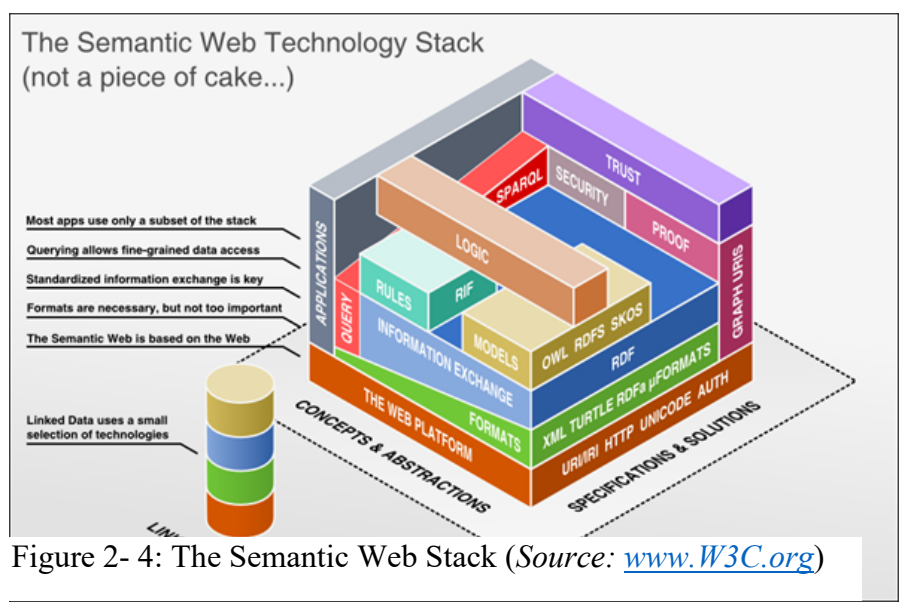

Figure 2- 4: The Semantic Web Stack (*Source: www.W3C.org*)

The middle layer of the Semantic Web stack describes the different formats for representing information in intelligent information systems, using technologies standardised by W3C for accurate knowledge representation. These technologies formally represent the meaning involved in information using languages that can be read by machines called machine-readable languages, well-known technologies are Resource Description Framework (RDF), Resource Description Framework Schema (RDFS), Web Language (OWL), SPARQL and Rule Interchange Format (RIF). Semantic technologies provide a new level of depth that offers more intelligent understanding of a knowledge base.

### 2.2.2.2.    Semantic Representation

The concept of how meaning and knowledge is represented has been a critical factor for effective communication since the dawn of humankind. According to (Vigliocco, Meteyard, Andrews & Kousta, 2009), the most important questions that arose from this concept are: (1) conceptual meanings related to conceptual structures? (2) How is the meaning of each word represented? (3) How are the meanings of different words related to one another? (4) Can the same principles of organisation hold in different content domains (e.g., words referring to objects, words referring to actions, words referring to properties). Researchers concur that a clear understanding and answers to these questions will maximize the utilisation of unstructured knowledge, such as indigenous knowledge and ensure effective integration and interoperability. This is achieved through appropriate knowledge management and transformation of the knowledge base into a model.

### 2.2.2.3.    Knowledge Management

Knowledge management is a vast discipline that deals with how people, process and technology come together for the utilisation of knowledge gathered or acquired. This entails the use of the right information and knowledge at the right time in the right context and the appropriate format. This is essential in this research study because knowledge management works by transforming data and information which comes from all available sources into reusable knowledge. For the sake of this research, various types of knowledge identified in the literature are explained briefly (Alavi & Leidner, 2001). Knowledge tends to come in pairs and often is the antithesis of each other:

a) *A Priori*: A priori is a term which means "from before" or "from earlier. It is a term that emaciated from epistemology (the study of knowledge). A priori knowledge is a knowledge that can be derived from the world without any form of experience. For example, a mathematical calculation of "2+5=?" can easily be derived without physically finding objects to count to get the answer. Mathematical equations are a typical example of *priori* knowledge.

b) *Posteriori*: This is the antithesis of *priori* and means "from what comes later" or "from what comes after." This type of knowledge experienced through the use of inductive reasoning to gain knowledge.

c) *Explicit Knowledge*: It is the knowledge an individual hold consciously in mental focus – knowledge identified in documents, images, audio-visual contents etc. This type of knowledge is easier to interpret and consumed externally.

d) *Implicit Knowledge*: This type of knowledge can be captured externally, it is based on experience and intuition, for example, capturing a domain knowledge be interviewing the domain expert in a particular domain.

e) *Tacit knowledge*: The knowledge gained from personal experience; it represents an internalised knowledge. This form of knowledge varies from individual to individual and comprises of experience and intuition; very difficult to express but can be captured externally.

The knowledge inferred or gathered in this research can be categorised based on the categories above, which indicates its lifecycle for use and application.

### 2.2.2.4. Knowledge Lifecycle

Typically, knowledge can be expressed in a two-dimensional life cycle, similar to software development (Studer, Benjamins & Fensel, 1998). The first phase is the innovation phase, and the second is the sharing phase. The innovation phase captures the lifecycle of the knowledge as it develops – how the new knowledge is created, represented and applied for use. On the other hand, there is the sharing phase, which involves identifying and capturing of the knowledge; organisation of unstructured knowledge in a structured format for consumption; dissemination of the structured knowledge in a form which is sharable externally to groups and used by intelligent information systems; and utilisation for decision-making processes.

### Why Capture Knowledge?

Knowledge access is important when needed and in the right format. Essentially, appropriate knowledge representation ensures the ease of information/knowledge search, access, share and reuse. Also, the capturing of knowledge is important because the cost of losing knowledge is great and significant (Van Vlaenderen, 2000). A typical example of this case is the indigenous knowledge (IK), which is currently going into extinction due to the adoption of modern methods. It is therefore, important to capture, organise and store this knowledge (IK) as it helps to make the utilisation of the knowledge more efficient and competitive for immediate and future use (Van Vlaenderen, 2000).

### 2.2.2.5. Knowledge Model

A knowledge model is similar to a mind map generated from human thoughts. These thoughts have been used during the course of human life to create what is called a mind map. A mind map is an illustration showing the interconnection of thoughts towards achieving an objective. The mind map is used to conceptualise concept and ideas by providing detailed relationships between concepts in the domain of discourse. This ensures a more meaningful interpretation of the concepts into something that is more interpretable. The possibility of breaking down information or ideas in a mind map into knowledge that is more interpretable by humans and machine is quite beneficial (Studer, Benjamins & Fensel, 1998), i.e. the information would be shared more easily by humans and offers a better way of sharing the information and meanings across machines (coded mind maps). This would ensure reasoning capabilities and more robust interaction between humans and machines based on the knowledge model.

In a nutshell, the knowledge modelling of information is called an *ontology*, where the knowledge representation is multivariate and multidimensional (Smith, 2003). An ontology is a mind map with an added structure that allows the representation of a domain and the meaning of concepts to be clearer. A knowledge model or ontology can be a visual representation (for human beings to view in the form of mind map) to understand and share, or a coded representation for the machine and intelligent systems' interpretation. A knowledge model allows the formalisation and capturing of the essence configuration and interrelationship of a subject matter.

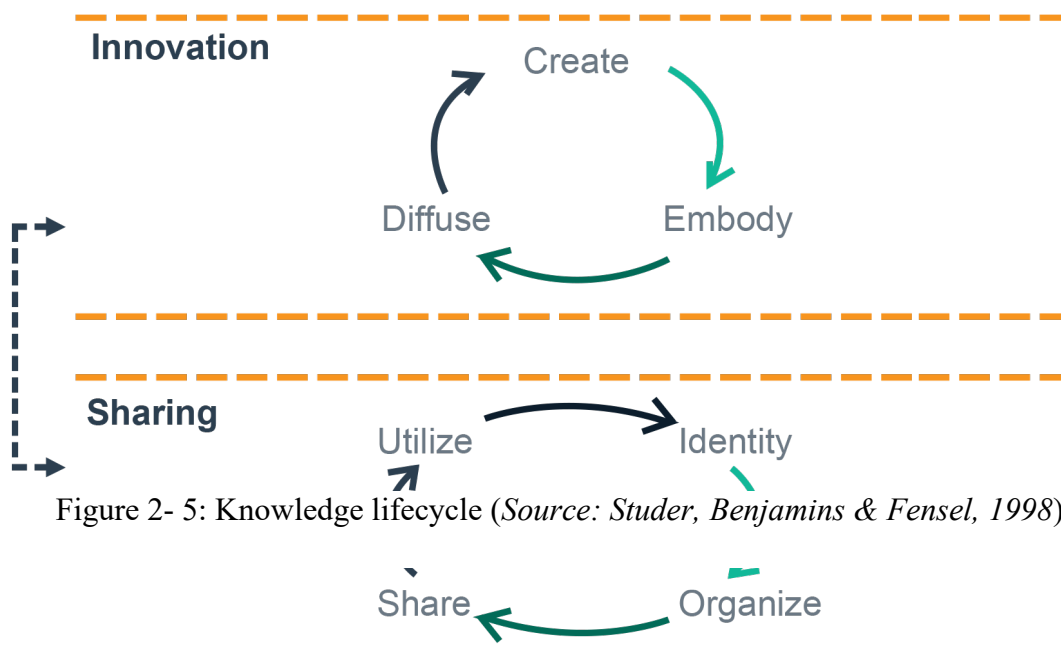

Figure 2- 5: Knowledge lifecycle (*Source: Studer, Benjamins & Fensel, 1998*)

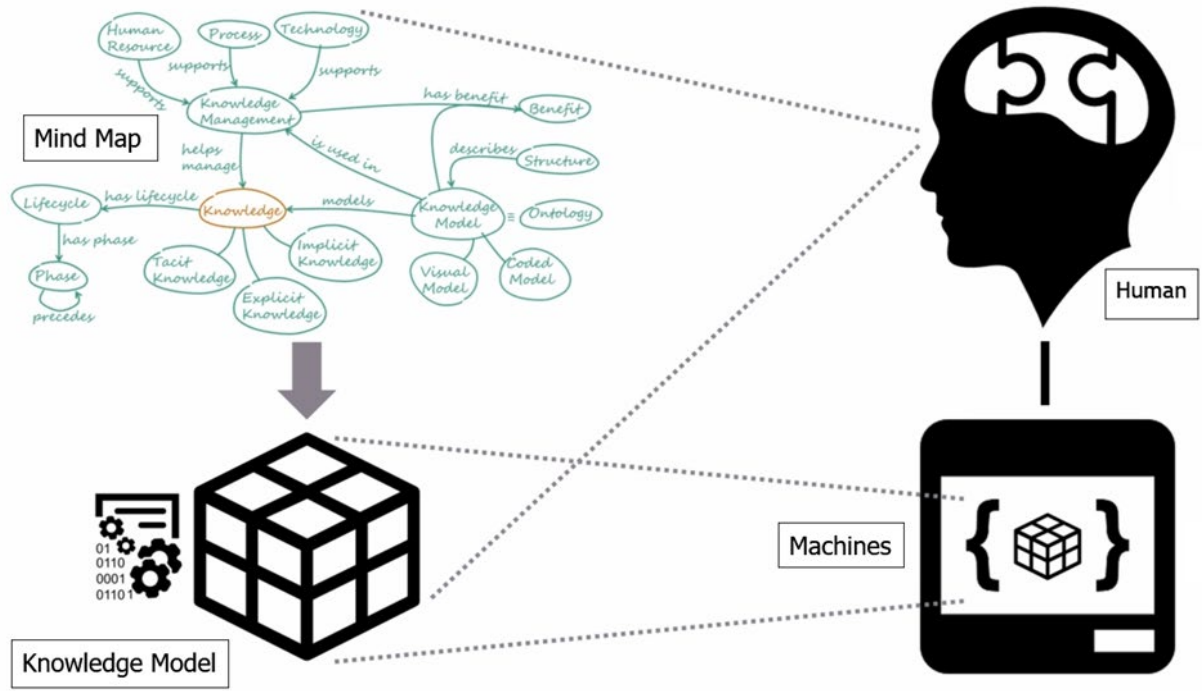

Figure 2- 6: Conceptual representation between Mind Map, Knowledge Model, Humans and Machines (*Source: Author*).

Knowledge access is important when needed and in the right format. Essentially, appropriate knowledge representation ensures the ease of information/knowledge search, access, share and reuse. Also, the capturing of knowledge is important because the cost of losing knowledge is significant. A typical example of this case is the local indigenous knowledge (IK), which is currently going into extinction owing to the adoption of modern methods. Therefore, it is important to capture, organise and store this knowledge (IK) as it helps to make the utilisation of the knowledge more efficient and competitive for immediate and future use.

### 2.2.2.6. Ontology

The concept of ontology in computer science is different from that in philosophy. According to Guarino, Oberle and Staab (2009), ontology is an explicit, formal specification of a shared conceptualisation to represent a specific domain of knowledge or discourse in a more typical way. An ontology defines terms with which to represent knowledge.

Ontology can provide formal semantic knowledge representation for the local indigenous knowledge. Moreover, since ontologies explicitly define the content of knowledge by formal sources, they ensure the integration and interoperability between these sources. Furthermore, ontologies can be used to detached domain knowledge from application-based knowledge in information-providing applications (Segaran, Evans & Taylor, 2009). The basic structural elements of ontology are namely:

a) *Class* is the collection of similar concepts related to a specific domain of knowledge; they can be a real object or abstract object concepts. Their attributes describe classes; meaning individuals populating a class shared common attributes. The class can be described in a formal, semi-formal or informal way, with preference given to formal ontologies. The formal description is a machine-understandable representation, for example class of animal, class of insect.

b) *Properties* are special attributes whose values are the object of (other) classes. It can be further divided into object properties and datatype properties.

c) *Instances* are the members (individuals) of the class and are the structural component of an ontology.

d) *Axioms* are rules that cannot be expressed with the help of other components.

In clear terms, an ontology can be an agreed blueprint for knowledge representation that has been designed to be interpretable by humans and machines. Ontology can be utilised and applied to meet the various needs, such as the perfect capturing of the meaning (semantics), domain representation, building controlled vocabulary, modelling etc. Several ontology languages have been developed with W3C standards, for example RDF, RDFS, OWL, DAML, and OIL.

### Creating a Domain Ontology (Informal Representation)

There are several types of ontology, ranging from upper ontology, application ontology, domain ontology to task ontology (Guarino, 1998; Noy & McGuinness, 2001). Figure 2-7 depicts these types of ontology. A domain or task ontology is built using an existing foundational ontology as a blueprint. In this thesis, a domain ontology will facilitate knowledge representation of the heterogeneous data sources. Hence, domain ontology provides vocabulary about the objects and concepts of a domain and their relationships (Berners-Lee, Hendler & Lassila, 2001). According to Guarino (1998) and Smith (2003), the ontology design or modelling approach is an iterative process that repeats continuously to improve the developed ontology; there are several stages involved which can be revisited if flaws detected are during the ontology design life cycle.

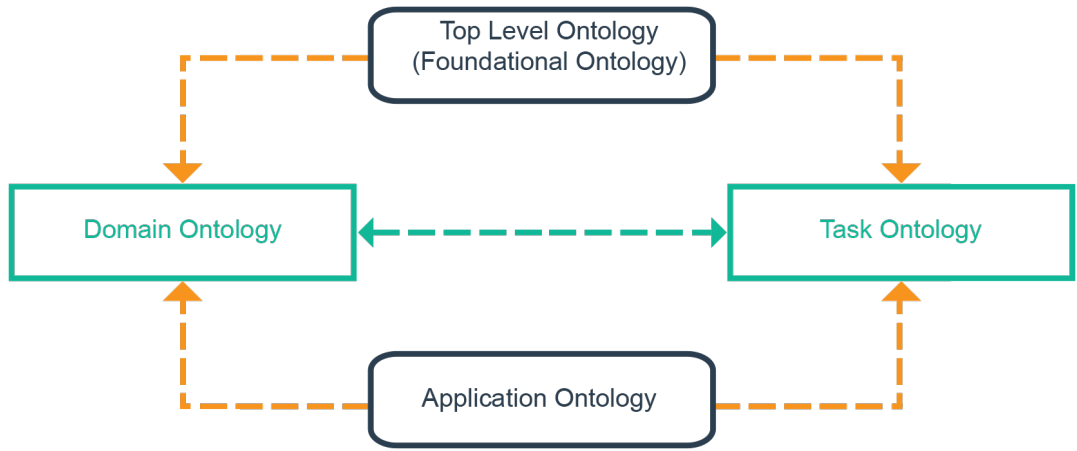

Figure 2- 7: Level of abstraction in ontology development (*Source: Guarino, 1998*).

The first step towards the development of a domain ontology is the determination of the ontology scope. This elaborates on what type of questions should be answered by the knowledge representation of the ontology and its re-usability. Reusing data or knowledge improves the quality of the development process. The next step in the iterative process is the development of the terminology about the domain; these are done by reviewing related published papers and interviewing the IK domain experts through questionnaires, workshops, and mobile apps.

## Semi-Formal Representation

The easiest way to develop semi-formal models for ontology is by applying logic, for example propositional logic (PL); first-order logic (FOL), descriptive logic (DL). The simplest type of logic is PL. In PL, the world consists of simple facts and nothing else, i.e. statement of assertions. An example of PL assertions and deductions based on local IK are:

1) *If Mugumo tree flowers, there would be bumper harvest;*
2) *If it does not flower, there won't be a bumper harvest.*

In PL, simple deductions can be made from the assertions. However, one problem in PL is that it only allows for making statements and assertions about a single object; it does not allow the summarisation of objects into a set of classes, or making a statement about a set of things. FOL is much more powerful than PL: in FOL, there are quantifiers/quantors that allow assertions about a set of objects, without naming the objects explicitly. This means there is the ability to make inductions out of a set of statements and infer implicit knowledge. For instance, considering the set of statements below, by understanding the assertion of the statements (1) and (2), implicit knowledge can be deduced from this statement to form statement (3).

1) *All crops need water to survive.*
2) *Lettuce is a crop.*
3) *Lettuce needs water to survive.*

FOL is a perfectly appropriate ontologies description, but the major disadvantages of FOL are that it is too expressive, too bulky for modelling because there are many interpretations that can be deduced from same knowledge in various forms, and too complex to prove the correctness or completeness of assertions.

## Formal Representation of Domain Ontology

The formal representation of a domain ontology knowledge base in detailed semantic annotation enables integration, interoperability, ease of data sharing among different platforms and eliminating data heterogeneity (Kuhn, 2005, Kuhn, 2009). It represents the unstructured data in a machine-readable language to facilitate effective use and integration. The manual ontology design process is costly and cumbersome. Therefore, an automated support system for ontology design is most often used. This involves the use of various software suites such as OntoEdit, KAON, and Protégé.

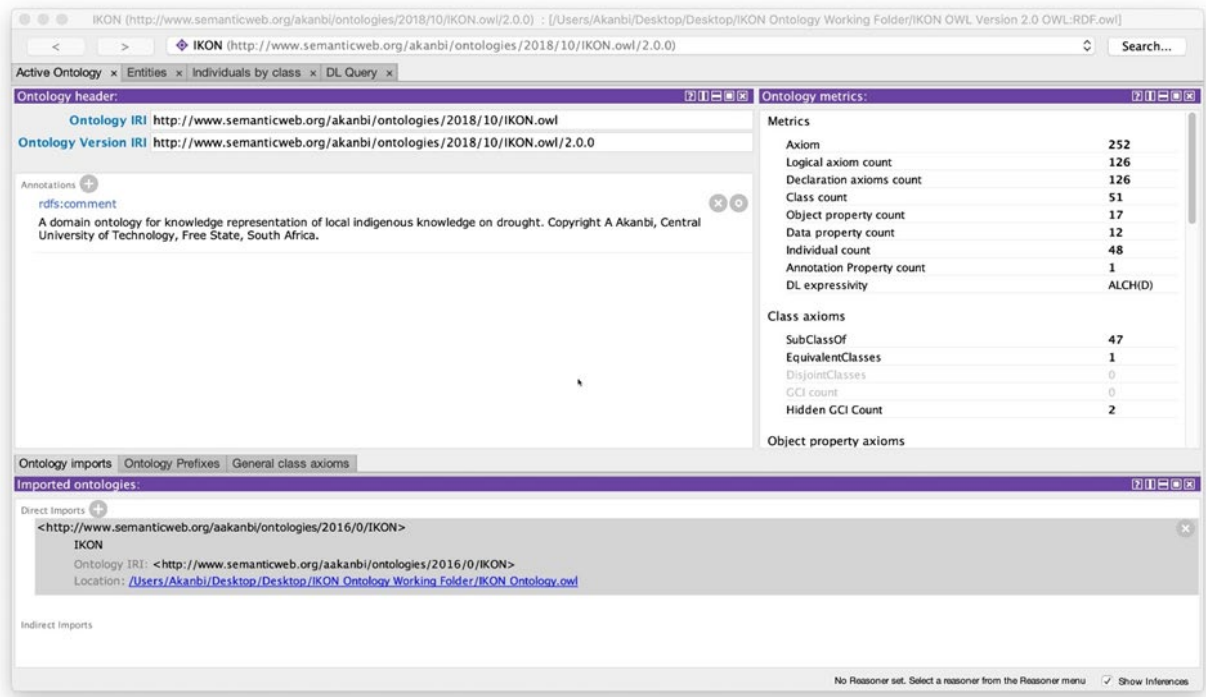

Figure 2- 8: A screenshot of Protégé IDE (*Source: Author*).

### 2.2.2.7. Knowledge Modelling of Heterogeneous Data Sources (D1 & D2)

Although modern scientific knowledge and methods are widely adopted in drought forecasting. Masinde (2014), Fogwill *et al.* (2012), Manyanhaire (2015), Coetzer, Moodley & Gerber (2014) and Akanbi & Masinde (2015b) all argue that modern science and technology, with the help of indigenous knowledge, will increase the level of accuracy of drought forecasting systems. Then, how can meaningful descriptions of environmental *events* be inferred from observations in the form of indigenous knowledge and sensor data? This research is tasked with identifying quality vocabularies that will facilitate the detailed understanding of the natural indicators associated with drought forecasting in the local indigenous domain (Akanbi & Masinde, 2015b). Currently, there is a lack of common definitions in terminology and semantically rich data representation models.

As stated earlier, there are two ways of representing a knowledge model: *visual representation* and *coded representation*. While visual representation is perfect for human interpretation and understanding, it is not suited for the machine and intelligent systems because visual representations are not encoded in a standardised format and well-defined languages that computer understand and interpret. A coded representation of indigenous knowledge and sensor readings using ontologies are necessary and important to make knowledge models meaningful and interpretable by computers. However, during the deliberation of the representation formalism for encoding knowledge models, detailed consideration was given to the *level of expressivity* of the standardised language, the *semantics* of the language and the *mathematical rigour* of the language, and hence, this research study has adopted the use of RDFa and OWL. Both standardised formal languages exhibit a high level of formality and expressivity, which are adequate for representing the heterogeneous data sources (**D1 & D2**) in knowledge models. Also, both standardised formal languages can be translated to JSON-LD for effective data communication between *functional groups* of the middleware without the loss of syntactic and semantic expressivity through the REST Manager.

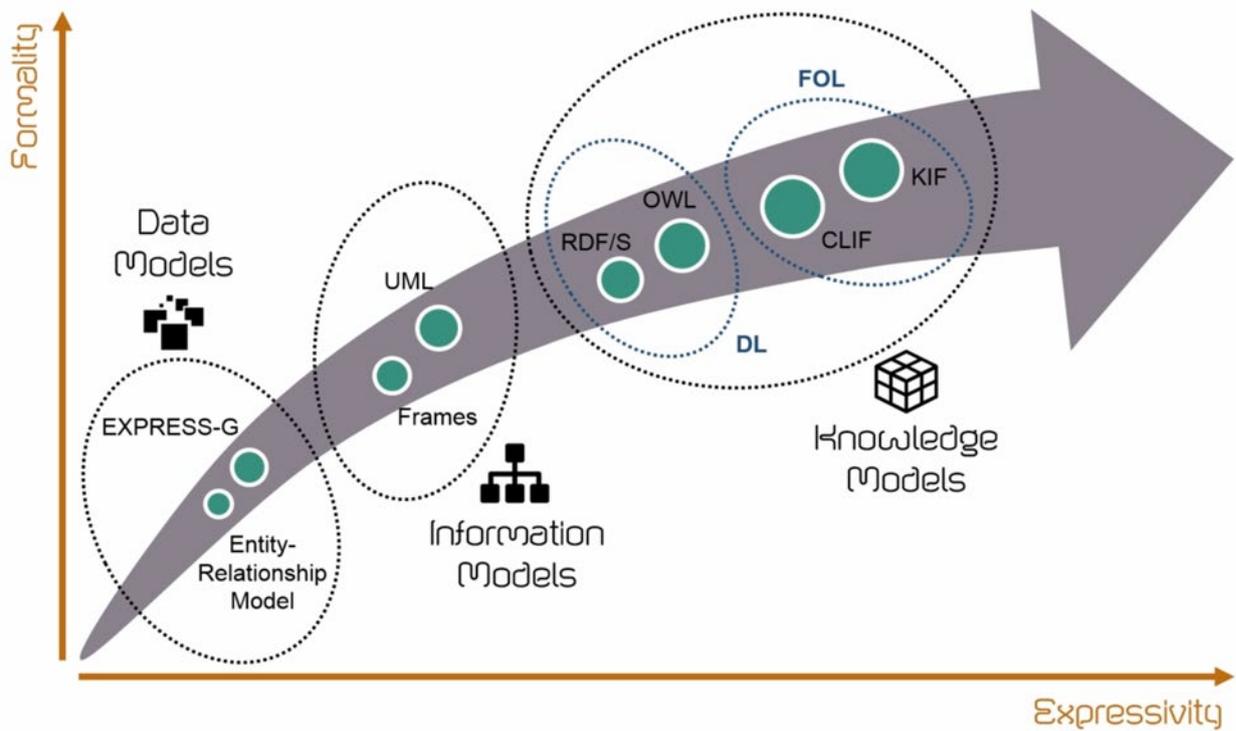

Figure 2- 9: Knowledge Representation Languages with a level of formality and degree of expressivity (*Source: Berners-Lee, Hendler & Lassila, 2001*)

Furthermore, the existing generic foundational ontology was used in the development of a domain ontology for local IK on drought and WSN sensor data. There are several top-level ontologies such as DOLCE, SUMO and BFO, which provide the standardised classification of very general concepts. This research tends to adopt DOLCE as the foundational ontology – because it provides relevant general notions under which the research domain concepts can be classified.

## DOLCE Foundational Ontology

The Descriptive Ontology for Linguistic and Cognitive Engineering (DOLCE) (Masolo, Borgo, Gangemi, Guarino, Oltramari & Schneider, 2003; Borgo & Masolo, 2010) is adopted as the foundational ontology for building the ontologies for the heterogeneous data sources (**D1 & D2**). DOCLE (Figure 2-10) embraces a pluralist perspective (Masolo *et al.*, 2003). The choice of DOLCE is because it provides most of the general notions for classifying the research domains concepts for the local indigenous knowledge on drought domain (**D1**) and to ensure ontology alignment with Semantic Sensor Network (SSN) ontology that will be adopted for the WSN domain (**D2**). Moreover, DOLCE has been widely adopted as the starting point for building an ontology in several ontology development initiatives (Kuhn, 2009; Probst, Gordon, Dornelas, 2006; Borgo, Cesta, Orlandini & Umbrico, 2016; Devaraju, 2009; Moreira, Pires, van Sinderen & Costa, 2017; Ludwig, 2016) in geospatial and sensing domains.

DOLCE aims to capture and represent the intuitive and cognitive bias underlying entities while recognising standard considerations. The top-level categories of DOLCE are **endurant**, **perdurant**, **quality** and **abstract** (Masolo *et al.*, 2003). Entities belonging to the **endurant** category are wholes at any time they are present, but at a certain instance of time, the same **endurant** may acquire or lose new parts and are subject to changes, for example, a **floral plant** such as flowering plant, and **blooming** or **withering** of the flowers (Masolo *et al.*, 2003). **Perdurant** is the category of entities that extends over time, at any time at which they exist they are only partially present, i.e., they can either be **eventive occurrences** such as drought and **stative occurrences** such as raining, etc. **Qualities** are physical or temporal (time-related) properties perceived or measure, for example, the temperature, duration of a rainfall, etc. Masolo et al. (2003) state "A participation relation holds between an **endurant** and a **perdurant**. A **physical-quality** is **inherent-in** a **physical-endurant,** whereas a **temporal-quality** is **inherent-in** a **perdurant**." The taxonomy of the domain concepts will be constructed using the DOLCE ontology classifications (Figure 2-10), and the knowledge is modelled and encoded using *Protégé*.

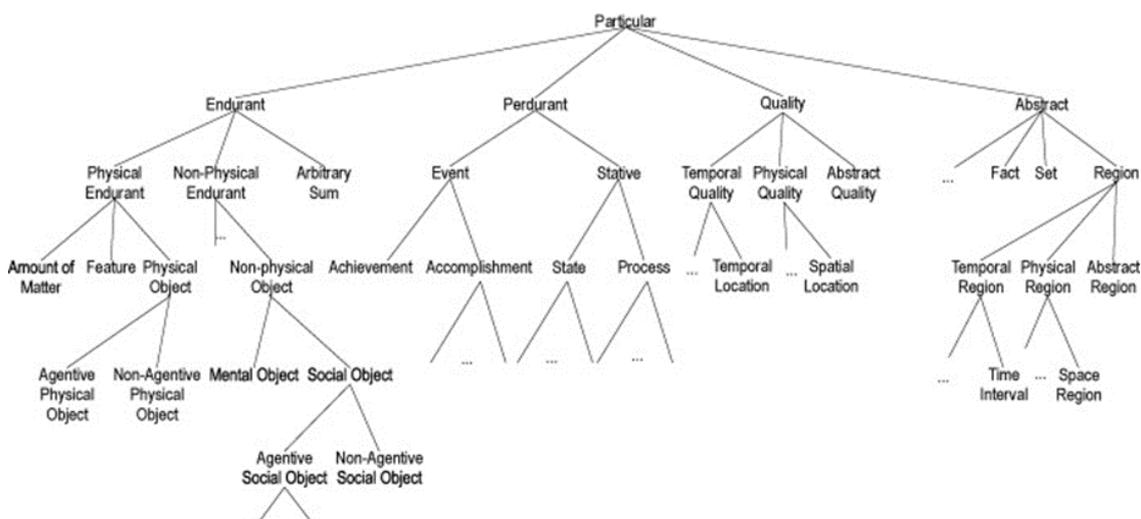

Figure 2- 10: Descriptive Ontology for Linguistic and Cognitive Engineering (DOLCE)
(*Source: Masolo et al., 2003*).

## Indigenous Knowledge on Drought ONtology (IKON)

The **I**ndigenous **K**nowledge on Drought **ON**tology (**IKON**) is part of the main contribution of this thesis, and it is a domain ontology that semantically represents the indigenous knowledge on drought based on DOLCE foundational ontology, and fully compatible with intelligent information systems also extendable for reuse. The detailed development is in Chapter Five.

## W3C Semantic Sensor Network (SSN) Ontology

The Semantic Sensor Network (SSN) ontology was developed by the W3C. It is an ontology for the formalisation, representation of sensors, their readings (observations), the methods, the features of interest, and the observed properties in a wireless network and IoT domain (Compton *et al.*, 2012). It also aligns other ontologies and standards such as OGC SensorML, SEMSOS and SWAMO (Ganzha, Paprzycki, Pawlowski, Szmeja & Wasielewska, 2016). It shares the same conceptualisation with DOLCE, which enhances perfect alignment with IKON. The SSN ontology provides a knowledge representation of the main concept of the domain, which is the **sensing device** (sensors) and models the event and temporal relationships. The sensors measure the environmental parameters and produce the measurements in real-time. However, while sensor data may be published as raw data, integrating and interpreting these data require more than just the observation results. Ontological representation of the sensors and their observations would enable the generation of deductive inference and improved reasoning capabilities (Poslad, Middleton, Chaves, Tao, Necmioglu & Bügel, 2015).

## 2.2.3. Inference Generation Systems and Reasoners

The fast development of the Internet of Things (IoT) sensors presents new challenges to Big Data platforms for performing real-time data analytics. For instance, in the environmental monitoring domain, deployed ubiquitous sensors forming Wireless Sensor Networks (WSN) generate huge streams of data that needs to be processed and analysed in real time to infer environmental events due to the time-sensitive nature of the data. Event processing of sensors data streams ensures enhanced analytic functionality, which provides a meaningful insight from IoT data and increases the productivity of processes for real-time utilisation of data (Cugola & Margara, 2012). In the domain of local IK on drought, the knowledge is in the form of indicators, rules and events, considering the practicability of implementing an inference system for this domain, where the only suitable option is using an inference engine of rule-based expert systems in performing a deductive inference based on the acquired rules. The two inference generation components for the proposed distributed middleware are presented below:

### 2.2.3.1. Stream Processing

Event Processing (EP) as an emergent research area is saddled with the goal of analysing a set of data either in batch – collected over a period of time or stream data fed to the processing engine – to extract meaningful insights, patterns and events in real-time without (the need of) committing this huge data stream to the database. This is achieved through the processing of raw data streams coming from diverse, heterogeneous data sources represented in a different data format in real-time through a processing engine based on predefined model or logic to identify likely events or future scenarios. For example, processing set soil moisture readings will automatically trigger a notification alert when it exceeds a certain threshold in real-time based on the specified limit. EP can be broadly addressed by Event Stream Processing (ESP) and Complex Event Processing (CEP) (Clemente & Lozano-Tello, 2018; Demers, Gehrke, Panda, Riedewald, Sharma & White, 2007; Flouris, Giatrakos, Deligiannakis, Garofalakis, Kamp & Mock, 2017). Irrespective of the category of the EP, EP uses time frames and use-case in the big data infrastructure to solve the problem using predictive and descriptive analytics.

Stream Processing (SP) is focused on analysing data streams from an event producer (for example, sensors) using a data analytics platform (engine and infrastructure) to detect and extract meaningful insights, patterns and events in real-time without (the need of) committing this huge data stream to the database. SP is important for real-time data analytics of continuous data streams from IoT sources (Demers *et al.*, 2007; Zhou, Simmhan & Prasanna, 2017). The huge volumes of data generated by IoT systems earned the title, 'Big Data'. These voluminous streams of sensor data are often characterised by the 5-Vs of Big Data – Volume, Variety, Value, Veracity and Velocity (Kao & Garcia-Molina, 1994). However, through efficient analysis of the diverse data from heterogeneous sources, the potential of the 5-Vs could be harnessed in providing meaningful insights for predictive analysis. This is achieved through online data stream processing, which takes into account the sensors' observations with temporal attributes in the form of time-value pairs for predicting events. This research used *Apache Kafka*; other common types of Event Stream Processing Engine are *Apache Samza*, *Apache Storm*, *Apache Flume*, *Amazon Kinesis*, and *Apache Flink*.

Complex Event Processing (CEP) on the other hand, is another side of the same coin used to analyse complex event rather than simple patterns from streams of sensor data. The capability of CEP engines over contemporary intelligent systems is the ability to carry out real-time analysis based on event pattern identification or matching from a data stream or sequence(s) of observations using initially specified models/logic. Events are triggered by multiple raw sensors data that are detected at the back-end server of the sensor-based systems. In this context, CEP is a form of stream processing technique which ingests raw data from several sensor data streams to detect various complex events through the use of declarative query language similar to SQL, called Event Processing Language (EPL). The EPL is used to continuously queries the incoming observations in real-time. The flow of unbounded data streams are aggregated in temporal bounds of data window, and the use of additional query constructs in EPL provides the ability to infer Complex Events (CE). Consecutively, the CE is identified through the occurrence of a sequence of raw observation which corresponds to a preset threshold of a sensor data. Examples include FiwareCEP (Rodriguez, Cuenca & Ortiz, 2018), KSQL, and Oracle EPL.

### Apache *Kafka*

Kafka is an open-source distributed event streaming processing engine by Apache. This streaming processing engine process sensor data streams in real-time to determine event patterns from incoming sensor's observation/readings and correlate the data with predefined/preset value threshold for prediction analysis.

The platform is similar to an enterprise messaging system based on the ability to process sensor data streams in a fault-tolerant way as they occur in a producer-publish and consumer-subscribe fashion (Figure 2-11). Apache Kafka provides real-time processing of streaming data pipelines using persistent querying systems (KSQL) without the need to commit the data stream to the database like conventional systems. This provides a huge benefit in IoT-enabled environmental monitoring systems for real-time monitoring of complex environmental phenomenon like drought.

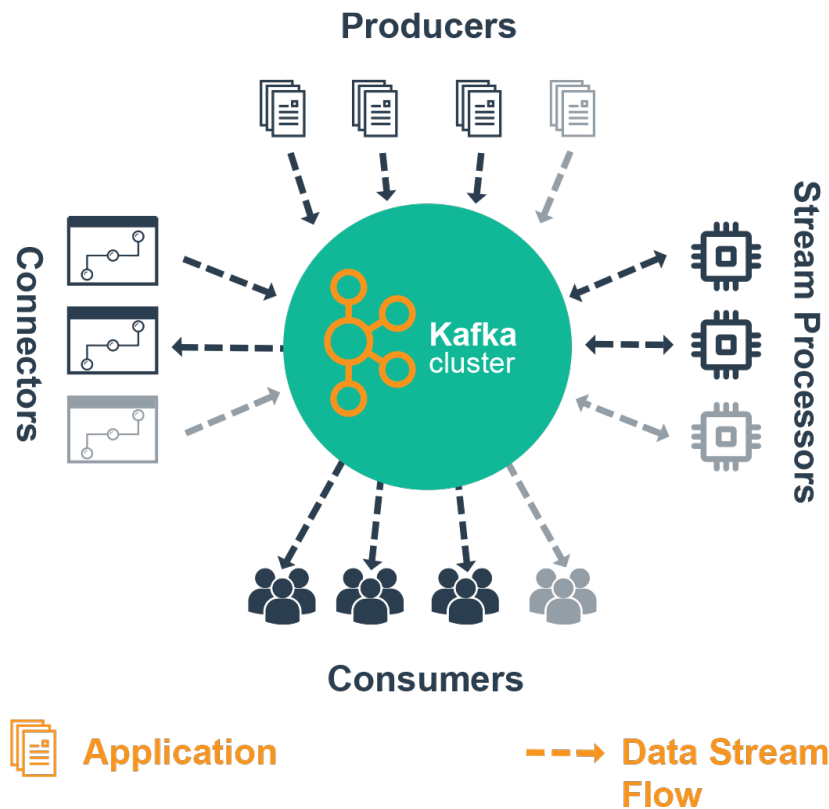

Figure 2- 11: Overview of Apache Kafka Ecosystem. (*Source: www.apache.org*)

## 2.2.3.2.    Rule-based Expert Systems (RBES)

RBES uses human expert knowledge to solve real-life challenges in a specific domain (Siler & Buckley, 2005). The domain-specific knowledge is stored in a knowledge base in the form of rules; and are usually created by the knowledge engineer in conjunction with the domain expert. Rules are expert knowledge in the form of *if-then* conditional statements. An inference engine component of the expert systems searches for a pattern in the input data that match patterns in the rule set to provide answers, predictions and suggestions in the way a human expert would. The *if* means when "*the condition is true*", the *then* means trigger a corresponding action. Hence, RBES require detailed information about the domain and the strategies for applying this information to problem-solving and generating inference.

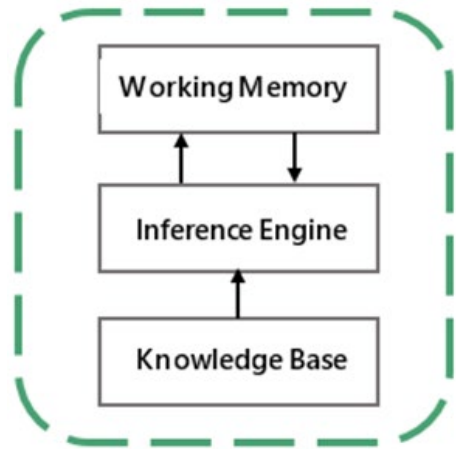

Figure 2- 12: Components of Rule-based System (*Source: Sasikumar et al., 2007*)

The knowledge of expert systems comes from experts (IK holder), and the representation of the domain knowledge is in the form of *rules*. A typical expert system consists of different sub-components. Each sub-component entails performing functionality for a specific purpose. The expert system integrates all sub-components and characterises the drought based on the knowledge base and generates inference in a form of drought forecasting information. The RBES consists of three basic components (Figure 2-12). They are:

- the rule base
- the working memory, and
- the inference engine.

a) *Rule base* (knowledge base): This is the set of rules which represent the knowledge of the domain (Sasikumar, Ramani, Raman, Anjaneyulu & Chandrasekar, 2007; Akanbi & Masinde, 2018a). The expert knowledge is represented in the form of "*if antecedents* then *consequent*". The rule base is used to generate inference from a sequence of a pattern from the input data. The general form of a rule is:

> IF    *Condition1 and*
>       *Condition2 and*
>       *Condition3*
>          *…*
> *THEN Action1, Action2, Action3….*

The conditions *Condition$(1-n)$* are known as antecedents. A rule is triggered if all antecedents (*Condition$(1-n)$*) are satisfied and consequents (*Action$(1-n)$*) are executed. However, some RBES allows the use of disjunctions such as 'OR' in the antecedents for complex scenarios before the *Action$(1-n)$* can be executed.

b) *Working memory (WM):* is typically used to store the data input or information about the particular instance of the problem or scenario. The WM is the storage medium in a rule-based system and helps the system focus its problem solving (Sasikumar *et al.*, 2007; Akanbi & Masinde, 2018a).

c) *Inference Engine:* The function of the inference engine is deriving information or generating reasoning from a given problem using the *rules* in the knowledge base. The inference engine must find the right facts, interpretations, and rules and assemble them correctly. The two basic methods for processing the *rules* are – Forward-Chaining (data-driven, antecedent-driven) and Backward-Chaining (Sasikumar *et al.*, 2007; Akanbi & Masinde, 2018a). In forward-chaining, all the facts are input to the systems and the system makes a deductive inference based on the rules available in the rule set. A system exhibits backward chaining if it tries to support a hypothesis by checking the facts in the rule base trying to prove that clauses are true in a systematic manner.

### 2.2.4. Distributed Middleware System

Middleware is a software layer composed of a set of sub-layers interposed between the application layer and the physical layer (Pietzuch & Bacon, 2002; Akanbi &

Masinde, 2015b). The whole idea of middleware is to facilitate interoperability between heterogeneous components (Pietzuch & Bacon, 2002). In distributed systems, it facilitates the integration and interoperability of heterogeneous components using a unified data pipeline eliminating data heterogeneity. One of the main challenges of developing a homogenised system with a heterogeneous component is developing a middleware between the user of the system and heterogeneous devices. Middleware ensures the ease of integrating heterogeneous devices while supporting interoperability within the diverse applications and services (Razzaque *et al.*, 2016).

The middleware for IoT acts like a bond joining heterogeneous domains of application community over heterogeneous interfaces. It also provides Application Programming Interface (API) for communication between layers or modules for easy usage and interoperability. Middleware provides seamless services and data integration for a plethora of heterogeneous devices making up the WSN to enable the various components of a WSN to communicate and manage data. Middleware supports application development, data integration, interoperability and service delivery. Middleware also enables interoperability between distributed applications that run on different platforms, by supplying services so the application can exchange data in a standardised way.

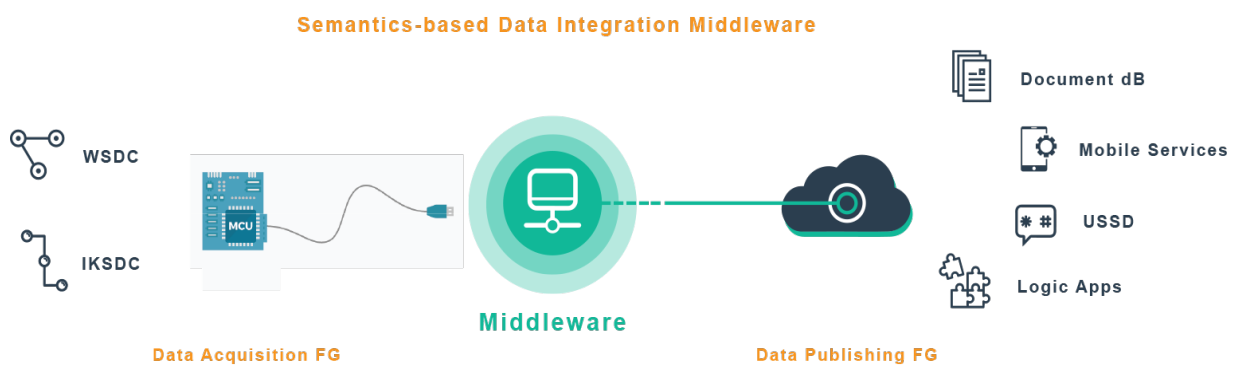

Figure 2- 13: Overview of the distributed semantics-based data integration Middleware (*Source: Author*)

Figure 2-13 depicted the Middleware structure of the proposed semantics-based data integration Middleware. This is a distributed three-tier system architecture that stretches across multiple systems or applications. Examples include

telecommunications software, messaging-and-queuing software (*Apache Kafka*), and transaction monitors. The proposed Middleware is implemented in the form of a DEWS called SB-DEWS and consist of the following sub-systems: *Data Acquisition FG*, Middleware, and the *Data Publishing FG*.

### 2.2.4.1.    Data Acquisition FG

This sub-system of the SB-DEWS performs the data acquisition for the heterogeneous data sources. The WSN measures the environmental parameters and transforms the observation into readings. Appropriate data collection instruments gather the local indigenous knowledge on drought for semi-formal representation. The heterogeneous data are transmitted to the next sub-system which is the semantic Middleware.

### 2.2.4.2.    Middleware

This sub-system is the core of the SB-DEWS. The Middleware is based on the developed proposed heterogenous data integration framework and comprises of *functional groups* such as *Data Storage FG*, *Stream Analytics FG* and *Inference Engine FG*. The Middleware sub-systems interact with the data from the *Data Acquisition FG* and publishes the output to the *Data Publishing FG* using embedded components that facilitates efficient integration of data and interoperability of services, namely: (1) interface protocols, (2) device abstraction, (3) content management, and (4) application abstraction.

a) *Interface Protocol*: The interface protocol component of the Middleware layer defines protocols for exchanging information among different networks based on different communications protocols. This component oversees providing technical interoperability. Enabling seamless connectivity using the same communication protocols ensures interoperability, for example *Apache Kafka* Connect and Sink APIs.

b) *Device Abstraction*: The device abstraction component is responsible for providing an abstract format to facilitate the interaction of application components with the heterogeneous devices. The abstraction layer ensures the integration of the devices by providing syntactic and semantic interoperability for the heterogeneous devices and communication networks using unified data pipelines. Veltman (2011) defines syntactic and semantic interoperability as follows:

- Semantic interoperability is creating a common understanding or knowledge of the various content (information) shared across the heterogeneous domain.
- Syntactic interoperability ensures the data (information) transferred by communication protocols must be represented using a well-defined syntax and encoding format such as JavaScript Object Notation.

Thus, the device abstraction provides the syntactic and semantic interoperability across the heterogeneous devices and communication networks in the domain Service Oriented Architecture (SOA) Model.

c) *Content Management*: The content management component of the middleware layer performs context-aware computation using data from various heterogeneous devices.

d) *Application Abstraction:* The application abstraction layer of the Middleware provides the interface for users to interact with devices.

### 2.2.4.3. Data Publishing FG

The output information from the Middleware is channelled to this FG for publishing and dissemination to the policymakers or system analyst for interpretation and use.

### 2.2.5. Service-Oriented Architecture

In this section, Service-Oriented Architecture (SOA) is presented, which is a software architecture used to develop the proposed distributed semantics-based data integration Middleware (SB-DIM). SOA, as a software architecture, allows functionality and is grouped around the related process and packaged as interoperable services (Nunavath, 2017). The basic principles of SOA are to achieve loose coupling among interacting and interconnected heterogeneous software components, *functional groups* (FG) or clusters within a distributed environment. SOA essentially allows the collection of services that communicates with each other using a unified data pipeline. Each service is a well-contained process that does not depend on the context or state of other services, allowing it to be independent of each other with the ability to function as a standalone application. To achieve a common task, services communicate with each other, requesting for input and output data in an orchestrated manner (Krafzig, Banke & Slama, 2005). Figure 2-14 presents the layered structure of SOA. The advantages of SOA are; it promotes scalability of individual component or FG and allows interaction between all interconnected components.

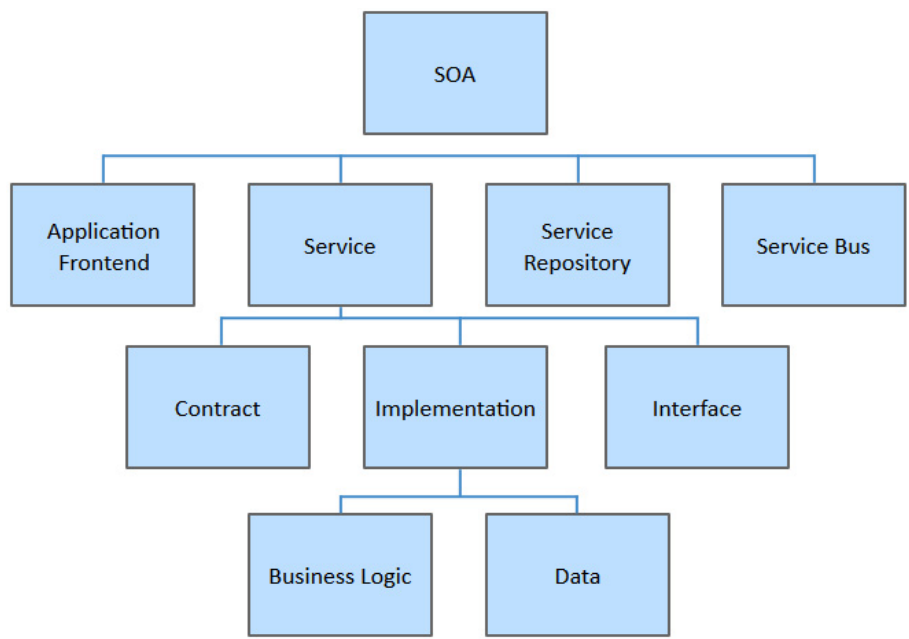

Figure 2- 14: Elements of SOA (*Source: Krafzig, Banke & Slama, 2005*).

## 2.3. Related Works

This section presents the existing efforts towards achieving heterogeneous data integration using standards or technologies related to environmental monitoring domain. In recent years, the amount of data such as computerised data and information available on the Web has spiralled out of control. Many different models and formats are being used that are incompatible with each other. Traditionally, approved standards are recommended to address interoperability (Llaves & Kuhn, 2014). Several standards have been created to cope with the data heterogeneities. Examples are data exchange such as EDXL Distribution Element (EDXL-DE), the Emergency Data Exchange Language Situation Reporting (EDXL-SitRep), the Customer Information Quality (CIQ), and National Information Exchange Model (NIEM). However, these standards provide data to a predefined application in a standardised format only and hence do not generally solve data heterogeneity.

Some of the mentions standards were developed by the Organisation for the Advancement of Structured Information Standards (OASIS) as a standard for representing and reporting an emergency in information systems. The current short-coming of these standards is incompatibility with other information systems. In the environmental monitoring domain, there are various geospatial standards such as Geography Markup Language (GML) standard for the representation and exchange of geographical information (OpenGeospatial, 2016). However, the background check of related works has shown existing standards do not solve the challenges of heterogeneous data integration in the environmental monitoring domain.

Furthermore, research efforts such as Masinde (2015) has primarily intend to utilise data from heterogeneous sources for forecasting and predicting drought. Also, for a more accurate drought prediction, Omidvar and Tahroodi (2019), recently propose a time-series modelling of precipitation data recorded from varieties of stations. Using the precipitation trends, the severity of the drought in the region are determined. The results of the model have acceptable accuracy in predicting annual precipitation.

## 2.4. Summary

This chapter presents the necessary technological background for addressing drought forecasting using heterogeneous data sources. The concept of drought, drought management, drought prediction models and indices were presented. Later, the representation of local indigenous knowledge and WSN data using semantic technology were discussed. Furthermore, the technologies that ensure automated inference generation from the unstructured indigenous knowledge and structured WSN data was described. In this research, the researcher selected the mediator-based data integration approach based on a SOA that allows loose coupling of services for achieving a common task. Lastly, some existing data integration standards and related works were reviewed as the background for the proposed solutions towards the integration of heterogeneous data sources in fulfilment of the research objectives listed in Chapter One.

# CHAPTER THREE

# RESEARCH DESIGN AND METHODOLOGY

## 3.1.    Introduction

This chapter focuses on describing the research design and methodology in detail. Firstly, the philosophical paradigms in which the methodology is grounded are discussed. It presents the framework design and methods adopted in this research; it includes research design type, data types, data collection, data pre-processing, and ethical considerations.

According to Murton (1998), a research design is the blueprint of a research project and provides the guideline for the execution of the design in a stepwise manner. Welman, Kruger and Mitchell (2005) defines a methodology as a system of methods, principles, and rules that govern a field of study. The methodology is the construction process using available methods and tools towards achieving the objective of the research (Ponterotto 2005; Cothran 2011; Houghton, Hunter & Meskell 2012; Creswell 2012). The research design to follow and the methodology of the research is chosen to support the outcome and importance of the result. Therefore, for every research, the underlying research design and research methodology of the research paradigm context needs to be discussed.

Initiation of research is often to find a solution – or a better solution than exists – to a problem or to contribute a novel idea or an invention. As mentioned in Chapter One, this thesis proffers a solution to the problem of lack of heterogeneous data integration and interoperability in the environmental monitoring domain.

In summary, in this chapter, the research design executed is, therefore, reported and the distributed semantic middleware framework is presented. This chapter is organised into eight (8) sections. Section 3.1 covers the introductory aspect of the research design and methodology; section 3.2 presents the research design. Section 3.3 focuses on data collection and analysis methods. Section 3.4 presents the semantic middleware data integration framework and its distributed *functional groups* (FG). The experimentation process is presented in section 3.5 and evaluation procedure in section 3.6. The ethical considerations are presented in section 3.7, and section 3.8 presents the summary of the chapter.

## 3.2. Research Design

Literature has shown that there are various methods and means by which to achieve the aim and objectives of the research. Straub, Gefen and Boudreau (2004) argued, however, that two principal forms of research are exploratory and confirmatory research. Exploratory research is appropriate for research projects with high levels of uncertainty (van Wyk, 2012). On the other hand, confirmatory research is used to test *a priori* alternative hypotheses about a subject of discourse, followed by the development of a research design to test and validate those hypotheses, the gathering of the data, data analysis and generation of deductive inference from the research (Jaeger & Halliday, 1998).

### 3.2.1. Qualitative vs Quantitative Techniques

This research is based on mixed research design where qualitative and quantitative techniques (Jaeger & Halliday, 1998) towards achieving the objectives were employed. A qualitative approach was used to gain a detailed understanding and opinion on the use of local IK on drought for drought prediction and forecasting, using unstructured or semi-structured data collection methods. The quantitative approach tested the hypothesis (see Section 1.2), examined the cause and effect and made predictions from it. Hence, formulating a research design for this research is important.

In 1999, Burstein and Gregor proposed action-based research design for system development in the field of information systems (IS). Multi-Methodological defines this action-based approach research cycle, which links conceptual and applied research approaches. The methodology involves three main steps: theory building, systems development, and the use of observation and/or experimentation for research evaluation. The first step is the theory- building or model-building studies, which involve the design of the conceptual framework for systems based on the research paradigm. The second step is the system development, which is based on the conceptual model to develop a prototype system for solving the IS problem. The last step is the use of observation and/or experimentation for research evaluation; this comprises five (5) distinct components, namely: significance, internal validity, external validity, objectivity/confirmability, and reliability/dependability /audibility. This criterion set is used to evaluate whether the proposed system successfully met the research objective and goals.

On the other hand, there are two spheres of research design (van Wyk, 2012), namely: (1) generating primary data – for example surveys, experiments, case studies, evaluation, ethnographic studies; and (2) analysing existing data – for example text data – content analysis, historical studies, or – numeric data – data analysis, statistical modelling.

Therefore, based on the objectives described in Chapter One, experimental and case study research design approaches were selected, which would involve the gathering of primary data, developing the middleware prototype, implementation and evaluation of the system. This is due to the context of the domain of sensor networks and the unstructured indigenous knowledge data collection. An experimental design is focused on constructing research with a high degree of validity. However, randomised experimental designs provide the highest levels of causal validity, which is important in validating sensor data readings used in this research (Mitchell, 2015). The case study design was applied to the validation of the research hypothesis.

### 3.2.2. Research Philosophy

In addition to being quantitative or qualitative, all research is executed either from a researcher's stance or philosophical, based on aspects such as truth and validity, and that determines acceptable research methods to be adopted (Derose, 2004; Myers, 1997; van der Merwe, Kotze & Cronje, 2004). According to Guba (1990), there is a need to comprehensively specify research design based on research philosophy, which is comprised of five choices on how to execute the research: 1) Ontology, 2) Epistemology, 3) Methodology, 4) Techniques (data gathering), and 5) Data Analysis Approaches. The terms and the relationship between them are graphically depicted in Figure 3-1.

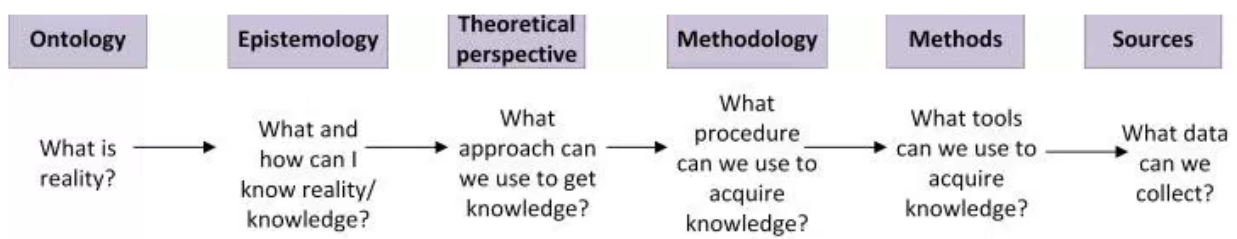

Figure 3- 1: Research design steps based on research philosophy (*Source: Guba, 1990*).

### 3.2.2.1.    Ontology

In this research, the ontological assumption is towards having an intricate understanding of the indigenous knowledge system on drought, identifying the meaningful indicators behind this century-old system and how this knowledge base can be integrated with modern knowledge for a more accurate drought forecasting system. Ontology (nature of reality) is the starting point of all research after which is the epistemological stance, and methodological positions logically follow (Denzin & Lincoln, 2011). For research in information systems, Myers (1997), outlines three paradigms, namely positivist, interpretive or critical. This research takes a positivist view, and from a positivist point of view has the notion that "*Truth*" exists and can be apprehended and measured. This research subscribed to a positivist view that the IK thus exists and can be documented for knowledge representation.

### 3.2.2.2.    Epistemology

Several epistemological stances are documented in the literature. According to Guba & Lincoln (1994), Bryman (2008), and Kovach (2010), epistemology is the branch of philosophy that deals with the origins, nature, methods and limits of knowledge. The researcher and the research paradigm are disconnected and independent of each other for the in-depth and unbiased study of the knowledge (Chalmers, 2002). This would ensure the attempts to distinguish between "what is true" in the knowledge and "what is false" are not influenced.

### 3.2.2.3.    Methodological

Methodological assumptions are the methods used which include experiments, content analysis, grounded theory, explanatory research, hypothesis-testing techniques and case study (Saunders, Lewis & Thornhill, 2007). Hence, there is no single 'right' way to undertake research.

### 3.2.3. System Development Methodologies

This section presents appropriate software systems methodologies adopted for developing various system modules in the overall systems design. The overall system (semantic middleware) incorporated several distributed *modules or* *functional groups* performing a different task using the same set of data, but semantically orchestrated to achieve the goal of effective data integration and systems interoperability.

### 3.2.4. Experimental Design

According to Teddlie & Tashakkori (2009), - "*an experiment is a blueprint of the procedure that enables the researcher to test his hypothesis by reaching valid conclusions about relationships between independent and dependent variables. It refers to the conceptual framework within which the experiment is conducted.*" The experimental design would allow the rigorous testing of the research hypothesis by reaching valid conclusions.

### 3.3. Data Collection and Analysis Methods

Data collection can be defined as the process of collecting information from all the relevant sources towards finding acceptable answers to the research problem, in an established systematic fashion to test the hypothesis and evaluate the outcomes (Gill, Stewart, Treasure & Chadwick, 2008). Extensive data collection improves the quality of data used for the data analysis and ensures the validity and reliability of research results (Cohen, Manion, & Morrison, 2013; Gill *et al.*, 2008). Data collection methods can be divided into two categories: primary methods of data collection and secondary methods of data collection.

According to Cohen *et al.* (2013), primary data collection methods can be divided into two groups: quantitative and qualitative. The quantitative data collection and analysis methods include interviews, closed-ended questionnaires, with methods of correlation and regression, mean, mode and median, and others. Qualitative research methods, on the other hand, aim to ensure a greater level of depth of understanding with data collection methods such as open-ended questionnaires, focus groups, observation and case studies. Qualitative research methods allow a better understanding of the scenarios, by providing details insights supported by data which are rich and holistic. Secondary data are readily available data already published in books, journals and online portals. The use of an appropriate set of criteria to select secondary data is crucial regarding increasing the levels of research validity and reliability.

Two forms of data were collected for this research — the sensor readings data from the wireless sensor networks and the local indigenous knowledge on drought. Hence, the research utilised the two data collection categories – primary and secondary – as necessary. The qualitative approach of primary data collection was adopted for the use of IK on drought. The quantitative methods were used for investigating the appropriate knowledge representation of the IK domain and the semantic integration with outputs from appropriate drought indices to predict drought.

### 3.3.1. Data Types

This research incorporates heterogeneous data for drought prediction. This data comes from two different domains – wireless sensor data and indigenous knowledge. The data for the wireless sensor network is structured and represented in data representation formats such as XML and JSON. On the other hand, IK is mostly unstructured data, available in the oral format. This type of data needed to be captured, documented, and represented in a form that can be used for knowledge representation, modelling and processing.

### 3.3.2. Data Sources

There are two heterogeneous data sources derived from the domain in this research study. The first domain (**D1**) is local indigenous knowledge on drought. The data obtained from this domain provides information on IK on drought, which is limited and varies from one geographic region to another. An indigenous community in a geographic area develops this knowledge system over the years and it is traditionally transmitted and shared orally across and within generations; it includes skills, technologies, practices and beliefs on the natural environment (World Bank, 2004). The data collection process involves the use of both primary and secondary data collections. The primary data collection involves the use of a participatory research approach involving interactive research methods such as in-depth interviews, questionnaire-based interviews, case studies, focus group discussions and participant observations. The secondary data involves the use of data and information available for the area under study in the literature. The data is categorised based on the scope of meteorological, astronomical, behaviour or living things (plants, such as flowers and trees.; animals, such as birds and insects), knowledge of seasons, and mythical beliefs.

The second data source is the sensor data (**D2**), which is obtained from deployed WSN in the area under study. This data is collected from the sensors that monitor various environmental parameters such as precipitation, soil moisture, temperature and humidity. The experimental prototype of the sensor networks provides data that would be used in the drought analysis model and integrated with local IK for more accurate drought forecasting systems. This involves primary data collection with the intention to obtain accurate readings, backed up with scientific validation.

### 3.3.2.1.    Pilot Study

A small-scale pilot study was conducted as a preliminary study to evaluate the feasibility, performance and effectiveness of the research study data collection tools and the research design. A selected domain expert in the area under study was recruited for the pilot study. An initial test questionnaire (Appendix A) was developed by the researcher with the help of the supervisor to compressively capture the demographics of respondents, knowledge of seasons, the indigenous knowledge locally indicated (astronomical, meteorological), the implication of event occurrences and behaviours based on the seasonal patterns. The test questionnaire was administered to the selected domain expert to provide feedback on the ease of use and practicability. The feedback received was beneficial and helped in the reformulation of the test questionnaire for the main study questionnaire.

### 3.3.2.2.    Use of Case Study

The use of a case-study provides an in-depth investigation of the intricate complexities of using local indigenous knowledge for forecasting and predicting a complex environmental phenomenon such as drought. This is possible through the use of a focus group, which are selected domain experts providing expert analysis and interpretation of environmental occurrence using indigenous knowledge.

The data collection and analysis method used for the primary data collection was to generate suitable data from respondents. The generated data in the form of local indigenous indicators on drought, relationships between indicators, the occurrence of ecological interactions with events and the expected weather outcomes were vetted and, verified by the focus group. The data collection tools used were a questionnaire and a developed Android application.

### 3.3.3. Target Population

This study took place at Swayimane, KwaZulu Natal province of South Africa and Mbeere district in Kenya. The data for **D1** (local indigenous knowledge on drought) were obtained in the two study areas for extensive qualitative data. The participants selected were local farmers and IK experts. The data for **D2** - WSN and weather station data was obtained from deployed sensors and installed weather stations in Swayimane, KwaZulu-Natal and Mbeere district in Kenya. This study took place in Swayimane from September 2017 to May 2018, and in Mbeere district from March 2018 to April 2018.

### 3.3.4. Sampling Techniques

Sampling is a statistical procedure in which a predetermined number of observations are taken from a larger population (Altmann, 1974). This research used a purposeful sampling technique (Patton, 2002) to select the indigenous knowledge domain experts (**DE**), which are mostly traditional farmers in Swayimane, KwaZulu Natal and Mbeere, Kenya. The selected farmers have relied on the use of their local IK for drought forecasts, weather predictions and farming-related decisions for generations. The selected respondents showed willingness and availability to participate in this research study. The data was collected through the use of questionnaires (see Appendix A), structured interviews, focus group meetings and ODK survey mobile application.

### 3.3.4.1.    Questionnaire

The survey's use of questionnaire was to measure the level of indigenous knowledge on drought application in the study area (Appendix – A). The questionnaire was used to gather each respondent's background information relevant to the context of the research. Also gathered was local indigenous knowledge on drought indicators such as the meteorological indicators, astronomical indicators, knowledge of seasons, ecological interaction of behaviours of birds, and insects and flowering and non-flowering plants based on seasonal patterns used by the local community in their IKS to predict and forecast drought and other environmental phenomena. The IK indicators collected and gathered from the respondents were summarised for further verification and detailed interpretation by the focus groups.

The questionnaire included 32 questions related to meteorological, astronomical, behavioural properties of local indicators, weather and climatic knowledge on drought. The questionnaire consisted of the following sections:

a) The first section of the questionnaire collected the biographical data of the respondents.
b) The second section is aimed to acquire the respondent's knowledge of weather forecasting and the area's indigenous knowledge system.
c) The third section aimed to gather and document the effectiveness and use of local indigenous knowledge for weather forecasting and cropping decisions.
d) The fourth section was aimed identifying and documenting the unstructured weather indicators for drought based on the categories such as knowledge of seasons, astronomical, and animal/plant behaviours with practical examples.

### 3.3.4.2. Survey Mobile Application

The adoption of mobile technology has tremendously improved the rate of data collection and gathering collation, and also helps remove ambiguities in responses. This research leverages on the benefit of mobile application through the use of Open Data Kit (ODK) – A mobile application coded for remote data collection and collation of the data in real time. The application is an android platform dependent on user-friendly Graphical User Interface (GUI) (Figure 3-2). The application is used to collect responses from text to pictures to location based on the questionnaire coded in the form of XML and support complex workflows via JavaScript customisation. It also supports complex branching, answer validation, multiple languages, and offline work. The data is uploaded to the database in the server in real time. This research adopts the use of Google Sheets as the database.

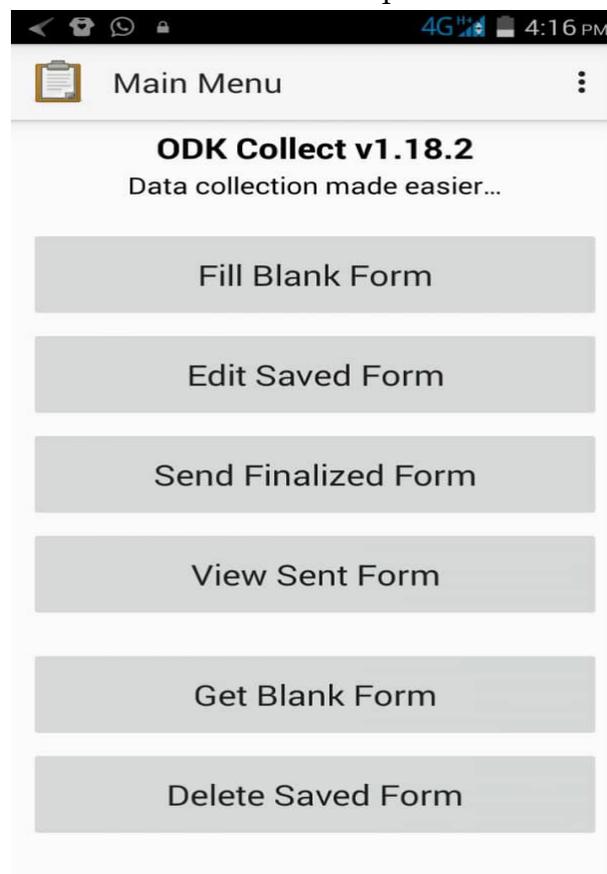

Figure 3- 2: Open Data Kit Collect GUI (*Source: ODK App*)

The ODK consist of a suite of tools for data/knowledge gathering and collection using mobile devices and data submission to an online server or phone cache. It consists of the frontend and the back-end to collect, use, and manage data. The front end is the ODK *Collect* open-source Android application; and the back-end is an online server or phone cache for offline saving. This research adopts the use of Google's free powerful hosting platform and services for back-end services. Google Sheets is used as the online database, and the data saved can be visualised on a map using Google Fusion Tables and Google Earth in real time. Fusion Tables is integrated with Google Sheets with some built-in geocoding functionality that allowed a seamless data analysis.

### 3.3.4.3. Focus Groups

The focus groups are selected IK experts used for verifying the authenticity and validating the obtained IK in the study areas through the sampling methods. The group consisted of five (5) elderly tribal farmers who are well-knowledgeable and expert in the use and application of IK in the study areas.

### 3.3.5. Data Analysis and Interpretation

Based on the research design, the target group consisted of indigenous knowledge domain experts (**DE**) and local farmers. The data analysis involved identifying the natural ecological indicators, ecological interaction scenarios, and interpretation of the scenarios in the form of *rules*. The ecological interaction of one or more natural indicators in a particular season is called an "*event*". These *event*(s) hold the clue to understanding an environmental phenomenon such as drought. The natural indicators with its corresponding *events* are gathered through the use of survey instrument of questionnaires, interviews, mobile applications and focus groups. The essence is to explicitly understand the IK domain for accurate knowledge representation. Hence, in order to develop an accurate knowledge representation of the domain, it is necessary to analyse data quantitatively. A section of the questionnaire (Appendix A) was designed to identify the natural indicators and events for this analysis. The two phases necessary for accurate data analysis and interpretation are data pre-processing and reliability.

### 3.3.5.1. Data Pre-processing

For this phase, data pre-processing of the responses gained through all forms of survey data collection instruments was undertaken to establish a reliable and useful information for accomplishing the research objectives. The research study adopted a mixed methodology; the data analysis process required different methods (Edmonds & Kennedy, 2012).

For **D1,** the collected data was collated using Statistical Package for Social Sciences (SPSS) software and Google Fusion Tables, for qualitative analysis, generation of descriptive statistics from the responses, and data visualisation. The data was analysed to identify the key natural indicators and to further understand the occurrence of *events* – astronomical *events*, meteorological *events* etc., albeit based on the period in the seasons – summer, autumn, winter and spring. The responses from all respondents were documented, digitalised and summarised, based on the section of the questionnaires towards providing answers to specific research objectives.

a) Study area and respondent's demographic information: This provides an understanding of the study area, the name of the village, the primary occupation of the respondents, age bracket, length of stay in the community. Analysis of this data category provides statistical data about the characteristics of a population, such as the age, gender, occupation and income of the respondents This information was necessary to understand the respondents' background, history of the use of IKS for drought forecasting and cropping decisions.

b) Respondent's knowledge on weather forecasting and prediction: Analysis of this category provides an understanding of IK by the respondent, the ways it is used in their daily activities, and most importantly for weather forecasts.

c) Types of weather forecasting used by the respondents: The interest here was to determine the frequency of use of IK for weather forecasts and it is used for cropping decisions. This analysis also provides an overview of sources of IK with an attributed confidence level of the sources;

d) Indigenous knowledge indicators: The analysis of this category provided a detailed list of the natural indicators of local indigenous knowledge on drought used in the study area. The indicators are categorised as astronomical indicators, meteorological indicators, behaviours of living things, the behaviour of non-living things etc.;

e) Indigenous knowledge events occurrences based on different seasonal patterns: The interpretation of this event provides an inference to likely weather outcomes, which help determine the level of correlation between the entire IKS of the area under study and the weather outcomes.

For **D2,** the sensors readings generated by different sensors (event producers) in the WSN are in structured formats, streamed wirelessly to the cloud repository for further processing in real-time. The sensor readings can be in various format and types. The pre-processing of streams of sensor readings performed in the cloud includes the average, median calculation as well as processing such as pattern matching and event forecasting and predictions.

### 3.3.5.2.    Reliability and Validity

Reliability and validity remain appropriate concepts for attaining rigour in qualitative and quantitative research (Morse, Barrett, Mayan, Olson, & Spiers, 2002; Guba and Lincoln, 1981)). This research ensures the accurate and truthful documentation of the local indigenous knowledge on drought, and, on the other hand, ensures the prevention of data delay and data denial and uncompromised integrity of sensors data. Opinion differs in the literature on the procedure to determine the validity of a research study. Wolcott (1994) stated that there is no distinction between procedures that determine validity during the course of a research study.

The calibration and validation of the instruments used are important in this research study. Drost (2011) stated that "*validity is the extent to which a research instrument reflects reality.*" The accuracy of the measurement would consecutively determine the truthfulness of the results. All data collection instruments were validated for reliability to remove errors. However, over the years, reliability and validity have been subtly substituted with criteria and standards.

### 3.3.6. Error Analysis

The basic principles for calibration of environmental monitoring sensors involve the use of a comparison method (Grykałowska, Kowal, & Szmyrka-Grzebyk, 2015). This principle is applied to all the sensors used in the experimental and field study of this research. There are two types of errors associated with an experimental research study: the "*precision*" and the "*accuracy*". According to Pugh and Winslow (1966) "*The word precision will be related to the random error distribution associated with a particular experiment or even with a particular type of experiment. Accuracy shall be related to the existence of systematic errors — differences between measurements.*" In this research, study effort was put in place to minimise errors of accuracy through calibration and determining the uncertainty of sensor measurement.

### 3.3.7. Data Collection Techniques

The data collection techniques for the pilot and case studies are based on the sub-framework of the semantics-based data integration framework. Both the structured and unstructured data sources are collected using the proposed data collection framework.

## 3.4.    Study Areas

### 3.4.1. KwaZulu-Natal

The Swayimane community – used as the case study – is located in the KwaZulu-Natal province, South Africa. KwaZulu-Natal (See Figure 3-3) is South Africa's third-smallest province with a total size of 92,100 km² in area. The province has two mountainous areas, the western Drakensberg Mountains and northern Lebombo Mountains. Tugela is the province's largest river and flows west to east across the centre of the province. The climate of the coastal regions is subtropical with the inland area becoming increasingly colder and summer temperature rising over 31ºC. KwaZulu-Natal is rich in biodiversity ranging from flora and fauna. The iSimangaliso Wetland Park and uKhahlamba Drakensberg Park host seasonal migratory species which provide a rich, in-depth avenue to study the biodiversity interactions.  The seasons are as follows: Summer: November – March; Autumn:

April – May; Winter: June – August; and Spring: September – October (Gouse, Pray, Schimmelpfennig & Kirsten, 2006). The average daytime temperature from January to March is 28°C and 23 °C from June to August with a minimum of 11 °C.

The KwaZulu-Natal Province is divided into eleven (11) municipalities – one (1) metropolitan municipality and ten (10) district municipalities, namely: eThekwini Metropolitan Municipality; Amajuba District, Zululand District, uMkhanyakude District, uThungulu District, uMzinyathi District, Uthukela District, uMgungundlovu District, iLembe District, Ugu District and Harry Gwala District municipality. The district municipalities have 48 local municipalities. The data collection took place in Swayimane village, which is located in the uMngeni local municipality of uMgungundlovu district of KwaZulu-Natal. The inhabitants are mostly Zulu by tribe with farming and livestock keeping the primary occupation of the study area. Swayimane terrain has undulating outcropping hills with an extensive altitudinal range of 2900m which influences the temperature changes in summer and winter (Ndlela, 2015).

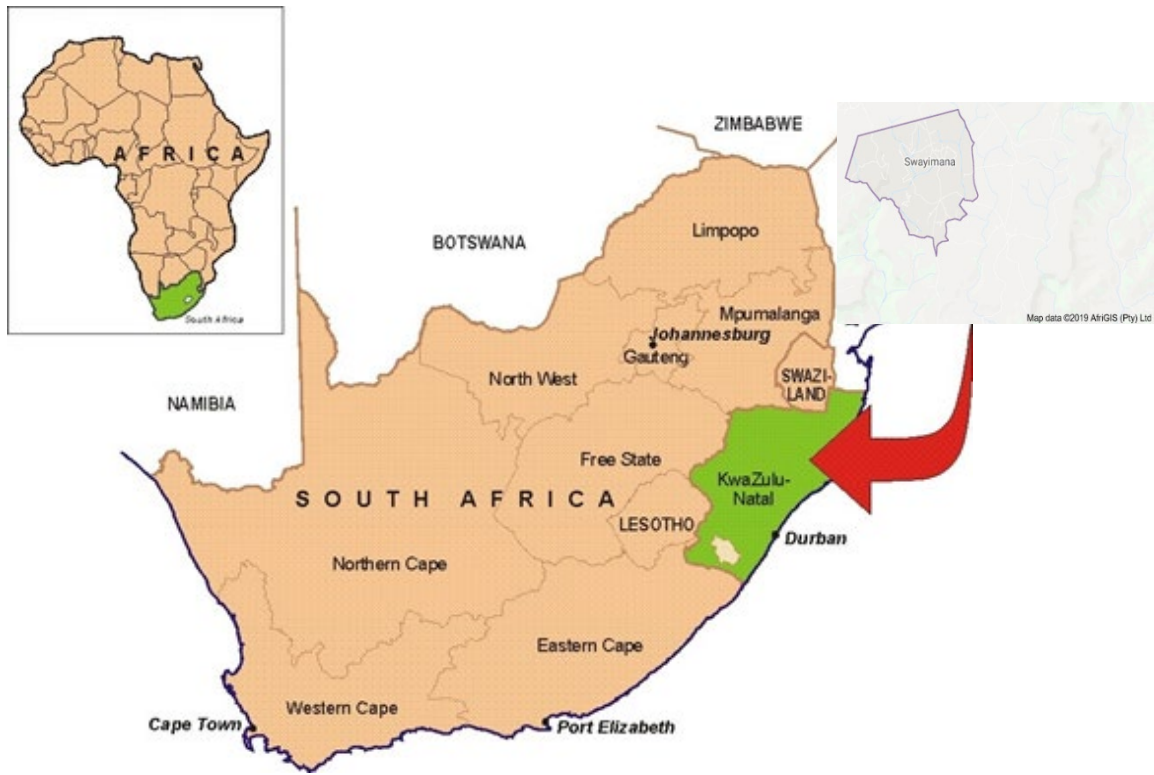

Figure 3- 3: Map of KwaZulu-Natal Province, South Africa and Swayimane. (*Source: Republic of South Africa, 2010*)

### 3.4.2. Mbeere District

Mbeere community is in Embu County in the Eastern province of Kenya (Republic of Kenya, 2001). Geographically, the Mbeere District lies between latitude 0° 20' and 0° 50' South and longitude 37°16' and 37°56' East, covering an area of 2,097 square kilometres (see Figure 3-4). Ambeeres/Mbeeres are predominantly farmers that specialise in growing a variety of crops such as melons, sorghum, maize, mangoes, pawpaws, millet, cowpeas, beans. (Kinuthia, Warui, & Karqanja, 2009).

The terrain is arid and classified as an Arid and Semi-Arid Lands (ASALs). The temperature varies from 20ºC to 32ºC due to several environmental factors and climatic conditions. The farmers have developed and use their indigenous knowledge systems based on local indicators and knowledge of seasons for the farming decision-making process and for predicting and forecasting environmental phenomena such as drought. Mbeere district experiences two main raining seasons: the March-April-May (MAM) long rains and the October-November-December (OND) short rains (Masinde, 2015).

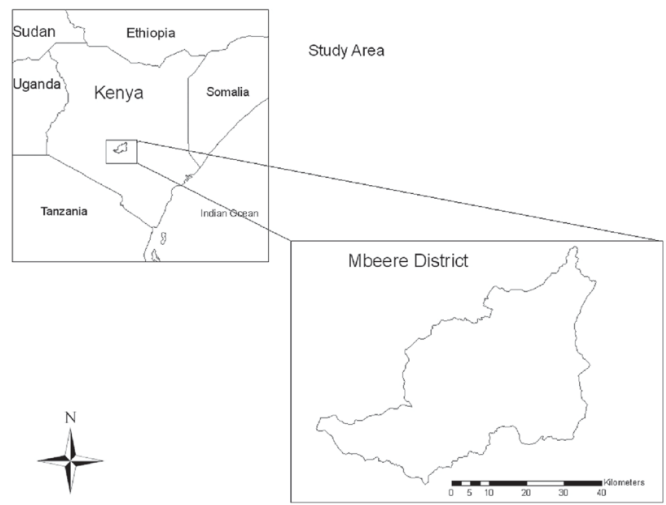

Figure 3- 4: Map of Kenya showing relative location and size of Mbeere district. (*Source: Republic of Kenya, 2001*).

Mbeere district has bimodal rainfall with annual averages of between 640 and 1110mm (Republic of Kenya, 2001). However, some parts receive less than 500mm per annum (Kinuthia *et al*., 2009). The erratic and irregular rainfall coupled with high temperature, make the district experience high evapotranspiration throughout the year (Kinuthia *et al.,* 2009).

## 3.5.    Semantics-based Data Integration Middleware Framework

This section presents a framework of a distributed semantic heterogeneous data integration middleware. The middleware aims to be implemented as Semantics-based Drought Early Warning System (SB-DEWS) that will enable semantic integration of heterogeneous data sources for drought forecasting and prediction in the study area. The system utilises local indigenous knowledge on droughts and data from wireless sensor network with weather station readings to generate deductive inference for drought forecasting and predictions. The semantic knowledge representation of local indigenous knowledge on drought and environmental readings will promote reuse of data, allow seamless integration and interoperability with intelligent information systems (Kuhn, 2005; Fogwill, Alberts & Keet, 2012; Akanbi & Masinde, 2015b).

The proposed middleware in the form of drought early warning systems semantically integrates the modern science with local indigenous knowledge using a middleware. This is important due to the complexity of environmental phenomena such as drought which necessitate the consideration of the localisation and variability of the environmental parameters of the area under study. The semantic-based data integration middleware (SM-DIM) framework provides a blueprint of the SB-DEWS. The middleware is a layered service-oriented architecture (SOA) which encompasses several distributed *functional groups* frameworks.

### 3.5.1. Framework Requirements

Based on the problem statements that motivated this research study and the research questions described in Chapter One, the chapter presents the framework requirements. The requirement is the criterion that project deliverables need to satisfy and verify how well the deliverable functions against the requirements. In this section, the essential basic requirements of the proposed framework that applies to solve these problems were elicited.

The system requirements are divided into two categories – functional requirements and non-functional requirements. The functional requirements (FR) describe what the framework should do, and the non-functional requirements (NFR) describe the properties of the framework (Rainardi, 2008; Nunavath, 2017).

#### Functional Requirements

- **FR1:** Due to drought complexity, accurate forecasting and prediction involve combining data from diverse sources. This heterogeneous data is often represented in abstruse terms, using different vocabulary and data representation format that causes data heterogeneity. This prevents seamless data exchange which impinges onto achieving interoperability. An introduction to the research problem indicates knowledge integration is limited by ontological divergence, and this could be solved by increasing the level of semantic expressivity. Therefore, the framework should provide a formal description and common understanding of the domain's concepts, relationships, constraints to eliminate semantic ambiguity based on a common ontology.

- **FR2:** The integration of data and interoperability of different systems is essential for an accurate information system. The framework should facilitate the semantic integration of data, data reuse, and exchange between various heterogeneous systems in an event-driven way using several clusters of functional groups.

- **FR3:** The framework should ensure the gathering and processing of the data, either structured data or unstructured in a timely event fashion.

- **FR4:** The middleware should be able to generate accurate deductive inference from the semantic integration of heterogeneous data sources for the area under study. The framework shall ensure the use of automated reasoning modules which infer events patterns and perform deductive inferences based on a set of syntactic derivation rules from indigenous knowledge and drought prediction model logic.

- **FR5:** The middleware framework must include a publishing system for publishing drought forecasting warnings in the form of drought forecasting advisory information (DFAI) across multiple channels for use by policymakers.

**Non-Functional Requirements**

- **NFR1:** The framework shall be flexible, distributed, offer reusability and extendable.

- **NFR2:** The framework shall be platform-independent and facilitate unified data communication via standard APIs.

### 3.5.2. The Middleware Framework Overview and Description

Integration and interoperability of heterogeneous data sources and systems respectively are critical in making efficient decisions and determining the accuracy of any EWS (Leonard, Johnston, Paton, Christianson, Becker, & Keys, 2008). However, due to the heterogeneity of data and information systems, it is quite difficult and challenging. This affects seamless data sharing and communication. Therefore, to have a common agreement in the terminologies and relationship between entities in different domains, the study has looked into the literature and found that the most suitable method is the adoption of ontology and semantic technologies (Llaves & Kuhn, 2014; Kuhn, 2005, Fogwill et al., 2012). Semantic technologies have a stronger approach to interoperability than contemporary standard-based approaches through detailed semantic referencing of metadata (Kuhn, 2005). Hence to address the requirements listed above in the development of an accurate EWS for drought forecasting, this middleware framework is based on the architecture proposed by Akanbi and Masinde (2018b).

The main fundamental characteristic of the presented semantics-based middleware framework is the ability to integrate both structured (sensors data) and unstructured data (indigenous knowledge). The study used ontology-based semantic annotation to deal with the integration and interoperability of heterogeneous data sources, and an automated reasoning system for the generation of accurate inference. The middleware is novel and revolutionary; it semantically integrates diverse legacy systems and diverse data sources like sensory data, weather station data and the local indigenous knowledge on drought by solving the semantic heterogeneity problem.

The presented framework provides the solutions to **FR1** and **FR2,** which is a semantic model that will facilitate the semantic integration and interoperability of systems. The semantic model will integrate different heterogeneous data sources (**FR3**); generate deductive inference from the semantic integration of data sources using automated systems – inference engines and CEP engines (**FR4**) and disseminate the output in the form of DFAI through various channels (**FR5**). The SB-DIM framework aims at improving the semantic interoperability among intelligent early warning systems (EWS) and their components.

A distributed layered SOA was adopted in which each layer consists of components (functional groups). Each *functional groups* (FG) consists of several modules that offer a high level of abstraction and functionalities suitable for each level (Akanbi & Masinde, 2018b). The middleware layer provides API for the communication and abstraction of complex modules and presenting the data in a machine-readable format for integration and interoperability (Akanbi, Agunbiade, Dehinbo & Kuti, 2014). The framework architecture is depicted in Figure 3-5. The framework consists of five *functional groups* (FG): *Data Acquisition FG*, *Data*

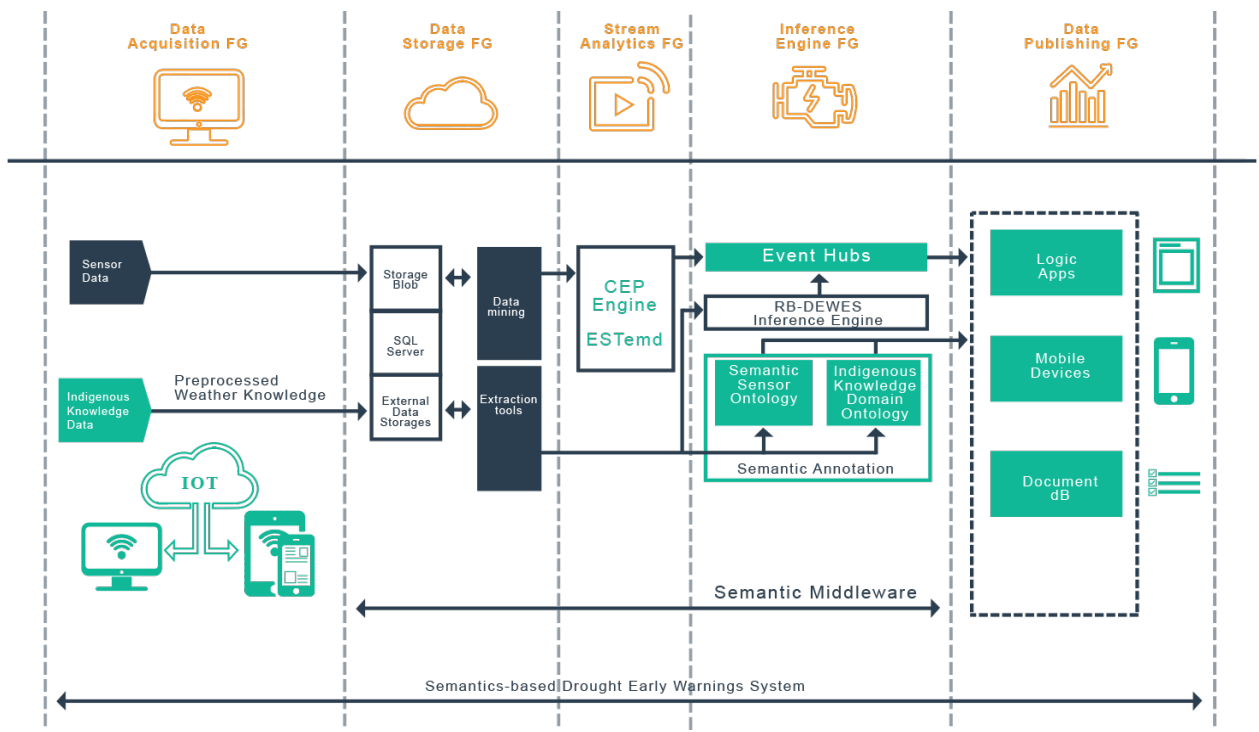

Figure 3- 5: The semantics-based data integration middleware framework (*Source: Author*).

*Storage FG*, *Stream Analytics FG*, *Inference Engine FG* and the *Data Publishing*

*FG*, with technologies and services that are based on a service-oriented approach (Akanbi & Masinde, 2018b) in fulfilment of the framework requirements.

### 3.5.2.1.     Data Acquisition FG

The data acquisition FG collects data from different data sources (structured and unstructured). The system utilises calibrated sensor data and local indigenous knowledge on drought. This FG encapsulates two functioning data collection modules: (i) Indigenous Knowledge System Data Collection (IKSDC) module, and (ii) Wireless Sensor Data Collection (WSDC) module. The results of the *data acquisition FG* fulfil the requirement **FR3** of the framework. The data collection and integration is based on Service Oriented Architecture (SOA) from heterogeneous data sources, and RESTful services are adopted for machine-to-machine data communication over the network.

### Indigenous Knowledge System Data Collection (IKSDC) Module

The IKS module of the Data Acquisition FG provides an abstraction for the collection, gathering and documentation of the IK data (**D1**) using appropriate data collection tools. Figure 3-6 depicts the architecture of the Indigenous Knowledge System Data Collection (IKSDC) module. The unstructured local indigenous knowledge on the drought of the area under study offers the desired level of scalability and variability is paramount to the realisation of the system on a micro-climatic level. The IK is obtained in the study area from the domain experts, farmers and focus groups through a series of oral consultation, questionnaires, interviews, field studies and meeting sessions. Furthermore, to achieve an updated collection of the IK from the IK experts, this research utilises a data collection application that captures the IK indicator (and its ecological interactions with detailed descriptions) and geographic coordination in the natural habitat. The IK data is temporarily stored in the indigenous Knowledge Database Server or Indigenous Knowledge Web App Server (backend) for further pre-processing and analysis. The acquired IK is pre-processed by the data mining tools into a form that is stored in the *Data Storage FG*.

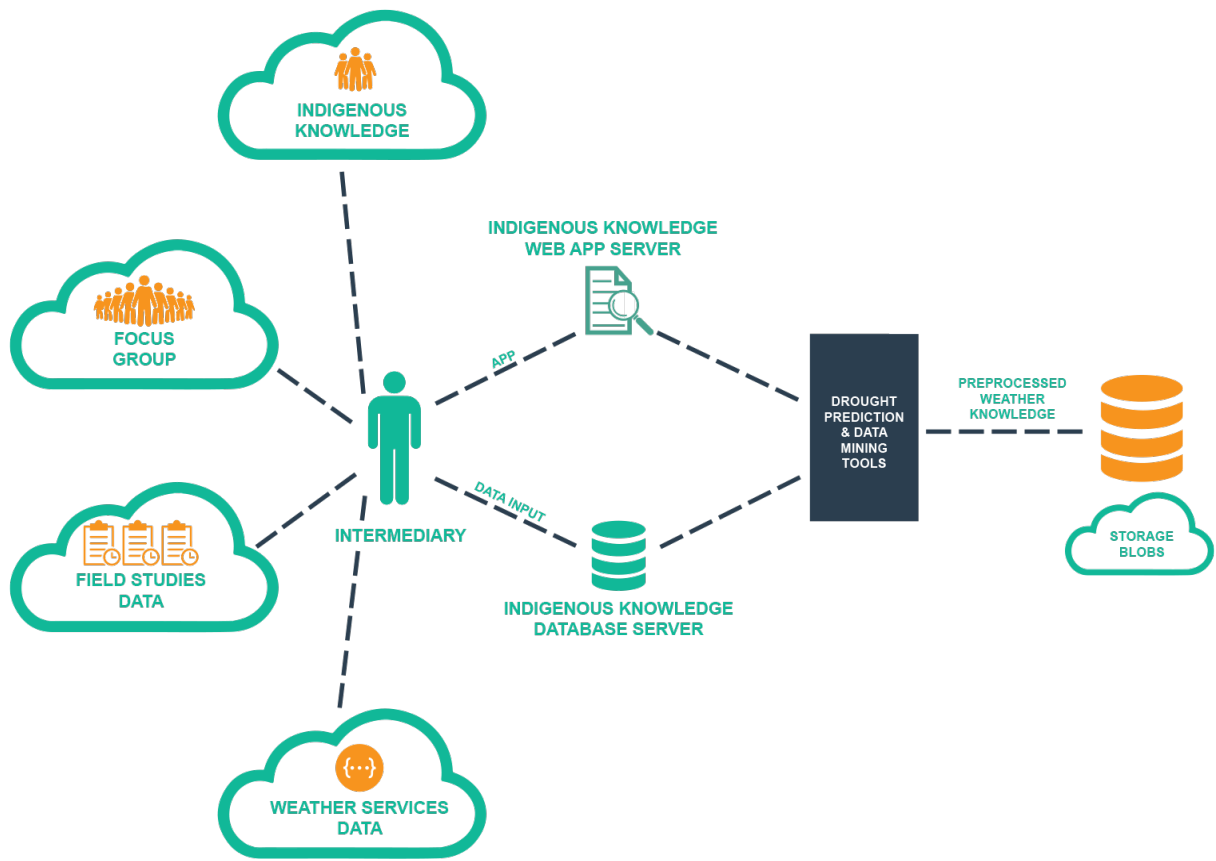

Figure 3- 6: Indigenous Knowledge System Data Collection (IKSDC) module framework (*Source: Author*).

## Wireless Sensor Data Collection (WSDC) Module

The Wireless Sensor Data Collection (WSDC) module architecture, as depicted in Figure 3-7 is a network of connected calibrated sensor devices for sensing atmospheric pressure, temperature, humidity, precipitation and soil moisture. The data (**D2**) are transmitted to the IoT hub (Microsoft Azure, Google Cloud, Sigfox Cloud) via the gateway.

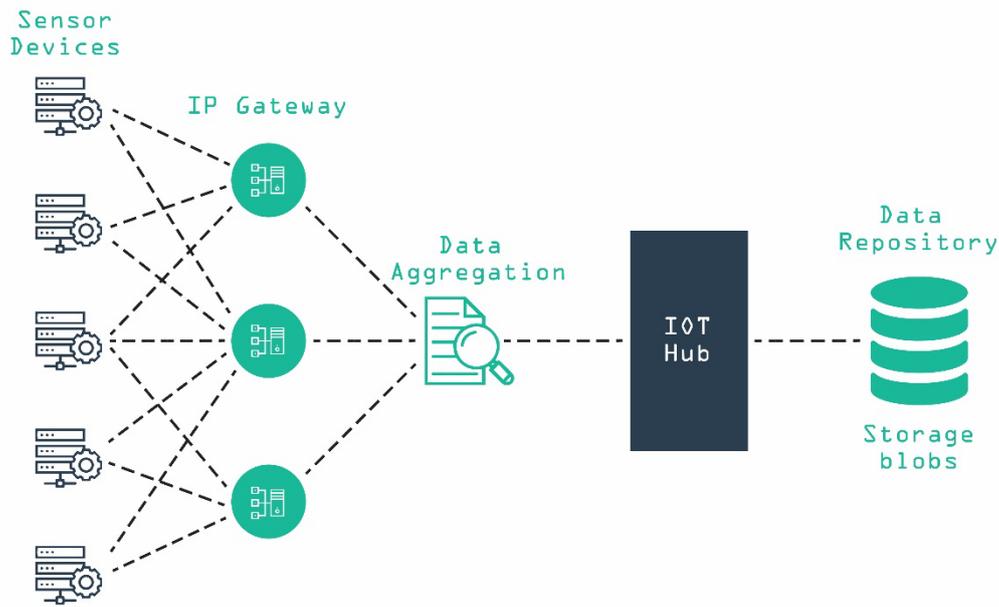

Figure 3- 7: Wireless Sensor Data Collection (WSDC) module framework (*Source: Author*).

The communication medium for transmitting the sensor readings from the sensors and the gateway to the cloud varies due to several factors. The communication medium ranges from the Bluetooth connection, ZigBee, MQTT, Sigfox network to HTTP protocol (Figure 3-8). The selection of an appropriate communication medium is based on the data necessity and secrecy factor of the transmission medium. This research used a Wi-Fi-enabled microcontroller board (Node MCU) mostly based on 6LoWPAN protocol. The time-series sensor readings are saved in the storage blobs and are retrieved in JSON-LD format using RESTful services.

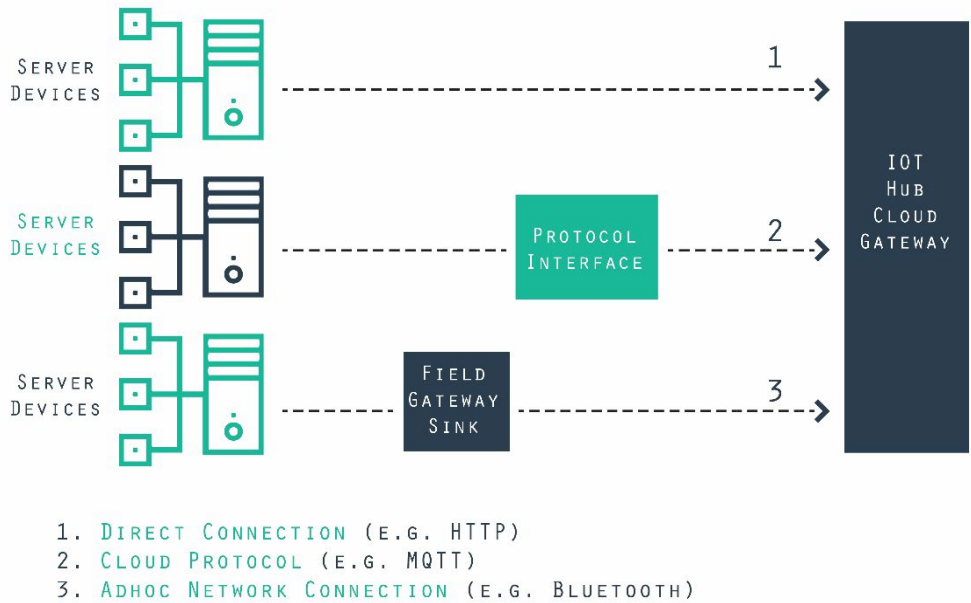

Figure 3- 8: Communication medium patterns (*Source: Author*).

### 3.5.2.2.    Data Storage FG

Pre-processed data collected from the *Data Acquisition FG* will be transferred to the *Data Storage FG*, where the data are stored in an internal context database (relation SQL or NoSQL). The *Data Storage FG* consists of modules that facilitate the storage and processing of structured and unstructured data types using online repository – Google Sheets, etc., and offline repository – storage media, phone cache etc.

The storage blobs filters and caches the streaming sensor data from the deployed WSN in a scalable real-time fashion. The raw sensor data is based on appropriate domain semantics is stored using the Open Geospatial Consortium Observation and Measurement (O&M) model (Botts, Percivall, Reed, & Davidson, 2008; Probst, 2006; Janowicz & Compton, 2010), which defines measurement units, concepts, values and uncertainty. The other set of data saved in the storage blobs is the IK from the domain experts which have been pre-processed and the knowledge extracted by semi-manual data mining techniques.

The dataset for **D2** is transferred to the *Stream Analytics FG* which transforms data into a consistent structure for the discovery of features and patterns to extract useful insights of real-world events for processing by the stream processing engine. The IK gathered for **D1** is mapped to a domain ontology specially developed for this research (See Chapter 5) to ensure common understanding and description of objects relationships and observed *events*.

### 3.5.2.3.    Stream Analytics FG

The *Stream Analytics FG* incorporates the implementation of Event Processing (EP) concepts to infer meaningful insights in the stream of sensor data in real time. The types of EP deployed are based on its application and are categorised under three sub-types: Event Processing Platforms (EPP), Distributed Stream Computing Platforms (DSCP) and Complex Event Processing (CEP) libraries (Dayarathna & Perera, 2018).

The EPPs type of EPs have functionalities such as event filtering and the ability to determine correlations of different scenarios. DSCPs incorporates the additional functionality of computation across multiple nodes in a distributed cluster. On the other hand, CEP engine (*or CEP libraries often used interchangeably in this thesis*) have the unique ability to infer meaningful patterns and relationships even in unrelated events. However, irrespective of the EPs, the suitability is based on the publish/subscribe patterns and compatibility with the use of RESTful services.

This research utilises the CEP engine (See Chapter Six) that detects composite events – specific patterns in the 'stream of time' series sensor data. The ability of the CEP engine to infer the pattern of the event is achieved through CEP rules that are embedded part of the application logic (Cugola, Margara, Pezzè, & Pradella, 2015). In this context, rules are in this form of general syntax:

$$CE\,(A1\,=\,J1\,(..)\,,...,An\,=\,Jn\,(...)\,)\,:=\,Pattern \ ...... \text{ (Equation 3-1)}$$

Where the symbol: = separates the rule head from the pattern. CE specifies the composite event captured by the rule and how its attributes $A_1....A_n$ are functionally defined by the attributes of the events that appear in the pattern. When a pattern is detected within the stream of input sensor data, the CEP engine knows that the corresponding composite event has occurred based on the specified CEP rule and notifies the interested components if the stream of input events satisfies the pattern (Dayarathna & Perera, 2018). For example, data from four sensors $S_1…S_4$ will serve as input to the CEP engine in the form of $S_1:=A_1 (T_1)$. The attribute value for the sensor is captured as well as the corresponding time stamp. A temperature sensor can capture four different reading within an hour period. Based on the drought forecasting model logic the average of those reading can trigger a pattern and used to infer an event such as "*High Temp*". Events inferred from the EPs component of the *Stream Analytics FG* are represented using the JSON-LD and transferred to the *Inference Engine FG*.

### 3.5.2.4.  Inference Engine FG

This FG of the middleware framework consists of the ontology modules for the semantic representation of the heterogeneous data sources (**D1 & D2**), automated reasoners and rule-based expert system modules that work in an event-driven fashion for drought prediction and forecasting. It addresses the requirement **FR1** and **FR2**. The *Inference Engine FG* implements the semantic representation of the heterogeneous data accordingly by using appropriate domain ontology; performs simple domain-specific reasoning on the IK in the RB-DEWES module. The domain ontologies in the *Inference Engine FG* address the need of a uniform representation for the data (structured and unstructured) in a way to be understood and processed by the reasoning engine module and support real-time persistent queries (Akanbi & Masinde, 2018b).

This research study adopted the W3C Semantic Sensor Network (SSN) ontology (Compton, Barnaghi, Bermudez, García-Castro, Corcho, Cox, Graybeal, Hauswirth, Henson, Herzog, & Huang, 2012) for the semantic representation and conceptualisation of the stream of sensor data and event inferred from it (**D2**). The ontology provides a comprehensive framework for the explicit description of sensor devices, observation, measurements, properties, etc., enabling reasoning of individual sensors or a WSN. The SSN ontology module represents the sensor data, properties of the data, and the events generated by the reasoners from the sequence of sensor reading (already represented in JSON-LD) in a machine-readable language – OWL based on the SSN ontology.

For the unstructured indigenous knowledge (IK) on drought (**D1**), the major challenge is the lack of an existing domain ontology that explicitly represents the local indigenous knowledge. The ontology module in the *Inference Engine FG* is a domain ontology that explicitly represents the local indigenous knowledge on drought. It is designed to semantically represent the entities and event (behavioural/observation) in the indigenous knowledge domain using a minimal number of classes, properties and restrictions (Akanbi & Masinde, 2018c). The SSN ontology and the IKON ontology are grounded on DOLCE as the foundational ontology. DOLCE provides a generic definition for conceptualisation, facilitating the perfect alignment between ontologies founded on it.

The semantic reasoner's module and the RB-DEWES module in the *Inference Engine FG* perform the generation of drought forecasting inference from the semantically represented **D1** data used in the middleware. Semantic reasoners module performs domain related reasoning based on the relationships and properties of the entities in the domain. The RB-DEWES module as a fully integrated expert system utilises *rules* derived from the knowledge representation of the IK to infer drought forecasting and prediction information with attributed *certainty factors*. Applying formal representation to all data using ontology ensures effective data exchange in the *Inference Engine FG* and high level of semantic expressivity in conjunction with the syntactic expressivity offered by the JSON-LD. Chapter 6 presents a completed overview of the reasoners and expert system component of the middleware.

## RB-DEWES Development Methodology

Harrison (1991) defined expert systems like "*computer programs, designed to make available some of the skills of an expert to non-experts*". Therefore, the development methodology starts with the use of expert's knowledge (skills) acquired in the *Data Acquisition FG* to system design, development and implementation. The development methodology consists of four (4) phases as depicted in Figure 8-1 below. Phase 1 starts with the knowledge engineering, knowledge categorisation, knowledge representation and rules ranking. Phase 2 of the methodology entails the system architecture, programming of the system's components etc. Phase 3 presents the system's design, development and implementation. Phase 4 presents

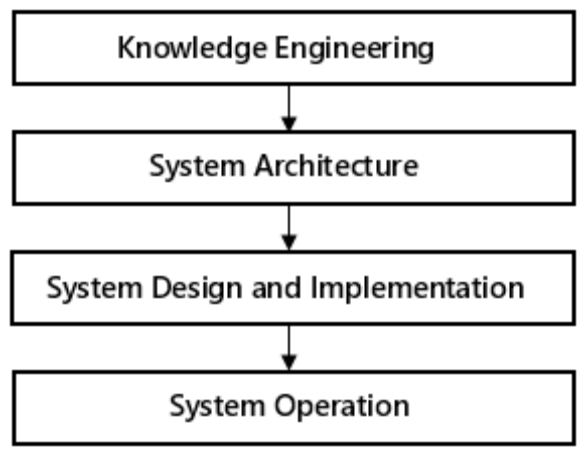

Figure 3- 9: RB-DEWES System Development Methodology. (*Source: Author*)

an illustration of the system operation and overall performance with evaluation.

a) *Knowledge Engineering:* Rule-based systems require that the expert's knowledge and thinking patterns be explicitly specified. Hence, the processes in this phase are knowledge acquisition from domain experts, categorisation of the knowledge and knowledge representation in the form of rules. Figure 3-10 depicts the processes involves in the knowledge

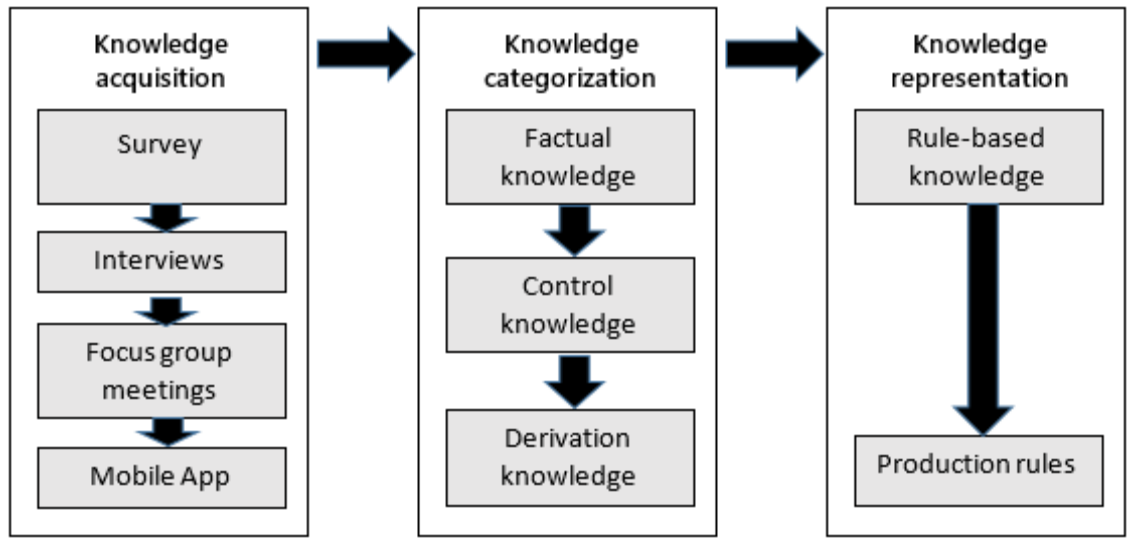

Figure 3- 10: Process Flowchart of Knowledge Engineering Phase. (*Source: Sasikumar et al., 2007*)

engineering phase.

The process of knowledge acquisition is the first step in the development of any knowledge-based systems. It is the process that facilitates the transfer of knowledge from a human expert to the knowledge base of an event-driven system through the construction of new, specific inference (production) *rules*. Obtaining this knowledge and writing proper rules is the main core of the knowledge acquisition phase (Scott *et al.*, 1991). The knowledge used in this process has been previously acquired through the implementation of the middleware's *Data Acquisition FG* and the *rules* were explicitly derived after the knowledge representation of the domain (Chapter Five) through the process of elicitation. The authors take the role of knowledge engineer and the local indigenous knowledge on drought was acquired through a series of structured interviews, conducted case studies, selected focus groups meetings and through the deployment of the developed data collection mobile application from the two study areas: KwaZulu-Natal, South Africa and Mbeere District, Kenya.

b) *System Architecture*: The nature of the middleware is taken into consideration for the requirement and specification criteria for the system architecture.

c) *System Design and Implementation*: This phase is achieved based on the middleware distributed architecture and implemented as a sub-system or component of the Inference Engine FG.

d) *System Operation*: The RB-DEWES can be implemented as a standalone system or as part of the distributed middleware DEWS.

### 3.5.2.5.    Data Publishing FG

The distributed semantic middleware framework seeks to automate and complement the existing drought alerts/weather forecast information for policy decision makers use in the study areas. This can include the application of modern technologies in the distribution and publishing of accurate inferred information. The inferred drought forecasting/prediction information is called 'drought forecasting advisory information' (DFAI) – presented in a standardised format with attributed *certainty value* to indicates the confidence level of the systems based on the set of inputs for use by policy decision makers. The DFAI can further be disseminated via mobile phones SMS, logic apps, notifications hubs, mobile services, web apps, document dB and also in a machine-readable format to promote reuse and integration with other third-party applications using REST APIs.

### 3.6.    Knowledge Modelling and Representation Methodology

Knowledge modelling and representation is carried out by the application of a methodology. A methodology is simply the organisation of some fundamental phases that ensure the correct completion of deliverables (Guarino, 1998; Gómez-Pérez & Benjamins, 1999). The methodology phases are planned towards achieving the heterogeneous data integration middleware requirements (**FR &**

**NFR**). This integrated bottom-up methodology consists of six phases, which allows seamless ontology development with system requirements at the centre of the development.

The methodology depicted in Figure 3.11 based on the data collected in section 3.5.2.1 starts by defining the high-level goals; which entails the type of ontology to be created, the foundational ontology to be adopted etc. This is followed by the information gathering (data collection) and elicitation phase. Elicitation is a term used in knowledge modelling that means fleshing out of the information, which typically means the extraction of knowledge from the domain experts or data source. The next phase is to start the preliminary modelling task, which is modelling in the form of light-weight ontologies. This helps generate more refined and encoded models in the formalisation phase. The initial structuring phases focus on the conceptual definition of ontology. The next phase is the formalisation – knowledge representation using machine-readable languages, then the deployment of the ontology which is the usage phase of the ontology and finally the ontology evaluation to determine the effectiveness of the ontology.

### 3.6.1. Phase One – Goal & Scope Definition

This phase is the starting point or preparatory stage of the KM methodology. It is the starting point of the ontology development cycle. In this research, the domain of interest is local indigenous knowledge on drought and WSN data sources. The focus is the development of a domain ontology for the local IK on drought and the sensor data from the WSN. The scope of ontology development is defined in terms of its boundaries and ontological requirements. The domain ontology for **D1 & D2** is be based on DOLCE as the foundational ontology for coherent ontology alignment between these heterogeneous knowledge bases as stated earlier.

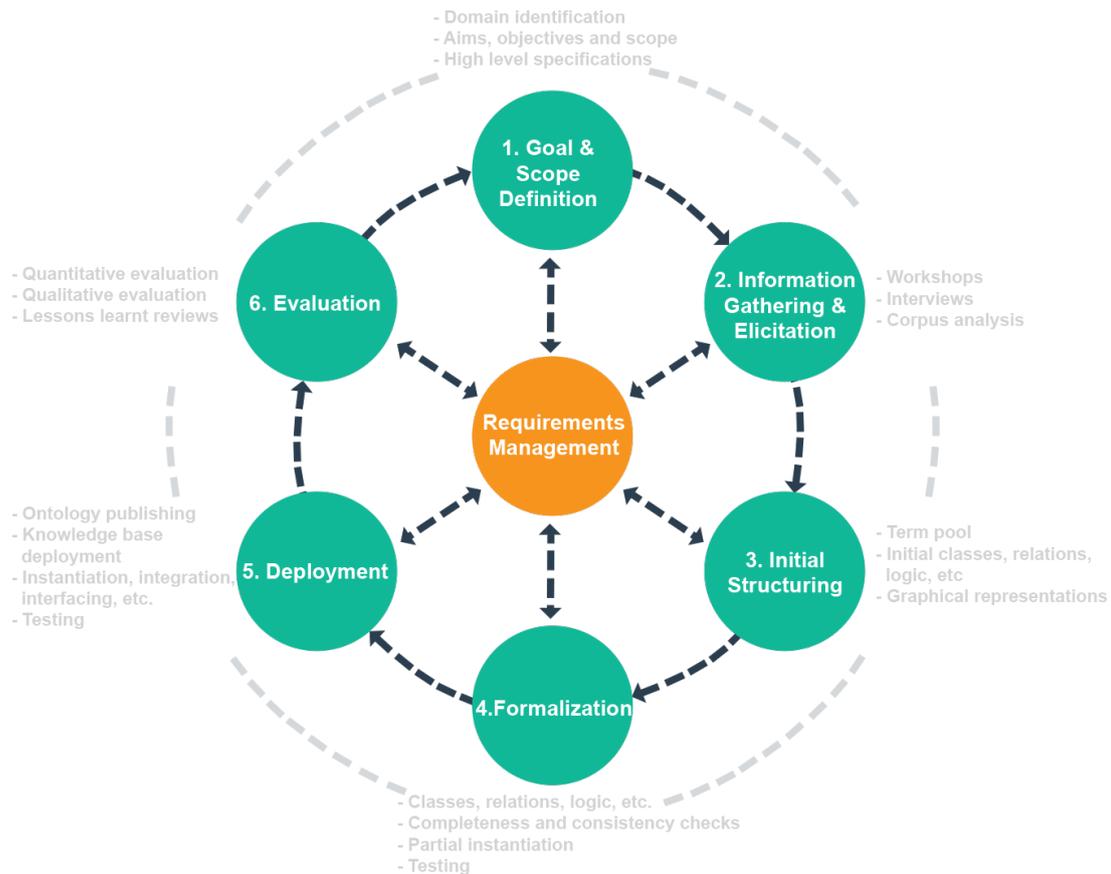

Figure 3- 11: Overview of Knowledge Modelling Methodology (*Source: Smith, 2003*)

### 3.6.2. Phase Two –Information Gathering & Elicitation

The information gathering and elicitation phase are about collecting information from a range of diverse sources (See Section 3.5.2.1). The capturing and understanding of information central to the domain of discourse are the key activities of the ontology development cycle. This entails the application of appropriate data collection and pre-processing tools of the *Data Acquisition FG*. For **D1** the data and information about the local indigenous knowledge are collected through a series of surveys, interview, focus groups and mobile application. The preliminary information is gathered using available tools and techniques such as simple documents, questionnaires, spreadsheets to more sophisticated means like mind maps and audio-visual recordings. Information is gathered from **D2** in the form of sensor readings and *events* inferred by the stream processing engine.

### 3.6.3. Phase Three – Initial Structuring

This phase of the methodology encompasses several tools and techniques for transforming the loosely organised information collected from the previous phase into a more refined, visually-represented lightweight model. In this phase, all the classes and the relationship are identified and mapped from the knowledge gathered (**D1 & D2**). The visual model representation at this phase is beneficial and helps provide an overview of the domain by providing a snapshot of the classes, sub-classes and relations. The visual ontologies generated in this phase are great for reviews and sharing purposes; this allows for easy updating of the knowledge model based on the feedback received. Visual lightweight representation makes it easier to build formal knowledge models for use in the next phase of the methodology; because the agreed visual lightweight model makes it easier during the encoding of the ontology. Hence, the process becomes more streamlined.

The first thing in this phase is the creation of the "term pool" – that captures the potential terms for inclusion into the knowledge model (Noy & McGuinness, 2001). The information and knowledge gathered from the domain experts are put in the form of statements about the things that make up and describe the domain, forexample statement of facts. These statements are analysed, and the *nouns* in the statements of facts are identified. After the identification of the nouns in the body of knowledge or statement of fact, the complex sentences and phrases are decomposed into several single statements (*rules*) that capture one or two simple ideas that are built around the nouns. These set of statements allows the basic understanding of the things that are relevant to the knowledge model — the spreadsheet term pool for capturing the domain specific terms in the knowledge base.

Graphical languages and notations such as UML are used to effectively model lightweight ontologies visually (Liepins *et al.*, 2012 ). In KM, the most important construct used are classes which represented the meaningful categorisation/classification that contains individuals, subsumptions such as inheritance between subclasses and classes; also, relations, which are the association between two or more classes. Then, association or relations are used to associate pairs of individuals which are instances of a particular class and are the most specific things in the universe of discourse (UoD) as illustrated below (Figure 3-12).

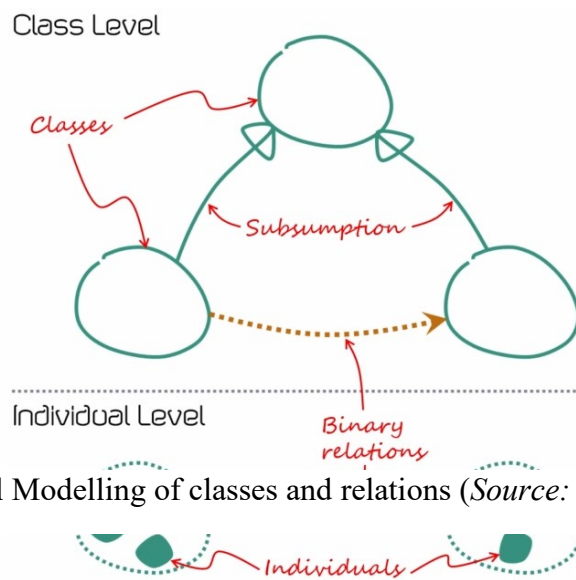

Figure 3- 12: Graphical Modelling of classes and relations (*Source: Author*).

Selecting the appropriate naming convention is the last step of the initial structuring phase of KM methodology (Noy & McGuinness, 2001). The naming convention ensures maintaining consistency in the manner or way of naming ontology entities, and this is enforced by following strict naming conventions. In KM, there are two widely used conventions: (i) camel case convention where words are written in such a way that the first or second word always begins with a capital letter while a first or single word starts with an uppercase or lowercase letter; (ii) the underscore convention, where underscore is used to separate words representing an entity. This research adopts the camel case conventions for naming entities, relations and individuals.

### 3.6.4. Phase Four – Formalization

As indicated in the previous phases of the methodology, visual lightweight representation or graphical notation are very useful for representation when sharing and covering meaning across human beings. For machine interpretation, reasoning and decision support at the systems level, the use of visual representation falls short of the ability to share meaning consistently with detailed semantics. Therefore, that shortcoming is overcome through formalisation, which means coding of knowledge models – ontology using formal, machine-readable languages and semantic technologies. The application of formalisation of knowledge bases allows computer and intelligent systems to be able to interpret, understand and generate reasoning from the knowledge model.

The formalisation phase deals with the encoding of knowledge models – which improves the ability to integrate diverse data sources, overcome data heterogeneity, enhance data integration and interoperability, search for information and use of heterogeneous knowledge base in information systems, intelligent schemas, etc. Formalisation requires using a formal knowledge representation language in representing and expressing knowledge models using appropriate software tools. There are various formal languages of description logic – RDF/s or OWL for encoding an ontology. In this research, OWL is adopted for the encoding and knowledge representation of our knowledge models due to the level of expressivity and the degree of formalism (Noy & McGuinness, 2001). OWL is a specification developed and maintained by the World Wide Web Consortium (W3C). There are several tools for encoding an ontology (Kapoor & Sharma, 2010; Stojanovic, 2004), *Protége* is the leading ontology editing tools with integrated add-ons to achieve reasoning capabilities of the developed ontology and also backed by an active community of users.

Classes identified during the initial structuring phases are represented and specified in a flexible hierarchy; the relation (called *properties*) is used for specifying axioms to define how classes and their individual components behave. The adoption of OWL allows reasoning facilities that automatically classifies concepts as well as to verify the effectiveness and consistencies of descriptions on the knowledge model. In *Protégé*, the first thing to specify is the ontology Internationalized Resource Identifier (IRI) and the ontology version of the IRI. It is appropriate to add a semantic annotation to the ontology being created. The semantic annotation provides a description of the ontology for the knowledge model. *Protégé* allows the ability to save the ontology in different knowledge representation formats like RDF/XML Syntax, Turtle Syntax, OWL/XML Syntax, OWL Functional Syntax, Manchester OWL Syntax, OBO format, LaTeX Syntax and lastly the JSON-LD format. Each of the file formats has a different level of syntactic and semantic expressivity. This research adopts the OWL/XML syntax and the JSON-LD for semantic representation of the domain and data integration respectively. JSON/JSON-LD is the standard output data format for all the FGs of the middleware due to compatibility with RESTful web services for scalability.

### 3.6.5. Phase Five – Deployment

After the successful formalisation and encoding of the ontology, this phase deals with the deployment of the developed ontology (Noy & McGuinness, 2001). The term "deployment" in this regard means the release of the ontology or knowledge model by publishing the ontology for use in intelligent information systems and ontology-driven information systems. The suitable way of deploying an ontology is dependent on the requirement management of the ontology in the context of its development. The deployment of the ontologies is about sharing the knowledge model with the wider audience or research community for download and reuse. Furthermore, the formal ontology can be exploited by integrating the knowledge model with another information system where the knowledge represented are used for decision-making processes.

The deployment phase also entails the ontology documentation of the entities as an important aspect in the deployment of the knowledge model. This ensures the representation of the encoded ontology in a natural language. For example, an ontology can be represented and view in HTML format for use by non-domain experts. The conversion of an OWL ontology file into HTML can be achieved through the use of the Live OWL Documentation Environment (LODE) tool developed by the University of Bologna, Italy (Peroni, Shotton & Vitali, 2012). The generated HTML files can be documented and shared with users for insight about the conceptualisation and formal representation of the domain.

The visual deployment or representation of the ontology can also be achieved through the use of OWLGrEd (Liepins, Cerans & Sprogis, 2012), OntoGraf, Visual Notation for OWL Ontologies – VOWL (Lohmann, Negru, Haag & Ertl, 2016) or OWLViz Import Graphs. Another method for the visual deployment of an ontology is through the use of Radial Diagram; at the centre of the radial diagram is the central concept of the ontology, i.e. the most important class to be emphasised using concentric shells with satellite classes relevant to the subject matter. The association between the classes are added using connectors which can be running outwards, inwards or centrally.

### 3.6.6. Phase Six – Evaluation

This phase involves assessing the goal and scope definition phase and determines the extent to which the aim and objectives of the project have been fulfilled and how the requirement has been met in the context of the established scope. This phase can be done iteratively during the ontology development life cycle. There are several methods used for ontology evaluation purposes. There is technical and specialist perspective for evaluation and ontology project through the use of ontology alignment or ontology comparison.

However, as part of the evaluation procedure, there is the need to ensure the use of appropriate ontology development methodology, because a perfect methodology provides the appropriate justification for ontology development from conception to implementation. Also, there exists the need to check for inconsistent naming conventions and typos, which are common mistakes in the ontology development and indicate a lack of attention to details. The evaluation of the developed ontology is similar to the initial data gathering phase; the major difference is that in this phase, the output of the domain formalisation is verified to be accurate and a true representation of the domain by the domain expert.

## 3.7.   Experimentation Process

The simulation was run using the implemented tool for short-term forecast and record the probability of accurate drought prediction or forecasting. The WSN provided a series of sensor data for the short term of forecasts. The accuracy of the drought prediction and forecasting information in the form of DFAI was verified during the evaluation stage.

## 3.8.   Middleware Evaluation Procedure

The implemented middleware in the form of drought early warning system tool was tested with usability specifications. This provides the ability to verify the effectiveness and ease of use of the implemented prototype. To evaluate the research methods used in this research study, a correlation between the forecasts/predictions and the actual weather data were analysed (Casati *et al.*, 2008). The evaluation procedure is presented in Chapter 7.

## 3.9.   Ethical Consideration

In this research study, participants/respondents were informed of their rights of ownership of the knowledge and that their privacy would be protected. Bryman and Bell (2007) stated ten principles related to ethical consideration in a research study; all were strictly adhered to, by ensuring full consent of each participant/respondents were obtained before data collection session through the completion of the "*Consent Form*" by each participant/respondent. The consent form contained clauses that must be approved by the participants/respondents; these clauses indicated that they had read and understood the information about the research; they had the free will to ask questions about participation in the research study; they voluntarily agree to participate in the research; they had the right to withdraw at any time without giving reasons or being penalised for doing so; and that adequate levels of confidentiality of the research data would be ensured.

The approval for conducting the research study was obtained from the Department of Information Technology's Departmental Research and Innovation Committee (DRIC); and the Faculty Research and Innovations Committee (FRIC) at Central University of Technology (CUT). The information collected from the participants/respondents remains the intellectual property of the participants/respondents of the area under study. The anonymity of participants/respondents participating in the research was ensured, detailed affiliations of researchers were declared, and all forms of communication in relation to the research were carried out with transparency through the chief/head of the community.

## 3.10. Summary

This chapter identified the research study design is a mixed research design where qualitative and quantitative techniques are used towards achieving the research objectives. Also, it included a description research paradigm, primary data sources, data collection methods of the heterogeneous data sources. The data pre-processing and analysis use case scenarios as well as the ethical consideration for the entire research study. The research was executed from a philosophical base on aspects such as truth and validity, which determines acceptable research methods to be adopted. A purposeful sampling process was followed, and the data collection instruments were the sensor devices, survey questionnaire, mobile application, structured interviews with a focus group and use of case study.

Furthermore, the chapter presents a vision of how the integration and interoperability of heterogeneous data sources can be achieved through a semantic middleware for drought forecasting and environmental monitoring systems. A distributed semantic middleware framework was presented, which acts as the main catalyst for heterogeneous data integration, providing the contrivance for the semantic data representation, annotation, generation of inference and reasoning. The methodology for the development of the RB-DEWS was also presented. The system generates levels of forecasting recommendation in the form of DFAI. This middleware takes processing, representation and dissemination of drought forecasting data where information will be shared in a machine-readable format for effective environmental monitoring or forecasting in the realm of this latest technology. The SBDIM can serve as the basis to provide other forms of integration among heterogeneous environmental data sources and interoperability of intelligent systems.

# CHAPTER FOUR

# HETEROGENEOUS DATA COLLECTION

## 4.1. Introduction

In Chapter Three, the form of the research methodology and outline of the semantic-based data integration middleware framework was presented. This chapter presents the implementation of the first *Functional Group* (FG) of the framework – *Data Acquisition* FG, which deals with the collection of data from two heterogeneous data sources – indigenous knowledge on drought (**D1**) and the wireless sensor data (**D2**) in this case. The indigenous knowledge on drought is mostly unstructured oral, with a historical knowledge base in the form of observation of the ecological interactions, natural indicators for predicting the occurrence of an environmental phenomenon such as drought. These natural indicators are identified and used in the future for prediction and forecasting purposes.

On the other hand, **D2** is a structured weather data collected from deployed sensor devices and calibrated weather stations in the area under study. The sensors are used to measure the environmental parameters in remote locations, while the professionally calibrated weather stations in the area under study are used as reference measurement model. These two data sources (**D1** and **D2**) are collected from the two areas under study: Swayimane in KwaZulu-Natal, South Africa and Mbeere in Embu County in Kenya.

For **D2**, five (5) weather data parameters that are crucial in this research are collected: (1) temperature; (2) humidity; (3) soil moisture; (4) atmospheric pressure; and (5) precipitation. Some of the weather data is observed using wireless sensors while other readings are observed from the weather stations. For example, temperature and humidity readings are remotely measured using the DHT22 sensor module on Arduino board; soil moisture is measured using the hygrometer sensor – SEN13322; the atmospheric pressure, precipitation, and rainfall are observed from the weather stations. The data from the sensor devices are pushed to the cloud for easy access and future analysis.

For the IK on drought domain, the preliminary task was to recognise the local indicators for the indigenous knowledge on drought. This is achieved through the review of existing literature presented in Chapter Two. Here, the indigenous drought forecasting indicator are categorised under: (1) patterns of seasons; (2) behavior of animals, insects and bird; (3) behavior of plant/trees; (4) meteorological; (5) astronomical; and (6) knowledge of seasons. Further, each of the local indicators has an attributed *certainty factor* (CF), which is the measure of belief/disbelief in the local indicator as determined by IK experts who are the custodian of the IK in the study areas. For example, if the sighting of *Phezukomkhono* (a migratory bird) indicating the onset of the raining season has a *CF* of 0.20, this might imply there is a 20% chance of onset of the raining season unless combined with other local indicators for accurate generation of inference from the set of local indicators.

## 4.2. Domain 1 – Local Indigenous Knowledge on Drought

This research study is focused on the indigenous knowledge system of Swayimane in KwaZulu-Natal, South Africa and Mbeere community in Embu County of Kenya. The local indigenous knowledge on drought knowledge collection and gathering is based on the Indigenous Knowledge System Data Collection (IKSDC) framework of the middleware's *Data Acquisition FG*. The IK on drought was gathered by the author with the help of facilitators. Both focus groups and questionnaires were used for the IK collection over a period of 12 months. IK data, previously collected in two related projects (Mwagha, 2017; Masinde, 2015) were also utilised.

### 4.2.1. Data Collection – Swayimane, KZN

In this case study, indigenous knowledge experts from Swayimane community participated in the knowledge acquisition process (Figure 4-1). Through the help of a local facilitator, the contents of the questionnaire and objective of the research were communicated in the isiZulu language. The surveying and the structured interview took place between September to March 2017 with the aim of using the questionnaire to measure the application level of indigenous knowledge on drought in the area under study.

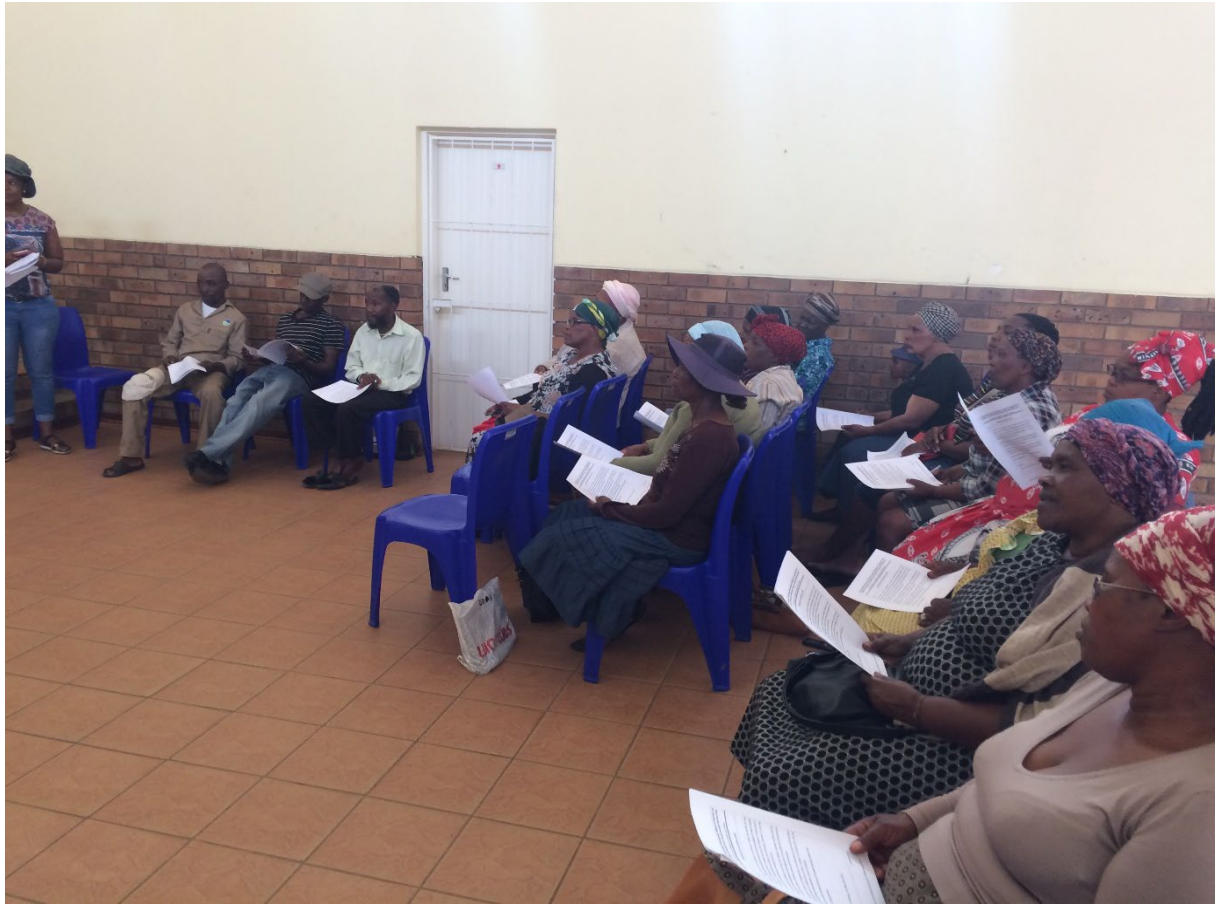

Figure 4- 1: Surveying and interviewing IK experts and local farmers at Swayimane, KZN, South Africa (*Source: Author*).

### 4.2.1.1. Demographics of Respondents

A sample of 61 respondents consisting of 82% females and 18% males participated, using the positive sampling technique (Duan & Hoagwood, 2013) from uMngeni local municipality of uMgungundlovu district of KwaZulu-Natal. All the respondents are active farmers utilising IKS for drought forecasting and cropping decisions and are from Swayimane village of the uMngeni municipality of uMgungundlovu district, KZN, South Africa.

The majority of the respondents were middle-aged females, with 8.1% falling in the age bracket of 18 to 35 years; 13.1% of the respondents were between the ages of 36 and 45 years; 32.7% of the respondents were in the age bracket of 46 and 55 years; 40.9% were between the ages of 56 and 65 years and only 4.9% fell in the age bracket of above 66 years.

Most of the respondents had a basic education, with 75.5% having some form of education, and 25.5% had none. The level of education distribution was 45.9% having primary education; 18% with a secondary qualification and 11.4% with a form of post-secondary qualification.

The main economic activity of the sample group was farming. The reason for this was obvious due to the fact that IK knowledge of forecasting and predicting drought was the criterion for selecting the respondents.

### 4.2.1.2.    Knowledge of Indigenous Knowledge System on Drought

During the interview and survey process, the respondents were asked about their knowledge and the significance of the IKS. Of these, 85.2% stated that they used one form of local indicator or another for forecasting drought and to determine when to prepare their crops or when to plant their crops, while 14.7% relied on drought forecasting information from the municipality weather services, radio or news channel.

When asked to categorise the indicators they use, 42.6% of the respondents indicated they use meteorological indicators such as knowledge of the seasons, 21.3% of the respondents use astronomical indicators such as moon phases or cloud patterns, 36.0% relied on behavioural indicators such as ecological interaction of animals and plants (Table 4-1).

Table 4- 1: Categories of IK used by the respondents – Swayimane, KZN.

|  | Frequency | Percent | Cumulative Percent |
|---|---|---|---|
| Meteorological | 26 | 42.6 | 42.6 |
| Astronomical | 13 | 21.3 | 63.9 |
| Behavioural | 22 | 36.0 | 100.0 |
| Myth and Religious Beliefs | 0 | 0 | 100.0 |
| Total | 61 | 100.0 | |

## 4.2.1.3.  Characteristics of Weather Seasons in Swayimane, KwaZulu-Natal

The findings from the survey and interviews indicated four (4) seasons in KwaZulu-Natal; this is further corroborated by existing research reported in Mwagha and Masinde (2016). The summer season is from October to February and locally called *ihlobo*; Autumn is from March to May and is called *intwasabusika*, Winter exists from May to July and is called *ubusika*, and Spring is called *intwasahlo* in the local language, isiZulu. Table 4-8 below shows the category of each season.

Table 4- 2: Onset and Cessation of Weather Seasons in KwaZulu-Natal.

| Season | Local Name | Onset signs | Cessation signs | Local indicators | Start | End |
|---|---|---|---|---|---|---|
| Summer | *ihlobo* | Hot weather Dry winds Less rain | Cold Winds Rain stops | *Magwababa, Inkojane, Ntuthwana ants, etc.* | Oct | Feb |
| Autumn | *intwasabusika* | Trees shed leaves | Very cold | *Inyosi bees, Mviti tree, etc.* | Mar | May |
| Winter | *ubusika* | Cold Mist | Warm weather | *Onogolantethe bird etc* | May | July |

| Spring | *intwasahlo* | Lot of winds | Hot weather | *Phezukomkhono bird, etc.* | Aug | Oct |
|--------|--------------|--------------|-------------|----------------------------|-----|-----|

### 4.2.1.4. Indigenous Knowledge Drought Indicator for KwaZulu-Natal

IK indicators are a critical component of the IKS. The observation or occurrence of the local indicators helps in making decisions about the likely occurrence of drought or related environmental phenomena. However, in most cases, several indicators are combined before reaching a likely interpretation of the local indicators or scenarios observed. Since observation of indicators is mostly in the form of sighting, observation or ecological interactions, listing the local indicators with their respective interpretation is paramount. IK holders, experts and local farmers provide the list of the indicators as well as the in-depth interpretation of the scenarios.

Table 4- 3: Swayimane KwaZulu-Natal Weather Indicators.

|  | SUMMER (Oct – Feb) | AUTUMN (March-May) | WINTER (May – July) | SPRING (Aug – Oct) |
|---|---|---|---|---|
| **Astronomical** | • Full moon | • The moon is small in size<br>• Full moon | • Half of small moon | • Half moon |
| **Meteorological** (*Knowledge of the Seasons*) | • Very hot weather<br>• High temperature during the day and night | | • Cold weather | • Its rains<br>• Presence of thunderstorm and lightning. |
| **Behaviors of Birds** | • *Magwababa* and *inkojane* flock in before the rain | | • *Onogolantethe* bird searching for small snakes and earthworms to eat | • Flocking in of *Phezukomkhono* which is a noisy yellow bird that flocks in during the spring |
| **Behaviors of Insects** | • Insects are present in the summer<br>• *Ntuthwana* ants are present. | • Insects are decreasing in Autumn.<br>• Present of I*nyosi* bees | • Insects are absent in the winter<br>• Ants are hiding<br>• No ants | • Absence of *Inyosi* bees<br>• Little insects are sighted |
| **Behaviors of Animals** | • The animals are beautiful and look well fed in summer<br>• Sighting of *Ingxangxa* frogs<br>• Cattles are gaining weight and getting fat<br>• Most animals are getting fat. | • The animals are thin<br>• Cows are fat | • Little trace of bush animals, because of the cold weather; activities<br>• The animals are thin | • The animals have average weight |
| **Flower, leaves and fruit productions** | • Mviti trees are flowering<br>• Peach trees are flowering | • Some plants leaves are withering | • Withering of leaves of some trees | • Blooming of Guava tree.<br>• Flowering of trees like Wattle, *Wiki-Jolo* and *Umphenjane.* |

| **by some Trees** | • *Amapetjies* trees are blooming | • *Miviti* tree is withering and loosing leaves | | |
|---|---|---|---|---|

## 4.2.2. Data Collection – Mbeere District

For Mbeere study area, the questionnaire, the ODK mobile application (see Appendix B) and focus groups were used. The data was collected through the application ODK *Collect* and saved to the database (Google Sheet). Figure 4-2 represents the structure of the database entry in the Google Sheet. The data distribution saved to the Google Sheets was visually analysed using Google Fusion Table for the data analysis.

Figure 4- 2: Mbeere's District Respondents entry in the database (*Source: Author*).

### 4.2.2.1. Data Analysis – Mbeere District

A sample of 1505 respondents' data was collected. The first set of data in the form of raw data is obtained by combining the digitalised respondents' information from the questionnaires with the data from the mobile application online database repository (Google Sheets). The combined data was processed to eliminate ambiguities and repetitions into a form compatible with Google Fusion for data visualisation.

By gender, the respondents consisted of 70.1% females and 29.9% males from Mbeere community (Figure 4-3). All the respondents were active farmers utilising IKS for drought forecasting and cropping decisions. Furthermore, 21.4% of the respondents fell in the age bracket of 18 to 35 years; 33.5% between the ages of 36 and 45 years; 36.1% of the respondents were between the ages of 46 and 55 years; and only 9% were above 66 years.

All the respondents had a basic education, with 62.5% having primary education, 31.2% with secondary qualification, and 6.3% with a form of post-secondary qualification.

Understanding the cropping practices in the area helped in determining the potential impact of drought on the crops. The response showed that most farmers engaged in mixed farming – where two or more crops are planted as displayed in the chart below (Figure 4-3). The chart indicates the quantity of crops produced by the respondents in the population sample in tonnes. For example, 10000Kg of maize, beans, sorghum and green grams were produced.

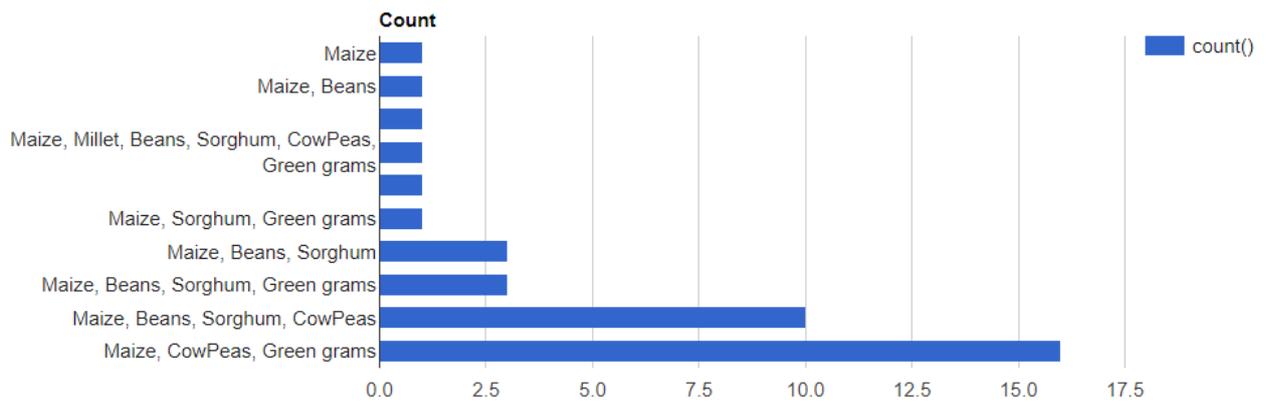

Figure 4- 3: Distribution of the Respondents by crops planted – Mbeere (*Source: Author*).

## 4.2.2.2.     Knowledge of Indigenous Knowledge System on Drought

To determine the knowledge of the respondents in IKS, respondents were asked about their level of understanding and usage of the IKS. Of these, 99.42% stated that they rely on one form of local indicator or another for forecasting drought and to determine when to prepare their crops or when to plant their crops, while only 0.58% relied on drought forecasting information from the weather services in the area (Figure 4-4).

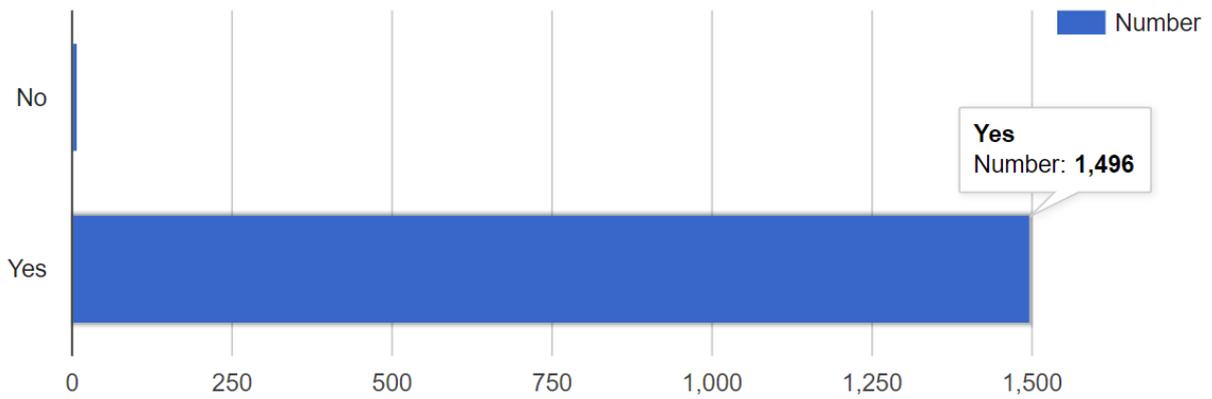

Figure 4- 4: Distribution of the Respondents by IK usage – Mbeere (*Source: Author*).

When asked to categorise the form of local indicators used, 47.92% of the respondents use meteorological indicators such as knowledge of the seasons, 29.7% used astronomical indicators such as moon phases or cloud patterns, 21.9% relied on behavioural indicators such as ecological interaction of animals and plants and 0.48% relied on myth and religious beliefs (Figure 4-5).

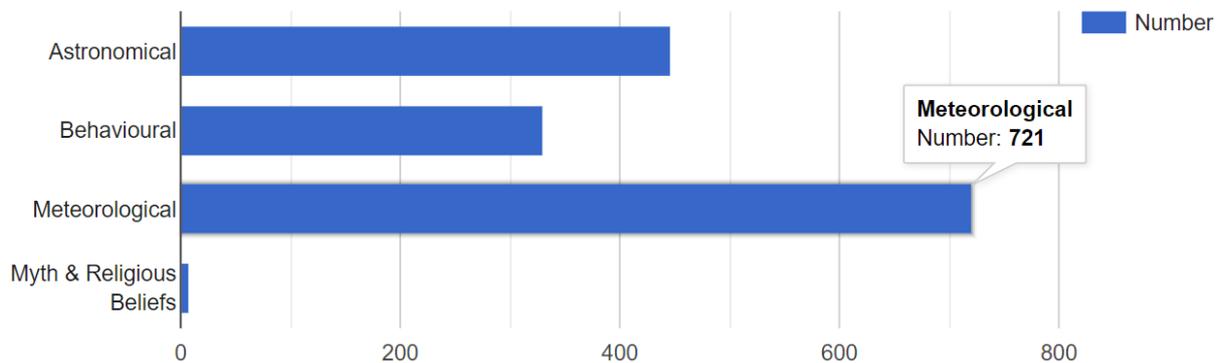

Figure 4- 5: Categories of IK used by the respondents – Mbeere (*Source: Author*)

### 4.2.2.3. Characteristics of Weather Seasons in Mbeere Community

IK among the Mbeeres has been extensively studied and served as a subject of research by Masinde (2013, 2015, 2018), which was used to validate the findings from this research. The onset and cessation of seasons in Mbeere community are stated in Table 4-4, which shows, the long rains, dry season and short rains (Masinde & Bagula, 2011).

Table 4- 4: Onset and Cessation of Weather Seasons in Mbeere Community.

| Season | Local Name | Onset signs | Cessation signs | Local indicators | Start | End |
|--------|-----------|-------------|-----------------|------------------|-------|-----|
| Transition | | High temperature | Rains | Temperature, lightning, *midithu* insects, etc. | Jan | Feb |
| Long Rains | *Mbura ya nihoroko* | Thunderstorms, lightning, etc. | Very cold & foggy | Thunderstorms, *Bugvare,* beetles, etc. | Mar | June |
| Dry Season | *Mbevo* (cold) & *Thano* (dry) | High temperature at night, windy, etc | Warm weather | *Kamutuanjiru, midithu* insects, etc. | June | Oct |
| Short Rains | *Mbura ya mwere* | Sharp lightning from the East, etc. | Hot weather | *Ngiri, Thari* bird, etc. | Oct | Dec |

### 4.2.2.4. Indigenous Knowledge Drought Indicator for Mbeere Community

IK indicators are a critical component of the IKS; the observation or occurrence of the local indicators helps in deciding the likely occurrence of drought or related environmental phenomena. Local indicators for Mbeere community are well documented by Masinde (2015). This helps in refining and listing the IK indicators in the study area. Table 4-5 below provides the list of the indicators on a seasonal basis as well as the interpretation or implication of the sighting and/or occurrence of the indicators.

Table 4- 5: Mbeere IK Weather Indicators (*Source: Masinde, 2015*)

| | January – February | Long Rains | Dry Season | Short Rains |
|---|---|---|---|---|
| **Astronomical** | | • Sighting on new moon<br>• Visible phases of the moon | | • Sighting on new moon<br>• Visible phases of the moon |
| **Meteorological** (*Knowledge of the Seasons*) | • Moderate daily temperature | • Drizzling in the evening<br>• Severe thunderstorms | • Sprouting of new leaves by cowpeas<br>• Cold temperature at night<br>• Whirling winds | • Raining daily<br>• Early morning dews |
| **Behaviors of Birds** | | • *Kivuta mbura* birds starts making sounds | • Nesting of *Ngoco* bird along water banks. | • Flocking of thari bird signifies onset of drought. |
| **Behaviors of Insects** | • Sighting of *Midithu* ants | • Croaking of frogs<br>• *Bugvare* birds are building their nests. | • *Mindithu* starts moving southwards | • *Ngiri* starts making noise |
| **Behaviors of Animals** | • Goats giving births | • Cows and bulls jumping up and down. | • Low nesting of ngoco bird near water banks | • Bulls behavior |
| **Flower, leaves and fruit productions by some Trees** | • Mango trees fruiting, Yield size of *Ngaa* | • Sprouting of *nthinuriu* and *mbaku* | • Blooming of *migaa* and *cowpeas*.<br>• Maturity and germination of *karamba ka nthi*<br>• Flowering of *mugaa, mutororo* | • Flowering of drought category mango tree |

### 4.2.3. Representation and Use of Aggregated Indigenous Knowledge

The gathering and collection of local IK on drought from the case studies help in documenting and understanding the local indicators used by the indigenous farmers in predicting drought (Manyanhaire, 2015). Each indicator is subjective to different interpretation based on the sighting or occurrence. However, in most cases, several indicators are combined to achieve a definite interpretation.

The aggregated IK data gathered from the two study areas are used for the semantic representation of the local indigenous knowledge on drought domain, using an ontology (see Chapter 5). Semantic modelling and knowledge representation of local IK on drought are fundamental in achieving **RO** – integration of the two heterogeneous data sources – IK and WSN data. The knowledge is formalised into the semantic structure using an ontology for machine readability, reusability, integration, and interoperability with another sub-system in the distributed FG of the middleware.

Also, interpretation of several indicators and observations identified from the indigenous knowledge gathered from the domain experts are constructed in the form of *rules* for use in the *Inference Engine* FG of the middleware. These *rule set* will be saved in the knowledge base and are used to infer and predict drought phenomenon by the inference engine. The expert system module generates inference by using the *rule set* derived from the IK and provides DFAI with attributed CF based on the set of user's inputs.

### 4.3.    Domain 2 – WSN & Weather Station Data

### 4.3.1. Wireless Sensor Data Collection

The WSN data gathering process started with the deployment of the sensors to remote locations in the area under study. The sensor boards transfer data from the sensors to the gateway/sink in the WSN. These time-critical sensor readings are sent to the *Sigfox* cloud (IoT Hub) using the *Sigfox* network. The *Squidnet* network module contains a unique identifier (UID) which gives access to the *Sigfox* cloud backend web interface. The *Sigfox* module is connected to the microcontroller board and all sensor readings based on the preset time frames are uploaded to the cloud accessible via the backend (Figure 4-7).

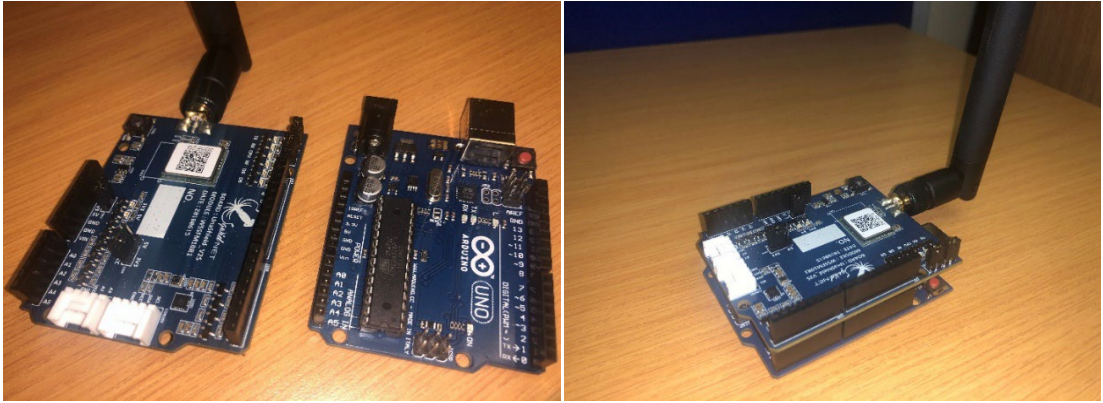

Figure 4- 6: Squidnet Network Module (*Source: Author*).

There are the capabilities to create callbacks to transfer data received from the devices associated with this device type to an IT infrastructure. The backend automatically forwards some *events* using the "callback" system. A callback is a custom HTTP request or routine that consists of the device(s') data and readings sent to a cloud server/platform. The callbacks are automatically triggered when a new message is received from the device, when a location has been computed, or when a device communication loss has been detected. In this research, the callback function will be used to push the streams of sensor data through the *Data Storage* FG to the *Stream Analytics* FG for data analytics and inference generation in real time and data streams in real time. The sensor readings are available in JSON, XML and CSV data format.

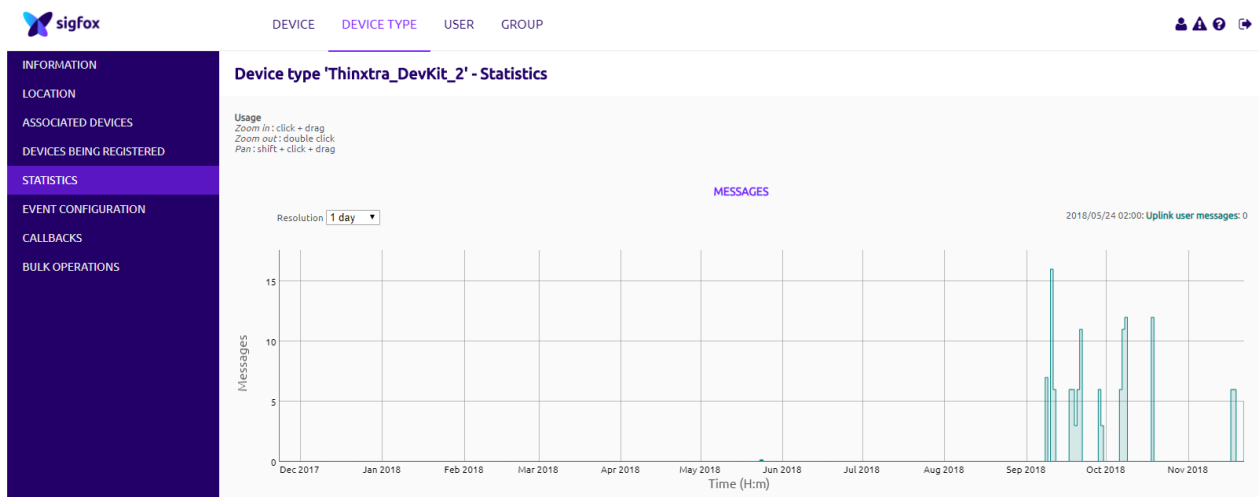

Figure 4- 7: Sigfox Cloud Web Interface (*Source: Author*).

### 4.3.2. Sensors

Miniature sensor modules were connected to the microcontroller boards to remotely measure the temperature, humidity, soil moisture and atmospheric pressure while the weather station was used for reference measurements. However, in some cases where the IoT devices could be damaged, the weather station is used. Table 4-6 below list the sensor modules used to measure temperature, humidity, soil moisture and atmospheric pressure.

Table 4- 6: List of sensor modules.

| Weather Parameter | Sensor Module |
|---|---|
| Humidity Temperature | Sensor module DHT22 is a sensor used to measure humidity and digital temperature. This sensor is a combination of the capacitive humidity sensor and a thermistor. The sensor measures the surrounding air and readings are channelled out in the form of a digital signal on the data pin. DHT22 is an improvement on 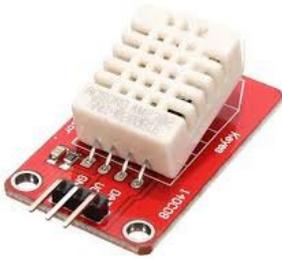 the previous version (DHT11), and compatible with most microcontroller boards. DHT22 (*Source: Author*) |

| Soil Moisture | The SEN13322 and the Irrometer 200SS-5PR Watermark sensor were used to measure the soil moisture. The SEN13322 is less fragile and can only be inserted to the depth of 5cm to prevent water/moisture from short-circuiting the exposed electronic component of the sensor. The most commonly known issue with soil moisture sensors is the exposure to moisture and water, which adversely shortens their lifespan. The 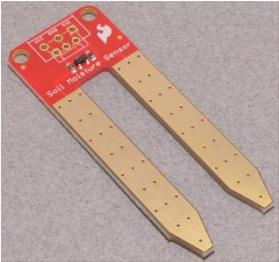 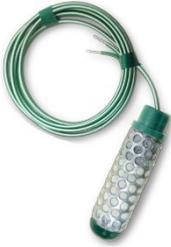 Watermark sensor is a probe can be embedded at a greater depth due to enclosed electronic components. SEN13322 (*Source: Author*)   Irrometer 200SS-5PR (*Source: Author*) |
|---|---|
| Atmospheric Pressure | The atmospheric pressure sensor used is the MPX4115A/MPXA4115A from Motorola. The sensor converts atmospheric pressure to an analogue voltage by using a silicon piezoresistive sensor element. 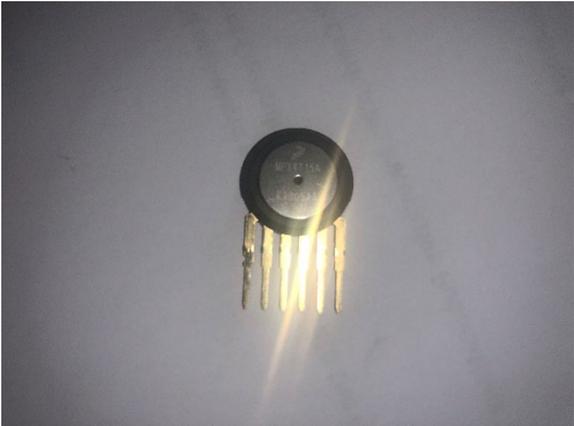 Pressure Sensor (*Source: Author*) |

The deployed sensors take the readings of the environmental parameters and streams the sensor readings to the CEP *engine* component of the *Stream Analytics* FG. The readings are uploaded to the *Sigfox* cloud at every interval.

### 4.3.3. Weather Station Data Collection

A weather station is the aggregation of instrument and equipment for measuring environmental conditions to provide information to understand and study the weather and climate. In the two areas under study, there is a weather station that monitors and record the precipitation, rainfall, relative humidity, air temperature and atmospheric pressure in real time. The weather stations are situated in an area free of obstruction in accordance with the manufacturer's specifications. The current readings and the historical data are stored in the repository accessible via

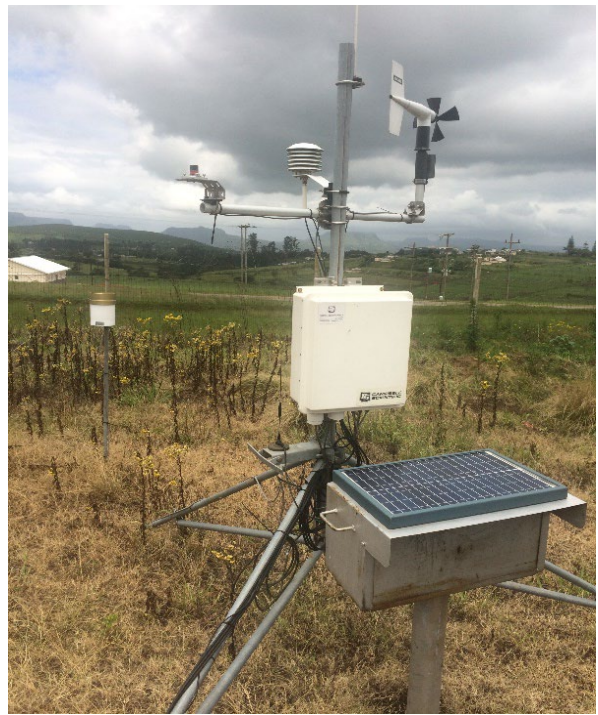

Figure 4- 8: Swayimane Weather Station (*Source: Author*).

the web interface.

The weather station consists of components that are used to measure and monitor the weather and climate, based on a programmable datalogger. The measuring instruments and sensors, measures, processes, stores, and transmits the data via multiple communications channel. In this instance, the readings and historical data are available and accessible online in real-time. Figure 4-9 shows the weather station readings at Swayimane, KwaZulu-Natal, South Africa through a web interface; the data are readily available for download in JSON, CSV and XML format.

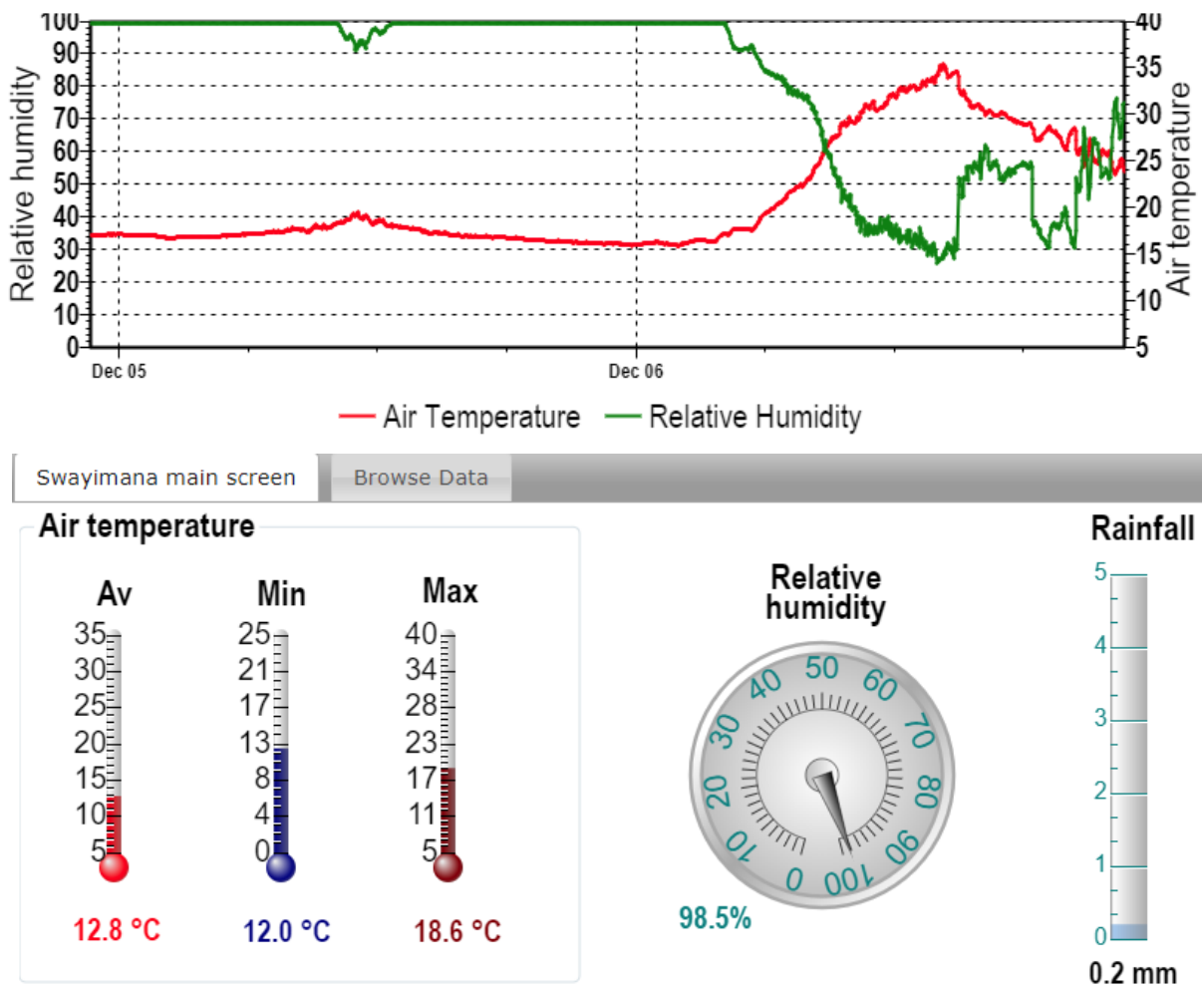

Figure 4- 9: Real-time readings of Swayimane Weather Station (*Source: http://agromet.ukzn.ac.za/*).

In the Kenya case study, the research study leverage on the partnership between the Central University of Technology, Free State and Trans-African Hydro-Meteorological Observatory (TAHMO) to have access to real-time weather station readings data. The current and historical weather data is available and accessible online in real time via the TAHMO web portal (Figure 4-10). TAHMO also offers access to backend data sources through the use of RESTful APIs for data integration and utilisation.

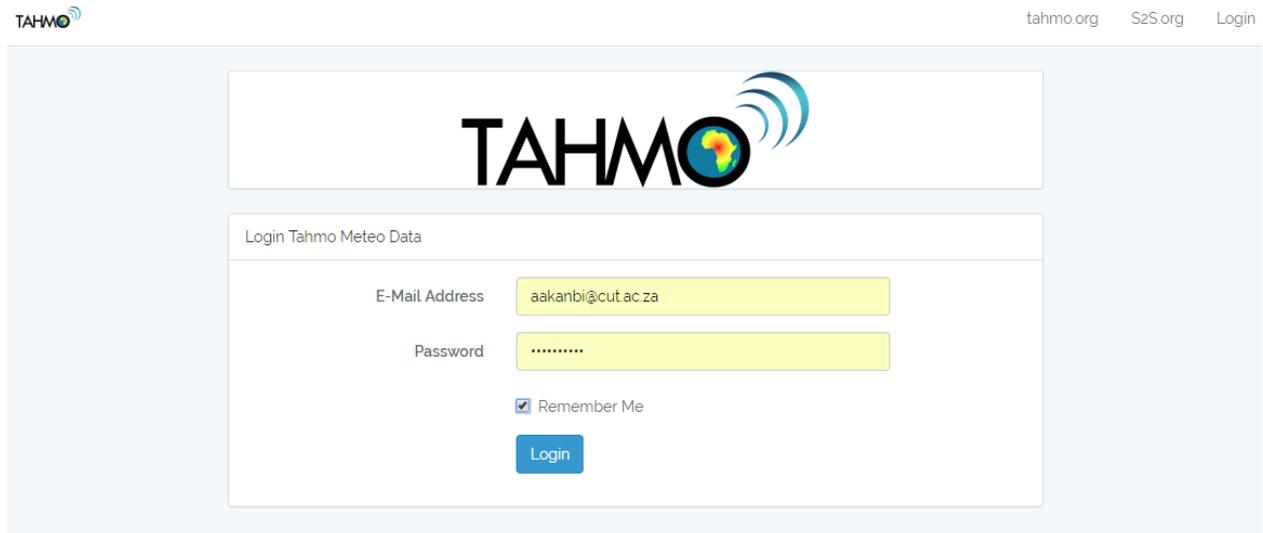

Figure 4- 10: TAHMO Web Portal (*Source: www.portal.tahmo.org*)

## 4.4. Summary

This chapter presents the data collection for the heterogeneous data sources through the *Data Acquisition FG*. The IK is gathered using qualitative interpretative methodology from the interviews with local farmers and indigenous knowledge domain expert using questionnaire, structured interviews, and survey mobile applications. The IK data are analysed and processed with the main focus to document the unstructured IK on drought knowledge. The structured WSN and weather station data are collected from the weather station and deployed sensors. The indigenous knowledge on drought obtained in the *Data Acquisition FG* serves as input for the *Data Storage FG*, *Stream Analytics FG* and *Inference Engine FG* towards the realisation of semantics-based data integration middleware for local indigenous knowledge and modern knowledge on drought.

# CHAPTER FIVE

# KNOWLEDGE MODELLING AND REPRESENTATION USING ONTOLOGIES

## 5.1. Introduction

In all forms of communication, the ability to share knowledge and information is often hindered because the meaning of information and the application of knowledge can be severely affected by the context in which it is interpreted. This notion is applicable in all spheres of human endeavours and subsequently now in intelligent information systems. The problem is most severe for application systems that must manage the data heterogeneity in various domains and integrate models of different domains into coherent frameworks (Ciocoiu, Nau & Gruninger, 2001). The lack of detailed meaning prevents or affects the full expressivity, use, reuse and application of knowledge and information irrespective of the context. When there is a lack of adequate meaning, the integration of several forms of information and knowledge towards a common goal of understanding become less desirable. Thus, the important aspect of interoperability is the mechanism to represent data. For example, in the environmental monitoring domain, the integration of different forms of knowledge in software applications have become central to improving the degree of accuracy of environmental monitoring systems due to the variability of environmental parameters. But this effort is hampered by different representations of the same information and use abstruse axioms and terminology in different contexts to mean different things (Kuhn, 2009; Akanbi & Masinde, 2015b; Devaraju, 2009), with the key challenge – how to infer accurate knowledge heterogenous environmental observations.

In this research, considering the challenges of integrating these heterogeneous data sources that were acquired and presented in Chapter 4, a solution to this problem was proposed through the use of semantic technologies and ontologies (Kuhn, 2005; Kuhn, 2009; Fogwill et al., 2012; Akanbi & Masinde, 2015b; Devaraju, 2009; Guarino, 1998; Walls, Deck, Guralnick, Baskauf, Beaman, Blum,

Bowers, Buttigieg, Davies, Endresen, & Gandolfo, 2014; Gómez-Pérez & Benjamins, 1999; Akanbi, Agunbiade, Kuti, & Dehinbo, 2014; Bally, Boneh, Nicholson & Korb, 2004; ) for a common understanding of concepts and terms. The application of ontology presents explicit semantics for the entities and concepts used, rather than relying just on the syntax used to encode those concepts. The adoption of ontological techniques helps with resolving ambiguity with abstruse terms, axioms and relationships for the local indigenous knowledge on drought domain (**D1**) and WSN domain (**D2**). This will ensure unifying the differences in how information and knowledge are conceptualised, and formal knowledge representation for translating those definitions and relationships into the specialised representation languages of intelligent systems.

This chapter presents the formal process of semantic representation of the heterogeneous data sources used in this research – the natural indicators, behavioural and ecological interactions of local IK on drought forecasting (**D1**) and the acquired sensor data from the WSN (**D2**). The knowledge is formalised into a semantic structure using an ontology for machine readability, reusability, reasoning, integration, and interoperability with intelligent systems in fulfilment of the research objectives (**RO**) and system requirement (**FR1**, **FR2** and **NFR1**). The objective is to model the acquired knowledge in an explicit form that is shareable and reusable for use by the *Inference Engine FG* subsystems. The chapter hence presents the *Inference Engine FG* component of the semantic middleware; it describes the ontology modules for the semantic representation of the heterogeneous data sources, the development processes of a domain ontology for local IK on drought and the adoption of an existing ontology for WSN sensors data.

## 5.2. Knowledge Modelling & Representation of Local Indigenous Knowledge on Drought (D1)

Before the commencement of knowledge modelling and formal representation of a domain using an ontology, effort have made towards reviewing the literature for any existing ontology that could be modified, extended or reuse. Currently, domain ontology that captures the context of local indigenous knowledge on drought explicitly using standardised languages for data exchange and semantic integration across software boundaries is missing (Akanbi & Masinde, 2015a). Domain ontology describes the properties, attributes and interrelationships of concepts, about a specific domain. Designing ontologies is the first step towards the integration and interoperability vision (Gerber *et al.*, 2015; Akanbi & Masinde, 2015b). Local IK on drought such as the behaviour of natural indicators, ecological interactions between different species of insects and animals, sighting of migratory birds, blooming and withering of floral and leaves – all pointing to the likely occurrence of an environmental phenomenon can be used to forecast drought accurately (Masinde & Bagula, 2011; Manyanhaire, 2015). The semantic knowledge representation of the domain using an ontology lead to richer processing of the concepts and knowledge through the use of a rule-based inferencing system as proposed in this research study (Akanbi & Masinde, 2018a).

Knowledge modelling and representation using an ontology has modernised the inference systems capability by permitting interoperability between heterogeneous knowledge systems and semantic web applications (Fahad & Qadir, 2008). Developed ontologies can also furnish the necessary semantics for inference generating capability required in intelligent systems (Fahad, Qadir & Shah, 2008). Ontological modeling of the indigenous knowledge on drought involves identifying the domain-controlled vocabularies, taxonomies, properties, and relationships; for adding of semantic (meaning) annotation to the data for an accurate inference generation from the knowledge base (KB) and to make them available in a structured form that can be processed by computers (Guarino, 1998). The methodology of knowledge modelling and representation has been outlined and presented in Chapter Three.

### 5.2.1. Ontology Development and Encoding of IKON – Knowledge Representation

The ontological representation and encoding of the local indigenous knowledge on drought in formal, machine-readable languages is a critical part of the knowledge modelling process for the domain. This phase comprises the initial knowledge gathering and formalisation phases, both intertwined, and led to the development and encoding of **I**ndigenous **K**nowledge on Drought Domain **ON**tology (**IKON**) (Akanbi & Masinde, 2018c) – domain ontology for local indigenous knowledge on drought. The method consists of the following steps a) enumerate terms in the ontology; b) define the classes and the class hierarchy; c) define the properties of classes; d) define the class instances.

a) E*numerate terms in the Ontology.*

This step involves the development of the terminology about the domain; this is done by reviewing related published and working papers and interviewing the indigenous knowledge domain experts through questionnaires and workshops with the focus groups. This allows the analysis of the domain data based on axioms and terms. Enumerating the terms in the domain provides an explicit knowledge of the domain. This is achieved by identifying the concepts that could become the classes of the domain. The phrase "*IS-A*" is a pointer to identify the class and sub-class relationship. For example, in the local indigenous domain, **Phezukomkhono** "*IS-A*" local bird sighted which is categorised under the **Bird** sub-classes under the **Vertebrate** sub-class etc. The "*IS-A*" indicate some form of inheritance between the class and its sub-classes. Another way of spotting an inheritance or sub-class of a class is through the direct mentioning of terms like "*KindOf*" and "*TypeOf*". This process is used to identify the classes and sub-classes in the domain before encoding *Protégé* ontology editor.

b) *Define class and hierarchy.*

Each class in the domain has a corresponding OWL class, since an OWL class represents a set of individuals that form the extension of the concept mapped by class. Based on DOLCE foundational ontology classification, the identified classes are categorised with on their attributes. For example, **DurationOfRainfall** and **StreamWaterLevel** are two physical qualities, i.e. subclasses of the **dolce:physical-quality** class. The activities or relationships of the entities are described ontologically as DOLCE processes. For example, **Blooming** is an **Event**, i.e., subclasses of **dolce:perdurant** based on DOLCE classification. These OWL classes form the basis of the **IKON** ontology as during the construction of taxonomy. The classes are organised in a taxonomy created by the *subsumption* relation.

All classes and sub-classes created are a subset of **Thing** which means **owl:Thing** is the class of set of all defined individuals. However, something inherent about the classes is that they can be disjoint which means expressing the logic that the individual cannot belong to more than one different classes, for example, an individual of the sub-class **Vertebrate** cannot belong to a sub-class **FloweringPlants** under any circumstances. In an ontology development project, it is critical to add the natural language description in the form of annotation to the classes and other ontology entities defined, this helps in the detailed documentation of the ontology. This is achieved by annotating each class or entity on the annotation section of the class or sub-classes. For example, the annotation of class **LivingThingsBehaviour** is depicted in Figure 5-1 below.

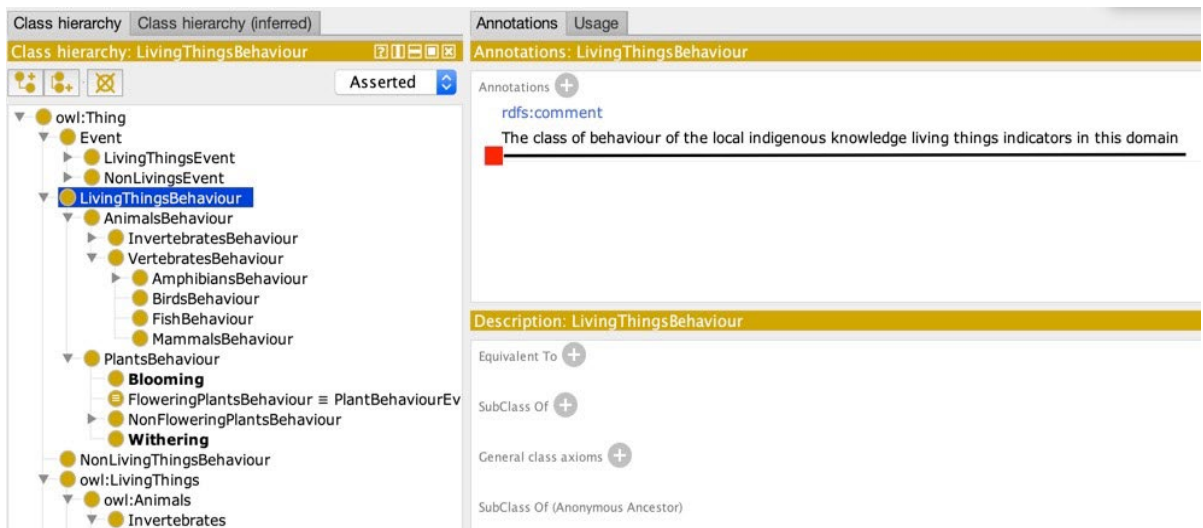

Figure 5- 1: Annotation of class `LivingThingBehaviour` (*Source: Author*).

Six main classes are identified and are subclasses of **Thing** as depicted in Figure 5-1. The six main classes were classified under the **owl:Things** into superclasses **owl:LivingThings**, **owl:NonLivingThings**, **owl:LivingThingsBehaviour**, **owl:NonLivingThingsBehaviour, owl:Event** and **owl:TimeAndPlace**. Each of these classes with subclass hierarchy. The domain was classified based on the expert knowledge, and the mapping of the domain classes to the ontology was achieved through object-oriented techniques using multiple inheritances. After the main superclasses of the local indigenous knowledge, the subclasses of each of the superclasses are defined. The **IKON** domain ontology represents all types of entities, relationships and events of the local indigenous knowledge on drought in the study areas. Through the use of natural indicators and relations to model the events or scenario in the domain. The domain ontology is reusable and fully extendable to accommodate additional indicators or drought-related events.

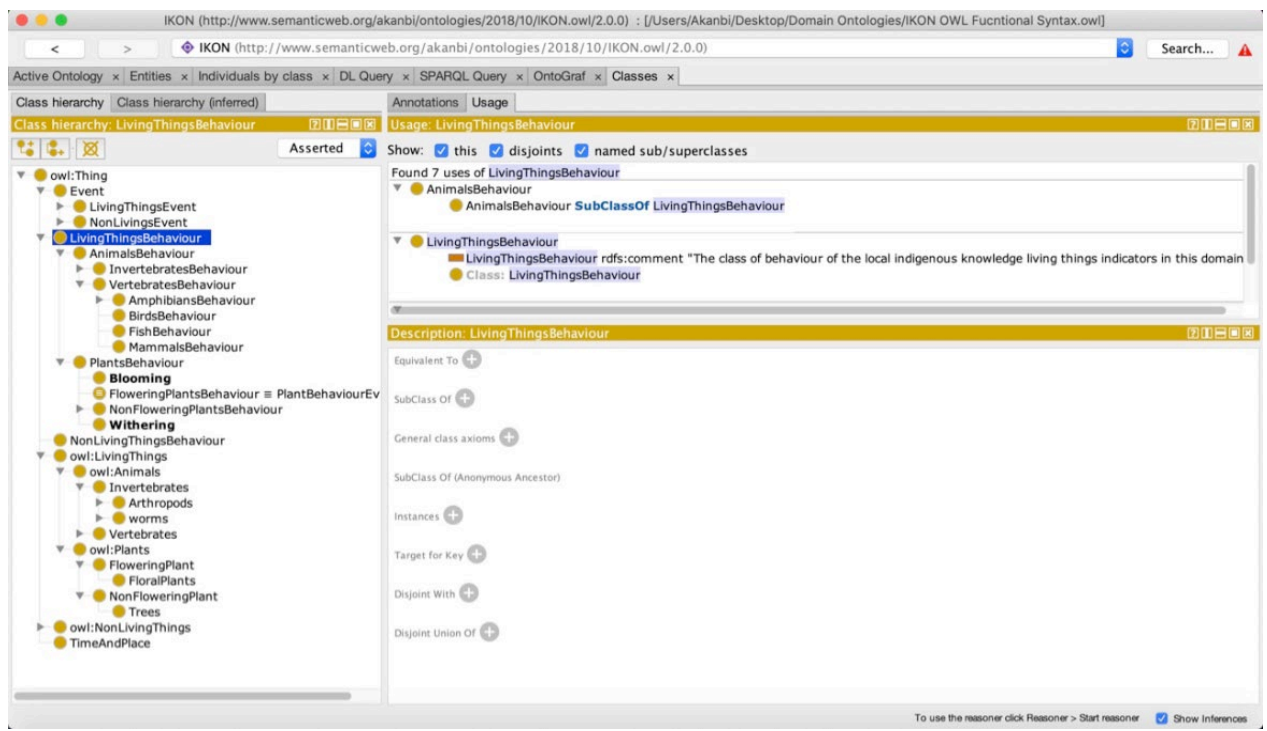

Figure 5- 2 The hierarchical representation of the IKON Ontology classes and subclasses (*Source: Author*).

The superclass **owl:LivingThings** will be classified into two subclasses of **owl:Animals** and **owl:Plants**, each with its own derived subclasses and individuals that are instances of the subclasses, for example, *Mugumo* tree, Wild figs, *Peulwane* bird, *Lehota* frog, *Sifenenefene* worms, etc. The **owl:NonLivingThings** class is used to capture the non-living entities of the IK domain, with individuals such as *Temperature*, *Rain*, *Humidity*, etc.

Behaviours (or observations) are represented as subclasses of **owl:LivingThingsBehaviour** and **owl:NonLivingThingsBehaviour**. An **owl:LivingThings** and **owl:NonLivingThings** provides a view on a set of entities which is consistent with a description. The **owl:LivingThingsBehaviour** and **owl:NonLivingThingsBehaviour** is used to model the corresponding behavioral activities of the **owl:LivingThings** and **owl:NonLivingThings** respectively, for example, sighting of migratory birds, blooming of flower, withering of plant, etc. The mapping of the semantically annotated behaviours (observations) to the entities is a formalisation of domain knowledge and allows deductive inference. The outcome of this phase for our proposed ontology helps to conceptualise and have a detailed understanding of the controlled vocabularies used, the class, properties, and relationship

c) *Define class properties.*

After identifying the classes and defining it, the next step is the class properties definition. The definition of class properties ensures the addition of semantics to the identified and defined concepts. Properties are used to describe attributes of the class, for example, characteristics of a class of **Animal**. In OWL the term used for relation is properties. OWL allows the specification of two main types of properties: *object property* and *data property*. In the IK domain, the relation (properties) capture the relation describing the objects in general. However, an *object property* is defined to relate classes and their objects. Further refinement could be added to the properties to include property constraints which describe or restrict the set of possible *property value*. Below are some of the object properties of this study's ontology. All the object properties (Figure 5-3) created are based on the ontology classes interrelationship. Few examples are stated below:

a) ***hasFlower*** relates a **Flower** plant with the **FloweringPlantBehaviour** which is the state of flowering depending on the seasonal changes.

b) ***OccursAt*** relates the occurrence of an **Event** with the corresponding class.

c) ***hasWithered*** relates the **Plant** with **PlantBehaviourEvent**.

d) ***BloomingOf*** is an object property that relates **Flower** with **Blooming** Behavior.

e) ***hasPhase*** relates **Moon** with phases of moon such as **FullMoon**, **HalfMoon** and **SmallMoon**.

f) ***hasTemp*** relates to the daily average *Temperature*. This can either be **High** or **Low** as assumed by the IK expert.

g) ***hasWindSpeed*** relates to the average **WindSpeed** for the day as determined by the IK expert.

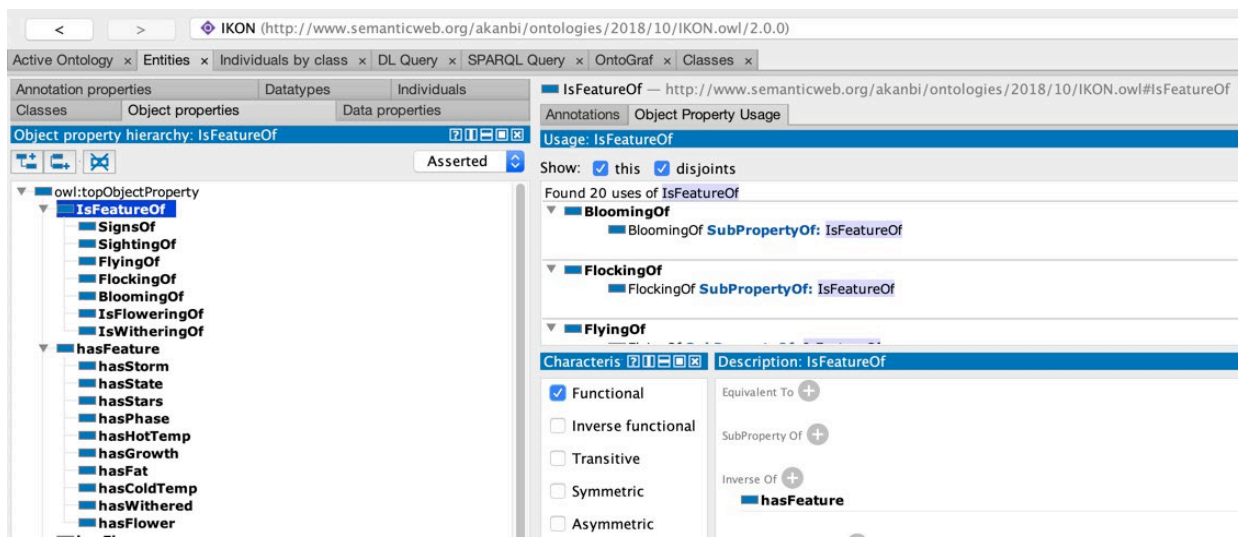

Figure 5- 3: The object properties of IKON Ontology classes and subclasses (*Source: Author*).

The *data property* can be simple or complex, this difference depends on the type of class, and are special attributes whose values are the object of (other) classes, or used to associate something to a *data value*. Figure 5-4 shows some of the captured data properties in *Protégé*.

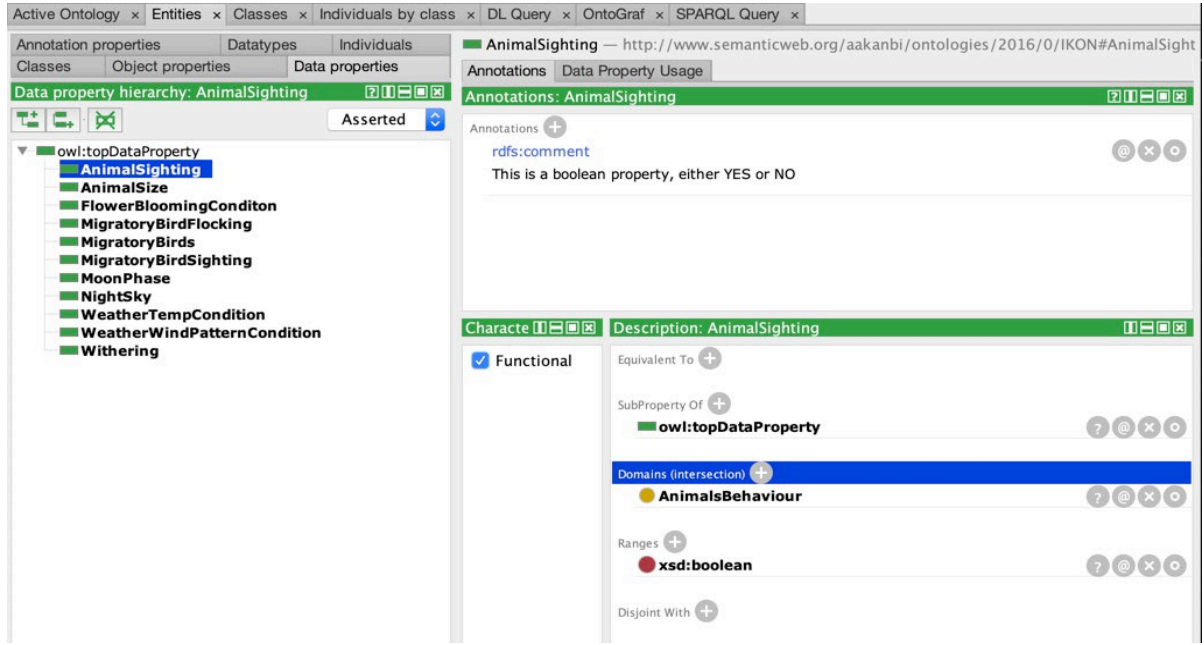

Figure 5- 4: The data properties of IKON Ontology classes and subclasses (*Source: Author*).

For example, during modelling, the data property **AnimalSighting** is a boolean property that has two data values, "Yes" or "No". This *data property* is used to represent the sighting of a particular local indicator, an animal in this instance. The type of animal sighted at an instance of time and period has an interpretation in the local indigenous knowledge on drought domain. Another example is **StreamWaterLevel**, **DurationOfRainfall** etc.

d) *Class instances.*

The class instances are the member (individuals) of the class and are the structural component of an ontology. The instances are "Individual" created after defining the classes, sub-classes, data and object properties of the domain. Individuals are added to the classes by *Protégé* by selecting the classes and click "Instance". This allows the add "individual button" to add a new instance to the class. Some individuals created are called asserted individuals if they declare explicitly the class they instantiate. In IKON

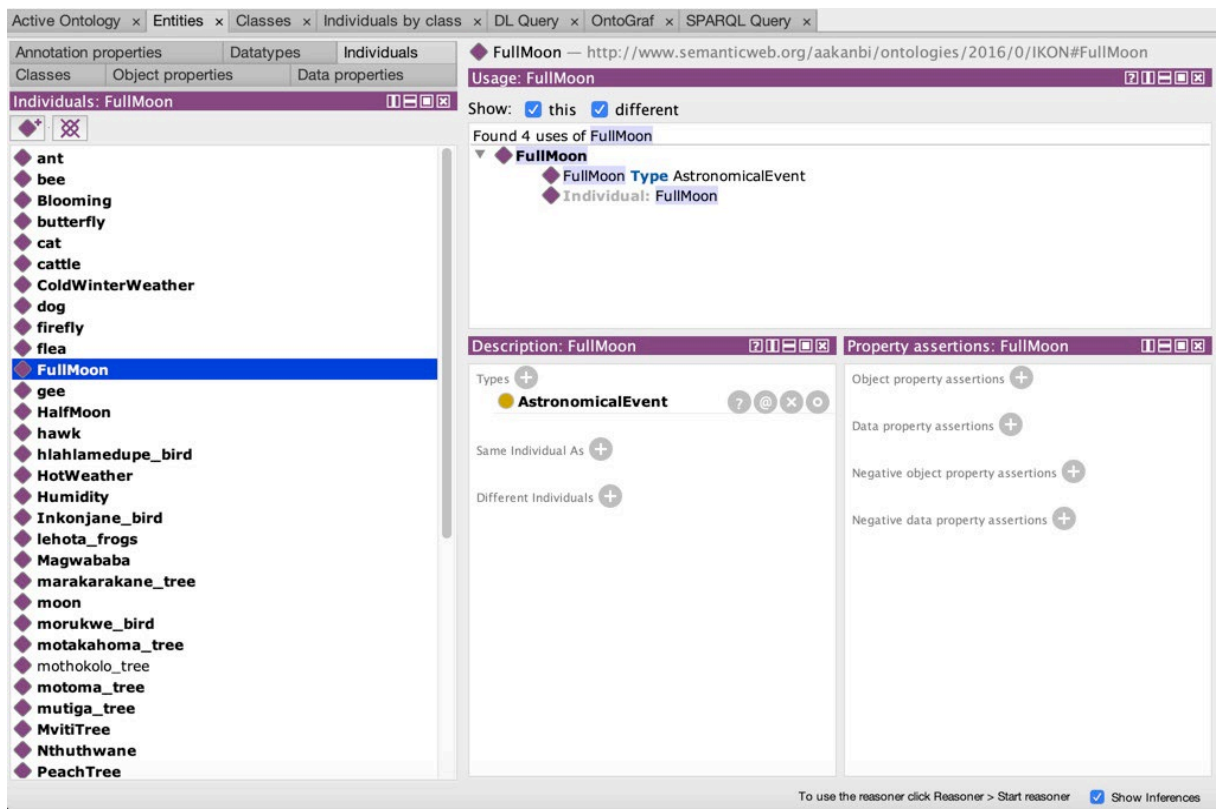

Figure 5- 5: Some Individuals of IKON Ontology (*Source: Author*).

ontology, each class has several individuals. Figure 5-5 depict some of the IKON *individuals*.

## 5.2.2. Lightweight Ontology Representation of IKON

The *lightweight ontology* representation of a domain is the visual representation of an ontology through the use of appropriate graphical notation. Lightweight ontology is tree-like structures where each node is labelled with corresponding natural language concept names. The lightweight ontology representation consists of backbone taxonomies of the domain.

In *Protégé*, there exist several plugins for the visual lightweight ontology representation of IKON ontology such as OntoGraf, OWLViz and VOWL. This research adopts the use of OntoGraf for the visual representation of the IKON ontology due to several inherent features that shows the detailed overview as well as the subsumptions relationships between the nodes (subclasses and the classes).

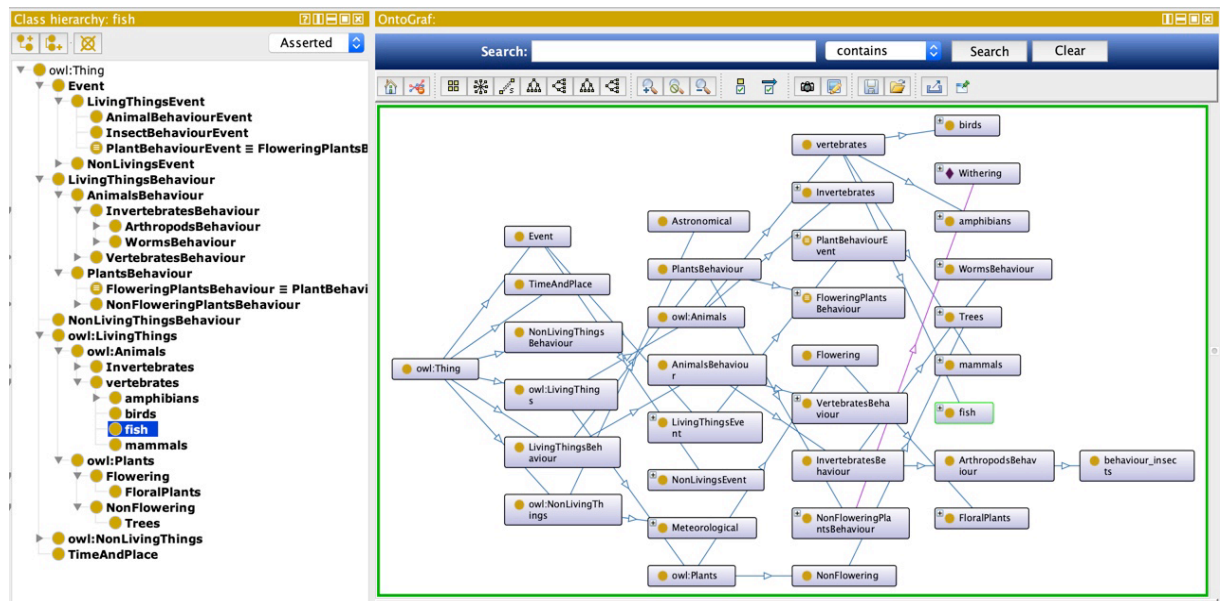

Figure 5- 6: Lightweight visual representation of IKON ontology using OntoGraf (*Source: Author*).

Figure 5-6 below shows the lightweight visual representation of IKON ontology.

### 5.2.3. Heavyweight Ontology Representation of IKON

A *heavyweight ontology* of **IKON** is an enriched version of the lightweight ontology encoded using OWL with necessary *axioms* to fix the semantic interpretation of concepts and relations. The inclusion of *axioms* is what differentiate *lightweight ontologies* from the *heavyweight ontologies*. For semantics-based information systems, a*xioms* are a critical component of the ontology module (Fürst & Trichet, 2006) and are in the form of statement, assertions and inference rules – which are used to perform deductive inference on the domain. The *heavyweight ontology* representation of IKON will allow the generation of deductive inference and automated reasoning. The heavyweight ontology representation of IKON includes axioms added to the domain ontology encoded using OWL in *Protégé*. The encoded IKON domain ontology is represented based on the OWL/XML Syntax. Figure 5-7 shows the code snippet of the representation of a class in IKON ontology. The

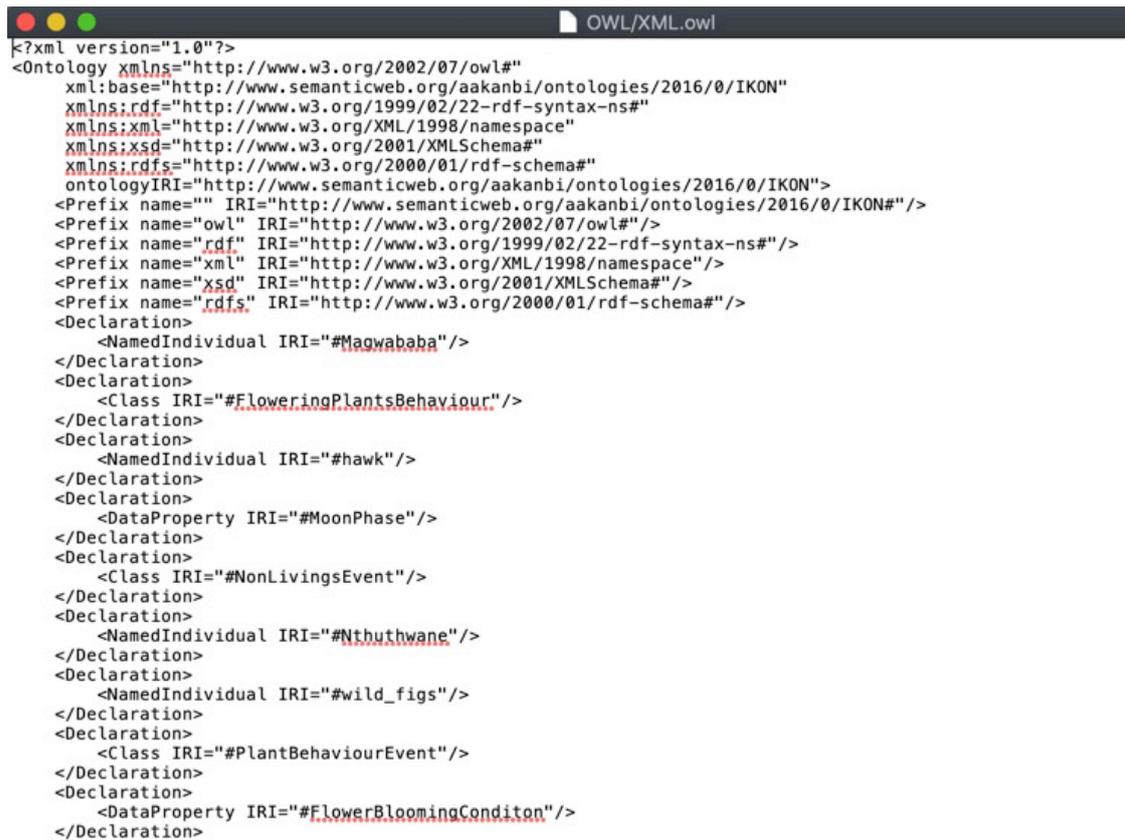

```xml
<?xml version="1.0"?>
<Ontology xmlns="http://www.w3.org/2002/07/owl#"
    xml:base="http://www.semanticweb.org/aakanbi/ontologies/2016/0/IKON"
    xmlns:rdf="http://www.w3.org/1999/02/22-rdf-syntax-ns#"
    xmlns:xml="http://www.w3.org/XML/1998/namespace"
    xmlns:xsd="http://www.w3.org/2001/XMLSchema#"
    xmlns:rdfs="http://www.w3.org/2000/01/rdf-schema#"
    ontologyIRI="http://www.semanticweb.org/aakanbi/ontologies/2016/0/IKON">
    <Prefix name="" IRI="http://www.semanticweb.org/aakanbi/ontologies/2016/0/IKON#"/>
    <Prefix name="owl" IRI="http://www.w3.org/2002/07/owl#"/>
    <Prefix name="rdf" IRI="http://www.w3.org/1999/02/22-rdf-syntax-ns#"/>
    <Prefix name="xml" IRI="http://www.w3.org/XML/1998/namespace"/>
    <Prefix name="xsd" IRI="http://www.w3.org/2001/XMLSchema#"/>
    <Prefix name="rdfs" IRI="http://www.w3.org/2000/01/rdf-schema#"/>
    <Declaration>
        <NamedIndividual IRI="#Magwababa"/>
    </Declaration>
    <Declaration>
        <Class IRI="#FloweringPlantsBehaviour"/>
    </Declaration>
    <Declaration>
        <NamedIndividual IRI="#hawk"/>
    </Declaration>
    <Declaration>
        <DataProperty IRI="#MoonPhase"/>
    </Declaration>
    <Declaration>
        <Class IRI="#NonLivingsEvent"/>
    </Declaration>
    <Declaration>
        <NamedIndividual IRI="#Nthuthwane"/>
    </Declaration>
    <Declaration>
        <NamedIndividual IRI="#wild_figs"/>
    </Declaration>
    <Declaration>
        <Class IRI="#PlantBehaviourEvent"/>
    </Declaration>
    <Declaration>
        <DataProperty IRI="#FlowerBloomingConditon"/>
    </Declaration>
```

complete OWL/XML code representation of IKON domain ontology is available on Appendix C.

### 5.2.4. Publishing and Deployment of IKON

The deployment and publishing of the developed **IKON** ontology involves the release of the ontology/knowledge model and publishing the ontology. Deployment of a domain ontology is about sharing the knowledge model with the research communities and published in a major ontology repository, for other users or researchers to download, reuse, extend or improve. **IKON** has been published as a research paper (Akanbi & Masinde, 2018c) and added to online ontology repository, available for download via Github (https://github.com/yinchar/Indigenous-Knowledge-on-Drought-Domain-Ontology) and in Appendix C and E as OWL/XML syntax and JSON-LD respectively.

Publishing and deployment of an ontology involve the ontology documentation of entities. This is as an important aspect in the development of the knowledge model, to ensure the documentation of the encoded ontology in a natural language. The HTML file of the developed IKON ontology is generated by using the Live OWL Documentation Environment (LODE) tool (Peroni, Shotton & Vitali, 2012) (Figure 5-8). The IKON OWL file is loaded to the LODE tool which automatically extracts the classes, properties, instances, axioms and namespace from the IKON OWL file and transformed the domain ontology into a human-readable HTML file with hyperlinks.

Figure 5- 7: A snippet of OWL/XML code representation of IKON (*Source: Author*).

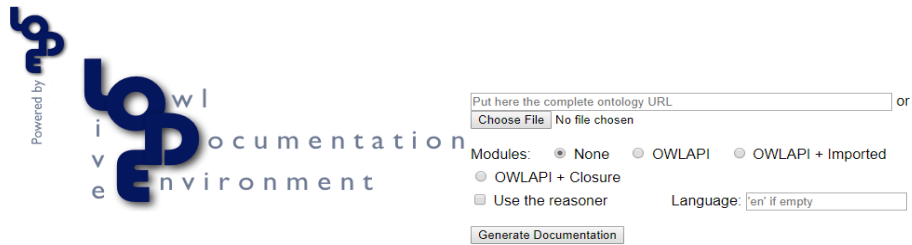

```
AddType application/rdf+xml .rdf

# Rewrite engine setup
RewriteEngine On

# Rewrite rule to serve HTML content
RewriteCond %{HTTP_ACCEPT} !application/rdf+xml.*(text/html|application/xhtml+xml)
RewriteCond %{HTTP_ACCEPT} text/html [OR]
RewriteCond %{HTTP_ACCEPT} application/xhtml+xml [OR]
RewriteCond %{HTTP_USER_AGENT} ^Mozilla/.*
RewriteRule ^ontology$ http://www.essepuntato.it/lode/http://www.mydomain.com/ontology [R=303,L]

# Rewrite rule to serve RDF/XML content if requested
RewriteCond %{HTTP_ACCEPT} application/rdf+xml
RewriteRule ^ontology$ ontology.owl [R=303]

# Choose the default response
RewriteRule ^ontology$ ontology.owl [R=303]
```

Figure 5- 8: Live OWL Documentation Environment (LODE) tool (*Source: https://essepuntato.it/lode/*).

## 5.3. Knowledge Representation of WSN (D2)

The sensor data and weather station data are in the form of raw data, formatted in binary without any metadata. This data lacks the formality and standardisation to ensure data integration with other datasets. Subsequently, this makes it difficult to generate meaningful inference and interpretation from the sensor readings. This is due to the lack of formal vocabularies to describe how observations (sensor readings) are related to the natural event (Devaraju, Kuhn & Renschler, 2015). The semantic annotation of these stream of sensor data will support data integration service interoperability and promote richer knowledge-driven use of data. The involves semantic representation of the sensor's' data using axioms that represents specific environmental property. The application of semantic model ensures the addition of variety of sensor through detailed semantic annotation of the concepts and data. This will enhance data integration and system interoperability when fusing heterogeneous sensor datasets.

In this research, the essence of the knowledge representation of **D2** through semantic-based ontology models is in two phases: a) to represent the sensor's data in machine-readable languages to enhance data and service interoperability, orchestration and extension in intelligent systems; b) to represent the *CEP engine* inference generated from the stream of time-sensitive sensors data in a shared knowledge – ontology. A semantic model incorporates explicit metadata definition and ontological concept definitions (Poslad, Middleton, Chaves, Tao, Necmioglu & Bügel, 2015). Example of a semantic model to represent these types of concepts is the Open Geospatial Consortium (OGC) Observations & Measurements Schema (O&M) (Botts, Percivall, Reed & Davidson, 2008) model and the World Wide Web Consortium (W3C) Semantic Sensor Network (SSN) ontology (Compton *et al.*, 2012). However, the OGC's O&G model is a lighter semantic model for representing concepts such as the Observed Properties, Features with the capability of a few reasoning mechanisms.

With the WSN domain (**D2**) there are existing domain ontologies for the semantic representation of the stream of sensor data, properties and inference outputs. Based on the methodology of ontology development (Noy & McGuinness, 2001), reviewing of existing ontologies and standards is paramount before developing a new ontology. Thus, this research adopted the W3C's SSN ontology for the ontological representation of sensors data and inference outputs due to the mathematical rigour, degree of expressivity and comprehensive reason capabilities and ontological alignments with DOLCE.

### 5.3.1. Axiomatisation of Semantic Sensor Network (SSN)

Currently, several conceptual modules are used to represent the sensor, actuation and sampling concepts. SSN ontology consists of eight (8) modules representing forty-one (41) concepts with thirty-nine (39) object properties. Eleven (11) concepts and fourteen (14) object properties are inherited from DOLCE-UltraLite (DUL), which is the foundational ontology (Compton *et al.*, 2012). Figure 5-9 below provides an overview of the modules.

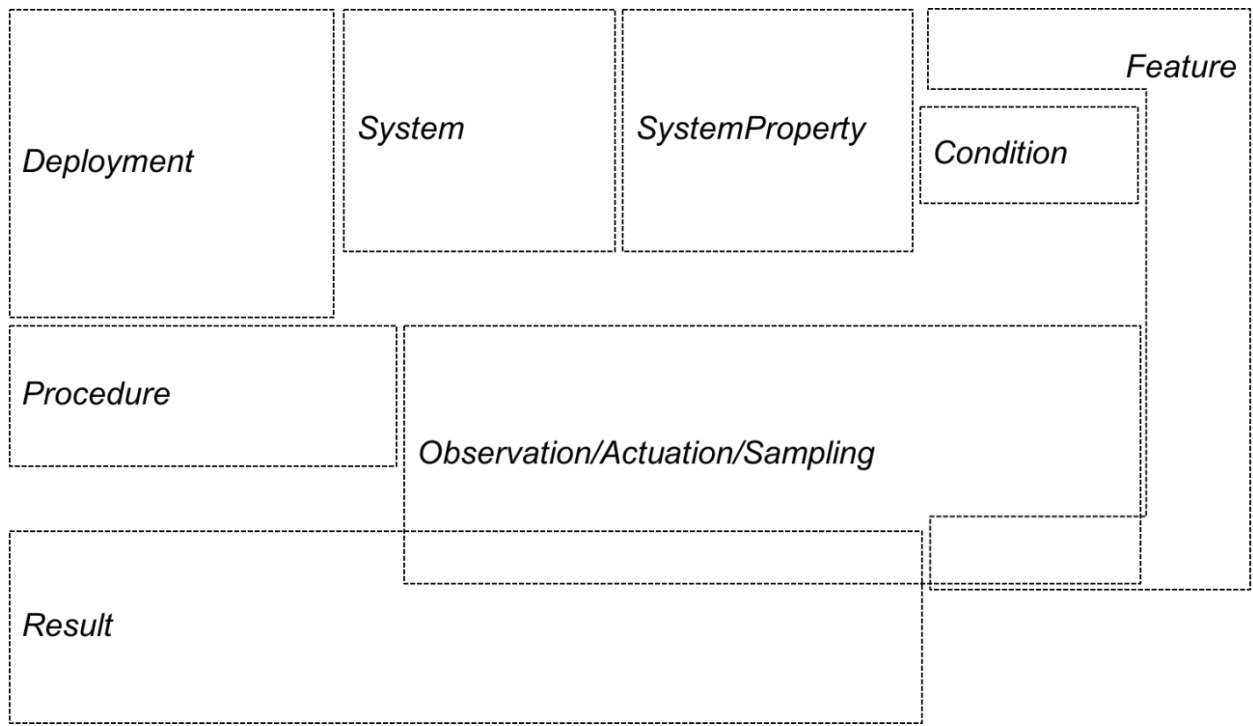

Figure 5- 9: Overview of the SOSA/SSN ontology modules (*Source:* Compton *et al.*, 2012)

The SSN ontology represents every **sensing device** as a function of the eight depicted modules. Each module contains several classes and properties inherent to it from the perspective of Observation, Actuation and Sampling. This research is only interested in the Observation paradigm of representing the **sensing device.** In other words, the *Deployment*, *System*, *SystemProperty*, *Condition*, *Feature*, *Procedure*, *Observation* and *Result* of each sensor is semantically annotated and represented using the SSN Ontology. Figure 5-10 below shows the classes and properties inherent for each module.

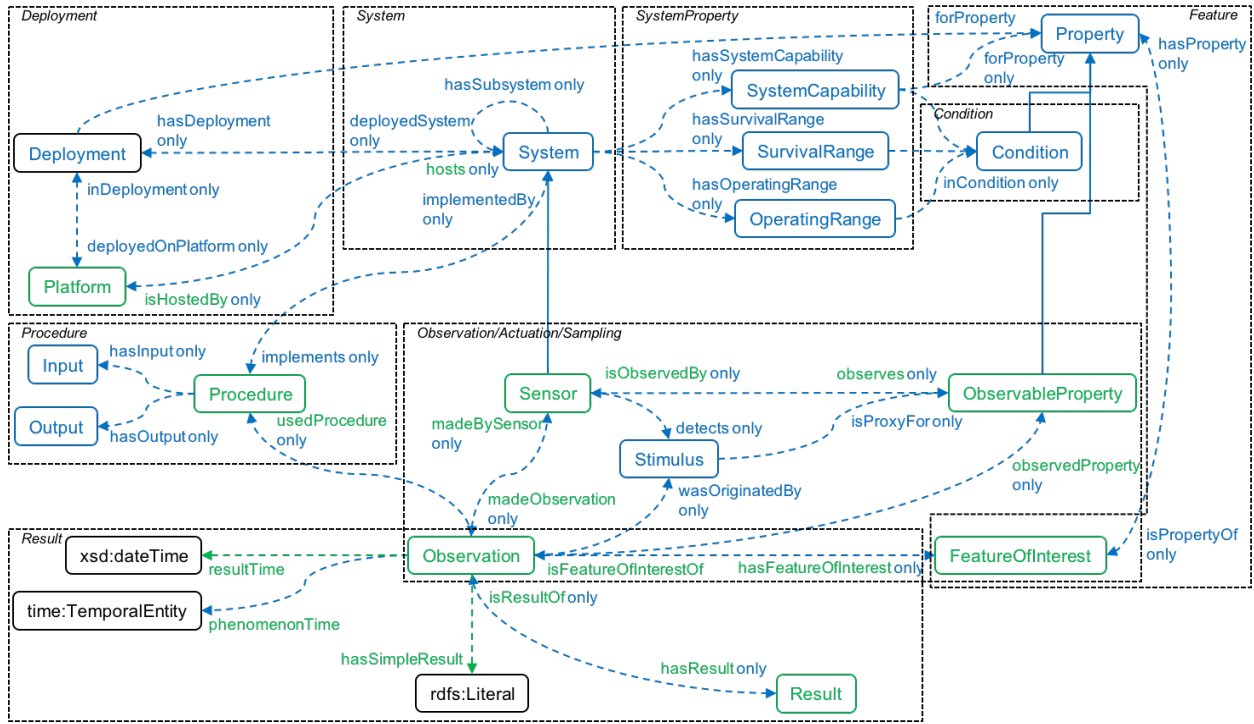

Figure 5- 10: Overview of the SSN classes and properties for the observation perspective, SSN only components in blue colour (*Source*: Compton *et al.*, 2012).

In this research we cover all the eight modules for the semantic annotation of the sensing device and its observation:

a) *Deployment module*: represents the Platform concept to indicate if the sensor is part of a platform or deployed alone. For example, a **Sensing Device** (measuring module) was hosted on a cloud repository platform (*Data Storage FG*) of the middleware. By utilising the properties *hasDeployment*, *inDeployment*, *deployedOnPlatform*, *isHostedBy*, the relationships of the concepts are modeled.

b) *System module*: represents the System concept composed by sub-systems (*hasSubsystems*) which are deployed (*deployedSystem*), hosted on a platform (*hosts*) and implemented (*implementedBy*) a procedural call or action.

c) *SystemProperty module*: covers the SystemCapability, SurvivalRange and OperatingRange using properties such as *hasSystemCapability*, *hasSurvivalRange* and *hasOperatingRange* to represent the property of the system or sensing device. For example, a DHT22 sensor deployed will have system capability to measure the temperature and humidity, with survival range as specified by the manufacturer and operating range of the minimum and maximum value that the sensor can measure correctly.

d) *Feature module*: covers the Property, the FeatureOfInterest and its Condition using properties such as *forProperty*, *hasProperty*, *inCondition* etc.

e) *Procedure module*: represents the procedural routine block of code that captures the Input and produces the Output using properties of *hasInput*, *hasOutput* and *implements*.

f) *Observation module*: represents the core concept of the SSN. The Stimulus is a core concept the Sensor is measuring after detection based on an Observation and must have an ObservableProperty. For example, the level of a water body. These classes are represented with properties such as *detects*, *observes*, *isProxyFor*, *wasOriginatedBy*, *madeObservation*, *observationResultTime*, *observationSamplingTime* etc.

g) *Result module*: covers the representation of the senosr's raw data output using annotation such as *resultTime*, *phenomenomTime*, *hasResult*, *isResultOf* with the appropriate data properties.

### 5.3.2. Application of SSN Ontology – Use Case

Succinctly, a sensor is an object that senses and measures the properties of the feature of interest. The ontological representation of the sensor and its related concepts using SSN ontology allows the generation of environmental events inference based on standardised rules expressed regarding observed properties.

This section explains the ontological representation of the soil moisture sensor (SEN13322) used in this research using the SSN ontology. The sensor and the observation are semantically represented as classes and properties of the SSN modules. The formalisation using SSN ontology provides a comprehensive specification to describe the ssn:Output, ssn:System, ssn-system:OperatingRange, ssn-system:Condition, ssn-system:SystemCapability, ssn-system:inCondition, ssn-system:hasSystemProperty, ssn-system:Accuracy, ssn-system:Sensitivity, ssn-system:Resolution, ssn:Property, ssn-system:Precision, ssn-

system:Frequency, ssn-system:qualityOfObservation. Figure 5-11 below presented the code snippet of the ontological representation of the SEN13322.

```
SEN13322 SSN Ontology Code.txt — Edited
@prefix rdf:  <http://www.w3.org/1999/02/22-rdf-syntax-ns#> .
@prefix rdfs: <http://www.w3.org/2000/01/rdf-schema#>.
@prefix xsd:  <http://www.w3.org/2001/XMLSchema#> .
@prefix qudt-1-1: <http://qudt.org/1.1/schema/qudt#> .
@prefix qudt-unit-1-1: <http://qudt.org/1.1/vocab/unit#> .
@prefix schema: <http://schema.org/>.
@prefix ex: <http://example.org/>.

@prefix sosa: <http://www.w3.org/ns/sosa/> .
@prefix ssn: <http://www.w3.org/ns/ssn/> .
@prefix ssn-system: <http://www.w3.org/ns/ssn/systems/> .

@prefix rdfp: <https://w3id.org/rdfp/>.

@base <http://example.org/data/> .

<SEN13322#Procedure> a sosa:Procedure ;
  ssn:hasOutput <SEN13322#output> .

<SEN13322#output> a ssn:Output , rdfp:GraphDescription ;
  rdfs:comment "The output is a RDF Graph that describes both the Soil moisture. It can be
validated by a SHACL shapes graph."@en ;
  rdfp:presentedBy [
    a rdfp:GraphDescription ;
    rdfp:validationRule <shacl_shapes_graph> ;
  ].

<SEN13322/4578> a ssn:System ;
  rdfs:comment "A sensing device contains a soil moisture sensor."@en ;
  rdfs:seeAlso <https://www.cdn.sparkfun.com/datasheets/Sensors/Biometric/
SparkFun_Soil_Moisture_Sensor.pdf> ;
  ssn:hasSubSystem <SEN13322/4578#SoilMoistureSensor>.

<SEN13322/4578#SoilMoistureSensor> a sosa:Sensor , ssn:System ;
  rdfs:comment "The embedded Soil Moisture sensor, a specific instance of Soil Moisture
sensor."@en ;
  ssn-system:hasOperatingRange <SEN13322/4578#SoilMoistureOperatingRange> ;
  ssn-system:hasSystemCapability <SEN13322/4578#SoilMoistureSensorCapability> ;
  ssn:implements <SEN13322#Procedure> .

<SEN13322/4578#SoilMoistureSensorOperatingRange> a ssn-system:OperatingRange ;
  rdfs:comment "The conditions in which the SEN13322 Soil Moisture sensor is expected to
operate."@en ;
  ssn-system:inCondition <NormalSoilMoistureCondition> .

<NormalOperatingCondition> a ssn-system:Condition , schema:PropertyValue ;
  rdfs:comment "A soil moisture range of 0 to 880."@en ;
  schema:minValue 0 ;
  schema:maxValue 880 ;
  schema:unitCode qudt-unit-1-1:value .
```

Figure 5- 11: A snippet of ontological representation of a SEN13322 sensor using SSN ontology (*Source: Author*)

The sensing unit's data are represented using SSN ontology. For example, in this case study of SEN13322 soil moisture sensor, the object property ssn:System provides the possibility to semantically annotate the description of the sensing device using the rdfs:comment and a sub-system ssn:hasSubSystem represents the relationship between the sensing device (microcontroller) and the soil moisture sensor (SEN13322). The ssn-system:OperatingRange represents the operating range of the sensor, in this particular instance, the range the SEN13322 Soil Moisture sensor is expected to operate. This is followed by the ssn-system:Condition an object property for the range of operation of the soil moisture, from ~0 to ~880 in accordance to the manufacturer's specification. The ssn-system:Accuracy annotates the accuracy of the Soil Moisture sensor which is 3% in all conditions. The accuracy value is represented with the schema:PropertyValue, using range schema:minValue 0 to schema:maxValue 3; and the unit value in percentage is represented as schema:unitCode qudt-unit-1-1:Percentage. The ssn-system:Sensitivity and ssn-system:Resolution of the Soil Moisture sensor is 0.1% VWC in normal conditions, represented with schema:PropertyValue of 0.1%.

Furthermore, the quality of the observation based on the existing parameters of the sensor can be represented using class ssn-system:qualityOfObservation (Figure 5-12) or subsequently use another quality ontology. The quality of the observation can be evaluated, and the attributed confidence value of the sensor observation declared as part of the ssn-system:qualityOfObservation.

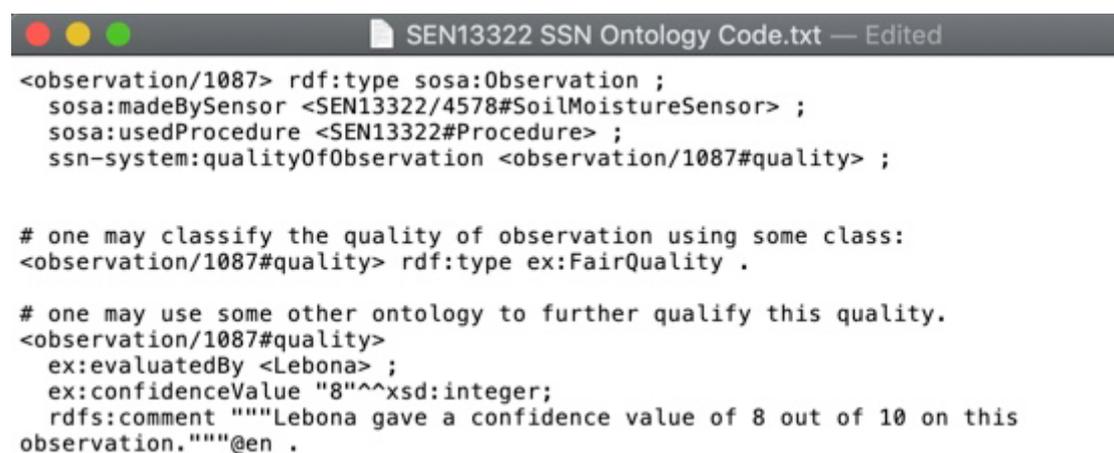

Figure 5- 12: A snippet of class ssn-system:qualityOfObservation (*Source: Author*).

## 5.4. Implementation Scenario

Figure 5-13 illustrate the integration scenarios of the semantic representation of data sources (**D1 & D2**) in the *Stream Analytics FG* and *Inference Engine FG* of the developed SB-DIM middleware. For **D1** – scenario **B** in the figure below, the inference is generated using an expert system inference engine module of the *Inference Engine FG*. The oral local indigenous knowledge gathered from the *Data Acquisition FG* is pre-processed at the *Data Storage FG* and represented in the digital format where the local indicators in the form of *rules* and interpretation of the rules are identified. These *rules* are saved in the knowledge base of the RB-DEWS module of the *Inference Engine FG* to deduct inference from the set of observations. The generated inference is semantically represented using the **IKON** domain ontology and also pushed to the *Data Publishing FG* of the middleware.

For domain data **D2** – scenario **A** in Figure 5-13. The inference is generated using stream processing engine of the *Stream Analytics FG* of the middleware. The stream of sensor readings/observation from the WSN are in the form of Machine-2-Machine (M2M) raw data generated at the *Data Acquisition FG* and is streamed through the storage blobs in the *Data Storage FG*. The data stream is then processed through the streaming platform and engine of the *Stream Analytics FG* to infer patterns from the sensor data based on the prediction model logic. The prediction logic is an EDI drought prediction or forecasting model represented in EP language. The data streams are queried in real-time, and the deductive inference generated by the *Stream Analytics FG* is semantically represented based on the ontology and also pushed to the *Data Publishing FG* of the middleware.

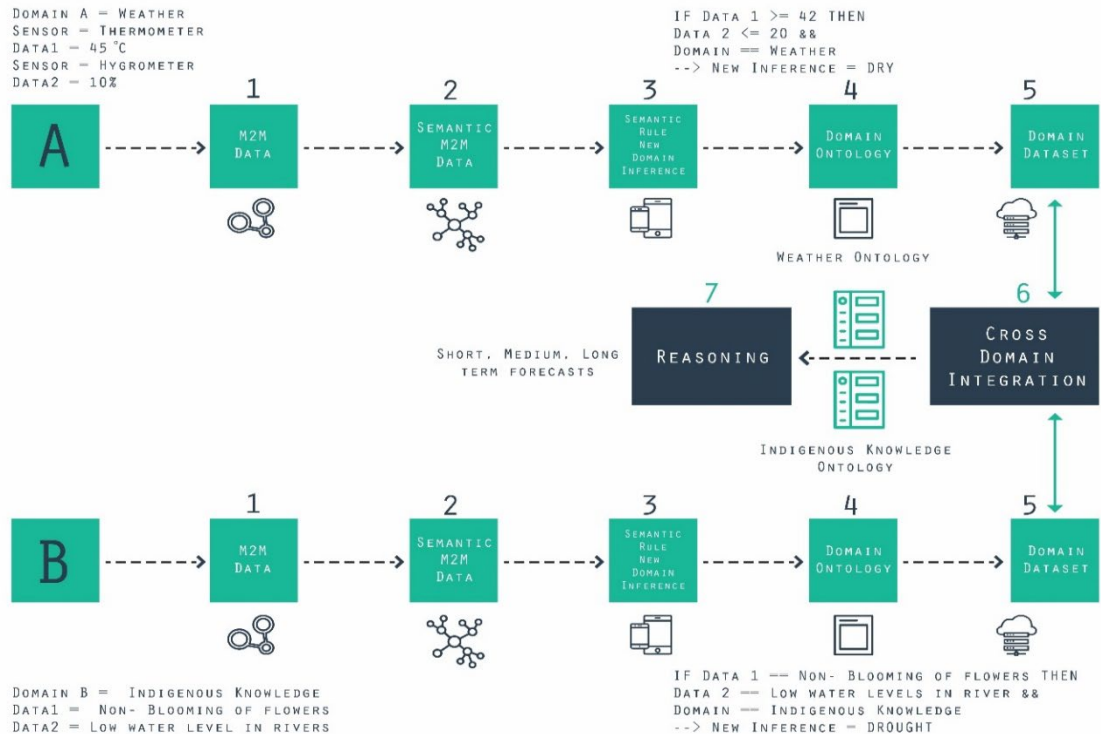

Figure 5- 13: Integration scenarios of semantic represented heterogeneous data sources (*Sources: Author*)

*Rule-based Reasoning.* The generation of inference in this domain requires additional reasoning techniques beyond that supported by the standard reasoning with OWL-DL semantics. This has been proven and adopted in several related research projects (Devaraju *et al.*, 2015; Borgo *et al.*, 2016; Patni, 2011). Hence, we employed a rule-based mechanism to perform the first set of deductive inference on the input data (See Chapter 6). A *rule* has the form "*IF Condition1 and Condition2; Then Action1, Action2,..*" (Figure 5-14).

In the case of the local IK on drought domain (**D1**), the oral documented knowledge is elicited for natural indicators in the form of rules for interpreting an observation or occurrences. The saved *rules* are for generating a new inference based on the set of inputs. The generated inference and the input data are semantically represented in a machine-readable language (OWL) based on the **IKON** ontology in the form of domain dataset for the cross-domain integration.

```
If      UID1==UmphenjaneIsBlooming        &&      CF==0.90
        UID2==MediumStreamLevel            &&      CF==0.50
        UID3==PhezukomkhonoIsSighted       &&      CF==0.80
        Domain==Local Indigenous Knowledge on Drought
```
Figure 5- 14: Example of an expert system rule definition for **D1** data (*Source: Author*).

*Stream Processing Reasoning.* Generating inference from the data streams **(D2)**. The sensing device measures the parameters every 1 min, and the readings are pushed to the *Stream Analytics FG* using RESTful services. The streaming engine of the *Stream Analytics FG* uses a persistent query system to infers patterns from the data streams based on a prediction model logic or set of pre-conditional rules. An example of the predictive logic query structure is depicted in Figure 5-15.

Figure 5- 15: Example of CEP persistent query logic for performing deductive inference from WSN data streams (*Source: Author*)

*Standard Ontological Reasoning.* Because both our domain ontologies share the same foundational ontologies, i.e. DOLCE, there is a perfect integration and

```
If        DHT22_Sensor>=45        &&
          SEN1322_Sensor<=10      &&
          Domain==weather

Then      New Inference = DRY
```

alignment of the semantically annotated domain datasets. Therefore, further standard ontological reasoning could be performed on the datasets to generate short, medium or long-term forecasts. The standard reasoning is done with Pellet OWL reasoners to check for the knowledge model consistency, deduce the forecast information and update the model with inferred information. This is outside the scope of this research work.

## 5.5.   Summary

This chapter presents the development and semantic representation of the heterogeneous data sources using ontologies. The ontology modules that perform the formalisation, semantic representation of the domains and data sources is a component of the *Inference Engine FG* of the middleware. The contribution of the chapter lies in the development and encoding of **I**ndigenous **K**nowledge on Drought Domain **ON**tology (**IKON**), which captures and models the description of local indicators related to drought forecasting in the area under study, using the entities, ecological interactions and behavioural relationships. The **IKON** ontology can be used to understand the overview of intricate indigenous knowledge on drought. The mapping of the semantic annotated observations or behaviours to the class entities results in the formalisation of domain knowledge and allows generating drought-related inference from events and sensor's data automatically.

This chapter presents the SSN ontology which uses declarative descriptions of sensors, networks and domain concepts to aid in searching, querying and managing the data sources. Both ontologies extend the functionalities of DOLCE, which aids cross-domain data integration and ontology alignment. The semantic annotations link the sensor data to more expressive ontological representations using reference models. This ensures that sensor data has semantic descriptions that would enhance heterogeneous data integration, and generation of accurate inference.

# CHAPTER SIX

# AUTOMATED INFERENCE GENERATION SYSTEMS

## 6.1. Introduction

In this chapter, the *Inference Engine FG* is presented that consists of the module to perform deductive inference from the local indigenous knowledge on drought (**D1**) and *Stream Analytics FG* that consists of the technological framework – **ESTemd** (**E**vent **ST**ream Processing Engine for **E**nvironmental **M**onitoring **D**omain) which is an event processing stack for the real-time data analytics of drought forecasting on the data streams from the deployed environmental monitoring sensors (**D2**) of the *Data Acquisition FG*. The RB-DEWES and ESTemd can be deployed in distributed mode as an FG of the distributed semantic middleware. In distributed mode, the *Stream Analytics FG* and *Inference Engine FG* consist of the deployed sensors, cloud-based infrastructure, stream processing engine using open-source *Apache Kafka*, JESS inference engine, notification system, adapters and APIs needed to perform the real-time data processing and analytics.

The inference generated from the local indigenous knowledge on drought (**D1**) inference engine *(Inference Engine FG)* is merged with the inference generated from the WSN data (**D2**) automated reasoners (*Stream Analytics FG*) for the creation of DFAI which is sent to the *Data Publishing FG* of the middleware for publishing. The published DFAI with attributed certainty factor is proposed to be used by the policymakers in their decision-making processes

## 6.2.    Rule-based Drought Early Warning Expert System (RB-DEWES)

The RB-DEWES is a software module and component of the *Inference Engine FG* of the distributed semantic-based data integration middleware aim at performing deductive inference from the acquired local indigenous knowledge on drought (**D1**) (Akanbi & Masinde, 2018a). This software module is tasked with the generation of drought forecasting inference from a set of input using the *rules* derived from the local indicators/observations on drought in the study areas. The sub-system utilises the domain indigenous knowledge and acquired facts stored in the *Data Storage FG*. The *rules* derived from the gathered knowledge indicators are saved in the knowledge base; and used by the inference engine for generating inference from a set of inputs.

In IK on drought, after the knowledge representation of the domain knowledge, natural indicators, their relationships, ecological interactions and interpretation of the scenarios are implicitly identified and is formulated in the form of *rules* making the adoption of an expert system with inference engine for reasoning suitable for automated generation of drought prediction inferences. A review of existing research projects and literature (Giarratano & Riley, 1998; Weiss & Kulikowski, 1991; Borgo *et al.*, 2014) emphasised the ability of the expert system in the reproduction of reasoning capabilities of the domain experts by formalising their knowledge for implicit reasoning through the emulation of human thoughts.

### 6.2.1.  Rules Ranking with Certainty Factor from Indigenous Knowledge Representation

Derivation knowledge, control knowledge and factual knowledge on drought acquired from the domain experts need to be represented and transformed into *rules* for use by the inference engine component of the RB-DEWES. Hence, knowledge representation process aims to encode the domain expert knowledge on drought. The researcher recognise that the study giving up an attractive feature of indigenous knowledge: a homogeneous definition of terms, concepts and events. However, the knowledge representation is a must and is achieved through the formalisation of local indicators and scenarios such as the sighting of a local indicator or ecological interactions into *rules;* using a rule-based programming style. Table 6-1 lists some of the main animals, plants, meteorological and astronomical indicators included in the expert system. Other includes the behavioural scenarios which are subjected to interpretations.

Table 6- 1: Indigenous animal, plants, meteorological, astronomical indicators included in the expert system.

| Animals | Plants | Meteorological | Astronomical |
|---|---|---|---|
| *Magwababa* bird | *Mviti* tree | Humidity | Full Moon |
| *Inkonjane* bird | *Wiki-jolo* tree | Soil Moisture | Half moon |
| *Ntuthwane* ant | *Umphenejane* tree | Weather temperature | Stars |
| *Ingxangxa* frog | Peach trees | Rainfall | Day Sky |
| *Onogolantethe* bird | *Amapetjies* tree | Thunderstorm | Night Sky |
| *Phezukomkhono* bird | *Tshi* tree | Sunlight intensity | Cloud patterns |
| Cows | *Motoma* tree | Windstorm | |
| *Inyosi* | *Marakarakane tree* | | |
| *Lehota* frog | *Mutiga* tree | | |
| All_animals | All_plants | | |

The *rule base* for RB-DEWES currently contains 33 natural indicators (behavioural observation, astronomical, meteorological), with the capability of adding additional indicators in the future. Each indicator has its corresponding *certainty factor (*CF*),* which is a measure of the indicator's relevance to natural occurrences, as determined by the focus group based on years of experience (Table 6-2) (Chu, Hwang, 2008).

Table 6- 2: Certainty Factor (CF) ranking scale.

| Percentage Scale (%) | Certainty Factor (CF) |
|---|---|
| 0 - 10 | 0.1 |
| 11 - 20 | 0.2 |
| 21 - 30 | 0.3 |
| 31 - 40 | 0.4 |
| 41 - 50 | 0.5 |
| 51 - 60 | 0.6 |
| 61 - 70 | 0.7 |
| 71 - 80 | 0.8 |
| 81 - 90 | 0.9 |
| 91 - 100 | 1.0 |

All *rules* comprise the natural indicator observation, or ecological scenario are represented as Object-Attribute-Value (O-A-V) by the expert system as shown in Table 6-3. Several indicators or observation scenarios can be combined in the expert system to improve the accuracy of the inference mechanism based on the user's input. Observation by a user is captured by the system with the user indicating the level of certainty (CF) of observing the captured scenario/observation. This helps the system to perform deductive inference using probabilistic forward-chaining method in calculating the overall CF attributed to the inferred output.

Table 6- 3: Representation of natural indicators and observation in O-A-V form.

| Rule condition | Object | Attribute | Value | *CF* |
|---|---|---|---|---|
| RC2 | *Umphenjane* | *Is* | *Blooming* | *0.40* |
| RC5 | *Soil moisture* | *Is* | *High* | *0.50* |
| RC6 | *Phezukomkhono* | *Is* | *Sighted* | *0.60* |
| RC10 | *Humidity* | *Is* | *High* | *0.60* |
| RC15 | *Mviti* | *Shows* | *Wilting* | *0.70* |
| RC15 | *Inyosibees* | *Is* | *Sighted* | *0.70* |
| RC15 | *Moon* | *Appears* | *Full* | *0.70* |
| RC17 | *All_animals* | *Appears* | *Thin* | *0.50* |
| RC17 | *All_plants* | *Shows* | *Withering* | *0.50* |

### 6.2.2. RB-DEWS Module Architecture

The architectural overview and components of the RB-DEWES are depicted in Figure 6-1 below. This sub-system consists of five (5) main components: (i) the Graphical User Interface (GUI), (ii) a database, (iii) inference engine, (iv) knowledge base, and (v) model base. The developed RB-DEWES module was implemented as a standalone distributed component of the middleware with the necessary GUI for interacting with the system while maintaining a uniform data pipeline for seamless integration with other FG components.

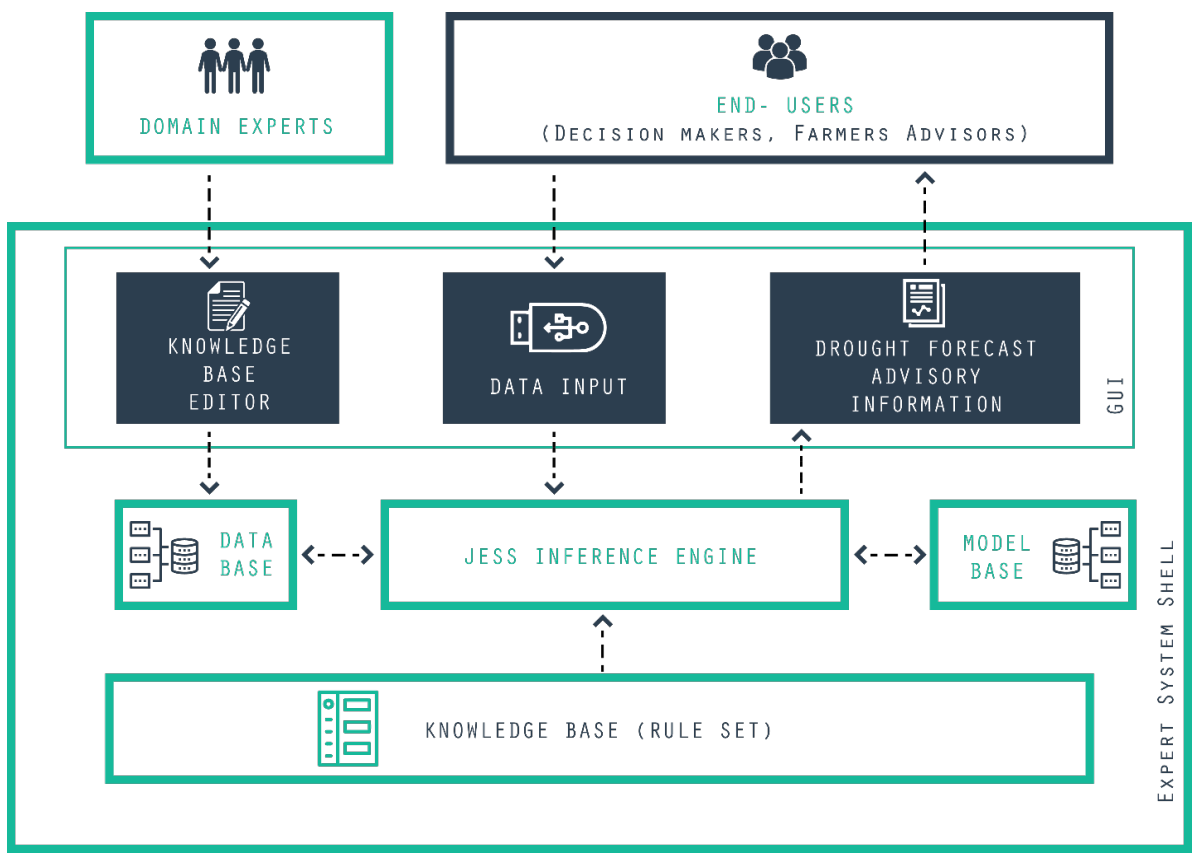

Figure 6- 1: The architecture of RB-DEWES (*Source: Author*).

### 6.2.2.1.    Graphical User Interface (GUI)

The GUI provides the interface that facilitates the communication with the frontend and the backend of the expert system module of the *Inference Engine FG*. Hence, there are two types of GUIs for accessing the system – the Frontend GUI and the Backend GUI. The frontend GUI (Figure 6-2) is designed to be user-friendly and achieve the desired usability. It provides the links that allow a non-registered user to create a profile and subsequently log-in to the system. On successful login to the system, the user can generate drought forecasting inference based on the response to a set of systems pre-programmed local indicator observation or scenarios. The user will click on the *Generate Inference* to start a new session of inference mechanism. Also, the *Reasoning History* provides an archive of previous inference outputs for record purposes. A different set of interfaces are designed for knowledge base editor, data input with CF and inference output in the form of Drought Forecast Advisory Information (DFAI) with attributed CF.

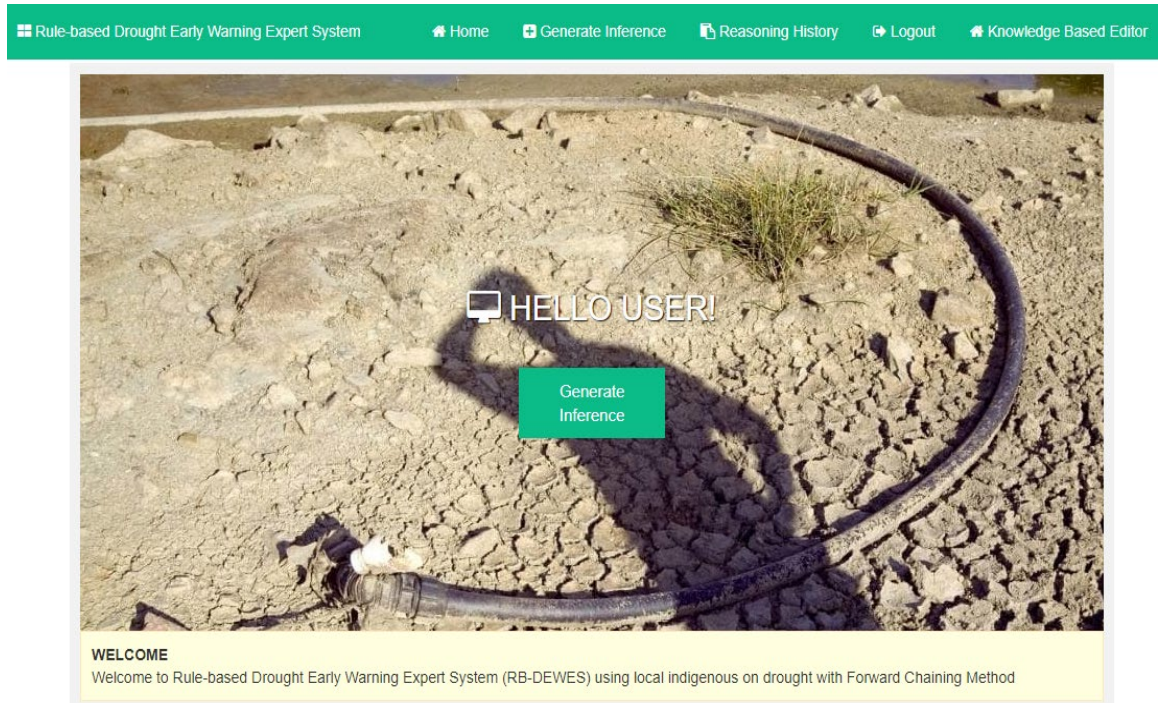

Figure 6- 2: RB-DEWES Frontend GUI (*Source: Author*).

The clicking of *Generate Inference*, the system interface displays a series of preconfigured local indicator occurrence or observation in a sequential fashion. The users have to select the appropriate option "Yes" or "No" as a response to the question. For instance, as displayed in Figure 6-3, the users have to reply to the first set of the question – "Do you experience observation/scenario like *Umphenjane* is blooming?

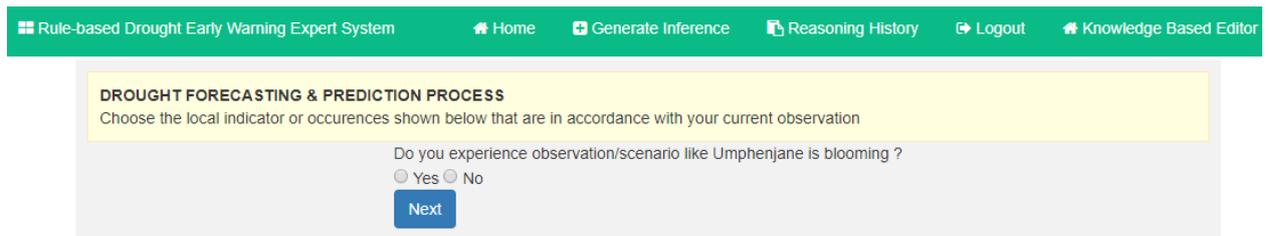

Figure 6- 3: Screenshot of RB-DEWES Inference Generation Process (*Source:*
*Author*).

At the end of the inference mechanism, the inference engine generates the inference and determine the classification and type of drought based on the severity using the EDI scale. The system captures the CF of each user's input observation/scenarios to calculate the CF or confidence level of the system's inferred output (Figure 6-4).

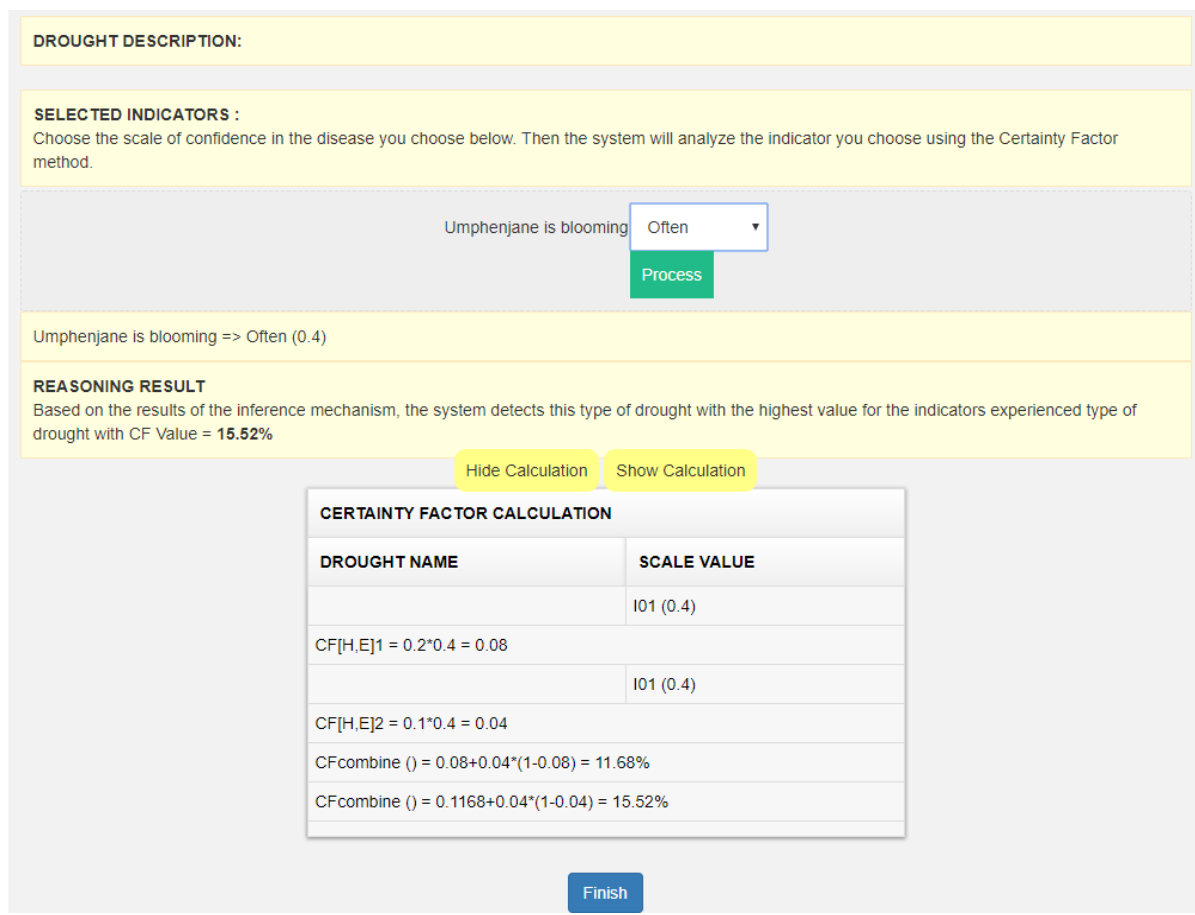

Figure 6- 4: A screenshot of Inference Output (*Source: Author*).

The backend GUI depicted in Figure 6-5 below provides the interface to the Knowledge Base Editor (KBE). This interface allows the knowledge engineer to add and edit the relevant section of the database through a user-friendly interface.

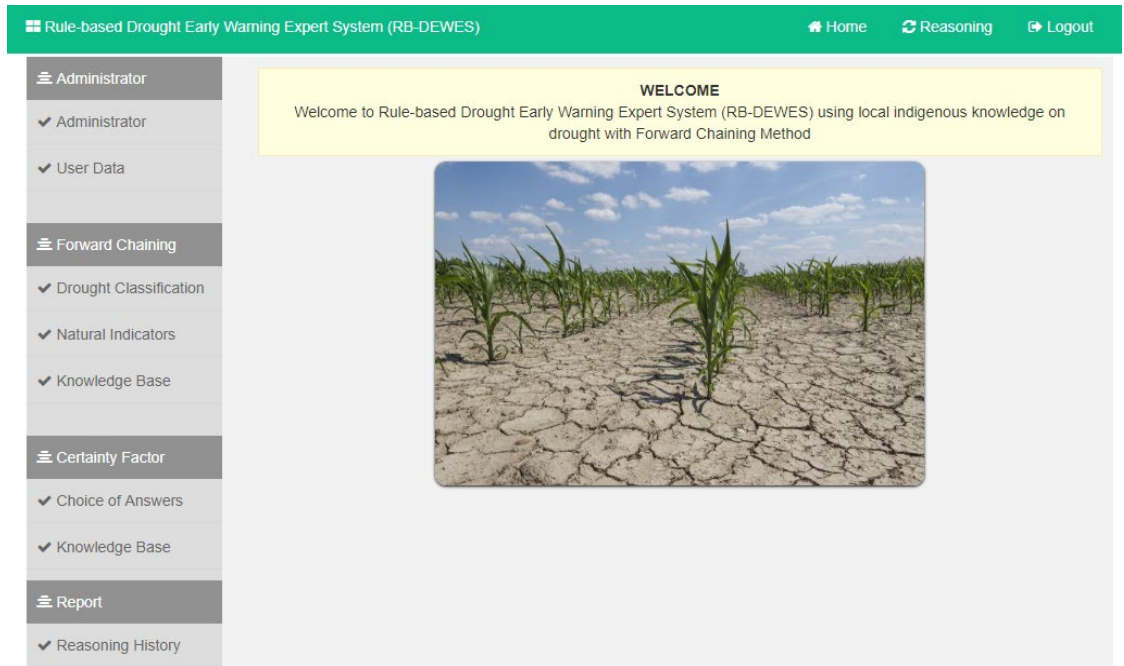

Figure 6- 5: Knowledge Editor Interface (*Source: Author*).

Through this interface the KE can perform knowledge base administration; add or edit the drought classification records (Figure 6-6); add, edit and delete from the natural indicators list; and specify the calculation for the *certainty factor*.

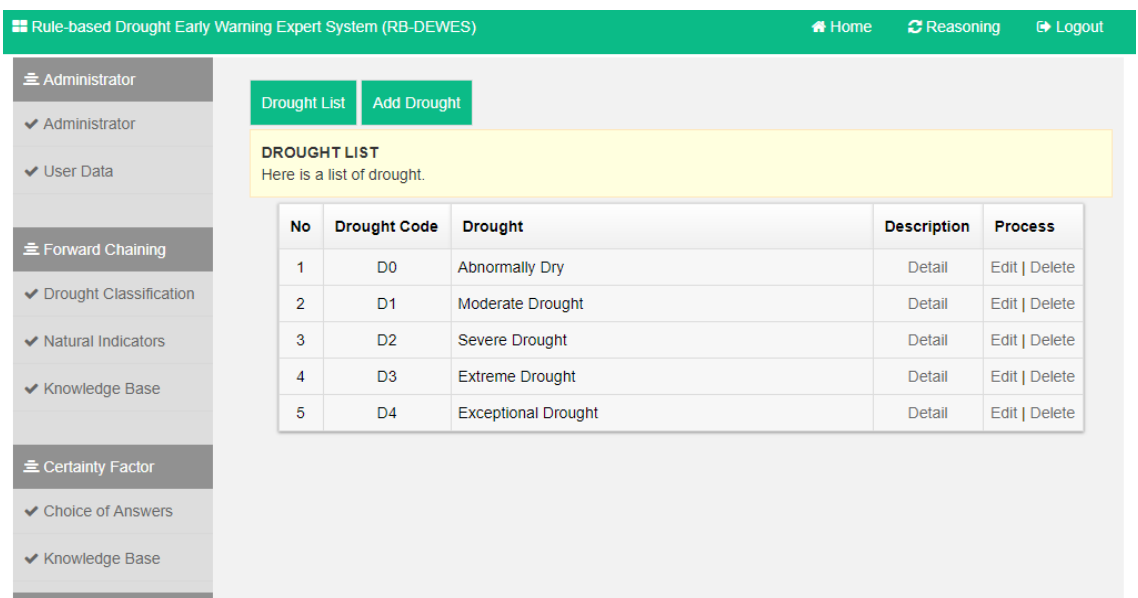

Figure 6- 6: Knowledge base administration interface (*Source: Author*).

### 6.2.2.2. Database

The database component of the expert system utilises a SQL-based relational database to store the indigenous knowledge on drought. The database is used to store the natural indicators, scenarios, CFs, classification of droughts and drought forecast advisory information. For the expert systems and database schema definition see Appendix F.

### 6.2.2.3. JESS Inference Engine

The function of the inference engine component is to perform the rule-based reasoning using forward chaining technique (Sasikumar, Ramani, Raman, Anjaneyulu & Chandrasekar, 2007). The engine is programmed and makes use of the Java Expert System Shell (JESS) (Hill, 2003). This component contains the software code that process the users selected local indicators/observation based on the rules derived from the domain expert knowledge. It predicts the onset of droughts based on the rule patterns experience stored in the knowledge base and generates part of the DFAI output.

### 6.2.2.4. Knowledge Base

The knowledge base is the repository to store the domain knowledge represented in the form of *rules*. This storage component is also used to save the inference output or interpretation from a combination of several *rules*. The interpretations or inference outputs are represented in O-A-V pattern and saved in the knowledge base. The following sample *rules* for local indicators and scenarios are listed below:

*RC18:  IF rainfall is High*
*AND soil moisture is high*
*AND soil temperature is moderate*
*THEN no evidence of drought (0.9)*
*RC21: IF phezukomkhono is sighted*
*AND Guavatree is flowering*
*AND Wiki-Jolo is blooming*
*AND Umphenjane is flowering*
*THEN No evidence of drought, onset of spring (0.85)*
*RC30: IF mviti tree is flowering*
*AND weather temperature is high*

AND ntuthwane ant was sighted
AND soil moisture is low
AND amapetjies is flowering
THEN No evidence of drought, onset of summer (0.70)

RC15: IF mviti shows wilting
AND Inyosibees is sighted
AND Moon appears full
THEN moderate evidence of drought, onset of autumn (0.75)

RC2:  IF umphenjane is blooming
THEN no evidence of drought (0.4)

RC5:  IF soil moisture is high
THEN no evidence of drought (0.5)

RC6:  IF phezukomkhono is sighted
THEN no evidence of drought (0.6)

RC10:     IF humidity is high
THEN no evidence of drought (0.6)

RC38: IF all_animals are thin
AND all_plants shows withering
AND humidity is high
AND rainfall is none
AND day sky appears clear
AND night sky is clear
AND stars are sighted
AND weather temperature is high
AND sunlight intensity is high
THEN evidence of drought (0.68)

Figure 6- 7: Production rules in the knowledge base (*Source: Author*).

### 6.2.2.5. Model Base

The model base component of the expert systems executes the probabilistic forward chaining algorithm that determines the certainty level of the output of the system. The qualitative probabilistic model is based on MYCIN (Shortliffe, Davis, Axline, Buchanan, Green & Cohen, 1975) and attributes calculated certainty factor to the inferred output.

### 6.2.3. RB-DEWS Module System Design and Implementation

The RB-DEWES is a modular sub-system of the *Inference Engine FG* of the distributed semantic middleware. The sub-system is compatible with the data representation and communication format of the middleware. The overall inference output is represented using JSON and merged with the inference from the streaming engine from the *Stream Analytics FG* to form the DFAI which is disseminated by the *Data Publishing FG* of the middleware. The DFAI output can also be integrated with other intelligent systems through the use of appropriate RESTful APIs.

As stated earlier, the RB-DEWES module was developed in a way that can be implemented as a standalone application for use independently of the middleware, in a situation where there are challenges obtaining drought prediction inference from **D2** – due to lack of data**,** or for quick inference generation based on a unique dataset. This will ensure a wider usage by policymakers for forecasting and predicting drought in the study areas. The RB-DEWES was implemented on Microsoft Windows and MacOS platform through the use of compatible web services. The minimum hardware and software requirement are as follows.

### 6.2.3.1. Software Component

RB-DEWS makes use of Java Expert System Shell (JESS) with SQL database for operation. The minimum requirement for the software either as a standalone or part of the middleware at runtime are:

- JAVA SE Runtime Environment 7
- SQL Server 2012
- Microsoft Windows OS 7

- MacOS Snow Leopard
- Web browser.

### 6.2.3.2.    Hardware Component

The hardware platform on which the RB-DEWS will be developed and, if different, where it will be run, is a major consideration when developing the module. Bearing in mind that the system is a component of the *Inference Engine FG* of the SBDIM Middleware, future reflection was considered for the use of the expert system as a standalone application-independent from the suite of *functional groups* of the middleware.   Hence, the system was developed to be compatible with five general platforms: personal computers, workstations, minicomputers, mainframe computers (servers), online cloud systems. However, the minimum hardware requirements are:

- A PC or Mac with Intel CPU processor, 4GB RAM and 2GB hard drive space.
- A VGA monitor.

### 6.2.4. RB-DEWS Module Implementation Operation

The system components were developed using a suite of programming languages such as JavaScript, PHP, HTML5, SQL etc. The frontend and backend GUI were developed using HTML5 – CSS, inference engine was based on Java Expert Shell Script (JESS) using JavaScript and PHP, while the knowledge base is a relational database – SQL.

### 6.2.4.1.    Module Execution

At the start of each drought forecasting and predicting session, a normal user is prompted to login into the system via GUI to commence the inference generation process. However, there are other available interfaces, such as the – knowledge base editor, data input and output. The user operates the system through the GUI and supplies data using push buttons, radio buttons, drop-down list, and text–field. The knowledge base editor interface allows the domain expert to add, edit, and delete rules and other contents in the knowledge base and database.

The data input interface displays a sequence of pre-defined observation and natural indicators to the end user. The user responds in affirmative to the sighting or observation of a scenario/local indicators. Multiple observation or occurrence(s) of natural indicators can be selected. The systems perform the deductive inference based on the user's responses using the *rules* stored the knowledge base. After each inference, the DFAI is generated as output with attributed CF; indicating the system level of certainty based on the users input.

### 6.2.5. Reasoning with Uncertainty

Determining the level of certainty in decision-making programs is very critical (Laudon & Laudon, 2000). In an expert system, the vagueness of expert rules and ambiguities in users' input are the major factors affecting the absolute certainty of system outputs. Hence, an expert system must exhibit a high level of modularity, and each rule may have associated with it a *certainty factor* (CF). The CF is a measure of the confidence in the piece of knowledge or observation of natural indicators (Juristo & Morant, 1998). However, there are many ways in which CFs can be defined and combined with the inference process. Our system incorporates the MYCIN model (Shortliffe *et al.*, 1975) for calculating the certainty factor (CF). The model ensures the rule probability is calculated by multiplying the domain expert implication probability by the user's input precondition probability. The domain expert implied probability is stated in the *rule* and expresses the expert confidence level based on a set of condition(s) (Akanbi & Masinde, 2018a). On the other hand, the user's input precondition probability determined by the user is also utilised. The CF value was calculated applying the formula:

$$P = P_{old} + (1 - P_{old}) * P_{new} \quad ......\text{Equation 6-1}$$

For example, the end-user input the following preconditions and their corresponding *certainty factors* (CF) of their observation through the system GUI (Table 6-4).

Table 6- 4: A random dataset of users input.

| User Input ID | Object | Attribute | Value | Relation | CF |
|---|---|---|---|---|---|
| UIID4 | *Umphenjane* | *is* | *Blooming* | && | 0.90 |
| UIID7 | *Soil moisture* | *is* | *High* | && | 0.50 |
| UIID8 | *Phezukomkhono* | *is* | *flocking* | && | 0.80 |
| UIID23 | *Relative humidity* | *is* | *High* | && | 0.70 |

The interpretation of the likely combination of several natural indicators/scenarios – UIID4 && UIID7 && UIID8 && UIID23, as obtained from the domain expert during the knowledge acquisition phase means "*No evidence of drought*" with the domain expert *certainty factor* (CF) of 0.80 as represented in Table 6-5 below.

Table 6- 5: Rule R28 in the knowledge base.

| Rule Number | IF | Relation | THEN | CF |
|---|---|---|---|---|
| R28 | UIID4 | && | | |
| | UIID7 | && | | |
| | UIID8 | && | | |
| | UIID23 | && | No evidence of drought | 0.80 |

Therefore, since the relation of all the preconditions is "AND". Using MYCIN model, the overall probability of the preconditions is given by the minimum CF of the precondition set, i.e. min[UIID4(CF), UIID7(CF), UIID8(CF), UIID22(CF)]. Therefore, the probability of the preconditions is: min(0.9,0.5,0.8,0.7) = 0.5. The CF of the inferred knowledge based on the RC28 will be as 0.8*0.5 = 0.4 = 40%. Therefore, the model base will attribute a CF value of 40% to the inferred output.

## 6.3. Streaming Analytics FG

In the novel approach of generating inference from the sensors data streams (**D2**), **ESTemd** framework is presented, which is an integrated method for knowledge reasoning and semantic annotation of the data streams using stream processor.

Accurate semantic annotation is achieved through adopted ontology – W3C SSN Ontology (Compton *et al.*, 2012), which is an effective way to associate meaning to raw data produced by sensors. In consideration of the major challenges of streaming data processing – (i) the requirement for a storage layer, and (2) a processing layer; there exist available cloud platforms that provide the infrastructure needed to build a streaming data application with sensor data integrated using APIs. Therefore, this research adopts the use of open source *Apache Kafka* as the real-time distributed stream processing system due to inherent ability to process complex queries on a stream of raw data in an efficient, highly scalable and easy to program manner. It also offers data durability, fault tolerance, lowered latency with increased high throughput and can be easily managed through a centralised platform – Confluent.

### 6.3.1. Event STream Processing Engine for Environmental Monitoring Domain (ESTemd)

Stream analytics as a Big Data technology has shown great promise and techniques in data analytics. Several analytics approaches and platform are already in existence to process data streams and detect simple or complex *events* using intelligent analytics methods. The application of stream analytics in this research is focused on identifying evidence of drought from the streams of sensor data/observation using appropriate drought prediction model and indices. In this context, the *Stream Analytics FG* of the SBDIM Middleware comprises of complex software modules and technologies where data streams are channelled using data pipelines from data sources (deployed sensors) to the stream processing engine, and processed output records are channeled to the data sinks in a real-time orchestrated manner. The inferred processing output is integrated with the data from other SBDIM Middleware *FGs* on the IK domain as part of the effort towards increasing the level of accuracy of drought forecasting systems using heterogeneous data sources.

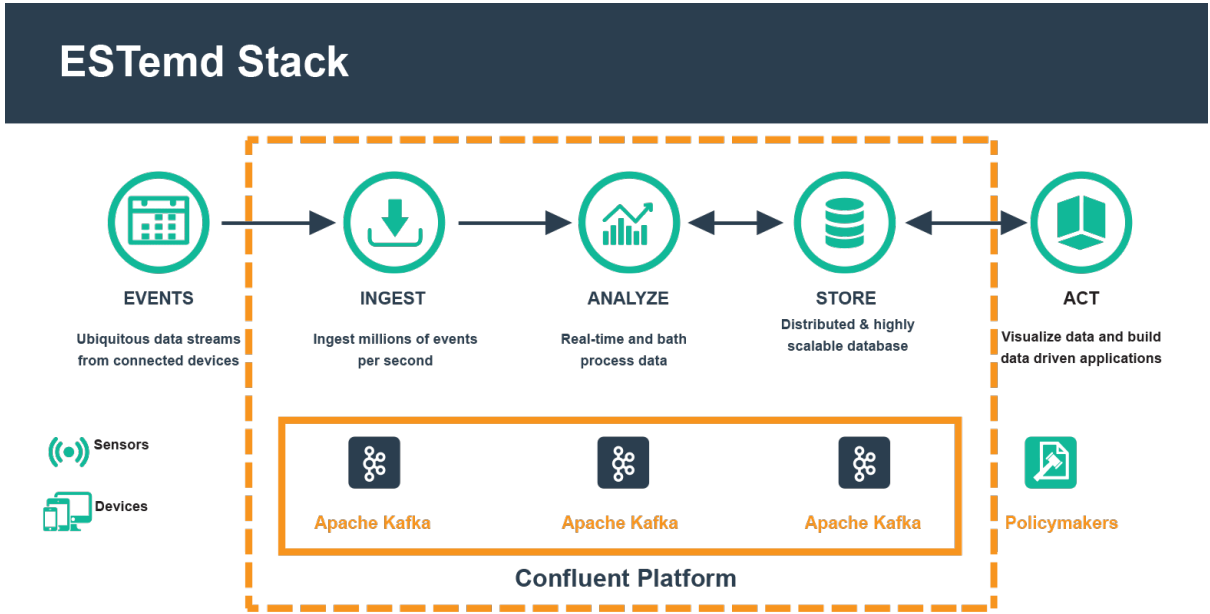

Figure 6- 8: ESTemd Stack (*Source: Author*).

To achieve the analytics functionality of the middleware's objective for a common conceptual representation of the heterogeneous data sources and outputs, the developed (IKON) and adopted ontologies (SSN) were incorporated for data annotation and semantic representation. This solution makes the system compatible with intelligent information systems and scalable for future extensions. The *Stream Analytics FG* design satisfies the requirement for efficient data processing for IoT applications and supports the extraction of insights from a stream of incoming sensors observation. The **ESTemd** stack and framework of the *Stream Analytics FG* for drought prediction and forecasting are depicted in Figure 6-8 and 6-9, respectively.

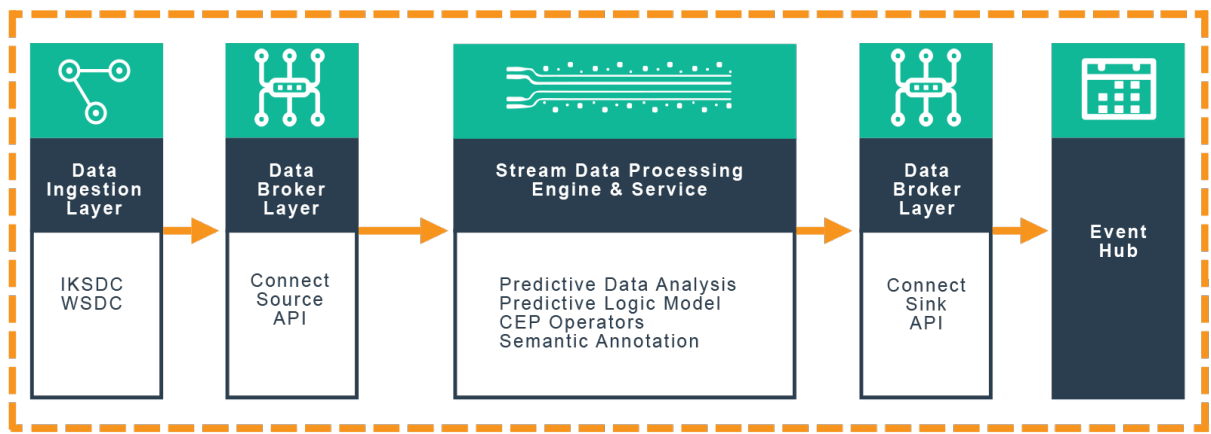

Figure 6- 9: Stream Analytics FG layered model (*Source: Author*).

### 6.3.1.1. Data Ingestion Layer

The data ingestion layer incorporates the data from the *Data Acquisition FG* (sensing device – called Producers) via the gateways in the form of messages. This layer must be a highly scalable using a publish-subscribe event bus which ensures that data streams are captured with minimal loss. *Apache Kafka* through the use of *Kafka* source connectors acting as a broker will buffer the incoming data streams from the producers and also helps to achieve better fault tolerance and load ...0

belov

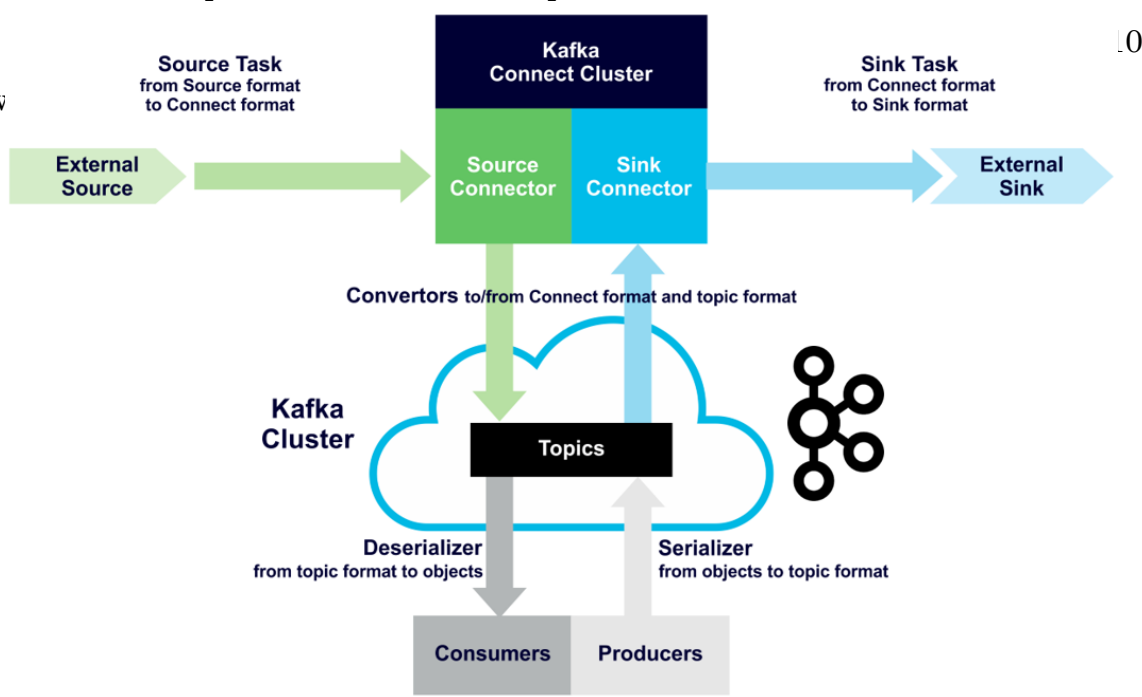

Figure 6- 10: Overview of the streaming engine – *Apache Kafka* (*Source: www.apache.org*)

The data are channelled from the producers – sensors (*Data Acquisition FG*) to the broker of the *Stream Analytics FG* via respective *Kafka* topics in the cluster – ready to be queried or utilised by the streaming engine. The data stream from the sensors in the *Data Acquisition FG* is responsible for feeding the system. The overall data flow to the system is driven by fixed sensor data acquisition from the

*Data Acquisition FG.* Figure 6-11 below depicts starting the *Kafka* broker through the Command Line Interface (CLI) on a local server.

Figure 6- 11: Starting *Apache Kafka* in the FG using CLI (*Source: Author*)

## 6.3.1.2.    Data Broker Layer (Kafka Connect Source)

The data broker layer performs the coordinated processing and transformation of the unbounded data stream coming from the data ingestion layer. The data is received from the *Data Storage FG* transformed using *Kafka* Connect protocol, with additional data preprocessing is performed, before the data is published to the next layer. We exploited the features of RESTful Web services and API to plug into *Sigfox cloud* infrastructure for seamless data flow through the use of appropriate adapters. An example of the several processes executed in this layer is the data cleaning process, where data are adjusted, normalised and inconsistencies resolved to attain a common structure through the *Kafka* Connect. This layer further employ the use of *Kafka* Connect to facilitate the onward data transmission and compatible data pipeline due to its compatibility with most technologies.

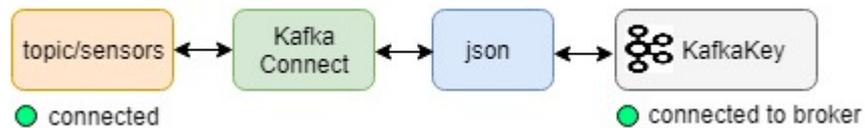

Figure 6- 12: Node-Kafka-broker data pipeline programming flow (*Source: Wang, 2016; Greco, Ritrovato & Xhafa, 2019*)

The programming flow of data is such that the data from the first node is fetched from the sensing devices encoded in a simple JSON format using *Kafka* Connect before being transmitted to other nodes, as illustrated in Figure 6-12 above. Subsequent nodes in the node chain plugged into the semantic repository are responsible for the parsing and converting of the JSON messages into JSON-LD for compatibility of the data pipeline in the middleware. The messages represented in the JSON-LD are transmitted to the next layer *node-red-contrib-Kafka-node* (Wang, 2016; Greco, Ritrovato & Xhafa, 2019). The *Apache Kafka* broker in the *Stream Analytics FG* host some *topics* for aggregating similar sensor data. For example, all *temperature sensors* can be assigned to *topics_temp*, which makes that categorisation and fetching of the all *temperature* data/event easier in a publish-subscribe manner.

### 6.3.1.3.       Stream Data Processing Engine and Service

The stream data processing layer is devoted to the stream processing of the semantically-enriched stream data collected by the *Kafka* broker. *Apache Kafka* offers *Kafka* stream processing engine with great throughput as an IaaS and higher-end API for seamless integration and interoperability using the Confluent platform. The stream of data flowing through several *Kafka* topics in *Kafka* broker is processed through the KSQL node to detect *events* in the time-attributed sensor data streams.

## Predictive Data Analytics

This layer consists of the data and processes analytics components of the *Stream Analytics FG,* that performs several analytics functionalities. The acquired data from the sensors is adequately enhanced with preprocessing techniques in the *Data Storage FG* to eliminate inconsistent observations before data analytics is performed on the sensors stream data set. The streaming dataset is queried from the integrated KSQL cluster with SQL-like operators based on the EDI drought model to gain drought prediction insights.

## Predictive Model Logic – Effective Drought Indices (EDI)

Several drought indices exist such as the PDSI, EDI or SPI – that serve as a measure to determine the onset of drought based on environmental observation of parameters like relative humidity, atmospheric pressure and soil moisture (see Chapter 2). The drought indices categorise the severity of a drought event at scale. EDI has been identified as a good index for determining and monitoring of both meteorological and agricultural drought (Byun and Wilhite, 1996). The EDI model is represented in the form of a logic using the EP language. Data from the deployed sensors would be used to calculate the EDI for profiling droughts in real time on a daily using the CEP engine. The EDI formula set, where precipitation is recorded is below.

$$EPi = \sum_{n-1}^{i}[(\sum_{m-1}^{n} Pm)/n] \qquad \text{(Equation 6-2)}$$

$$DEP_n = EP_n - MEP_n \qquad \text{(Equation 6-3)}$$

$$EDI_n = DEP_n / SD (DEP_n) \qquad \text{(Equation 6-4)}$$

where, $EP_i$ represents the valid accumulations of precipitation of each day, accumulated for $n$ days, $P_m$ is the precipitation for $m$ days, $m = n$. In Equation 1, if $m/n = 365$, then, EP becomes the valid accumulation of precipitation for 365 days divided by 365. $DEP_n$ in Equation 6-3 represents a deviation of $EP_n$ from the mean of $EP_n$ (MEP) – typically 30-year average of the EP. $EDI_n$ in Equation 3 represents the Effective Drought Index, calculated by dividing the DEP by the standard deviation of DEP – $SD (DEP_n)$ for the specified period. In order to detect the onset of drought based on the EDI prediction model, analysis and manipulation were performed on the datasets using *Kafka* operators – Filter (), Map (), FlatMap (), Aggregation (), Sum (), Average () used to represent the EDI model in KSQL. The sensors streams in the *Kafka* topics are queried in real-time using the EDI model in KSQL. The historical precipitation data will be read from a file system to a *Kafka* topic. The output of the persistent query is committed to the output *Kafka* topic in the form of drought indices belonging to one of the four classes of the EDI.

The drought levels are categorised into four classes in  EDI (Table 2-1). After computation using Equation 6-3, the output value of the EDI which ranges from negative to positive determines the category of the drought, which indicates the intensity of the drought, giving a clear definition of the onset, end and duration of drought. For example, a value of -1.05 indicates near normal drought. The interpretation and classification of the drought based on the output values of the EDI calculation are published by the event publisher component of the *Stream Analytics FG*. The output is represented in JSON format to be used by the next FG, which is the *Inference Engine FG*.

### *Kafka* CEP Operators

A stream processing engine utilises the use the CEP operators to identify meaningful patterns,  relationships and gain weather-related insights from streams of unbounded sensor data. *Kafka* streaming processing engine primitive operators such as Filter (), Map (), FlatMap (), Aggregation (), Projection (), Negation () are used for various combination and permutation of parameters of the stream sensor data. These operations are invoked on the *Kafka* topics in the cluster(s) using KSQL. Once a pattern(s) is/are identified and extracted, the KSQL will encapsulate it into a composite (derived) event to be published into an output *Kafka* output topic saved in the cluster or in the form of a message to a secondary index by the event publishers.

The *Selection* filter selection is based on atrributes values. For example, the following pseudocode which selects DHT22 Sensor messages from the message queue to detect temperature readings between $31 - 45$ (Celsius).

**Pattern 1**:

*Select DHT22 (temp >= 31.0 and temp <= 45.0)*
*From DataSource*

*Projection* operator extracts a subset of attributes of the event. For example, Pattern 2 select the humidity attributes of the DHT22 events.

**Pattern 2**:

*Select DHT22 (humidity)*
*From DataSource*

The *Conjunction* operator determines the occurrence of two or more events, either simultaneously or consecutively within a window time frame. As an example, the following pattern can be used to determine in real time a hypothetical onset of a near normal drought event where high temperature and low soil moisture events are notified within the window frame of 4320 hours (6 months).

**Pattern 3**:

*Within 4320hr. SoilMoisture(Value < 10%) and Thermometer(temp > 35)*
*From DataSource*

The *Aggregation* operator is used to perform a calculation to determine aggregated attributes values. For example, Pattern 4 computes the average value of temperature from the DHT22 Sensor events.

**Pattern 4**:

*Select Avg(DHT22.humidity)*
*From DataSource*

*Disjunction* operator determines the occurrence of either one or more event in a predefined set.

*Repetition* operator determines some occurrence observation of a particular event in the messages queue. As an example, Pattern 5 detects the number of occurrence of high temperature.

<u>**Pattern 5**</u>:

*Select DHT22(temp > 35) as Temp*
*From DataSource*
*Where count(Temp) > 10*

*Sequence* operator is useful to determine ordering relations or sequence of corresponding events of a pattern which is satisfied when all the events have been detected.

*Negation* operator usually considers the non-occurrence/absence of an event, used to further strengthen an inference generation or assertion. For example, additional credence could be given to Pattern 6 for the onset of drought by introducing the absence of Rain events.

<u>**Pattern 6**</u>:

*Within 4320hr. SoilMoisture(Value < 10%) and Thermometer(temp > 35) and not Rain ()*
*From DataSource*

The use of several and combination of primitive CEP operators to perform CEP query ensure the identification of complex patterns and determination of composite events. The queries are matched against data streams and get triggered whenever the queries condition have been fulfilled (Lam, & Haugen, 2016). *Apache Flink* will chain the operators together to form a single task.

## Semantic Annotation Layer

This layer deal performs the enrichment of the data with metadata and semantic annotation using the available ontology with Semantic Technologies. Semantic annotation of the data stream with well-defined knowledge will ensure contextual representation, analysis and integration. Lightweight semantics are added and linked via the SSN ontology repository in the *Inference Engine FG* to annotate the data for further enhanced inferencing procedure semantically. The semantic service is responsible for analysing the data or information to predict the conceptual states of the entities or event occurrences.

### 6.3.1.4.      Data Broker Layer (*Kafka* Connect Sink)

The output from the stream data processing and service system is represented by the data broker layer *Kafka* Connect Sink connection protocol for the transformation of the data into a middleware data pipeline compatible format.

### 6.3.1.5.      Data Sink (Event Publishers)

The output from the stream processing engine is made available to other clusters using *Kafka* Connect sink connectors and standard APIs. The data sink acts as a buffer to output from the streaming engine. The output can be saved in *Kafka* topic or other secondary indexes such as MongoDB, Cassandra, NoSQL databases for an offline longer time series analysis or immediate visual analysis using *AKKA* to get further insights.

### 6.3.2. Experimental Implementation and Use Case Discussion

For testing the **ESTemd** framework – *Stream Analytics FG*, events records from the sensors deployed in the study area are feed into the system. Data are captured at a constant stipulated interval from the sensors and the weather station. Each reading entry is in the form of a key-value pair containing the information and the time when data was collected critical for the stream processing.

The hardware used for this experimental implementation was provided by the Unit for Research and Informatics for Drought in Africa (URIDA) of Information Technology Department at Central University of Technology, Free State, South Africa. The entire *Stream Analytics FG* clusters and infrastructure could be deployed as docker containers and managed by kebenetics in the cloud, Virtual Machine (VM), bare-metal computer or local servers depending on the requirement and scale of the ecosystem. For this *FG* implementation, the physical machine employ is Intel Core i7 Quad-Core 3.1GHz running macOS Mojave; the VM is running Ubuntu Linux with Intel Core-based processor as a base machine of the distributed middleware module.

The infrastructure is composed of two clusters: (1) a cluster running on a local machine with a quad-core Intel CPU and 16GB RAM hosts the ZooKeeper, an instance of *Kafka* broker, an active controller and *Kafka* broker; (2) *Kafka* client hosting the *Kafka* streaming engine API and the KSQL for persistent querying of the streams in real time, both clusters monitored and managed through the Confluent streaming platform.

### 6.3.2.1.    Central Streaming Platform

In order to achieve a fully streaming architecture of sensors in the context of IoT, a central streaming platform is required to monitor and manage the data pipelines of deployed sensors and devices in remote locations. This research leverages on the compatibility of Confluent Platform with Apache *Kafka*. Confluent is an enterprise streaming platform based on open-source Apache *Kafka*. It is a central platform which ensures the real-time monitoring of streams through the infrastructure clusters from producers to consumers as depicted in Figure 6-13. It provides the ability to build contextual event-driven applications with Apache *Kafka* using a variety of connectors for different native clients and process event streams in real time using Confluent KSQL.

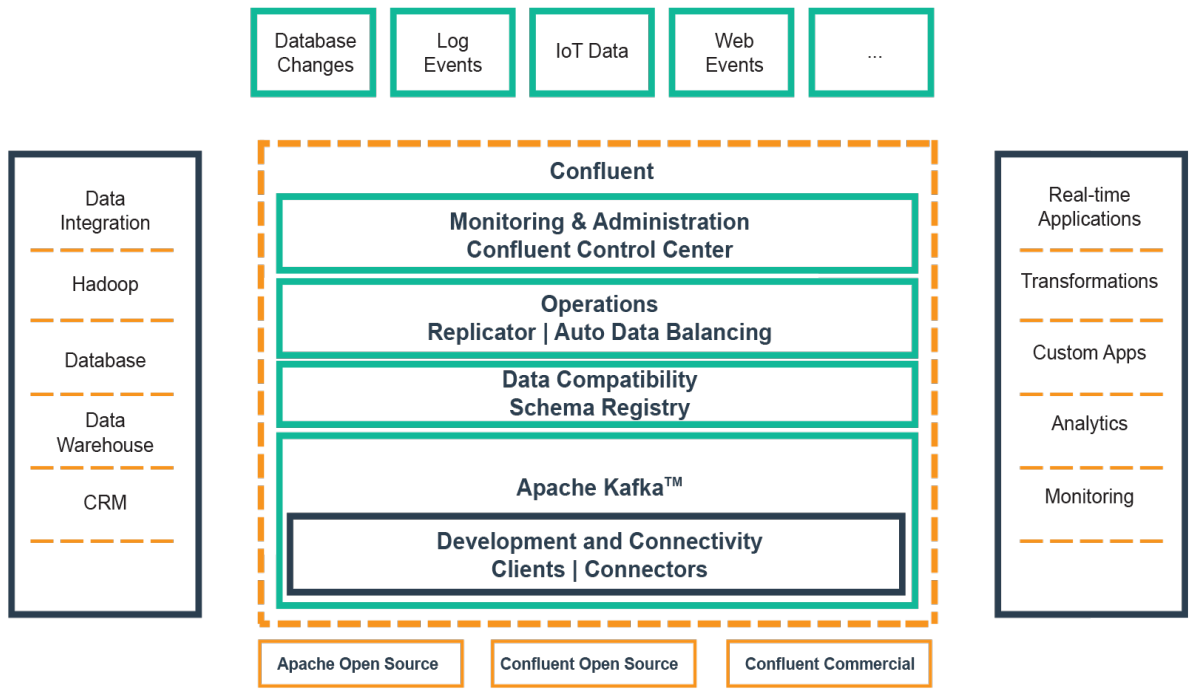

Figure 6- 13: Confluent Enterprise Streaming Framework (*Source:*
*www.confluent.io*)

Unique topics can be created for each type of sensor streams in the system. This allows the grouping of a particular type of sensor data in the same topic, and consumers can retrieve the right data through the sensor group. Confluent Platform is started through the Terminal (Figure 6-14) by invoking the bash file to start an array of services such as zookeeper, *Kafka*, schema-registry, *Kafka*-rest, *Kafka* connect, KSQL-server and the control-center services all in a sequence.

```
control-center-reset                kafka-streams-application-reset
control-center-run-class            kafka-topics
control-center-set-acls             kafka-verifiable-consumer
control-center-start                kafka-verifiable-producer
control-center-stop                 ksql
kafka-acls                          ksql-datagen
kafka-api-start                     ksql-print-metrics
kafka-avro-console-consumer         ksql-run-class
kafka-avro-console-producer         ksql-server-start
kafka-broker-api-versions           ksql-server-stop
kafka-configs                       ksql-stop
kafka-console-consumer              replicator
kafka-console-producer              schema-registry-run-class
kafka-consumer-groups               schema-registry-start
kafka-consumer-perf-test            schema-registry-stop
kafka-delegation-tokens             schema-registry-stop-service
kafka-delete-records                security-plugins-run-class
kafka-dump-log                      sr-acl-cli
kafka-log-dirs                      support-metrics-bundle
kafka-mirror-maker                  windows
kafka-mqtt-run-class                zookeeper-security-migration
kafka-mqtt-start                    zookeeper-server-start
kafka-mqtt-stop                     zookeeper-server-stop
kafka-preferred-replica-election    zookeeper-shell
IoT:bin Akanbi$ /Users/Akanbi/Desktop/Confluent/confluent-5.2.1/bin/confluent start
This CLI is intended for development only, not for production
https://docs.confluent.io/current/cli/index.html

Using CONFLUENT_CURRENT: /var/folders/1w/zl4_x60d1fz5y7b491gtcn7w0000gn/T/confluent.2wRXgGSv
Starting zookeeper
zookeeper is [UP]
Starting kafka
kafka is [UP]
Starting schema-registry
schema-registry is [UP]
Starting kafka-rest
kafka-rest is [UP]
Starting connect
connect is [UP]
Starting ksql-server
ksql-server is [UP]
Starting control-center
control-center is [UP]
IoT:bin Akanbi$ 
```

Figure 6- 14: Starting Confluent Platform in the Terminal (*Source: Author*).

After starting Confluent, the streaming platform interface can be accessed through the localhost server on Port 9021 (Figure 6-15). The dashboard provides an integrated approach to monitor the health of the clusters, brokers, topics, measure the system load, performance operations and even aggregated statistics at a broker or topic level. Confluent Platform provides a broker-centric view of the clusters, used to perform end-to-end stream monitoring, configure the data pipeline using *Kafka* Connect and query the data streams, also with the ability to inspect streams, measure latency and throughput.

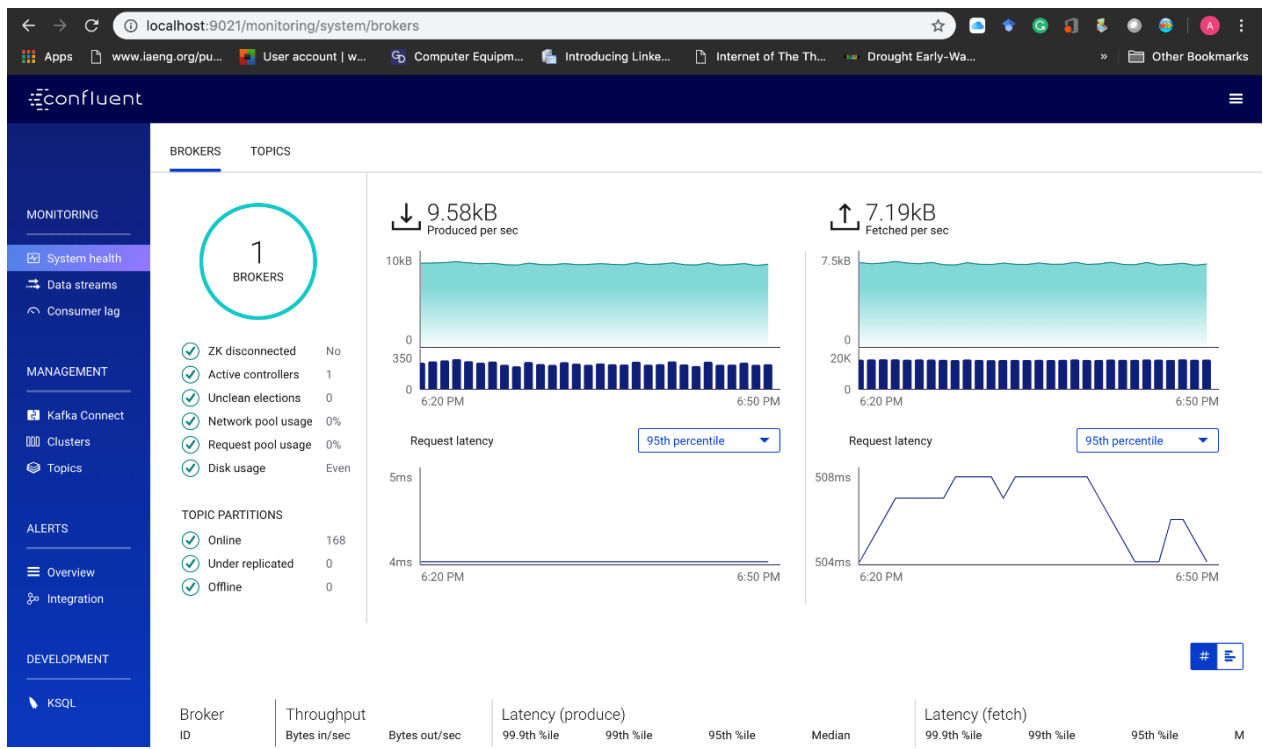

Figure 6- 15: Confluent Platform Interface (*Source: Author*).

## 6.3.2.2.     Configuring data pipelines using *Kafka* Connect

The Confluent Platform ensures the integration of all services and managing of the data connectors to connect data emanating from heterogenous *FG* in one place. The integration of heterogeneous data sources is made possible through *Kafka* Connectors; it provides meaningful data abstractions to pull or push data to *Kafka* brokers (*Kafka* Connect — Confluent Platform, 2019). *Kafka* connectors are forward and backward compatible with vast data representation formats such as XML, JSON, AVRO etc. The configuration of the *Kafka* connector is through the *Kafka* Connect management console. There are two major types of *Kafka* connectors – the *Kafka* Source Connector for connecting to the producers and the *Kafka* Sink Connector for connecting to the secondary data storage indexes. In the *Kafka* Connect management console, the connector class, key converter class, value converter class are defined for the data formats for the *Kafka* Source Connector and the *Kafka* Sink Connector to achieve common serialization format and ecosystem compatibility. This will specify the Kafka messages and convert it based on the key-value pairs using key.converter and value.converter configuration settings. In this research, the entire data pipeline in the middleware infrastructure is represented in JSON. Hence, for JSON, the key.converter will be represented as "key.converter": "org.apache.kafka.connect.json.JsonConverter". If we want *Kafka* to include the schema we insert "key.converter.schemas.enable=true". The same will be applicable for the value.converter.

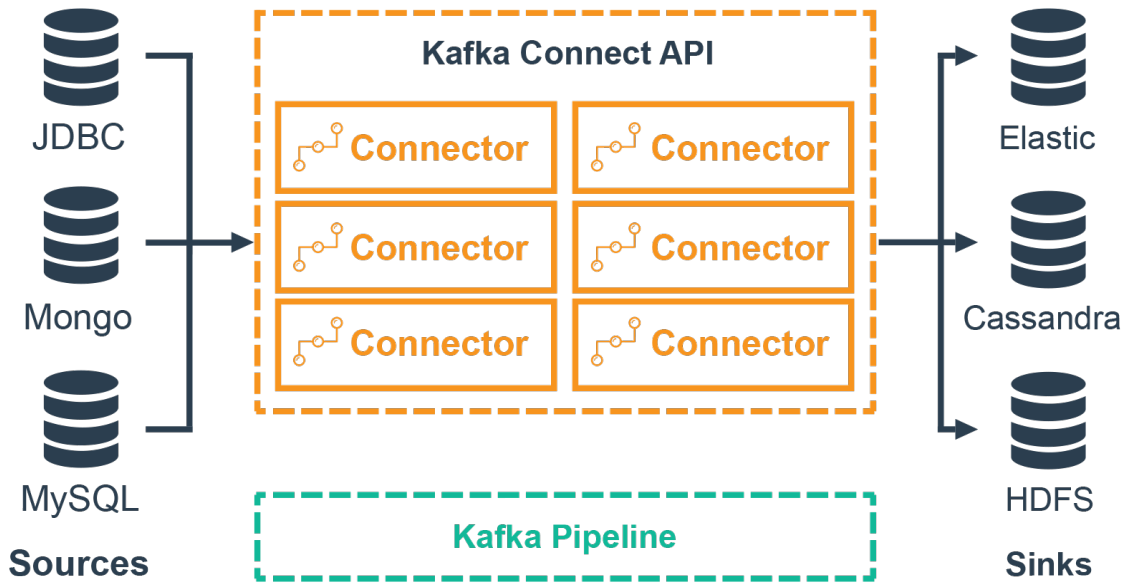

Figure 6- 16: Overview of Kafka Connect. (*Source: www.apache.org*)

## Kafka Source Connector

*Kafka* Connect (Figure 6-16) provides the set of API classes based on different messaging protocols to facilitate stream messages from the producers (sensors) gateways channels to the *Kafka* broker. The *Kafka* Source Connectors broker buffers the incoming messages, kept it in a queue and are replicated across all the brokers in the cluster. The connectors automatically perform data transformations on the messages to make it easier to process. The source connectors ingest the data streams table or entire database and pass it on to the appropriate *Kafka* topics in the broker.

The *Kafka* Source Single Message Transform makes real-time light-weight modifications to the raw messages before publishing to *Kafka* stream engine. There are several source connectors available on the *Kafka* platform, depending on the native language of event producers. For example, *Kafka* Connect MQTT, *Kafka* Connect RabbitMQ, *Kafka* Connect JDBC, *Kafka* Connect CDC Microsoft SQL and many more.

## Kafka Sink Connector

*Kafka* Sink Connector streams the data out of *Kafka* clusters to other secondary indexes such as Elasticsearch or Cassandra using *Kafka* Source Single Message Transform to make light-weight modifications to *Kafka* messages before writing the output to an external repository. The stream processed outputs are delivered from the *Kafka* topics to the secondary indexes for visual representation and analysis or offline batch analysis with Hadoop. In the context of this research, the output data will be consumed and used by policymakers as a critical output of the middleware. Configuration of *Kafka* Connect for MQTT-JSON (Figure 6-17), other relevant examples of *Kafka* Sink Connectors are *Kafka* Connect Neo4j, and *Kafka* Connect HDFS, *Kafka* Connect HTTP.

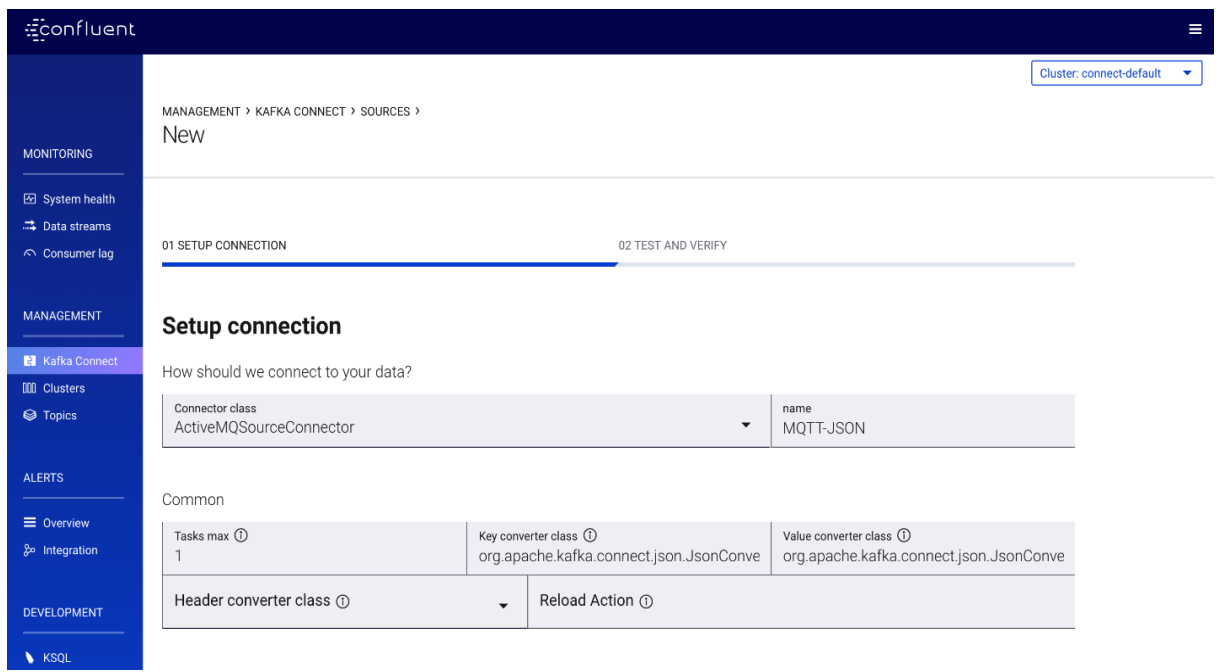

Figure 6- 17: Configuration of Kafka Connect in Confluent Platform (*Source: Author*).

### 6.3.2.3.  *Kafka* Topics

Topics in *Kafka* are similar to RSS feeds that allow users to access updates in a standardised format. Hence, a *Kafka* topic is a feed that stores similar messages or event records. The messages or event records are generated from the Producer – data sources (sensor) and are written to the appropriate topic. Several topics can be created to categorise similar types of messages belonging to a broker. Consumers make use of the messages by reading the messages from the topics. New topics (Figure 6-18) can be created to store the output of manipulation performed on an existing topic within the same cluster and infrastructure.

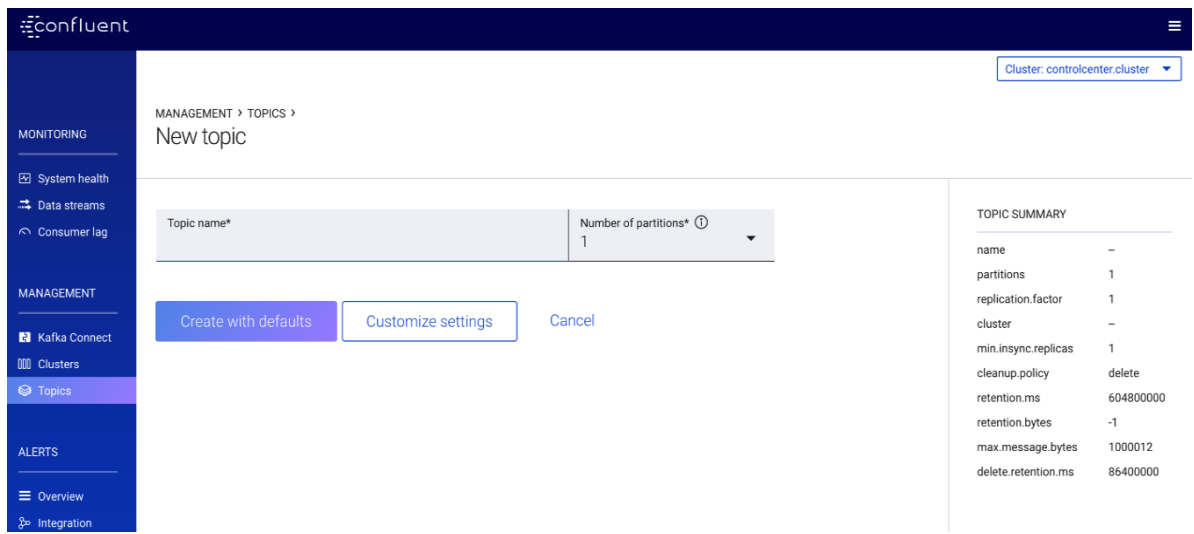

Figure 6- 18: Creating a new topic in Confluent platform (*Source: Author*).

In this research, five unique topics will be created to cater for and specifically categorise the temperature readings, humidity readings, atmospheric pressure readings, precipitation readings and the soil moisture readings from the producers (sensors). Table 6-6 below shows the grouping of the sensor readings to a specific topic.

**Table 6-6: Categorisation of the Sensors Readings to *Kafka* Topics.**

| Type of Readings | *Kafka* Topic |
|---|---|
| Temperature | TemperatureSensors |
| Humidity | HumiditySensors |
| Precipitation | PrecipitationSensors |
| Atmospheric Pressure | AtmosPressureSensors |
| Soil Moisture | SoilMoistureSensors |

| EDI Output | EDIOutput |
|------------|-----------|

Further manipulation of the *Kafka* topics messages using CEP operators based on the EDI model formula will yield new topics to store the processed messages. Performing the average operator (Avg ()) on the topics will create five (5) new additional topics namely: TemperatureSensors ➔ Avg_Temperature; HumiditySensors ➔ Avg_Humidity; AtmosPressureSensors ➔ Avg_AtmosPressure; SoilMoistureSensors ➔ Avg_SoilMoisture; PrecipitationSensors ➔ Avg_Precipitation. Additional six (6) *Kafka* topics will be created to further store the output of the EDI computations, namely: DEP, Standard deviation of DEP, EP, Mean of Effective Precipitation (MEP), Sum of precipitation (Sum_Precipitation) and EDI. Lastly, a new topic that stores the historical precipitation data from file – "HistoricalPrecipitation" will be created for calculating the MEP. Therefore, there are 17 *Kafka* topics in our broker, all created with the same number of partition and replication factor across the cluster (Figure 6-19).

Figure 6- 19: Available topics in the *Kafka* broker (*Source: Author*).

**6.3.2.4.       Workflows**

In this case study, a couple of producers deployed in the area under study send sensor readings (messages) to four (4) different *Kafka* topics. The data streams generated by the sensors (producers) are passed on to the *Kafka* topics in the *Kafka* broker for stream processing. The *Kafka* cluster is composed of two (2) nodes having similar setting running Intel-based processors. *Kafka* broker runs operators and user-defined functions inside the JVM. EDI computational process performed on the data streams using KSQL will generate new tables that will be committed to the appropriate topics in the broker. KSQL performs persistent line queries, filtering and aggregation of data records for drought predictions and forecasting over a period of time.

### 6.3.2.5. Persistent Querying of the Data Streams using KSQL

Each record or message from a producer is typically represented as a *key-value* pair, and the streams of record are processed in real-time with the smallest amount of latency through the help of *Kafka*-SQL (KSQL). KSQL is a streaming SQL engine for *Kafka*, with almost identical syntax and mode of operations to normal SQL, the only difference is that SQL queries a relational database while KSQL queries data streams. KSQL allows the stream processing of data streams using operators such as data filtering using WHERE clause to filter data that comes from streams and meet certain requirements and save the filtered output to other topics in the broker. As depicted in Figure 6-20. KSQL Server consists of the KSQL engine and the REST API. KSQL Server routines communicate with the *Kafka* cluster through the KSQL UI.

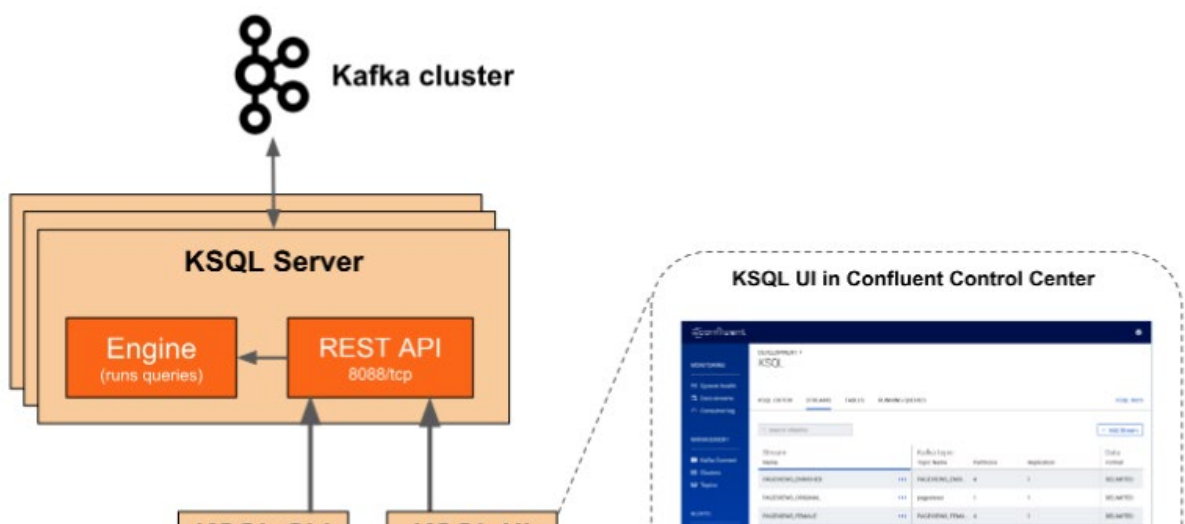

Figure 6- 20: KSQL cluster interfacing with the *Kafka* broker (*Source: www.apache.org*).

Data transformation are performed with JOIN or SELECT operator for data enrichment or scalar functions; while data analysis with stateful processing, aggregation and windowing operation for time-series analysis are also possible. KSQL consumes streams of sensor data stored in *Kafka* topics – TemperatureSensors, HumiditySensors, AtmosPressureSensors and SoilMoistureSensors; which are mostly structured data set in JSON but could be in a format like AVRO or delimited formats (CSV) by using the appropriate *Kafka* Connect API for the data pipeline. Queries are performed through the use of KSQL cluster connected to the *Kafka* broker. KSQL supports standard Data Definition Language (DDL) and Data Manipulation Language (DML) statements.

### KSQL Querying Algorithm

*Generate KSQL (DStream)*

(1)     FOR historical precipitation dataset

IF dataset is Filesystem WHERE file format is .xslv

READ file (.csv)

CREATE Table "HistoricalPrecipitation"

SAVE file (.csv) to Table "HistoricalPrecipitation"

(2)     FOR $Sum\_Precipitation = SUM (PrecipitationSensors)$

CREATE Table "Sum_Precipitation"

SAVE "Sum_Precipitation" to Table "Sum_Precipitation"

(3)     FOR $EP = {Sum\_Precipitation}/{Time\ Frame}$

CREATE Table "EP"

SAVE "EP" values to Table "EP"

(4)     FOR $MEP = Mean (HistoricalPrecipitation)$

CREATE Table "MEP"

SAVE "MEP" values to Table "MEP"

(5)     FOR $DEP = EP - MEP$

CREATE Table "DEP"

SAVE "DEP" values to Table "DEP"

(6)     FOR SD(DEP) = Standard deviation (DEP)

CREATE Table "SD(DEP)"

SAVE "SD(DEP)" values to Table "SD(DEP)"

(7)     FOR $EDI = {DEP}/{SD(DEP)}$

CREATE Table "EDI"

SAVE "EDI" values to Table "EDI"

(8)     RETURN persistent KSQL query

## Prediction Model Logic Codes

The detailed KSQL code for querying the data streams based on the EDI model is available on https://github.com/yinchar/KSQL-Code-for-EDI-Model-Logic.

## 6.4.     Inferences Outputs as Drought Forecast Advisory Information (DFAI)

The inference output for the *Inference Engine FG* from **D1** and inference output from the *Stream Analytics FG* from **D2** are merged together to form the DFAI with attributed CF. The higher the CF attributed to the inferred output, the higher the certainty level of the system. Hence, the certainty of the systems is dependent on the number of input data and the attributed CF of each observation/scenarios. The final DFAI output contains a categorisation of the predicted drought based on the EDI scale. In this case study, the DFAI is meant to be interpreted and used by policymakers in the study areas for their drought-related decision-making processes.

## 6.5.     Integration of the Stream Analytics and Inference Engine FGs to the Middleware

The development tools and data input/outputs format adopted for the *Stream Analytics FG* and *Inference Engine FG* ensures the easy integration with other existing functional groups of the distributed semantic middleware as well as making it forward compatible with conventional software environments. This is achieved through the consistent use of compatible data representation format throughout the middleware's data pipeline. In the Middleware, effective data sharing and communication is important and achieved through the semantic representation of the data flow using uniform JSON/JSON-LD machine-readable language in all the FG. This ensures ease of data integration and interoperability of the distributed FGs. The inferences outputs from the RB-DEWES and EStemd are passed to the *Eventhub* and merged for the creation of the DFAI, which will be subsequently published by the *Data Publishing FG* of the middleware.

## 6.6. Summary

This chapter presents the inference generation systems of the middleware. The inference from the heterogeneous data sources is achieved in the *Stream Analytics FG* and *Inference Engine FG* of the semantic middleware from the indigenous knowledge on drought and sensors data, respectively. The overview of the *Stream Analytics FG* is outlined using the **ESTemd** framework. The key technological components of the FG that facilitates the effective stream processing of the sensor streams in the *FG* cluster are *Apache Kafka, Kafka* Connect, *Kafka* Streaming Engine, KSQL and Confluent Platform. *Apache Kafka* provides a lightweight stateful streaming operation of records from the data sources by storing and replicating the data across several nodes in the cluster, using *Kafka* Connect which provides the necessary API to ensure data compatibility in the middleware pipeline. The KSQL that queries the data streams in real time through the use of *Kafka* Streaming Engine API was also presented through the Confluent streaming platform.

The RB-DEWES – is an expert system component of the *Inference Engine FG* of the semantic Middleware for drought forecasting and prediction using *rules* identified from the local indigenous knowledge acquired in the areas under study was presented. The sub-system utilises a rule-based methodology and probabilistic reasoning technique using *rules* derived from the IKS. This approach enabled the generation of inference from the IK acquired from the domain experts. RB-DEWES allows the ascription of CFs with the input and output information, which vastly helps with evaluating the quality and confidence level of the user's observation and the system's inferred output. The inference outputs of the automated inference generation systems of the middleware are merged in the *Eventhub* to form the DFAI which uses the EDI index to categorise the severity or onset of drought.

# CHAPTER SEVEN

# EVALUATION OF SEMANTICS-BASED DATA INTEGRATION MIDDLEWARE

## 7.1. Introduction

In the following paragraphs, the evaluation of all the *Functional Groups* (FG) of the experimental system and the middleware prototype is presented. This is used for software verification and validation (V&V) processes of each of the semantic middleware FGs modules. This chapter also reports on the data flow in the semantic middleware geared towards achieving a semantics-based data integration for drought forecasting and prediction systems. There are several V&V approaches for software modules evaluation; however, the evaluation of the semantic middleware will be based on the core five categories of V&V (Ferreira, Collofello, Shunk & Mackulak, 2009).

Also, during the implementation procedure, the data pipeline is uniformly represented in JSON/JSON-LD for continuous data flows in the data plane to connect various part of the middleware infrastructure irrespective of the schema or specification. The middleware is intelligently capable of data transformation at dedicated FG nodes or edge/gateway in a cloud or standalone environment. This eliminates data heterogeneity and provides efficient data integration with service interoperability in the middleware with strict adherence to the principles of SOA.

The core objective of the V&V of the semantic middleware is for building and quantifying confidence in the software development process through adequate testing of the modules. The functioning walkthrough of the aggregated FGs with test results of the middleware services are presented below. This is ascertained through a series of experimental test, V&V of the FGs, presenting of results and user experience (UX) evaluation.

## 7.2.    FGs Verification and Validation (V&V)

The core aspect of the middleware's FGs V&V is to determine the semantic middleware performs the intended functions correctly based on the **NFRs** and **FRs** (see Section 3.5.1); and as a measure of middleware quality and reliability. The verification involves evaluating the middleware to ensure it meets the middleware initial requirements; and the validation involves testing of each middleware FGs during the implementation to ensure the initial requirements are indeed met against the system requirements. The FGs V&V gives the incremental preview of the middleware FGs performance as required by the IEEE 1012-2012 – IEEE Standard for System and Software Verification and Validation (Freund, 2012). The V&V is performed during the implementation of each of the middleware's FGs and requires minimum input and output requirements for the V&V task (Wallace & Fujii, 1989; Wallace, Ippolito & Cuthill, 1996).

During the development process, different forms of execution and non-execution-based testing were performed to ensure conformity. For example, reviews, audits, document-driven walkthroughs were performed for each FG of the semantic middleware. Subsequently, detailed inspections were performed for each FG during the implementation to ensure it satisfies the five behavioural properties of utility, robustness, reliability, performance and correctness for in-depth evaluation.

## 7.3.    Overview of the SB-DIM Middleware Implementation

To start with, tests were presented on the *data acquisition FG* of the middleware – IK data representation and the WSN data transformation, as well as all other FGs. The outcomes of a verification and validation processes based on the comparison of weather forecasts to actual weather observation are presented. To make evident the validity of SB-DIM middleware, experiments and results using actual data acquired from study area during the month of September and October 2017 are presented. The structure of the FGs is based on the middleware's framework presented in Chapter Three.

The procedural components of the distributed middleware are programmed in parallel. This ensures the middleware was developed using an incremental software development life cycle model and was continuously enhanced during and after development. The realisation of individual components consisted of experimental tests, coding, and execution to gauge the sensor devices outputs with the acceptable outputs from the weather station. After the development, real tests were run transmitting and uploading the data through the middleware FGs. The middleware process that data across all the FGs with a disjointed interactive platform for input and output visualisations.

## 7.4. Data Acquisition FG Phase

### 7.4.1. Configuration of the Wireless Sensor Network and Professional Weather Station

The microcontroller and sensors were deployed in a simple start network topology for effective transmission of sensor readings data to the sink. The sensors connected are the DHT22 – for temperature and relative humidity, atmospheric pressure sensor. Each sink was equipped with *Sigfox* module/Wi-Fi module acting as the to transmit the data to the Sigfox cloud. Each sink is powered by a 3.3V 18600 battery pack for the microcontroller. These battery packs are rechargeable and could be easy replaced when the voltage is low. All components are encased in a Pyrex box for prevention against weather effects (Figure 7-1).

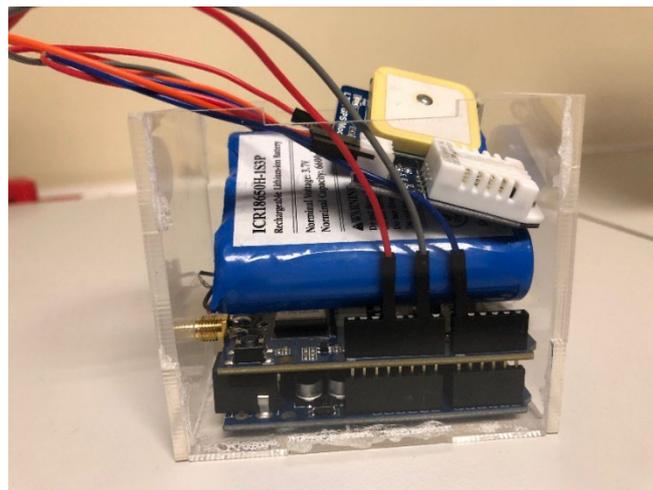

Figure 7- 1: Micro-controllers, sensors with a battery in a Pyrex casing (*Source: Author*).

The wireless weather station was used as a reference model for the wireless sensor network and also for accurate monitoring of weather conditions. The sensor probes – some embedded in the soil – are directly connected to the weather station. The Campbell Scientific WxPRO™ research-grade equipment is a programmable datalogger used for the reliable monitoring enhanced with several components that are used to measure, monitor, and study the weather and climate. The weather station has been comprehensively calibrated, validated, and ISO 9001:2015 Certified. The wireless weather station gathered in real time the temperature, precipitation, relative humidity, atmospheric pressure, wind direction, soil moisture, wind speed, among other parameters. The weather station used in this research is depicted below (Figure 7-2).

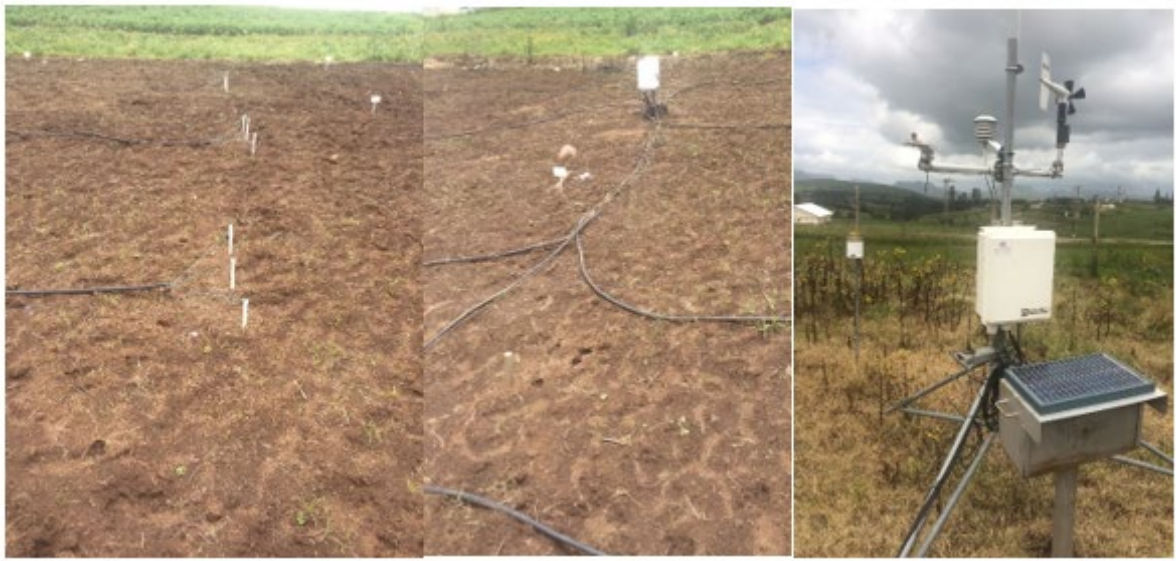

Figure 7- 2: Campbell Scientific Research Grade Weather Station (*Source: Author*).

### 7.4.1.1.    Data Representation Formats

The WSN sensors motes directly send the messages to the *Sigfox* Cloud through the sink/gateway where it's available for offline processing. *Sigfox* Cloud provides the ability to download or export the sensor readings in .csv formats for further analysis from the *Sigfox* cloud as shown in Figure 7-3 and Figure 7-4. The data format can be further converted from CSV format to JSON format for compatibility with other FGs in the middleware.

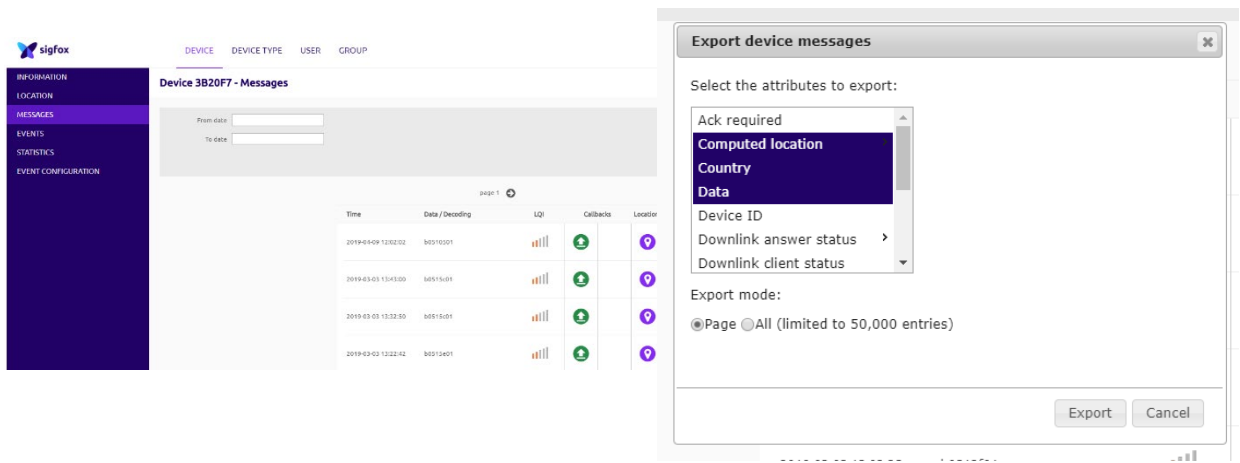

Figure 7- 3: Exporting sensor device messages from the *Sigfox* Cloud (*Source: Author*).

| 1 | Latitude (Computed location);"Status (Computed location)";"Longitude (Computed location)";"Radius |
|---|---|
| 2 | ;;;;;"";"b0510501";"Limit";"N/A";"SIGFOX_South_Africa_Sqwidnet";"2019-04-09 12:02:02" |
| 3 | ;;;;;"";"b0515c01";"Limit";"N/A";"SIGFOX_South_Africa_Sqwidnet";"2019-03-03 13:43:00" |
| 4 | ;;;;;"";"b0515c01";"Limit";"N/A";"SIGFOX_South_Africa_Sqwidnet";"2019-03-03 13:32:50" |
| 5 | ;;;;;"";"b0515e01";"Limit";"N/A";"SIGFOX_South_Africa_Sqwidnet";"2019-03-03 13:22:42" |
| 6 | ;;;;;"";"b0512f01";"Limit";"N/A";"SIGFOX_South_Africa_Sqwidnet";"2019-03-03 13:02:25" |
| 7 | ;;;;;"";"4e6f77204f6e6c696e65";"Limit";"N/A";"SIGFOX_South_Africa_Sqwidnet";"2019-03-03 13:02:06" |
| 8 | ;;;;;"";"b0514301";"Good";"N/A";"SIGFOX_South_Africa_Sqwidnet";"2018-12-04 14:42:11" |
| 9 | ;;;;;"";"b0513601";"Good";"N/A";"SIGFOX_South_Africa_Sqwidnet";"2018-12-04 14:32:01" |
| 10 | ;;;;;"";"4e6f77204f6e6c696e65";"Good";"N/A";"SIGFOX_South_Africa_Sqwidnet";"2018-12-04 14:31:41" |
| 11 | ;;;;;"";"b051f900";"Limit";"N/A";"SIGFOX_South_Africa_Sqwidnet";"2018-11-22 06:23:16" |
| 12 | ;;;;;"";"b051fb00";"Limit";"N/A";"SIGFOX_South_Africa_Sqwidnet";"2018-11-22 06:13:07" |
| 13 | ;;;;;"";"b051f300";"Limit";"N/A";"SIGFOX_South_Africa_Sqwidnet";"2018-11-22 06:03:00" |
| 14 | ;;;;;"";"b051f300";"Limit";"N/A";"SIGFOX_South_Africa_Sqwidnet";"2018-11-22 05:52:51" |
| 15 | ;;;;;"";"4e6f77204f6e6c696e65";"Limit";"N/A";"SIGFOX_South_Africa_Sqwidnet";"2018-11-22 05:52:30" |
| 16 | ;;;;;"";"b0516101";"Limit";"N/A";"SIGFOX_South_Africa_Sqwidnet";"2018-11-18 12:53:24" |
| 17 | ;;;;;"";"b0514a01";"Limit";"N/A";"SIGFOX_South_Africa_Sqwidnet";"2018-11-18 12:43:16" |
| 18 | ;;;;;"";"b0514801";"Limit";"N/A";"SIGFOX_South_Africa_Sqwidnet";"2018-11-18 12:33:09" |
| 19 | ;;;;;"";"b0515201";"Limit";"N/A";"SIGFOX_South_Africa_Sqwidnet";"2018-11-18 12:23:00" |
| 20 | ;;;;;"";"b0513201";"Limit";"N/A";"SIGFOX_South_Africa_Sqwidnet";"2018-11-18 12:12:50" |

Figure 7- 4: Sensor device messages in CSV format (*Source: Author*).

The weather station data are represented and are downloaded in an array of formats such as HTML, JSON, TOA5, XML depending on the suitability and requirement using a custom data query. The readings are available through - http://143.128.64.9:5355/Sw_weather/index.html (Figure 7-5) where the historical data are downloaded in JSON format (Figure 7-6).

Figure 7- 5: Sensor device messages in CSV format (*Source: Author*).

Figure 7- 7: NPM conversion code (*Source: Author*).

Figure 7- 6: Weather station readings in JSON format (*Source: Author*).

## 7.4.1.2.    Conversion and Representation of Sensor Data in JSON files

This section presents the method of converting the sensor readings data in CSV format to JSON files. Data files in CSV format are converted using NPM package installed on a LAMP localhost server. The installation command takes the form below; it is self-contained without dependencies.

```
> npm i csvjson-csv2json
```

After installation, the command **csv2json** was called to reliably convert the CSV files to JSON; the command will auto-detect the separator although you may override or force it via the separator option. The converted sensor devices messages in JSON format is depicted below. The outputs show the process conforms with the **NFR** and the **FR** initially specified.

### 7.4.2. Indigenous Knowledge on Drought Component

```
[
  {
    "Latitude (Computed location);\"Status (Computed location)\";\"Longitude
(Computed location)\";\"Radius (Computed location)\";\"Source (Computed
location)\";\"Country\";\"Data\";\"Link Quality Indicator\";\"Link Quality Indicator
repeaters\";\"Operator\";\"Timestamp":
";;;;;\"\";\"b0510501\";\"Limit\";\"N/A\";\"SIGFOX_South_Africa_Sqwidnet\";\"2019-04-
09 12:02:02"
  },
  {
    "Latitude (Computed location);\"Status (Computed location)\";\"Longitude
(Computed location)\";\"Radius (Computed location)\";\"Source (Computed
location)\";\"Country\";\"Data\";\"Link Quality Indicator\";\"Link Quality Indicator
repeaters\";\"Operator\";\"Timestamp":
";;;;;\"\";\"b0515c01\";\"Limit\";\"N/A\";\"SIGFOX_South_Africa_Sqwidnet\";\"2019-03-
03 13:43:00"
  },
  {
    "Latitude (Computed location);\"Status (Computed location)\";\"Longitude
(Computed location)\";\"Radius (Computed location)\";\"Source (Computed
location)\";\"Country\";\"Data\";\"Link Quality Indicator\";\"Link Quality Indicator
repeaters\";\"Operator\";\"Timestamp":
";;;;;\"\";\"b0515c01\";\"Limit\";\"N/A\";\"SIGFOX_South_Africa_Sqwidnet\";\"2019-03-
03 13:32:50"
  },
  {
```

Figure 7- 8: Converted sensor readings in JSON format (*Source: Author*).

#### 7.4.2.1. Overview of Indigenous Knowledge Indicators

This section presents the verification experiments using various forecast skill metrics in determining the level of confidence are presented. The transformation processes applied to the data set in transforming the data in a structured format with the final output in JSON.

#### 7.4.2.2. Data Collection Tool

The data collection tool used was an Android application for smart devices – ODK Tool Version 2.3.1. The application latest version (APK) could be downloaded from the Google Play Store and is based on a free and open-source framework for collecting data from respondents. It allows the collection of data offline and submission of the data when internet connectivity is available. The application was a configurable and programmable survey tool that could be customised to meet the survey requirements, and in this instance, for collecting the IK. It consists of a programmable frontend and the backend database that saves each respondent response entry in the database. The questions are prepared in XML format and uploaded to the smart device for use. Each entry was saved by clicking the submit button and are automatically saved to the database in real time. Figure 7-8 shows the code snippet of the developed questionnaire in XML format; complete code is available in Appendix B.

Figure 7- 8: Sample questionnaire in XML format (*Source: Author*).

### 7.4.2.3. Indigenous Knowledge Verification and Confidence Level

Two IK input data set were obtained for verification purposes. The primary IK data set was gathered from the local farmers using ODK IK Collector. The reference data set was obtained from a focus group comprising of ten (10) IK domain experts uniquely selected to perform verification and validation of the knowledge sample. These two data sets are crucial in the validation of the IK component of the data sources. The gathered knowledge was further refined.

## 7.5.    Data Storage FG Phase

### 7.5.1. Data Pipeline Data Format

The data in the data pipeline has been transformed and stored in a unified JSON format, making it compatible for processing by other *Functional Groups* (FGs). The data represented in JSON will be ingested or consumed by other services within the context of the middleware. The outputs shows the process conforms with the **NFR** and the **FR** initially specified.

## 7.6.    Stream Analytics Phase

### 7.6.1. Overview of the Stream Analytics FG

This FG process the streams of sensor data from the wireless sensor network (WSN) component of the *Data Acquisition FG* in real-time. Through the use of persistent query of the data streams, the inference is generated in real-time without committing the data to the database. This phase consists of several stacked layers producing services in a unified manner.

### 7.6.2. Implementation Scenario

For implementation, the technical specifications of the entire *Stream Analytics FG* clusters and infrastructure bare-metal computer with a localhost server. The physical machine employ is a MacBook Pro Intel Core i7 Quad-Core 3.1GHz running MacOS Mojave; the VM is running Ubuntu Linux with Intel Core-based processor as the base machine. Full experimental implementation is available in Chapter 5.

The infrastructure is composed of 2 clusters: a cluster running on a local machine with a quad-core Intel CPU and 16GB RAM. The cluster hosts the ZooKeeper, instance of *Kafka* broker, an active controller and *Kafka* broker; the second cluster is a *Kafka* client hosting the *Kafka* streaming engine API and the KSQL for persistent querying of the streams in real time, both clusters monitored and managed through the Confluent streaming platform.

The streaming platform is started up through the Terminal by first navigating to the location of the installation folder and by calling the associated bash script file *./confluent start*, which will invoke and start the streaming platform on the dedicated port – 9021. This starts up the Zookeeper, *Apache Kafka*, Schema-Registry, *Kafka*-rest, *Kafka* connect, ksql-server and the streaming control center (Figure 7-9).

Figure 7- 9:Starting up the streaming platform and associated services in Terminal (*Source: Author*).

After startup, the streaming platform could be assessed through a web browser using port 9021. However, through the central platform, the configurations for creating Topics, *Kafka* Connect, KSQL and metrics for monitoring the cluster's health are accessible through the interface in real time.

Figure 7- 10: Querying the data stream through the KSQL CLI (*Source: Author*).

### 7.6.3. Persistent Query Output Data Format.

The data streams are queried in real time through the KSQL CLI (Figure 7-11). The query is structured and based on the EDI formula algorithm (see Appendix G). The output created from the real-time persistent querying are saved and committed to the output topic EDI in JSON format. The output is represented as a category of EDI and can be viewed in the output topic using the SHOW TABLE or SHOW STREAM command with the stream/table name, *Kafka* topic name and the data format (Figure 7-11).

```
CLI v5.0.0, Server v5.0.0 located at http://ksql-server:8088

Having trouble? Type 'help' (case-insensitive) for a rundown of how things work!

ksql> SHOW STREAMS;

Stream Name | Kafka Topic | Format
```

Figure 7- 11: Querying output stream format using SHOW command in KSQL CLI (*Source: Author*).

## 7.7.    Inference Engine FG Phase

### 7.7.1. Overview of the Inference Engine FG

This *Inference Engine FG* consists of various sub-systems for the generation of accurate inference from the heterogeneous data sources. It consists of the semantic annotation, the event hub and the reasoner's subsystem.

### 7.7.2. Semantic Annotation Sub-System - Transformation of IK into Structured Machine-Readable Format

The verified IK gathered were analysed using a top-down approach to identify the indicators, the relationship between the indicators, the occurrence of an indicator with the significance. The entire IK domain was modelled and represented in a domain ontology – capturing the core objects (indicators), mappings with the relationships. This is carried out in the Semantic Annotation sun-system. The domain ontology was transformed and represented in a machine-readable format that can be used by intelligent information systems and part of the web of linked data such as RDF, OWL, XML and JSON. The knowledge representation process was presented in Chapter Six.

For representing the data as JSON Files – using Protégé, the IK domain ontology is transformed and exported in JSON format for use or integration with other machine-readable ontology and intelligent information systems. JSON format provides seamless data integration and service interoperability through the utilisation of RESTful web services. Figure 7 – 12 below shows the JSON format of the IK domain ontology. The complete JSON code is available in Appendix D.

```
][ {
    "@id" : "http://www.semanticweb.org/aakanbi/ontologies/2016/0/IKON#AnimalSize",
    "http://www.w3.org/2000/01/rdf-schema#range" : [ {
      "@id" : "http://www.w3.org/2002/07/owl#real"
    } ]
}, {
    "@id" : "http://www.semanticweb.org/aakanbi/ontologies/2016/0/
IKON#FlowerBloomingConditon",
    "http://www.w3.org/2000/01/rdf-schema#range" : [ {
      "@id" : "http://www.w3.org/2001/XMLSchema#string"
    } ]
}, {
    "@id" : "http://www.semanticweb.org/aakanbi/ontologies/2016/0/
IKON#MigratoryBirdSighting",
    "http://www.w3.org/2000/01/rdf-schema#range" : [ {
      "@id" : "http://www.w3.org/2001/XMLSchema#boolean"
    } ]
}, {
    "@id" : "http://www.semanticweb.org/aakanbi/ontologies/2016/0/IKON#MigratoryBirds",
    "http://www.w3.org/2000/01/rdf-schema#range" : [ {
      "@id" : "http://www.w3.org/2001/XMLSchema#string"
    } ]
}, {
    "@id" : "http://www.semanticweb.org/aakanbi/ontologies/2016/0/IKON#WeatherTempCondition",
    "http://www.w3.org/2000/01/rdf-schema#range" : [ {
      "@id" : "http://www.w3.org/2001/XMLSchema#float"
    } ]
}, {
    "@id" : "http://www.semanticweb.org/aakanbi/ontologies/2016/0/IKON#Wiki-Jolo",
    "http://www.semanticweb.org/aakanbi/ontologies/2016/0/IKON#FlowerBloomingConditon" : [ {
      "@type" : "http://www.w3.org/2001/XMLSchema#boolean",
      "@value" : "true"
    } ]
}, {
    "@id" : "http://www.semanticweb.org/aakanbi/ontologies/2016/0/IKON#Withering",
    "http://www.w3.org/2000/01/rdf-schema#range" : [ {
      "@id" : "http://www.w3.org/2001/XMLSchema#boolean"
    } ]
}, {
    "@id" : "http://www.semanticweb.org/aakanbi/ontologies/2016/0/IKON#cattle",
    "http://www.semanticweb.org/aakanbi/ontologies/2016/0/IKON#AnimalSize" : [ {
      "@type" : "http://www.w3.org/2001/XMLSchema#integer",
      "@value" : "150"
    } ]
}, {
    "@id" : "http://www.semanticweb.org/aakanbi/ontologies/2016/0/IKON#hasFlower",
    "http://www.w3.org/2000/01/rdf-schema#domain" : [ {
      "@id" : "http://www.semanticweb.org/aakanbi/ontologies/2016/0/IKON#FloralPlants"
    } ],
    "http://www.w3.org/2000/01/rdf-schema#range" : [ {
      "@id" : "http://www.semanticweb.org/akanbi/ontologies/2018/10/IKON.owl#Blooming"
    } ],
```

Figure 7- 12: Indigenous Knowledge Ontology in JSON format (*Source: Author*).

### 7.7.3. Expert System Event Hub

The expert system event hub is a component of the *Inference Engine FG* of the SBDIM Middleware called RB-DEWES. The event hub is deployed on the local server and provides a tool for drought forecasting and prediction using local IK acquired in the study area. The sub-system employs rule-based methodology and probabilistic reasoning technique using *rules* derived from the IKS. The derived *rules* which are based on different scenarios and interpretation are saved in the knowledge base of the expert system event hub. The hub has an interactive interface accessed through the localhost, where the end user can select their current observation and the inference engine of the expert system event hub is fired using deductive mechanism from a *rule* or combination of *rules* with certainty factors. The output with attributed *certainty factors* is represented in JSON for use by the reasoners. Complete code in JAVA is available in Appendix E.

### 7.7.4. Reasoners

The task of augmenting the service output from the Semantic Annotation and Expert System Event Hub Sub-System is the responsibility of the reasoners. Several semantics reasoners exist as a plugin for achieving reasoning services. The middleware utilises the *FACT++* reasoners. The reasoner's leverage on the semantic representation of the sub-systems' outputs in JSON/JSON-LD for merging and aggregation of the outputs with a simple generation of information to be published by the *Data Publishing FG*.

### 7.8. Data Publishing FG Phase

The final output of the middleware is called Drought Forecast Advisory Information (DFAI). This information is made available to policymakers for decision-making processes and dissemination to the farmers. The system analyst interacts with the middleware using data input sources from the WSN and the IK as shown in Figure 7-13, and the middleware processes the data through the FGs and also factored in the current IK observation, and a final inferred output is generated. The output is published via Web apps, notifications hubs, mobile services or saved to document repository for offline storage.

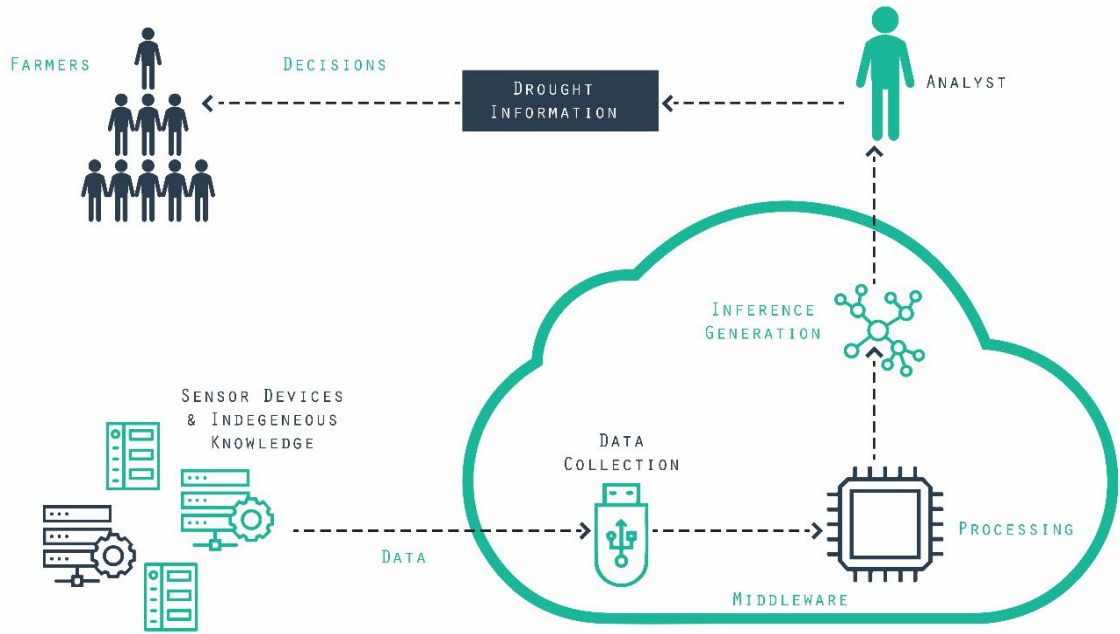

Figure 7- 13: SBDIM Middleware Process Flow Chart (*Source: Author*).

## 7.9.   Review of Software Verification and Validation (V&V) Process

The system evaluation in terms of the V&V were performed during the implementation processes with minimum inputs and outputs to ensure strict adherence with the initial requirements. The is carried out during the unit evaluation of each FGs, with the V&V outcome indicated that the distributed FGs of the middleware conforms perfectly with the FRs and NFRs of the semantic middleware.

## 7.10.  UX Evaluation of Prototype

After the unit evaluation of each FGs phase, this section further presents the results UX evaluation of the developed distributed semantics-based data integration middleware – an intermediary distributed middleware infrastructure that integrate heterogeneous data sources. The aim is to test the applicability of the distributed FGs of the middleware prototype from an end users' point of view. The evaluation procedure adopted the human-centred design process method (Mabanza, 2018). After developing the prototype, a UX evaluation of the semantic middleware was done to determine the ease of use.

For UX evaluation of the middleware prototype, a focus group comprising of twelve (12) participants (SQA testers and proposed users) – six (6) literate farmers and six (6) software developers were tasked to rate the UX experience at a workshop session. The number of participants for the evaluation was relatively small, but according to Nielsen (1994) – "a small number of participants can be sufficient for having a valid result for testing a developed system". Hence, the result of the evaluation process was accepted to be a valid result. The workshop started with a background explanation, demonstration of the distributed prototype to the participants, and a simple hand-on interaction of each FG phases by the participants through the middleware's FGs GUI.

After the participant's interactions with the middleware prototype, the participants were tasked to rate the usability experience through a given System Usability Scale (SUS) (Brooke, 1996). The SUS questionnaire (Appendix F) provides a measure to determine how efficiently and easily users can utilise a software product or service.

### 7.10.1.    Performance and Usability Evaluation

Using the System Usability Scale (SUS) as mentioned above, the SUS consists of ten (10) statements with 5 points each on the Likert scale of agreement or disagreement (Brooke, 1996). To calculate the overall SUS score – a cumulative of the statements points was performed using the division of the overall scores as follows: score of 0-25: worst, score of 25-39: poor, score of 39-52: ok, score of 52-85 excellent, and score of 85-100: best imaginable (Brooke, 1996). Hence, SUS scores have a range of 0 to 100. The results of the SUS scores are shown in Figure 7-14 below.

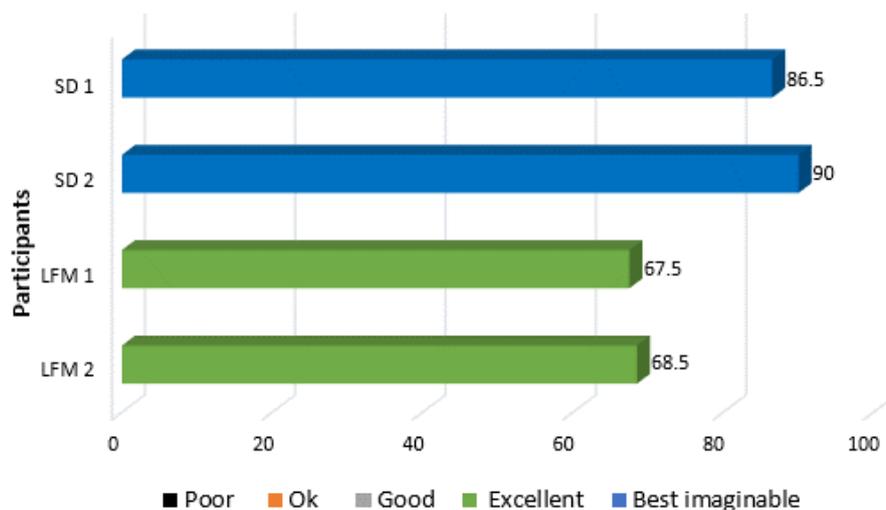

Figure 7- 14: SUS Scores (*Source: Author*).

The participants were divided into group of three each – SD 1, SD 2, LFM 1, and LFM 2. From Figure 7-15, the results indicate an approval rating of above 65%. It is observed that the LFM 1 (Literate Farmers Group 1), LFM 2 (Literate Farmers Group 2) rated the middleware prototype as "excellent"; while the SD 1 (Software Developer Group 1) and SD 2 (Software Developer Group 2) rated the systems as "best imaginable". Therefore, the Middleware prototype attained an "Excellent and Best Imaginable" SUS score.

## 7.10.2.    Recommendation from the Participants

Despite achieving positive evaluation feedback from the study's distributed middleware infrastructure prototype, few recommendations were received from the participants towards improving the overall usability of the system. The most important recommendation received was about the unification of the entire distributed FGs of the middleware as a unified system in the form of IaaS (Infrastructure-as-a-Service) accessible through the cloud.

## 7.11.  Summary

This chapter has presented the evaluation of all the *Functional Groups (FG)* of the middleware in the form of V&V during implementation from a holistic point of view. The *FG* service(s) output data format from the implementation was presented and the data flow from the first FG (*Data Acquisition FG*) to the last FG (*Data Publishing FG*). The V&V evaluation is a way of ensuring the initial requirements have been satisfied, and effective at uncovering basic design assumption errors and deviation from research objectives.

# CHAPTER EIGHT

# DISCUSSION AND CONCLUSIONS

## 8.1. Introduction

This chapter summarises the evaluation of the thesis objectives together. The chapter also presents the main contribution, innovative aspects of the research, conclusion and future research directions.

## 8.2. Evaluation of Thesis Objectives

In this the thesis, all objectives, which were described in the introduction, were achieved.

### 8.2.1. Weather Prediction based on Integration of Heterogeneous Data Sources

The complex nature of drought demands a complete understanding of all knowledge spheres for a holistic integration, analysis and inference generation. While it was a difficult requirement, especially considering the heterogeneity of data and technology, there is a gap in providing efficient and scalable methods towards achieving this – and it is a vital objective of this research – towards more accurate drought early warning systems (DEWS). An investigation was accomplished on the most effective exploit in achieving a perfect integration of IK and WSN data for accurate drought forecasting (Akanbi & Masinde, 2015b). The investigation established that ontologies and Semantic Web technologies might facilitate the integration of heterogeneous data and interoperability of services. The identification proved usable and resulted in the development of several frameworks for the semantic integration of different data sources.

The first of the series of frameworks developed was the IKSDC module framework, which facilitated the collection of indigenous knowledge. From the IK data collected, over 90% stated that they knew and applied IK to predict likely rainfall and onset of drought in their area. The second framework is the WSDC framework for the deployment of IoT/WSN sensors in the study area for collecting accurate localised data. The two heterogeneous data sources were semantically integrated towards creating a more accurate drought early warning system (DEWS) using the SB-DIM framework. An analysis of the SB-DIM framework as presented (Akanbi & Masinde, 2018b) was found to enhance effective data collection, integration and development of a semantics-based data integration middleware.

## 8.2.2. The Semantic Representation of Heterogeneous Weather Data (IK & WSN Data)

The problem of information integration and interoperability of the two different data sources was encountered, discussed; with the semantic representation of the data as the main solution. Representation of the knowledge base using semantic representation was described. As the first requirement for resolving data heterogeneity, a domain ontology was developed. There exists no semantic ontological framework for the local indigenous knowledge on drought currently in existence (Akanbi & Masinde, 2018c). Hence, it is a primary objective to develop from scratch a domain ontology for the representation of local indigenous knowledge.

Detailed attention was paid to the use of foundational ontologies (mainly DOLCE) for supporting the task of knowledge representation. Next, the development and encoding of **I**ndigenous **K**nowledge on Drought Domain **ON**tology (**IKON**), which captures and models the description of local indicators related to drought forecasting in the study area, using the entities, ecological interactions with behavioural relationships were described (Akanbi & Masinde, 2018c). The proposed solution for sensor data was built on SSN ontology – which was extended by the required concepts.

The main benefit of the ontology utilisation is the capability of unambiguous identification of natural indicators used in the context of indigenous knowledge on drought and sensor devices' data which may not be misinterpreted even without a given context. The employment of ontologies for the knowledge representation of the heterogeneous knowledge bases eliminates data heterogeneity and ensures a unified approach for representing the data models and seamless use of the data in the proposed semantics-based data integration Middleware or other intelligent information systems. Moreover, the proposed solution also offers additional benefits:

a) The application of the developed domain ontologies in Semantic Web and Web of Linked Data.
b) Ontology matching methods for accurate identifications of objects or entities (natural indicators or sensors), irrespective of the representation format.
c) The use of IK on drought domain ontology, which is publicly available and an extension. The adoption of DOLCE ontology (upper ontology) ensures the ease of reusability and compatibility.

### 8.2.3. Using IoT/WSN in Real-Time Monitoring of Drought Parameters

To achieve this objective, a wireless sensor network was deployed in the study area, using different varieties of sensors devices and weather station all with different data representation format. At first, the sensors were calibrated before deployment, to factor in instrument error against a standard instrument. The calibrated sensors were deployed to a remote part of the study area with the gateway/sink at the centre of the star network topology, ensuring complete coverage. Accurate readings were taken every 15 minutes and the data streamed and saved to the cloud.

The study result has proven that the use of IoT/WSN in environmental monitoring provides an accurate *in-situ* measurement of the parameters that could be committed to the cloud in real-time. However, several challenges do exist – such as powering the sensors and keeping them safe from environmental conditions. The pros outweigh the cons and have proven dependable towards achieving a reliable and accurate dataset.

Through the implementation of a CEP engine like *Apache Kafka* in this study's cluster, IoT/WSN, sensor readings in the form of data streams are processed in real time using filters, aggregations, joins on a set of window data based on different predefined patterns in the cloud. The streaming platform facilitates the end-to-end enterprise stream monitoring of the entire cluster's health with the ability to receive alerts or set triggers, measure system loads and network utilisation, determine latencies and throughput for each broker per cluster.

### 8.2.4. Application of Semantic Middleware in Solving Integration and Interoperability of Different Entities.

The SM-DIM framework was formulated as an overview of the semantics-based data integration middleware based on a service-oriented architecture (SOA). The semantic middleware comprises various *Functional Groups* (FG) already discussed in earlier chapters, working in an orchestrated way towards achieving seamless data integration and interoperability. This is achieved through the representation of the inputs/output data in a unified machine-readable language. The ease of a unified language in the data pipeline compatible with the plethora of sub-systems in the middleware eliminates data heterogeneity, which hampers the integration of data and interoperability of services.

### 8.2.5. Implementing the Middleware as a DEWS for Creating Accurate Drought Prediction and Forecasting

After the design and development of the semantics-based data integration middleware, the semantic middleware was implemented as a form of Drought Early Warning System (DEWS) to ensure the feasibility of the middleware. The middleware integration is based on semantic technologies, and the inference generation is based on the use of CEPs, inference engines and reasoners as encompassed by the SB-DIM framework. The proposed solution incorporates several inference generation mechanisms in different FGs of the middleware to provides adequate flexibility and optimal inference generation capability. The CEP engine is an open-source *Apache Kafka* in the streaming platform – Confluent. The inference engine is JESS, with various reasoners in Protégé.

## 8.3.    Innovative Contributions of the Research Thesis

In order to improve the accuracy level of drought prediction and forecasting systems, this thesis investigates the possibility of integrating available heterogeneous data sources by solving the challenges of data integration and interoperability. The main contributions to the knowledge of the research in this thesis are summarised below.

a) Development encoding of **I**ndigenous **K**nowledge on Drought Domain **ON**tology (**IKON**) – In this research, a domain ontology for the local indigenous knowledge on drought was developed. This ontology provides a machine-readable format of the domain. The model is developed in Protégé and available in RDF and OWL format. This domain ontology is based on DOLCE, making it more easily reusable and extendable for future research purposes. More details can be found in Chapter Five and Paper D.

b) The conceptualisation of semantics-based data integration middleware framework – A model semantics-based data integration middleware framework has been proposed and implemented to solve the challenges of heterogeneous data integration and interoperability. The proposed framework facilitated the semantic representation of the data sources eliminating data heterogeneity and created a model with a unified data format. The framework is presented in Chapter Three. The details of the framework can be found in papers B and C.

c) Implementation of semantic middleware for the integration and interoperability of heterogeneous data sources for drought forecasting and prediction – A semantically-enhanced distributed middleware approach has been utilised for integrating the heterogeneous data. Using this approach, the structured and unstructured data sources are transformed and represented in a machine-readable language for seamless integration and inference generation. This contribution is presented in Chapter Four and paper C.

d) A streaming processing engine based on *Apache Kafka* for real-time processing of sensor data – Streams of data from the deployed sensors are channelled through a streaming platform; using a drought prediction model; the streaming engine determines patterns in the data streams, and inference are generated as outputs. More details can be found in Chapter Six.

e) RB-DEWES sub-system that could be implemented as a standalone system – A component of the entire system can be implemented as a standalone system with customisable GUI for end-users to specify current indigenous knowledge observation of occurrences. The inference engine of the RB-DEWS will fire and determine the likely implication of the scenarios using expert knowledge saved in the knowledge base. Details of the system can be found in Chapter Six and paper E.

f) Implementation of a more accurate semantics-based DEWS based on the semantic middleware – The middleware is implemented as a DEWS for the study area. More details can be found in the thesis and published papers.

## 8.4. Conclusion and Future Work

This thesis proposed a semantics-based data integration middleware for drought forecasting and prediction. The aim of the research was to develop a framework for semantic middleware that facilitates integration of heterogeneous data sources (IK on drought and sensors data) and interoperability of services towards achieving more accurate drought early warning systems (DEWS); using heterogeneous data from various places through mediator-based data integration approach would be beneficial and increase the level of reliability and variability. In the requirement elicitation phase of this study's approach, the researcher conducted a survey, interviewed IK domain experts, collected and documented the IK on drought in the study areas. The study also reviewed the literature on the most suitable IoT/WSN based systems that would facilitate useful measurement of the required environmental parameters and determine the challenges of integrating the heterogeneous data sources.

Based on the requirements identified and the research gap, in this thesis, the solution for achieving an accurate drought forecasting and prediction system using different data sources has been presented, and subsequently, this work proved that the integration and interoperability using Semantic Web technologies are feasible and reliable. The presented semantic middleware performs semantic representation and metadata annotation of input data and knowledge base to create unified machine-readable data for use in various functional groups that perform aggregation and computational analysis based on forecasting models and current indigenous observations. As this thesis has shown, heterogeneous data integration and interoperability could be solved.

Through the thesis, the study introduced the proposition, conceptualised framework and system design, and explained all detailed implementations in stages based on the presented semantics-based data integration frameworks. The multitude of sub-systems in the semantic middleware produces a service(s) as a combined output – enabling other services to be created – with drought forecast advisory information (DFAI) as an output of the middleware. The DFAI as an output of the middleware is based on the EDI drought severity index – which categorises the severity of the drought. This serves as advisory information to policy-makers or system analyst for interpretation and recommendation to the farmers (end-users). Accurate risk perception and knowledge needed to interpret the advisory information by the policy-makers is essential.

Heterogeneous data integration and interoperability were fascinating but challenging subjects to study. Nevertheless, this research has made a meaningful contribution to the challenging task of solving the data integration and interoperability problems of the data-driven solution towards achieving a more accurate inference in the environmental monitoring domain – for drought forecasting and prediction. The results of this research are focused mainly on drought forecasting and prediction. Also, it applies to the challenges of integration and interoperability will eliminate the bottlenecks hampering the full realisation of IoT potentials.

The presented work is the first step for achieving seamless integration, interoperability and improving the accuracy of drought forecasting and prediction systems. Constant improvement of warning systems is challenging and necessary to reflect the trend and improving the systems accuracy (Twigg & Lavell, 2006; Leonard, Johnston, Paton, Christianson, Becker & Keys, 2008). Future research and development will be aimed to complement the developed system and suggests to explore the following:

a) Improving the mechanism of drought early warning system through the application of an ontological-based reasoning technique.

b) The semantic representation and integration of inferences generated from heterogeneous knowledge bases with other intelligent information systems for a more accurate drought forecasting and prediction system.

c) The implementation of the proposed middleware FGs approach could be further improved by a formalisation how to utilise cloud-based services as an IaaS (Infrastructure-as-a-Service); currently – it is a distributed service with some of the FGs residing on a local server environment and others in the cloud. Primary experiment with the local servers and sub-systems of the *Inference Engine FG* in the cloud were conducted, but proper methodology and formalisation of these servers together could exploit web-based capabilities – to promote ease of use.

d) Indigenous knowledge component (and its developed domain ontology) of this research is currently limited to the study areas. More case studies could be done to document the indigenous knowledge on the drought of other communities, expand the knowledge base and extend the domain ontology for extensive reuse purposes.

e) Even though the evaluation model presented here has been developed centred on drought forecasting and prediction early warning systems, the developed framework and middleware apply to other warning systems that need to integrate heterogeneous data sources (structured and unstructured).

f) In this research, to integrate the heterogeneous data sources, manual and semi-automatic methods were used in the semantic middleware in a distributed manner. For future work, complex algorithms for automatic data integration could be developed. This can include the adoption of more complex streaming techniques, mapping, reasoning methods.

g) The security of the data pipeline was not taken into consideration for the data exchange and communication amongst all the devices, sub-systems, clusters and all the *functional groups* in the middleware. It was assumed all communication and data exchanges are handled using secure channels. As future work, securing the entire data pipeline could be carried out.

# The Development of Semantic-based Data Integration Middleware for Integrating Local Indigenous Knowledge and Scientific Data for Drought Forecasting/Monitoring System

*Questionnaire for Local Indigenous Knowledge Data Gathering*

## RESEARCH INVITATION LETTER

Dear ____________________,

I am pleased to invite you to participate in an interview to identify and document the local indigenous knowledge weather indicators based on the following categories (1) patterns of seasons (cold, dry, hot, raining and so on); (2) animals, insects and bird's behaviors; (3) astronomical; (4) meteorological; (5) human nature and behavior; and (6) behaviors of plant/trees. No more than thirty minutes would be required to complete the interview.

Be assured that any information you provide will be treated in the strictest confidence and your participation will not be identifiable in the resulting report. You are entirely free to discontinue your participation at any time or to decline to answer particular questions.

I will seek your consent, on the attached form, to record the interview and to use the recording in preparing the report, on condition that your name or identity is not revealed, and to make the recording available to other researchers on the same conditions.

Direct any enquiries concerning this study to the **main** Researchers contacts below.

Thank you for your assistance.

Researcher

Central University of Technology, Free State, South Africa

# APPENDIX A

Questionnaire page 1

**The Development of Semantic-based Data Integration Middleware for Integrating Local Indigenous Knowledge and Scientific Data for Drought Forecasting/Monitoring System**

*Questionnaire for Local Indigenous Knowledge Data Gathering*

## INTERVIEW/QUESTIONNAIRE GUIDE

The purpose of the interview is to gather the local indigenous knowledge on drought forecasting and environmental monitoring using indicators.

The researcher/research assistant will: -

1. Introduce the interview session by explaining the purpose of the interview, welcome the respondent(s) and make clear why they were chosen.

2. Explain the presence and purpose of any recording equipment and give the option for respondent(s) to opt out of recording.

3. Outline ground rules and interview guidelines such as participants can end the interview at any time or refuse to answer any questions,

4. Inform the respondent(s) that a break will be provided if time goes beyond 30 minutes.

5. Address the issue of privacy and confidentiality and inform the respondent(s) that information gathered will be analyzed aggregately and respondent's personal details will not be used in any report. The researcher will also make it clear that respondents' answers and any information identifying the respondent(s) as a participant of this research will be kept confidential.

6. Inform the respondent(s) that they must sign consent forms before the interview begins.

7. Inform the respondent(s) that the interview consists of 19 questions, some with sub sections.

8. Inform the respondent**(s)** how to provide answers to questions by either putting a mark on a check box for optional questions or by giving a short answer to open ended questions.

9. Inform the respondent**(s)** that during or after the interview additional questions can be asked to clarify the respondent**(s)** answer.

10. Inform respondent**(s)** that they may choose not to answer a particular question; in that event, he will need to inform the researcher or research assistant.

11. Inform the respondent**(s)** that oral interview will be recorded to ensure responses are captured and transcribed accurately.

12. Inform the respondent**(s)** that they are allowed ask questions before, during and after the interview

13. Go through the process of completing a questionnaire with the respondent(s) through as an example

14. Inform the respondent(s) of follow-up activities and that they should provide their contact details at the end of the questionnaire if they may wish to be involved in the implementation phase of the research.

15. Assist the respondent(s) to properly fill the questionnaires to competition.

16. Collect all the questionnaire from the respondent(s)

17. Close the interview by thanking the respondent(s), maintaining on privacy and confidentiality considerations;

Questionnaire page 3

**The Development of Semantic-based Data Integration Middleware for Integrating Local Indigenous Knowledge and Scientific Data for Drought Forecasting/Monitoring System**

*Questionnaire for Local Indigenous Knowledge Data Gathering*

### CONSENT FORM

**I, the undersigned, confirm that (please tick box as appropriate):**

| | | |
|---|---|---|
| [1] | I have read and understood the information about the research, | ☐ |
| [2] | I have been given the opportunity to ask questions about the research and my participation. | ☐ |
| [3] | I voluntarily agree to participate in the research. | ☐ |
| [4] | I understand I can withdraw at any time without giving reasons and that I will not be penalized for withdrawing | ☐ |
| [5] | The procedures regarding confidentiality have been clearly explained to me. | ☐ |
| [6] | If applicable, separate terms of consent for forms of data collection have been explained and provided to me. | ☐ |
| [7] | The use of the data in research, publications, sharing and archiving has been explained to me. | ☐ |
| [8] | I understand that other researchers will have access to this data only if they agree to preserve the confidentiality of the data and if they agree to the terms I have specified in this form. | ☐ |
| [9] | Select only **ONE** of the following: | |
| | • I would like my name used and understand what I have said or written as part of this research will be used in reports, publications and other research outputs so that anything I have contributed to this project can be recognised. | ☐ |
| | • I do not want my name used in this research. | ☐ |
| [10] | I agree to sign and date this informed consent, along with the Researcher. | ☐ |

______________________ _______________ ________________
**Name of Respondent**          **Signature**                          **Date**

______________________ _______________ ________________
**Name of Researcher**          **Signature**                          **Date**

**The Development of Semantic-based Data Integration Middleware for Integrating Local Indigenous Knowledge and Scientific Data for Drought Forecasting/Monitoring System**

**QUESTIONNAIRE FOR LOCAL INDIGENOUS KNOWLEDGE DATA GATHERING**

**[September 2017]**

**PART A: INTRODUCTION**

The local Indigenous knowledge has been built over the years from an understanding of local weather, climate, interpretations of animals, insects, birds, and plants behaviour of a particular geographical area. The major strength of IK lies in long time-series of observations in a particular region. The veracity of the knowledge is based on diachronic data (long time-series) as opposed to synchronic data (short time-series over a large area) obtained from modern weather monitoring devices. The two kind of data when semantically integrated would provide accurate and reliable drought forecasting input.

The Department of Information Technology at the Central University of Technology, Free State in conjuction with the University of KwaZulu Natal is conducting a research to identify and document the unstructured weather indicators based on the following categories (1) patterns of seasons (cold, dry, hot, rainy and so on); (2) animal, insects and bird's behaviour; (3) astronomical; (4) meteorological; (5) human nature and behaviour; and (6) behaviour of plants/trees.

Phase I of this research seeks to collate the Indigenous Knowledge (IK) from natives, local farmers, IK holders at KwaZulu-Natal province of South Africa and Ndau people, Muchenedze District of Mozambique. The results of this research will be used to develop an ontology that captures all the entities and relationship among the entities in the weather monitoring domain. This knowledge base will be useful in refining and development of an accurate IoT-based drought forecasting system.

You are requested to participate in this research by completing this questionnaire. You are required to put a mark ($\sqrt{}$ or X) in the check box for the appropriate option or write down your response in the area provided.

Questionnaire page 5

# The Development of Semantic-based Data Integration Middleware for Integrating Local Indigenous Knowledge and Scientific Data for Drought Forecasting/Monitoring System

## PART B: DEMOGRAPHIC INFORMATION

Q1  **Names:** ________________________________________ (Optional)

Q2  **Gender?**  ❑ Male ❑ Female

Q3  **Age bracket?**

❑ Under 18  ❑ 18-35  ❑ 36-45  ❑ 46-55  ❑ 56-65  ❑ above 66

Q4  **Highest Education Level:**

❑ None ❑ Primary ❑ Secondary ❑ Post-Secondary

Q5  **What is the name of your community?**
_______________________________________________________

Q6  **What is the main economic activity in your community?**
_______________________________________________________

Q7  **How long have you stayed in this community?**

❑ 5- 10 years        ❑ 10-20 years        ❑ over 20 years

Q8  **Do you own a phone or have access to a phone?**

❑ Yes        ❑ No

Q9  **Do you own a smart phone?**

❑ Yes        ❑ No

# PART C: KNOWLEDGE ON WEATHER FORECASTING

Q10 **Do you check the weather forecast?**

❑ Yes          ❑ No

**If Yes, how often do you check it?**

❑ Daily          ❑ Weekly          ❑ Monthly          ❑ Seasonal

Q11 **Do you regularly check for the weather forecast during your cropping decisions?**

❑ Yes          ❑ No

Q12 **Where do you get your weather forecast information? (You may tick more than one box).**

**Do you have confidence in the accuracy of information you get from these options? Please, tick on a scale 1 – 5, with one (1) being the lowest level of confidence and five (5) being the highest.**

|  | 1 | 2 | 3 | 4 | 5 |
|---|---|---|---|---|---|
| ❑ Radio | ❑ | ❑ | ❑ | ❑ | ❑ |
| ❑ TVs | ❑ | ❑ | ❑ | ❑ | ❑ |
| ❑ Newspapers | ❑ | ❑ | ❑ | ❑ | ❑ |
| ❑ Local observations e.g. the clouds and behavior of animals | ❑ | ❑ | ❑ | ❑ | ❑ |
| Others please specify _____________ | ❑ | ❑ | ❑ | ❑ | ❑ |

Q13 **Do you use the information from the weather forecast to plan your work?**

❑ Yes ❑ No

**If Yes, what kind of decisions do you make based on the weather forecast?**

❑ Planting date selection ❑ Crop selection

❑ Planting method ❑ Weeding

❑ Harvesting ❑ Marketing

Others, please specify: _______________________________________

Q14 **Does the weather forecast provide you with the kind of information you need to make decisions for planting and managing your crops?**

❑ Yes ❑ No

Q15 **What other information would you want to get from the forecast that could help you to make decisions on your farm?**

_________________________________________________________________
_________________________________________________________________
_________________________________________________________________
_________________________________________________________________
_________________________________________________________________
_________________________________________________________________
_________________________________________________________________
_________________________________________________________________
_________________________________________________________________
_________________________________________________________________
_________________________________________________________________
_________________________________________________________________
_________________________________________________________________
_________________________________________________________________
_________________________________________________________________
_________________________________________________________________
_________________________________________________________________
_________________________________________________________________
_________________________________________________________________
_________________________________________________________________
_________________________________________________________________
_________________________________________________________________

## PART D: EXAMPLES OF INDIGENOUS/LOCAL INDICATORS FOR WEATHER

Q16  **Which of the following cropping decisions do you use indigenous knowledge to reach?** You can tick more than one one**)**

❑ When/if to plant; for example. *decide not to plant at all based on very prolonged rains onset*

❑ What to plant; e.g. *to decide to plant sweet potatoes instead if maize based on the anticipated rainfall*

❑ How to plant; e.g. decision to practice mixed cropping

❑ When to harvest; e.g. *if I know there will be frost next week, I can decide to harvest all my crop before*

❑ Disposal/selling of produce; e.g. *when I know that a drought is imminent, I conserve all my produce stead of selling it*

Q17  **List some of the indigenous indicators that you commonly use in order to make to make decisions in Q12 above.**

| | Summer | Autumn | Winter | Spring |
|---|---|---|---|---|
| ❑ Meteorological | _________ | _________ | _________ | _________ |
| | _________ | _________ | _________ | _________ |
| | _________ | _________ | _________ | _________ |
| ❑ Behaviors of birds | _________ | _________ | _________ | _________ |
| | _________ | _________ | _________ | _________ |
| | _________ | _________ | _________ | _________ |
| ❑ Behaviors of insects; e.g. *ants moving in a straight line indicate a dry spell is imminent* | _________ | _________ | _________ | _________ |
| | _________ | _________ | _________ | _________ |
| | _________ | _________ | _________ | _________ |
| ❑ Behaviors of animals, e.g. *cattle coming home jumping with their tails up is a sign of a good season* | _________ | _________ | _________ | _________ |
| | _________ | _________ | _________ | _________ |
| | _________ | _________ | _________ | _________ |
| ❑ Flower, leave and Fruit Production by some trees | _________ | _________ | _________ | _________ |
| | _________ | _________ | _________ | _________ |
| ❑ Astronomical, e.*g. Visible phases of full moon signifies drier period* | _________ | _________ | _________ | _________ |
| | _________ | _________ | _________ | _________ |
| | _________ | _________ | _________ | _________ |

| ❑ Myths and religious beliefs, e.g. *an extreme drought is a curse for the offences to the gods* | ____________ | ____________ | ____________ | ____________ |
| --- | --- | --- | --- | --- |
| | ____________ | ____________ | ____________ | ____________ |
| | ____________ | ____________ | ____________ | ____________ |
| ❑ Knowledge of seasons e.g. *it always rains during summer* | ____________ | ____________ | ____________ | ____________ |
| | ____________ | ____________ | ____________ | ____________ |
| | ____________ | ____________ | ____________ | ____________ |

Questionnaire page 8

# APPENDIX B

```xml
<h:html xmlns="http://www.w3.org/2002/xforms" xmlns:h="http://www.w3.org/1999/xhtml" xmlns:xsd="http://www.w3.org/2001/XMLSchema"
xmlns:jr="http://openrosa.org/javarosa">
  <h:head>
    <h:title>Data Collection Tool - KZN</h:title>
    <model>
      <instance>
        <data id="build_Data-Collection-Tool-KZN_1558795977">
          <meta>
            <instanceID/>
          </meta>
          <Name/>
          <Gender>
            Please select
          </Gender>
          <Age>
            Select Age Bracket
          </Age>
          <HighestEducationLevel/>
          <Village/>
          <DurationInVillage/>
          <NoInHousehold/>
          <HouseholdSourceOfIncome/>
          <HouseholdIncome-IF-Others/>
          <CropsGrown/>
          <CropGrownOthers/>
          <FarmPlotSize/>
          <IncomeFromFarming/>
          <DifficultyInGrowingCrops/>
          <BiggestChallengesLastSeason/>
          <DoYouOwnAPhone/>
          <PhoneSmartness/>
          <PhonePrimaryFunction/>
          <CheckingWeatherForecasts/>
          <FrequencyOfWeatherForecastCheck/>
          <WhereDidYouGetYourWeatherForecastInfo/>
          <OtherSourceOfWeatherForecast/>
          <UsageOfIndigenousKnowledge/>
          <TypeOfIndigenousKnowledgeUsed/>
          <OtherIndigenousKnowledge/>
          <ExactUsageOfIndigenousKnowledge/>
          <BenefitsOfCUTDroughtPredictionTool/>
          <WillingnessToPayForTheDroughtPredictionTool/>
          <CostForDroughtPredictionTool/>
          <InterestInDroughtServiceSubscription/>
          <RespondentPhoneNumber/>
          <ONDPlantPeriod/>
          <MaizeFarmSize/>
          <MaizeLandOwnership/>
          <AppliedCroppingMethods/>
          <QuantityOfProducedMaizeOneSeasonAgo/>
          <QuantityOfProducedMaizeTwoSeasonAgo/>
          <QuantityOfProducedMilletOneSeasonAgo/>
          <QuantityOfProducedMilletTwoSeasonsAgo/>
          <QuantityOfProducedBeansOneSeasonAgo/>
          <QuantityOfProducedBeansTwoSeasonsAgo/>
          <QuantityOfProducedCowpeasOneSeasonAgo/>
          <QuantityOfProducedCowpeasTwoSeasonsAgo/>
          <QuantityOfProducedSorghumOneSeasonAgo/>
          <QuantityOfProducedSorghumTwoSeasonsAgo/>
          <GPSLocation/>
        </data>
      </instance>
      <itext>
        <translation lang="English">
          <text id="/data/Name:label">
            <value>What is your name?</value>
          </text>
          <text id="/data/Name:requiredMsg">
            <value>Please type the name of the respondent</value>
          </text>

          <text id="/data/Gender:label">
            <value>Gender</value>
          </text>
          <text id="/data/Gender:requiredMsg">
            <value>Please select Gender</value>
          </text>
          <text id="/data/Gender:option0">
            <value>Male</value>
          </text>
          <text id="/data/Gender:option1">
            <value>Female</value>
          </text>
          <text id="/data/Age:label">
            <value>Age Bracket</value>
          </text>
          <text id="/data/Age:requiredMsg">
            <value>Please select Age Bracket</value>
          </text>
          <text id="/data/Age:option0">
            <value>Under 18</value>
          </text>
          <text id="/data/Age:option1">
            <value>18-35</value>
          </text>
          <text id="/data/Age:option2">
            <value>36-45</value>
          </text>
          <text id="/data/Age:option3">
            <value>46-55</value>
          </text>
          <text id="/data/Age:option4">
            <value>Above 66</value>
          </text>
          <text id="/data/HighestEducationLevel:label">
            <value>Highest Education Level</value>
          </text>
```

```xml
107      <text id="/data/HighestEducationLevel:requiredMsg">
108        <value>Please Select Highest Education Level</value>
109      </text>
110      <text id="/data/HighestEducationLevel:option0">
111        <value>None</value>
112      </text>
113      <text id="/data/HighestEducationLevel:option1">
114        <value>Primary</value>
115      </text>
116      <text id="/data/HighestEducationLevel:option2">
117        <value>Secondary</value>
118      </text>
119      <text id="/data/HighestEducationLevel:option3">
120        <value>Post-Secondary</value>
121      </text>
122      <text id="/data/Village:label">
123        <value>What is the name of your village?</value>
124      </text>
125      <text id="/data/DurationInVillage:label">
126        <value>How long have you stayed in this village?</value>
127      </text>
128      <text id="/data/DurationInVillage:option0">
129        <value>5-10 Years</value>
130      </text>
131      <text id="/data/DurationInVillage:option1">
132        <value>10-20 Years</value>
133      </text>
134      <text id="/data/DurationInVillage:option2">
135        <value>Over 20 Years</value>
136      </text>
137      <text id="/data/NoInHousehold:label">
138        <value>How many people in your household?</value>
139      </text>
140      <text id="/data/HouseholdSourceOfIncome:label">
141        <value>How does your household earn money/source of income?</value>
142      </text>
143      <text id="/data/HouseholdSourceOfIncome:option0">

144        <value>Farming</value>
145      </text>
146      <text id="/data/HouseholdSourceOfIncome:option1">
147        <value>Business</value>
148      </text>
149      <text id="/data/HouseholdSourceOfIncome:option2">
150        <value>Others</value>
151      </text>
152      <text id="/data/HouseholdIncome-IF-Others:label">
153        <value>Others?</value>
154      </text>
155      <text id="/data/CropsGrown:label">
156        <value>What crops do you grow?</value>
157      </text>
158      <text id="/data/CropsGrown:option0">
159        <value>Maize</value>
160      </text>
161      <text id="/data/CropsGrown:option1">
162        <value>Millet</value>
163      </text>
164      <text id="/data/CropsGrown:option2">
165        <value>Beans</value>
166      </text>
167      <text id="/data/CropsGrown:option3">
168        <value>Sorghum</value>
169      </text>
170      <text id="/data/CropsGrown:option4">
171        <value>Cow Peas</value>
172      </text>
173      <text id="/data/CropsGrown:option5">
174        <value>Others</value>
175      </text>
176      <text id="/data/CropGrownOthers:label">
177        <value>Others?</value>
178      </text>
179      <text id="/data/FarmPlotSize:label">
180        <value>How large is your farm?</value>
181      </text>
182      <text id="/data/FarmPlotSize:hint">

183        <value>In Hectares</value>
184      </text>
185      <text id="/data/IncomeFromFarming:label">
186        <value>How much do you make in income from farming?</value>
187      </text>
188      <text id="/data/DifficultyInGrowingCrops:label">
189        <value>What do you find most difficult about growing your crops?</value>
190      </text>
191      <text id="/data/DifficultyInGrowingCrops:option0">
192        <value>Unexpected Rainfall</value>
193      </text>
194      <text id="/data/DifficultyInGrowingCrops:option1">
195        <value>Less Rainfall</value>
196      </text>
197      <text id="/data/DifficultyInGrowingCrops:option2">
198        <value>Cost of inputs (seeds, fertilizer etc)</value>
199      </text>
200      <text id="/data/DifficultyInGrowingCrops:option3">
201        <value>Cost of Labour</value>
202      </text>
203      <text id="/data/DifficultyInGrowingCrops:option4">
204        <value>Unavailability of Machinery</value>
205      </text>
206      <text id="/data/DifficultyInGrowingCrops:option5">
207        <value>Transportation to market</value>
208      </text>
209      <text id="/data/DifficultyInGrowingCrops:option6">
210        <value>Spoilage</value>
211      </text>
212      <text id="/data/DifficultyInGrowingCrops:option7">
213        <value>Attacks by Pests</value>
214      </text>
215      <text id="/data/BiggestChallengesLastSeason:label">
216        <value>What was the biggest challenge you faced last season (October-November-December rains)</value>
217      </text>
218      <text id="/data/BiggestChallengesLastSeason:option0">
219        <value>Unexpected rainfall</value>
220      </text>
```

```xml
221  <text id="/data/BiggestChallengesLastSeason:option1">
222    <value>Less than expected rainfall</value>
223  </text>
224  <text id="/data/BiggestChallengesLastSeason:option2">
225    <value>Costs of inputs (seeds, fertilizer etc)</value>
226  </text>
227  <text id="/data/BiggestChallengesLastSeason:option3">
228    <value>Cost of Labour</value>
229  </text>
230  <text id="/data/BiggestChallengesLastSeason:option4">
231    <value>Unavailability of machinery</value>
232  </text>
233  <text id="/data/BiggestChallengesLastSeason:option5">
234    <value>Transportation to market</value>
235  </text>
236  <text id="/data/BiggestChallengesLastSeason:option6">
237    <value>Spoilage</value>
238  </text>
239  <text id="/data/BiggestChallengesLastSeason:option7">
240    <value>Attacks by pests</value>
241  </text>
242  <text id="/data/DoYouOwnAPhone:label">
243    <value>Do you own a phone or have access to a phone?</value>
244  </text>
245  <text id="/data/DoYouOwnAPhone:option0">
246    <value>Yes</value>
247  </text>
248  <text id="/data/DoYouOwnAPhone:option1">
249    <value>No</value>
250  </text>
251  <text id="/data/PhoneSmartness:label">
252    <value>Is the phone a smart phone?</value>
253  </text>
254  <text id="/data/PhoneSmartness:requiredMsg">
255    <value></value>
256  </text>

257  <text id="/data/PhoneSmartness:option0">
258    <value>Yes</value>
259  </text>
260  <text id="/data/PhoneSmartness:option1">
261    <value>No</value>
262  </text>
263  <text id="/data/PhonePrimaryFunction:label">
264    <value>What do you use your phone for primarily?</value>
265  </text>
266  <text id="/data/PhonePrimaryFunction:option0">
267    <value>To call friends and family</value>
268  </text>
269  <text id="/data/PhonePrimaryFunction:option1">
270    <value>To check news</value>
271  </text>
272  <text id="/data/PhonePrimaryFunction:option2">
273    <value>To get customers for crops/animals</value>
274  </text>
275  <text id="/data/CheckingWeatherForecasts:label">
276    <value>Did you regularly check for the weather forecast prior to the current rain season?</value>
277  </text>
278  <text id="/data/CheckingWeatherForecasts:option0">
279    <value>Yes</value>
280  </text>

281  <text id="/data/CheckingWeatherForecasts:option1">
282    <value>No</value>
283  </text>
284  <text id="/data/FrequencyOfWeatherForecastCheck:label">
285    <value>How often did you check for the weather forecasts</value>
286  </text>
287  <text id="/data/FrequencyOfWeatherForecastCheck:option0">
288    <value>Daily</value>
289  </text>
290  <text id="/data/FrequencyOfWeatherForecastCheck:option1">
291    <value>Weekly</value>
292  </text>
293  <text id="/data/FrequencyOfWeatherForecastCheck:option2">
294    <value>Monthly</value>
295  </text>
296  <text id="/data/FrequencyOfWeatherForecastCheck:option3">
297    <value>Seasonal</value>
298  </text>
299  <text id="/data/WhereDidYouGetYourWeatherForecastInfo:label">
300    <value>Where did you get your weather forecast information?</value>
301  </text>
302  <text id="/data/WhereDidYouGetYourWeatherForecastInfo:option0">
303    <value>Radio</value>
304  </text>
305  <text id="/data/WhereDidYouGetYourWeatherForecastInfo:option1">
306    <value>TVs</value>
307  </text>
308  <text id="/data/WhereDidYouGetYourWeatherForecastInfo:option2">
309    <value>Newspaper</value>
310  </text>
311  <text id="/data/WhereDidYouGetYourWeatherForecastInfo:option3">
312    <value>Local Indigenous Knowledge</value>
313  </text>
314  <text id="/data/WhereDidYouGetYourWeatherForecastInfo:option4">
315    <value>Others?</value>
316  </text>
```

```xml
419          <text id="/data/InterestInDroughtServiceSubscription:label">
420            <value>Are you interested in subscribing for this service?</value>
421          </text>
422          <text id="/data/InterestInDroughtServiceSubscription:requiredMsg">
423            <value>Please answer</value>
424          </text>
425          <text id="/data/InterestInDroughtServiceSubscription:option0">
426            <value>Yes</value>
427          </text>
428          <text id="/data/InterestInDroughtServiceSubscription:option1">
429            <value>No</value>
430          </text>
431          <text id="/data/RespondentPhoneNumber:label">
432            <value>Please provide your phone (cell) number</value>
433          </text>
434          <text id="/data/RespondentPhoneNumber:hint">
435            <value>Type the phone number</value>
436          </text>
437          <text id="/data/ONDPlantPeriod:label">
438            <value>At what point during the October-November-December rain season did you plant?</value>
439          </text>
440          <text id="/data/ONDPlantPeriod:option0">
441            <value>Before the rains started</value>
442          </text>
443          <text id="/data/ONDPlantPeriod:option1">
444            <value>Within 2Weeks after the rains had started</value>
445          </text>
446          <text id="/data/MaizeFarmSize:label">
447            <value>What is the size (in hectares) of the land under which you have planted maize?</value>
448          </text>
735            <value>Others</value>
736          </item>
737        </select>
738        <input ref="/data/CropGrownOthers">
739          <label ref="jr:itext('/data/CropGrownOthers:label')"/>
740        </input>
741        <input ref="/data/FarmPlotSize">
742          <label ref="jr:itext('/data/FarmPlotSize:label')"/>
743          <hint ref="jr:itext('/data/FarmPlotSize:hint')"/>
744        </input>
455          <text id="/data/MaizeLandOwnership:label">
456            <value>Do you own this piece of land?</value>
457          </text>
458          <text id="/data/MaizeLandOwnership:requiredMsg">
459            <value>Please answer</value>
460          </text>
461          <text id="/data/MaizeLandOwnership:option0">
462            <value>Yes</value>
463          </text>
464          <text id="/data/MaizeLandOwnership:option1">
465            <value>No</value>
466          </text>
467          <text id="/data/AppliedCroppingMethods:label">
468            <value>Which of the following two cropping method have you applied?</value>
469          </text>
470          <text id="/data/AppliedCroppingMethods:requiredMsg">
471            <value>Please answer</value>
472          </text>
473          <text id="/data/AppliedCroppingMethods:option0">
474            <value>Pure cropping (Only maize)</value>
475          </text>
476          <text id="/data/AppliedCroppingMethods:option1">
477            <value>Mixed  cropping (Maize planted with other crops)</value>
478          </text>
479          <text id="/data/QuantityOfProducedMaizeOneSeasonAgo:label">
480            <value>Please provide the quantity (in Kilograms) of MAIZE produced on your piece of land last season (one season ago)</value>
481          </text>
482          <text id="/data/QuantityOfProducedMaizeOneSeasonAgo:hint">
483            <value>In kilogram, type zero (0) if none.</value>
484          </text>
485          <text id="/data/QuantityOfProducedMaizeTwoSeasonAgo:label">
486            <value>Please provide the quantity (in Kilograms) of MAIZE produced on your piece of land two (2) seasons ago?</value>
487          </text>
488          <text id="/data/QuantityOfProducedMaizeTwoSeasonAgo:hint">
489            <value>In kilogram, type zero (0) if none.</value>
490          </text>
491          <text id="/data/QuantityOfProducedMaizeTwoSeasonAgo:requiredMsg">
492            <value>Please answer</value>
493          </text>
494          <text id="/data/QuantityOfProducedMilletOneSeasonAgo:label">
495            <value>Please provide the quantity (in Kilograms) of MILLET produced on your piece of land ONE (1) season ago?</value>
496          </text>
497          <text id="/data/QuantityOfProducedMilletOneSeasonAgo:hint">
498            <value>In kilogram, type zero (0) if none.</value>
499          </text>
500          <text id="/data/QuantityOfProducedMilletTwoSeasonsAgo:label">
501            <value>Please provide the quantity (in Kilograms) of MILLET produced on your piece of land TWO (2) seasons ago?</value>
502          </text>
503          <text id="/data/QuantityOfProducedMilletTwoSeasonsAgo:hint">
504            <value>In kilogram, type zero (0) if none.</value>
505          </text>
506          <text id="/data/QuantityOfProducedMilletTwoSeasonsAgo:requiredMsg">
507            <value>Please answer</value>
508          </text>
509          <text id="/data/QuantityOfProducedBeansOneSeasonAgo:label">
510            <value>Please provide the quantity (in Kilograms) of BEANS produced on your piece of land ONE (1) season ago?</value>
511          </text>
512          <text id="/data/QuantityOfProducedBeansOneSeasonAgo:hint">
513            <value>In kilogram, type zero (0) if none.</value>
514          </text>
515          <text id="/data/QuantityOfProducedBeansOneSeasonAgo:requiredMsg">
516            <value>Please answer</value>
517          </text>
518          <text id="/data/QuantityOfProducedBeansTwoSeasonsAgo:label">
519            <value>Please provide the quantity (in Kilograms) of BEANS produced on your piece of land TWO (2) seasons ago?</value>
520          </text>
```

```xml
<text id="/data/QuantityOfProducedBeansTwoSeasonsAgo:hint">
  <value>In kilogram, type zero (0) if none.</value>
</text>
<text id="/data/QuantityOfProducedCowpeasOneSeasonAgo:label">
  <value>Please provide the quantity (in Kilograms) of COWPEAS produced on your piece of land ONE (1) season ago?</value>
</text>
<text id="/data/QuantityOfProducedCowpeasOneSeasonAgo:hint">
  <value>In kilogram, type zero (0) if none.</value>
</text>
<text id="/data/QuantityOfProducedCowpeasTwoSeasonsAgo:label">
  <value>Please provide the quantity (in Kilograms) of COWPEAS produced on your piece of land TWO (2) seasons ago?</value>
</text>
<text id="/data/QuantityOfProducedCowpeasTwoSeasonsAgo:hint">
  <value>In kilogram, type zero (0) if none.</value>
</text>
<text id="/data/QuantityOfProducedCowpeasTwoSeasonsAgo:requiredMsg">
  <value>Please naswer.</value>
</text>
<text id="/data/QuantityOfProducedSorghumOneSeasonAgo:label">
  <value>Please provide the quantity (in Kilograms) of SORGHUM produced on your piece of land ONE (1) season ago?</value>
</text>
<text id="/data/QuantityOfProducedSorghumOneSeasonAgo:hint">
  <value>In kilogram, type zero (0) if none.</value>
</text>
<text id="/data/QuantityOfProducedSorghumOneSeasonAgo:requiredMsg">
  <value>Please answer.</value>
</text>
<text id="/data/QuantityOfProducedSorghumTwoSeasonsAgo:label">
  <value>Please provide the quantity (in Kilograms) of SORGHUM produced on your piece of land TWO (2) seasons ago?</value>
</text>
<text id="/data/QuantityOfProducedSorghumTwoSeasonsAgo:hint">
  <value>In kilogram, type zero (0) if none.</value>
</text>
<text id="/data/QuantityOfProducedSorghumTwoSeasonsAgo:requiredMsg">
  <value>Please answer.</value>
</text>

<text id="/data/GPSLocation:label">
  <value>GPS Location</value>
</text>
<text id="/data/GPSLocation:hint">
  <value>Allow app to capture GPS coordinates</value>
</text>
    </text>
  </translation>
</itext>
<bind nodeset="/data/meta/instanceID" type="string" readonly="true()" calculate="concat('uuid:', uuid())"/>
<submission method="form-data-post" action="https://docs.google.com/spreadsheets/d/15SEVc2Ivt4SW5exvbGlwwxxYuC-OiuCPUQWlL7uJgPw/edit?usp=sharing"/>
<bind nodeset="/data/Name" type="string" required="true()" jr:requiredMsg="jr:itext('/data/Name:requiredMsg')"/>
<bind nodeset="/data/Gender" type="select1" required="true()" jr:requiredMsg="jr:itext('/data/Gender:requiredMsg')"/>
<bind nodeset="/data/Age" type="select1" required="true()" jr:requiredMsg="jr:itext('/data/Age:requiredMsg')"/>
<bind nodeset="/data/HighestEducationLevel" type="select1" required="true()" jr:requiredMsg="jr:itext('/data/HighestEducationLevel:requiredMsg')"/>
<bind nodeset="/data/Village" type="string"/>
<bind nodeset="/data/DurationInVillage" type="select1" required="true()"/>
<bind nodeset="/data/NoInHousehold" type="int" required="true()" constraint="(. > '0') and (. <= '100')" jr:constraintMsg="Value must be between
<bind nodeset="/data/HouseholdSourceOfIncome" type="select"/>
<bind nodeset="/data/HouseholdIncome-If-Others" type="string" relevant="(selected(/data/HouseholdSourceOfIncome, 'Others'))"/>
<bind nodeset="/data/CropsGrown" type="select"/>
<bind nodeset="/data/CropsGrown0thers" type="string" relevant="(selected(/data/CropsGrown, 'Others'))"/>
<bind nodeset="/data/FarmPlotSize" type="int" required="true()" constraint="(. >= '0') and (. <= '500')" jr:constraintMsg="Value must be between
<bind nodeset="/data/IncomeFromFarming" type="string" required="true()"/>
<bind nodeset="/data/DifficultyInGrowingCrops" type="select"/>
<bind nodeset="/data/BiggestChallengesLastSeason" type="select"/>

<bind nodeset="/data/DoYouOwnAPhone" type="select1" required="true()"/>
<bind nodeset="/data/PhoneSmartness" type="select1" required="true()" jr:requiredMsg="jr:itext('/data/PhoneSmartness:requiredMsg')" relevant="(sele
<bind nodeset="/data/PhonePrimaryFunction" type="select"/>
<bind nodeset="/data/CheckingWeatherForecasts" type="select1" required="true()"/>
<bind nodeset="/data/FrequencyOfWeatherForecastCheck" type="select1" relevant="(selected(/data/CheckingWeatherForecasts, 'Yes'))"/>
<bind nodeset="/data/WhereDidYouGetYourWeatherForecastInfo" type="select"/>
<bind nodeset="/data/OtherSourceOfWeatherForecast" type="string" relevant="(selected(/data/WhereDidYouGetYourWeatherForecastInfo, 'Others'))"/>
<bind nodeset="/data/UsageOfIndigenousKnowledge" type="select1" required="true()" jr:requiredMsg="jr:itext('/data/UsageOfIndigenousKnowledge:require
<bind nodeset="/data/TypeOfIndigenousKnowledgeUsed" type="select" required="true()" jr:requiredMsg="jr:itext('/data/TypeOfIndigenousKnowledgeUsed
<bind nodeset="/data/OtherIndigenousKnowledge" type="string" required="true()" relevant="(selected(/data/TypeOfIndigenousKnowledgeUsed, 'Others'))
<bind nodeset="/data/ExactUsageOfIndigenousKnowledge" type="select"/>
<bind nodeset="/data/BenefitsOfCUTDroughtPredictionTool" type="select1" required="true()" jr:requiredMsg="jr:itext('/data/BenefitsOfCUTDroughtPredi
<bind nodeset="/data/WillingnessToPayForTheDroughtPredictionTool" type="select1" required="true()" jr:requiredMsg="jr:itext('/data/WillingnessToPay
<bind nodeset="/data/CostForDroughtPredictionTool" type="string" required="true()" jr:requiredMsg="jr:itext('/data/CostForDroughtPredictionTool:req
<bind nodeset="/data/InterestInDroughtServiceSubscription" type="select1" required="true()" jr:requiredMsg="jr:itext('/data/InterestInDroughtServic
<bind nodeset="/data/RespondentPhoneNumber" type="string" required="true()"/>
<bind nodeset="/data/ONDFlantPeriod" type="select1"/>
<bind nodeset="/data/MaizeFarmSize" type="int" required="true()" jr:requiredMsg="jr:itext('/data/MaizeFarmSize:requiredMsg')"/>
<bind nodeset="/data/MaizeLandOwnership" type="select1" required="true()" jr:requiredMsg="jr:itext('/data/MaizeLandOwnership:requiredMsg')"/>
<bind nodeset="/data/AppliedCroppingMethods" type="select1" required="true()" jr:requiredMsg="jr:itext('/data/AppliedCroppingMethods:requiredMsg')".
<bind nodeset="/data/QuantityOfProducedMaizeOneSeasonAgo" type="int" required="true()"/>
<bind nodeset="/data/QuantityOfProducedMaizeTwoSeasonAgo" type="int" required="true()" jr:requiredMsg="jr:itext('/data/QuantityOfProducedMaizeTwoSea
<bind nodeset="/data/QuantityOfProducedMilletOneSeasonAgo" type="int" required="true()"/>
<bind nodeset="/data/QuantityOfProducedMilletTwoSeasonsAgo" type="int" required="true()" jr:requiredMsg="jr:itext('/data/QuantityOfProducedMilletTw
<bind nodeset="/data/QuantityOfProducedBeansOneSeasonAgo" type="int" required="true()" jr:requiredMsg="jr:itext('/data/QuantityOfProducedBeansOneSe
<bind nodeset="/data/QuantityOfProducedBeansTwoSeasonsAgo" type="int" required="true()"/>
<bind nodeset="/data/QuantityOfProducedCowpeasOneSeasonAgo" type="int" required="true()"/>
<bind nodeset="/data/QuantityOfProducedCowpeasTwoSeasonsAgo" type="int" required="true()" jr:requiredMsg="jr:itext('/data/QuantityOfProducedCowpeas
<bind nodeset="/data/QuantityOfProducedSorghumOneSeasonAgo" type="int" required="true()" jr:requiredMsg="jr:itext('/data/QuantityOfProducedSorghumOn
<bind nodeset="/data/QuantityOfProducedSorghumTwoSeasonsAgo" type="int" required="true()" jr:requiredMsg="jr:itext('/data/QuantityOfProducedSorghumO
<bind nodeset="/data/GPSLocation" type="geopoint"/>
</model>
</h:head>
<h:body>
  <input ref="/data/Name">
    <label ref="jr:itext('/data/Name:label')"/>
  </input>

  <select1 ref="/data/Gender">
    <label ref="jr:itext('/data/Gender:label')"/>
    <item>
      <label ref="jr:itext('/data/Gender:option0')"/>
      <value>Male</value>
    </item>
    <item>
      <label ref="jr:itext('/data/Gender:option1')"/>
      <value>Female</value>
    </item>
  </select1>
  <select1 ref="/data/Age">
    <label ref="jr:itext('/data/Age:label')"/>
    <item>
      <label ref="jr:itext('/data/Age:option0')"/>
      <value>Under18</value>
    </item>
    <item>
      <label ref="jr:itext('/data/Age:option1')"/>
      <value>18-35</value>
    </item>
    <item>
      <label ref="jr:itext('/data/Age:option2')"/>
      <value>36-45</value>
    </item>
    <item>
      <label ref="jr:itext('/data/Age:option3')"/>
      <value>46-55</value>
    </item>
    <item>
      <label ref="jr:itext('/data/Age:option4')"/>
      <value>Above66</value>
    </item>
  </select1>
  <select1 ref="/data/HighestEducationLevel">
    <label ref="jr:itext('/data/HighestEducationLevel:label')"/>
    <item>
      <label ref="jr:itext('/data/HighestEducationLevel:option0')"/>
```

```xml
657        <value>None</value>
658      </item>
659      <item>
660        <label ref="jr:itext('/data/HighestEducationLevel:option1')"/>
661        <value>Primary</value>
662      </item>
663      <item>
664        <label ref="jr:itext('/data/HighestEducationLevel:option2')"/>
665        <value>Secondary</value>
666      </item>
667      <item>
668        <label ref="jr:itext('/data/HighestEducationLevel:option3')"/>
669        <value>PostSecondary</value>
670      </item>
671    </select1>
672    <input ref="/data/Village">
673      <label ref="jr:itext('/data/Village:label')"/>
674    </input>
675    <select1 ref="/data/DurationInVillage">
676      <label ref="jr:itext('/data/DurationInVillage:label')"/>
677      <item>
678        <label ref="jr:itext('/data/DurationInVillage:option0')"/>
679        <value>5-10</value>
680      </item>
681      <item>
682        <label ref="jr:itext('/data/DurationInVillage:option1')"/>
683        <value>10-20</value>
684      </item>
685      <item>
686        <label ref="jr:itext('/data/DurationInVillage:option2')"/>
687        <value>20years</value>
688      </item>
689    </select1>
690    <range ref="/data/NoInHousehold" start="1" end="100" step="1" appearance="picker">
691      <label ref="jr:itext('/data/NoInHousehold:label')"/>
692    </range>
693    <select ref="/data/HouseholdSourceOfIncome">
694      <label ref="jr:itext('/data/HouseholdSourceOfIncome:label')"/>
695      <item>
        ...
1011      <hint ref="jr:itext('/data/CostForDroughtPredictionTool:hint')"/>
1012    </input>
1013    <select1 ref="/data/InterestInDroughtServiceSubscription">
1014      <label ref="jr:itext('/data/InterestInDroughtServiceSubscription:label')"/>
1015      <item>
1016        <label ref="jr:itext('/data/InterestInDroughtServiceSubscription:option0')"/>
1017        <value>Yes</value>
1018      </item>
1019      <item>
1020        <label ref="jr:itext('/data/InterestInDroughtServiceSubscription:option1')"/>
1021        <value>No</value>
1022      </item>
1023    </select1>
1024    <input ref="/data/RespondentPhoneNumber">
1025      <label ref="jr:itext('/data/RespondentPhoneNumber:label')"/>
1026      <hint ref="jr:itext('/data/RespondentPhoneNumber:hint')"/>
1027    </input>
1028    <select1 ref="/data/ONDPlantPeriod">
1029      <label ref="jr:itext('/data/ONDPlantPeriod:label')"/>
1030      <item>
1031        <label ref="jr:itext('/data/ONDPlantPeriod:option0')"/>
1032        <value>BeforeRainStarted</value>
1033      </item>
1034      <item>
1035        <label ref="jr:itext('/data/ONDPlantPeriod:option1')"/>
1036        <value>Within2WeeksAfterRain</value>
1037      </item>
1038    </select1>
1039    <input ref="/data/MaizeFarmSize">
1040      <label ref="jr:itext('/data/MaizeFarmSize:label')"/>
1041      <hint ref="jr:itext('/data/MaizeFarmSize:hint')"/>
1042    </input>
1043    <select1 ref="/data/MaizeLandOwnership">
1044      <label ref="jr:itext('/data/MaizeLandOwnership:label')"/>
1045      <item>
1046        <label ref="jr:itext('/data/MaizeLandOwnership:option0')"/>
1047        <value>Yes</value>
1048      </item>
1049      <item>
1050        <label ref="jr:itext('/data/MaizeLandOwnership:option1')"/>
1051        <value>No</value>
1052      </item>
1053    </select1>
1054    <select1 ref="/data/AppliedCroppingMethods">
1055      <label ref="jr:itext('/data/AppliedCroppingMethods:label')"/>
1056      <item>
1057        <label ref="jr:itext('/data/AppliedCroppingMethods:option0')"/>
1058        <value>PureCropping</value>
1059      </item>
1060      <item>
1061        <label ref="jr:itext('/data/AppliedCroppingMethods:option1')"/>
1062        <value>MixedCropping</value>
1063      </item>
1064    </select1>
1065    <input ref="/data/QuantityOfProducedMaizeOneSeasonAgo">
1066      <label ref="jr:itext('/data/QuantityOfProducedMaizeOneSeasonAgo:label')"/>
1067      <hint ref="jr:itext('/data/QuantityOfProducedMaizeOneSeasonAgo:hint')"/>
1068    </input>
1069    <input ref="/data/QuantityOfProducedMaizeTwoSeasonAgo">
1070      <label ref="jr:itext('/data/QuantityOfProducedMaizeTwoSeasonAgo:label')"/>
1071      <hint ref="jr:itext('/data/QuantityOfProducedMaizeTwoSeasonAgo:hint')"/>
1072    </input>
1073    <input ref="/data/QuantityOfProducedMilletOneSeasonAgo">
1074      <label ref="jr:itext('/data/QuantityOfProducedMilletOneSeasonAgo:label')"/>
1075      <hint ref="jr:itext('/data/QuantityOfProducedMilletOneSeasonAgo:hint')"/>
1076    </input>
1077    <input ref="/data/QuantityOfProducedMilletTwoSeasonsAgo">
1078      <label ref="jr:itext('/data/QuantityOfProducedMilletTwoSeasonsAgo:label')"/>
1079      <hint ref="jr:itext('/data/QuantityOfProducedMilletTwoSeasonsAgo:hint')"/>
1080    </input>
1081    <input ref="/data/QuantityOfProducedBeansOneSeasonAgo">
1082      <label ref="jr:itext('/data/QuantityOfProducedBeansOneSeasonAgo:label')"/>
1083      <hint ref="jr:itext('/data/QuantityOfProducedBeansOneSeasonAgo:hint')"/>
1084    </input>
1085    <input ref="/data/QuantityOfProducedBeansTwoSeasonsAgo">
```

# APPENDIX C

Prefix(:=<http://www.semanticweb.org/akanbi/ontologies/2018/10/ IKON.owl#>)

Prefix(owl:=<http://www.w3.org/2002/07/owl#>) Prefix(rdf:=<http://www.w3.org/1999/02/22-rdf-syntax-ns#>) Prefix(xml:=<http://www.w3.org/XML/1998/namespace>)

Prefix(xsd:=<http://www.w3.org/2001/XMLSchema#>)

Prefix(rdfs:=<http://www.w3.org/2000/01/rdf-schema#>)

Ontology(<http://www.semanticweb.org/akanbi/ontologies/2018/10/ IKON.owl>

<http://www.semanticweb.org/akanbi/ontologies/2018/10/IKON.owl/ 2.0.0>

Import(<http://www.semanticweb.org/aakanbi/ontologies/2016/0/IKON>) Annotation(rdfs:comment "A domain ontology for knowledge representation of local indigenous knowledge on drought. Copyright A Akanbi, Central University of Technology, Free State, South Africa.")

Declaration(Class(:Blooming))Declaration(Class(:Withering))

Declaration(ObjectProperty(:IsFeatureOf))

Declaration(ObjectProperty(:IsFloweringOf))

Declaration(ObjectProperty(:IsWitheringOf))

Declaration(ObjectProperty(:hasFeature))

###########################

\#      Object  Properties

###########################

\# Object Property: <http://www.semanticweb.org/aakanbi/ontologies/2016/0/IKON#BloomingOf>

(<http://www.semanticweb.org/aakanbi/ ontologies/2016/0/IKON#BloomingOf>)

SubObjectPropertyOf(<http://www.semanticweb.org/aakanbi/ontologies/ 2016/0/IKON#BloomingOf>   :IsFeatureOf)

\# Object Property: <http://www.semanticweb.org/aakanbi/ontologies/2016/0/IKON#FlockingOf>

(<http://www.semanticweb.org/aakanbi/ ontologies/2016/0/IKON#FlockingOf>)

SubObjectPropertyOf(<http://www.semanticweb.org/aakanbi/ontologies/ 2016/0/IKON#FlockingOf> :IsFeatureOf)

\# Object Property: <http://www.semanticweb.org/aakanbi/ontologies/2016/0/IKON#FlyingOf>

(<http://www.semanticweb.org/aakanbi/ ontologies/2016/0/IKON#FlyingOf>)

SubObjectPropertyOf(<http://www.semanticweb.org/aakanbi/ontologies/ 2016/0/IKON#FlyingOf> :IsFeatureOf)

\# Object Property: <http://www.semanticweb.org/aakanbi/ontologies/2016/0/IKON#SightingOf>

(<http://www.semanticweb.org/aakanbi/ ontologies/2016/0/IKON#SightingOf>)

SubObjectPropertyOf(<http://www.semanticweb.org/aakanbi/ontologies/ 2016/0/IKON#SightingOf> :IsFeatureOf)

\# Object Property: <http://www.semanticweb.org/aakanbi/ontologies/2016/0/IKON#SignsOf>

(<http://www.semanticweb.org/aakanbi/ ontologies/2016/0/IKON#SignsOf>)

SubObjectPropertyOf(<http://www.semanticweb.org/aakanbi/ontologies/ 2016/0/IKON#SignsOf> :IsFeatureOf)

\# Object Property: <http://www.semanticweb.org/aakanbi/ontologies/2016/0/IKON#hasColdTemp>

(<http://www.semanticweb.org/aakanbi/ ontologies/2016/0/IKON#hasColdTemp>)

SubObjectPropertyOf(<http://www.semanticweb.org/aakanbi/ontologies/ 2016/0/IKON#hasColdTemp>  :hasFeature)

# Object Property: <http://www.semanticweb.org/aakanbi/ontologies/ 2016/0/IKON#hasFat>

(<http://www.semanticweb.org/aakanbi/ontologies/2016/0/IKON#hasFat>)

SubObjectPropertyOf(<http://www.semanticweb.org/aakanbi/ontologies/ 2016/0/IKON#hasFat>
:hasFeature)

# Object Property: <http://www.semanticweb.org/aakanbi/ontologies/2016/0/IKON#hasFlower>

(<http://www.semanticweb.org/aakanbi/ ontologies/2016/0/IKON#hasFlower>)

SubObjectPropertyOf(<http://www.semanticweb.org/aakanbi/ontologies/ 2016/0/IKON#hasFlower>
:hasFeature) ObjectPropertyDomain(<http://www.semanticweb.org/aakanbi/ontologies/
2016/0/IKON#hasFlower> <http://www.semanticweb.org/aakanbi/
ontologies/2016/0/IKON#FloralPlants>)

ObjectPropertyRange(<http://www.semanticweb.org/aakanbi/ontologies/ 2016/0/IKON#hasFlower>
:Blooming)

# Object Property: <http://www.semanticweb.org/aakanbi/ontologies/2016/0/IKON#hasGrowth>

(<http://www.semanticweb.org/aakanbi/ ontologies/2016/0/IKON#hasGrowth>)

SubObjectPropertyOf(<http://www.semanticweb.org/aakanbi/ontologies/ 2016/0/IKON#hasGrowth>
:hasFeature)

# Object Property: <http://www.semanticweb.org/aakanbi/ontologies/2016/0/IKON#hasHotTemp>

(<http://www.semanticweb.org/aakanbi/ ontologies/2016/0/IKON#hasHotTemp>)

SubObjectPropertyOf(<http://www.semanticweb.org/aakanbi/ontologies/
2016/0/IKON#hasHotTemp  :hasFeature)

# Object Property: <http://www.semanticweb.org/aakanbi/ontologies/2016/0/IKON#hasPhase>

(<http://www.semanticweb.org/aakanbi/ ontologies/2016/0/IKON#hasPhase>)

SubObjectPropertyOf(<http://www.semanticweb.org/aakanbi/ontologies/ 2016/0/IKON#hasPhase>
:hasFeature)

# Object Property: <http://www.semanticweb.org/aakanbi/ontologies/2016/0/IKON#hasStars>

(<http://www.semanticweb.org/aakanbi/ ontologies/2016/0/IKON#hasStars>)

SubObjectPropertyOf(<http://www.semanticweb.org/aakanbi/ontologies/ 2016/0/IKON#hasStars>
:hasFeature)

# Object Property: <http://www.semanticweb.org/aakanbi/ontologies/2016/0/IKON#hasState>

(<http://www.semanticweb.org/aakanbi/ ontologies/2016/0/IKON#hasState>)

SubObjectPropertyOf(<http://www.semanticweb.org/aakanbi/ontologies/ 2016/0/IKON#hasState>
:hasFeature)

# Object Property: <http://www.semanticweb.org/aakanbi/ontologies/2016/0/IKON#hasStorm>

(<http://www.semanticweb.org/aakanbi/ ontologies/2016/0/IKON#hasStorm>)

SubObjectPropertyOf(<http://www.semanticweb.org/aakanbi/ontologies/ 2016/0/IKON#hasStorm>
:hasFeature)

# Object Property: <http://www.semanticweb.org/aakanbi/ontologies/2016/0/IKON#hasWithered>

(<http://www.semanticweb.org/aakanbi/ ontologies/2016/0/IKON#hasWithered>)

SubObjectPropertyOf(<http://www.semanticweb.org/aakanbi/ontologies/
2016/0/IKON#hasWithered> :hasFeature)

ObjectPropertyDomain(<http://www.semanticweb.org/aakanbi/ontologies/
2016/0/IKON#hasWithered> owl:Plants)

ObjectPropertyRange(<http://www.semanticweb.org/aakanbi/ontologies/
2016/0/IKON#hasWithered>  :Withering)

# Object Property: :IsFeatureOf (:IsFeatureOf) InverseObjectProperties(:IsFeatureOf

:hasFeature)

\# Object Property: :IsFloweringOf  (:IsFloweringOf)

SubObjectPropertyOf(:IsFloweringOf :IsFeatureOf) ObjectPropertyDomain(:IsFloweringOf
<http://www.semanticweb.org/aakanbi/ontologies/2016/0/IKON#FloweringPlant>)

ObjectPropertyRange(:IsFloweringOf  :Blooming)

\# Object Property: :IsWitheringOf  (:IsWitheringOf)

SubObjectPropertyOf(:IsWitheringOf :IsFeatureOf)ObjectPropertyDomain(:IsWitheringOf
owl:Plants) ObjectPropertyRange(:IsWitheringOf  :Withering)

\# Object Property: :hasFeature  (:hasFeature)

TransitiveObjectProperty(:hasFeature)

###########################

\#       Data  Properties

###########################

\# Data  Property: <http://www.semanticweb.org/aakanbi/ontologies/2016/0/IKON#AnimalSize>
(<http://www.semanticweb.org/aakanbi/ ontologies/2016/0/IKON#AnimalSize>)

DataPropertyRange(<http://www.semanticweb.org/aakanbi/ontologies/ 2016/0/IKON#AnimalSize>
owl:real)

\# Data  Property: <http://www.semanticweb.org/aakanbi/ontologies/
2016/0/IKON#FlowerBloomingConditon> (<http://www.semanticweb.org/
aakanbi/ontologies/2016/0/IKON#FlowerBloomingConditon>)

DataPropertyRange(<http://www.semanticweb.org/aakanbi/ontologies/
2016/0/IKON#FlowerBloomingConditon>  xsd:string)

\# Data  Property: <http://www.semanticweb.org/aakanbi/ontologies/
2016/0/IKON#MigratoryBirdSighting> (<http://www.semanticweb.org/
aakanbi/ontologies/2016/0/IKON#MigratoryBirdSighting>)

DataPropertyRange(<http://www.semanticweb.org/aakanbi/ontologies/
2016/0/IKON#MigratoryBirdSighting>  xsd:boolean)

\# Data  Property: <http://www.semanticweb.org/aakanbi/ontologies/2016/0/IKON#MigratoryBirds>
(<http://www.semanticweb.org/aakanbi/ontologies/2016/0/IKON#MigratoryBirds>)

DataPropertyRange(<http://www.semanticweb.org/aakanbi/ontologies/
2016/0/IKON#MigratoryBirds>  xsd:string)

\# Data  Property: <http://www.semanticweb.org/aakanbi/ontologies/
2016/0/IKON#WeatherTempCondition> (<http://www.semanticweb.org/
aakanbi/ontologies/2016/0/IKON#WeatherTempCondition>)

DataPropertyRange(<http://www.semanticweb.org/aakanbi/ontologies/
2016/0/IKON#WeatherTempCondition> xsd:float)

\# Data  Property: <http://www.semanticweb.org/aakanbi/ontologies/2016/0/IKON#Withering>
(<http://www.semanticweb.org/aakanbi/ ontologies/2016/0/IKON#Withering>)

DataPropertyRange(<http://www.semanticweb.org/aakanbi/ontologies/ 2016/0/IKON#Withering>
xsd:boolean)

###########################

\#       Classes

###########################

\# Class: :Blooming  (:Blooming)

SubClassOf(:Blooming <http://www.semanticweb.org/aakanbi/ontologies/
2016/0/IKON#PlantsBehaviour>)

\# Class: :Withering  (:Withering)

SubClassOf(:Withering <http://www.semanticweb.org/aakanbi/
ontologies/2016/0/IKON#PlantsBehaviour>)

```
##############################
#      Named Individuals
##############################
# Individual:  <http://www.semanticweb.org/aakanbi/ontologies/2016/0/IKON#Wiki-Jolo>
(<http://www.semanticweb.org/aakanbi/ontologies/  2016/0/IKON#Wiki-Jolo>)

DataPropertyAssertion(<http://www.semanticweb.org/aakanbi/
ontologies/2016/0/IKON#FlowerBloomingConditon> <http://
www.semanticweb.org/aakanbi/ontologies/2016/0/IKON#Wiki-Jolo>  "true"^^xsd:boolean)

# Individual:     <http://www.semanticweb.org/aakanbi/ontologies/2016/0/    IKON#cattle>
(<http://www.semanticweb.org/aakanbi/ontologies/2016/0/IKON#cattle>)

DataPropertyAssertion(<http://www.semanticweb.org/aakanbi/
ontologies/2016/0/IKON#AnimalSize>  <http://www.semanticweb.org/
aakanbi/ontologies/2016/0/IKON#cattle>  "150"^^xsd:integer)

AnnotationAssertion(rdfs:comment <http://www.semanticweb.org/
aakanbi/ontologies/2016/0/IKON#LivingThingsBehaviour> "The  class  ofbehaviour of the local
indigenous knowledge living things  indicators in this  domain")

)
```

```
1   [ {
2     "@id" : "http://www.semanticweb.org/aakanbi/ontologies/2016/0/IKON#AnimalSize",
3     "http://www.w3.org/2000/01/rdf-schema#range" : [ {
4       "@id" : "http://www.w3.org/2002/07/owl#real"
5     } ]
6   }, {
7     "@id" : "http://www.semanticweb.org/aakanbi/ontologies/2016/0/IKON#FlowerBloomingConditon",
8     "http://www.w3.org/2000/01/rdf-schema#range" : [ {
9       "@id" : "http://www.w3.org/2001/XMLSchema#string"
10    } ]
11  }, {
12    "@id" : "http://www.semanticweb.org/aakanbi/ontologies/2016/0/IKON#MigratoryBirdSighting",
13    "http://www.w3.org/2000/01/rdf-schema#range" : [ {
14      "@id" : "http://www.w3.org/2001/XMLSchema#boolean"
15    } ]
16  }, {
17    "@id" : "http://www.semanticweb.org/aakanbi/ontologies/2016/0/IKON#MigratoryBirds",
18    "http://www.w3.org/2000/01/rdf-schema#range" : [ {
19      "@id" : "http://www.w3.org/2001/XMLSchema#string"
20    } ]
21  }, {
22    "@id" : "http://www.semanticweb.org/aakanbi/ontologies/2016/0/IKON#WeatherTempCondition",
23    "http://www.w3.org/2000/01/rdf-schema#range" : [ {
24      "@id" : "http://www.w3.org/2001/XMLSchema#float"
25    } ]
26  }, {
27    "@id" : "http://www.semanticweb.org/aakanbi/ontologies/2016/0/IKON#Wiki-Jolo",
28    "http://www.semanticweb.org/aakanbi/ontologies/2016/0/IKON#FlowerBloomingConditon" : [ {
29      "@type" : "http://www.w3.org/2001/XMLSchema#boolean",
30      "@value" : "true"
31    } ]
32  }, {
33    "@id" : "http://www.semanticweb.org/aakanbi/ontologies/2016/0/IKON#Withering",
34    "http://www.w3.org/2000/01/rdf-schema#range" : [ {
35      "@id" : "http://www.w3.org/2001/XMLSchema#boolean"
36    } ]
37  }, {
38    "@id" : "http://www.semanticweb.org/aakanbi/ontologies/2016/0/IKON#cattle",
39    "http://www.semanticweb.org/aakanbi/ontologies/2016/0/IKON#AnimalSize" : [ {
40      "@type" : "http://www.w3.org/2001/XMLSchema#integer",
41      "@value" : "150"
42    } ]
43  }, {
44    "@id" : "http://www.semanticweb.org/aakanbi/ontologies/2016/0/IKON#hasFlower",
45    "http://www.w3.org/2000/01/rdf-schema#domain" : [ {
46      "@id" : "http://www.semanticweb.org/aakanbi/ontologies/2016/0/IKON#FloralPlants"
47    } ],
48    "http://www.w3.org/2000/01/rdf-schema#range" : [ {
49      "@id" : "http://www.semanticweb.org/akanbi/ontologies/2018/10/IKON.owl#Blooming"
50    } ],
51    "http://www.w3.org/2000/01/rdf-schema#subPropertyOf" : [ {
52      "@id" : "http://www.semanticweb.org/akanbi/ontologies/2018/10/IKON.owl#hasFeature"
53    } ]
54  }, {
55    "@id" : "http://www.semanticweb.org/aakanbi/ontologies/2016/0/IKON#hasWithered",
56    "http://www.w3.org/2000/01/rdf-schema#domain" : [ {
57      "@id" : "http://www.w3.org/2002/07/owl#Plants"
58    } ],
59    "http://www.w3.org/2000/01/rdf-schema#range" : [ {
60      "@id" : "http://www.semanticweb.org/akanbi/ontologies/2018/10/IKON.owl#Withering"
61    } ],
62    "http://www.w3.org/2000/01/rdf-schema#subPropertyOf" : [ {
63      "@id" : "http://www.semanticweb.org/akanbi/ontologies/2018/10/IKON.owl#hasFeature"
64    } ]
65  }, {
66    "@id" : "http://www.semanticweb.org/akanbi/ontologies/2018/10/IKON.owl",
67    "@type" : [ "http://www.w3.org/2002/07/owl#Ontology" ],
68    "http://www.w3.org/2000/01/rdf-schema#comment" : [ {
69      "@value" : "A domain ontology for knowledge representation of local indigenous knowledge on drought. Copyright A Akanbi, Central University of Tech
70    } ],
71    "http://www.w3.org/2002/07/owl#imports" : [ {
72      "@id" : "http://www.semanticweb.org/aakanbi/ontologies/2016/0/IKON"
73    } ],
74    "http://www.w3.org/2002/07/owl#versionIRI" : [ {
75      "@id" : "http://www.semanticweb.org/akanbi/ontologies/2018/10/IKON.owl/2.0.0"
76    } ]
77  }, {
78    "@id" : "http://www.semanticweb.org/akanbi/ontologies/2018/10/IKON.owl#Blooming",
```

```
78      "@id" : "http://www.semanticweb.org/akanbi/ontologies/2018/10/IKON.owl#Blooming",
79      "@type" : [ "http://www.w3.org/2002/07/owl#Class" ],
80      "http://www.w3.org/2000/01/rdf-schema#subClassOf" : [ {
81        "@id" : "http://www.semanticweb.org/aakanbi/ontologies/2016/0/IKON#PlantsBehaviour"
82      } ]
83    }, {
84      "@id" : "http://www.semanticweb.org/akanbi/ontologies/2018/10/IKON.owl#IsFeatureOf",
85      "@type" : [ "http://www.w3.org/2002/07/owl#ObjectProperty" ],
86      "http://www.w3.org/2002/07/owl#inverseOf" : [ {
87        "@id" : "http://www.semanticweb.org/akanbi/ontologies/2018/10/IKON.owl#hasFeature"
88      } ]
89    }, {
90      "@id" : "http://www.semanticweb.org/akanbi/ontologies/2018/10/IKON.owl#IsFloweringOf",
91      "@type" : [ "http://www.w3.org/2002/07/owl#ObjectProperty" ],
92      "http://www.w3.org/2000/01/rdf-schema#domain" : [ {
93        "@id" : "http://www.semanticweb.org/aakanbi/ontologies/2016/0/IKON#Flowering"
94      } ],
95      "http://www.w3.org/2000/01/rdf-schema#range" : [ {
96        "@id" : "http://www.semanticweb.org/akanbi/ontologies/2018/10/IKON.owl#Blooming"
97      } ],
98      "http://www.w3.org/2000/01/rdf-schema#subPropertyOf" : [ {
99        "@id" : "http://www.semanticweb.org/akanbi/ontologies/2018/10/IKON.owl#IsFeatureOf"
100     } ]
101   }, {
102     "@id" : "http://www.semanticweb.org/akanbi/ontologies/2018/10/IKON.owl#IsWitheringOf",
103     "@type" : [ "http://www.w3.org/2002/07/owl#ObjectProperty" ],
104     "http://www.w3.org/2000/01/rdf-schema#domain" : [ {
105       "@id" : "http://www.w3.org/2002/07/owl#Plants"
106     } ],
107     "http://www.w3.org/2000/01/rdf-schema#range" : [ {
108       "@id" : "http://www.semanticweb.org/akanbi/ontologies/2018/10/IKON.owl#Withering"
109     } ],
110     "http://www.w3.org/2000/01/rdf-schema#subPropertyOf" : [ {
111       "@id" : "http://www.semanticweb.org/akanbi/ontologies/2018/10/IKON.owl#IsFeatureOf"
112     } ]
113   }, {
114     "@id" : "http://www.semanticweb.org/akanbi/ontologies/2018/10/IKON.owl#Withering",
115     "@type" : [ "http://www.w3.org/2002/07/owl#Class" ],
116     "http://www.w3.org/2000/01/rdf-schema#subClassOf" : [ {

116     "http://www.w3.org/2000/01/rdf-schema#subClassOf" : [ {
117       "@id" : "http://www.semanticweb.org/aakanbi/ontologies/2016/0/IKON#PlantsBehaviour"
118     } ]
119   }, {
120     "@id" : "http://www.semanticweb.org/akanbi/ontologies/2018/10/IKON.owl#hasFeature",
121     "@type" : [ "http://www.w3.org/2002/07/owl#ObjectProperty", "http://www.w3.org/2002/07/owl#TransitiveProperty" ]
122   } ]
```

# APPENDIX E

---

```
1     --phpMyAdminSQL
Dump 2 -- version 3.2.4
3 --http://www.phpmyadmin.net 4 --
5    -- Host: localhost
6    -- Waktu pembuatan: 02. November 2018 jan17:17
7    --VersionServer:51.41 8 --
Versi PHP: 5.3.1
9
10SET
SQL_MODE="NO_AUTO_VALUE_ON_ZERO"; 11
13   /*!40101 SET@OLD_CHARACTER_SET_CLIENT=@@CHARACTER_SET_CLIENT */;
14   /*!40101 SET@OLD_CHARACTER_SET_RESULTS=@@CHARACTER_SET_RESULTS*/;
15   /*!40101 SET
@OLD_COLLATION_CONNECTION=@@COLLATION_CONNECTION */; 16 /*!40101
SET NAMES utf8 */;
19 --Database:`db_expert_drought`
28   CREATE TABLE IF NOT EXISTS `tbl_admin` (
29     `id_admin` int(5) NOT NULL AUTO_INCREMENT,
30     `username` varchar(30) NOT NULL,
31     `password` varchar(32) NOT NULL,
32     `para` varchar(50) NOT NULL,
33     PRIMARY KEY (`id_admin`)
34   )ENGINE=MyISAM DEFAULTCHARSET=latin1
AUTO_INCREMENT=5;
37 --Dumpingdata label`tbl_admin` 38
40 INSERTINTO `tbl_admin`(`id_admin`,`username`,`password`,
       `para`) VALUES
41(3,'admin','21232f297a57a5a743894a0e4a801fc3','admin');
46 --Strukturdaritabel`tbl_drought`
49   CREATE TABLE IF NOT EXISTS ` tbl_drought`(
50     `id_diagnosis` int(5) NOT NULL AUTO_INCREMENT,
51     `id_member` int(5) NOT NULL,
52     `kd_penyakit` char(3) NOT NULL,
```

---

```
53     `tanggal_diagnosa` varchar(30) NOT NULL,
54     PRIMARY KEY (`id_diagnosa`)
55   )ENGINE=MyISAM DEFAULTCHARSET=latin1
AUTO_INCREMENT=7;
58 --Dumpingdatafrom table` tbl_drought `
61 INSERTINTO` tbl_drought`(`id_diagnosis`,`id_member`,`kd_penyakit`,
       `tanggal_diagnosa`) VALUES 62
(6, 4, 'D01','02-11-2018'),
63 (5, 4, 'D01','02-11-2018');
68 --Strutureof table tbl_indicator`
71   CREATE TABLE IF NOT EXISTS `tbl_indicator`(
72     `val1` char(3) NOT NULL,
```

```sql
73      `val2` text NOT NULL,
74      PRIMARY KEY (`kd_Val1`)
75    )ENGINE=MyISAM DEFAULT
CHARSET=latin1;

78  -- Dumping data from table `tbl_indicator`

81  INSERT INTO `tbl_1`(`kd_gejala`,`gejala`) VALUES 82 ('I03', 'Indicator
3'),
83  ('I02', 'Indicator2'),
84  ('I01', 'Indicator1');

89  -- Struktur dari tabel `tbl_member`

92    CREATE TABLE IF NOT EXISTS `tbl_member` (
93      `id_member` int(5) NOT NULL AUTO_INCREMENT,
94      `username` varchar(30) NOT NULL,
95      `password` varchar(32) NOT NULL,
96      `email` varchar(50) NOT NULL,
97      `nama_lengkap` varchar(40) NOT NULL,
98      `jenis_kelamin` enum('L','P') NOT NULL,
99      `alamat` text NOT NULL,
100     PRIMARY KEY (`id_member`)
101   )ENGINE=InnoDB  DEFAULT CHARSET=latin1
AUTO_INCREMENT=5;

104 -- Dumping data untuk tabel `tbl_member`
```

---

```sql
107 INSERT INTO `tbl_member`(`id_member`,`username`,`password`,`email`,
              `nama_lengkap`, `jenis_kelamin`, `alamat`) VALUES
108 (4,'member1','c7764cfed23c5ca3bb393308a0da2306','member1@gmail.com', 'member1','L','-');

113 -- Struktur dari tabel `tbl_penyakit` -

116   CREATE TABLE IF NOT EXISTS `tbl_penyakit` (
117     `kd_penyakit` char(3) NOT NULL,
118     `nama_penyakit` varchar(250) NOT NULL,
119     `keterangan` text NOT NULL,
120     `gambar` varchar(255) NOT NULL,
121     PRIMARY KEY (`kd_penyakit`)
122   )ENGINE=MyISAM DEFAULT
CHARSET=latin1;

125 -- Dumping data untuk tabel `tbl_penyakit`

128 INSERT INTO `tbl_penyakit`(`kd_penyakit`,`nama_penyakit`,`keterangan`,
        `gambar`) VALUES
129 ('D01','Example Drought 1', '-', '1.jpg'),
130 ('D02','Example Drought 2', '-', '2.jpg'),
131 ('D03','Example Drought 3', '-', '3.jpg');

136 -- Struktur dari tabel `tbl_responsed`
138
139   CREATE TABLE IF NOT EXISTS `tbl_responsed` (
140     `kd_rule_fc` char(3) NOT NULL
141   )ENGINE=MyISAM DEFAULT
CHARSET=latin1;

144 -- Dumping data untuk tabel `tbl_responsed`

147 INSERT INTO `tbl_responsed`(`kd_rule_fc`) VALUES 148 ('R01'),
```

```sql
149  ('R02');
154  --Struktur dari tabel `tbl_rule_cf`

157  CREATE TABLE IF NOT EXISTS `tbl_rule_cf` (
158      `id_rule_cf` int(5) NOT NULL AUTO_INCREMENT,
159      `id_admin` int(5) NOT NULL,
160      `kd_penyakit` char(3) NOT NULL,
161      `kd_gejala` char(3) NOT NULL,
162      `nilai_cf` varchar(20) NOT NULL,
163      PRIMARY KEY (`id_rule_cf`)
164  )ENGINE=MyISAM DEFAULT CHARSET=latin1
AUTO_INCREMENT=27;
167  --Dumping data untuk tabel `tbl_rule_cf` -
170  INSERT INTO `tbl_rule_cf` (`id_rule_cf`,`id_admin`,`kd_penyakit`,
         `kd_gejala`,`nilai_cf`) VALUES 171
(26, 0, 'D03', 'I03', '0.3'),
172  (25, 0, 'D02', 'I01', '0.1'),
173  (24, 0, 'D01', 'I02', '0.3'),
174  (23, 0, 'D01', 'I01', '0.2');
175
179  --Struktur dari tabel `tbl_rule_fc`
182  CREATE TABLE IF NOT EXISTS `tbl_rule_fc` (
183      `kd_rule_fc` char(3) NOT NULL,
184      `kd_gejala` char(3) NOT NULL,
185      `jika_ya` char(3) NOT NULL,
186      `jika_tidak` char(3) NOT NULL,
187      `id_admin` int(5) NOT NULL,
188      PRIMARY KEY (`kd_rule_fc`)
189  )ENGINE=MyISAM DEFAULT
CHARSET=latin1;
192  --Dumping data untuk tabel `tbl_rule_fc`
195  INSERT INTO `tbl_rule_fc` (`kd_rule_fc`,`kd_gejala`,`jika_ya`,
         `jika_tidak`,`id_admin`) VALUES 196
('R03', 'I03', 'D03', '0', 0),
197  ('R02', 'I02', 'D01', 'D02', 0),
198  ('R01', 'I01', 'R02', 'R03', 0);
203  --Struktur dari tabel `tbl_skala`

206  CREATE TABLE IF NOT EXISTS `tbl_skala` (
207      `id_skala` int(5) NOT NULL AUTO_INCREMENT,
208      `skala` varchar(30) NOT NULL,
209      `bobot` varchar(10) NOT NULL,
210      PRIMARY KEY (`id_skala`)
211  )ENGINE=InnoDB DEFAULT CHARSET=latin1
AUTO_INCREMENT=6;
214  --Dumping data untuk tabel `tbl_skala`
217  INSERT INTO `tbl_skala` (`id_skala`,`skala`,`bobot`) VALUES 218 (2, 'Often',
'0.4'),
219  (3, 'Sometimes','0.3'),
220  (4, 'Rarely','0.2'),
```

```sql
221 (5, 'Very rare', '0.1');
226 -- Struktur dari tabel `tbl_tmp`
229 CREATE TABLE IF NOT EXISTS `tbl_tmp` (
230     `logic` char(3) NOT NULL
231 ) ENGINE=MyISAM DEFAULT
CHARSET=latin1;
234 -- Dumping data untuk tabel `tbl_tmp`
237 INSERT INTO `tbl_tmp` (`logic`) VALUES 238
('D01');
243 -- Struktur dari tabel `tbl_tmp_cf`
246 CREATE TABLE IF NOT EXISTS `tbl_tmp_cf` (
247     `id_tmp_cf` int(5) NOT NULL AUTO_INCREMENT,
248     `kd_gejala` char(3) NOT NULL,
249     `id_skala` int(5) NOT NULL,
250     PRIMARY KEY (`id_tmp_cf`)
251 ) ENGINE=InnoDB DEFAULT CHARSET=latin1
AUTO_INCREMENT=1;
254 -- Dumping data untuk tabel `tbl_tmp_cf`
258 /*!40101 SET CHARACTER_SET_CLIENT=@OLD_CHARACTER_SET_CLIENT */;
```

---

```sql
259 /*!40101 SET CHARACTER_SET_RESULTS=@OLD_CHARACTER_SET_RESULTS */;
260 /*!40101 SET
COLLATION_CONNECTION=@OLD_COLLATION_CONNECTION */; 261
```

# APPENDIX F

**System Usability Scale**

|   | Strongly disagree | | | | Strongly agree |
|---|---|---|---|---|---|

1. I think that I would like to use this system frequently

| 1 | 2 | 3 | 4 | 5 |

2. I found the system unnecessarily complex

| 1 | 2 | 3 | 4 | 5 |

3. I thought the system was easy to use

| 1 | 2 | 3 | 4 | 5 |

4. I think that I would need the support of a technical person to be able to use this system

| 1 | 2 | 3 | 4 | 5 |

5. I found the various functions in this system were well integrated

| 1 | 2 | 3 | 4 | 5 |

6. I thought there was too much inconsistency in this system

| 1 | 2 | 3 | 4 | 5 |

7. I would imagine that most people would learn to use this system very quickly

| 1 | 2 | 3 | 4 | 5 |

8. I found the system very cumbersome to use

| 1 | 2 | 3 | 4 | 5 |

9. I felt very confident using the system

| 1 | 2 | 3 | 4 | 5 |

10. I needed to learn a lot of things before I could get going with this system

| 1 | 2 | 3 | 4 | 5 |

# APPENDIX G

1. FOR historical precipitation dataset
   a. IF dataset is Filesystem WHERE file format is .xslv
   b. READ file (.xlsx)
   c. CREATE Table "HistoricalPrecipitation"
   d. SAVE file (.xlsx) to Table "HistoricalPrecipitation"
2. FOR Sum_Precipitation = SUM (PrecipitationSensors)
   a. CREATE Table "Sum_Precipitation"
   b. SAVE "Sum_Precipitation" to Table "Sum_Precipitation"
3. FOR EP= (Sum_Precipitation)/(Time Frame)
   a. CREATE Table "EP"
   b. SAVE "EP" values to Table "EP"
4. FOR MEP = Mean (HistoricalPrecipitation)
   a. CREATE Table "MEP"
   b. SAVE "MEP" values to Table "MEP"
5. FOR DEP = EP - MEP
   a. CREATE Table "DEP"
   b. SAVE "DEP" values to Table "DEP"
6. FOR SD(DEP) = Standard deviation (DEP)
   a. CREATE Table "SD(DEP)"
   b. SAVE "SD(DEP)" values to Table "SD(DEP)"
7. FOR EDI= DEP/(SD(DEP))
   a. CREATE Table "EDI"
   b. SAVE "EDI" values to Table "EDI"